\documentclass[11pt]{report} 

\usepackage{hyperref}
\usepackage{lipsum}
\usepackage[british]{babel}
\usepackage[utf8]{inputenc}
\usepackage{amsmath}
\usepackage{amssymb}
\usepackage{mathtools}
\usepackage{braket}
\usepackage{graphicx}
\usepackage{xcolor}
\usepackage[inner = 4cm, outer = 2.8cm, bottom = 2.8cm]{geometry}
\usepackage{setspace}
\usepackage{mathrsfs}
\usepackage{fancyhdr}
\begin{document}

\begin{titlepage}
\centering
\vspace{1cm}
{\Large \textbf{A Theory of Local Photons\\
 with Applications in Quantum Field Theory}}\\
\vspace{1cm}
\begin{figure}[h]
\centering
\includegraphics[width = 0.64\textwidth]{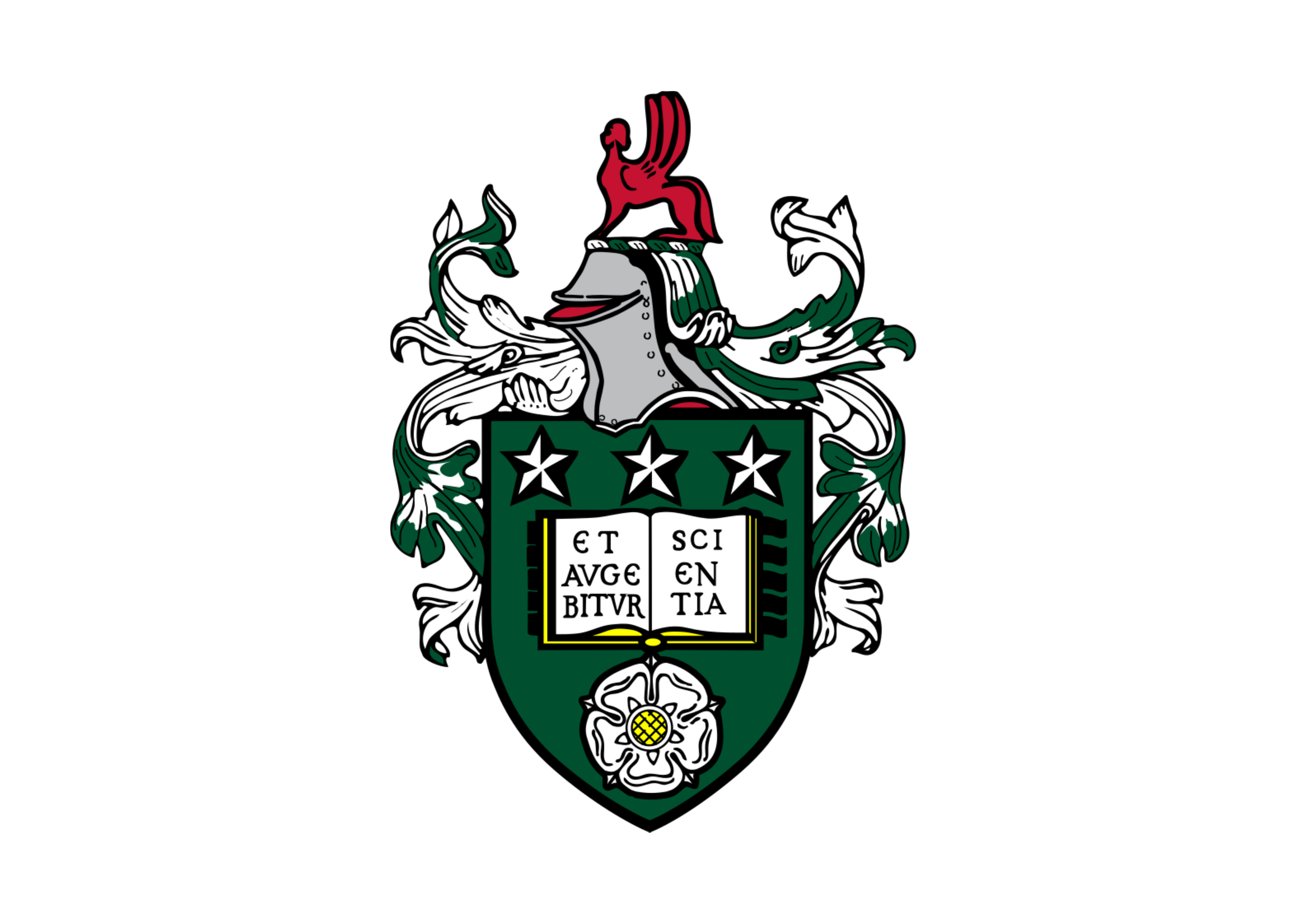}
\end{figure}
\vspace{1cm}

{\Large Daniel Richard Ernest Hodgson}\\
\vspace{0.5cm}
{\Large The University of Leeds}\\
{\Large School of Physics and Astronomy}\\
\vspace{5cm}
{\large Submitted in accordance with the requirements for the degree of\\
Doctor of Philosophy\\}
\vspace{0.5cm}
{\large September, 2022}
\end{titlepage}

\begin{onehalfspace}

The candidate confirms that the work submitted is his own, except where work which has formed part of jointly-authored publications has been included. The contribution of the candidate and the other authors to this work has been explicitly indicated below. The candidate confirms that appropriate credit has been given within the thesis where reference has been made to the work of others.
\\ 

Chapter \ref{Chapter:1D} of this thesis includes work appearing in 

\begin{itemize}
\item Hodgson, D., Southall, J., Purdy, R. and Beige, A. Local Photons. Frontiers in Photonics. 2022, \textbf{3}: 978855.
\end{itemize}

In this publication I carried out the research into the relevant literature, and performed the calculations and derivations that contributed to the main result of this paper.  I wrote the first draft of the publication and was involved in all aspects of this research.  All authors contributed to discussions on the research, checked calculations and provided feedback on the writing of the paper. Robert Purdy and Almut Beige designed the project and posed the original research question.
\\

The derivation of the one-dimensional Casimir effect appearing in Chapter \ref{Chapter:Casimir} of this thesis includes work from

\begin{itemize}
\item Hodgson, D., Burgess, C., Altaie, M. B., Beige, A. and Purdy, R. An intuitive picture of the Casimir effect. arXiv:2203.14385. 2022.
\end{itemize}

In this publication I performed the original calculations that contributed to the final results of the paper, researched the relevant literature, and was involved in all aspects of this research.  Robert Purdy wrote the first draft of the paper.  I contributed to the writing of later versions of the paper, especially those parts concerning the writing of calculations and providing references.  All authors contributed to discussions on the research, checked calculations and provided feedback on the writing of the paper. Robert Purdy and Almut Beige designed the project and posed the original research question.
\\

This copy has been supplied on the understanding that it is
copyright material and that no quotation from the thesis may be published without proper
acknowledgement.

\newpage
\begin{center}
\textbf{Acknowledgements}
\end{center}

I would like to express my sincerest gratitude to my Ph.D.~supervisors Rob Purdy and Almut Beige.  Over the past four years I have received their unending help and support in all parts of this degree, and through their kind and constant mentorship I have learned and achieved more than I could have hoped for.  Their advice and knowledge has always been invaluable to me, and because of them I don't think there has been a single moment during my Ph.D.~where I haven't thoroughly enjoyed myself.

I would like to thank my fellow Ph.D.~students, in particular Jake Southall and Matthew Horner, whose friendship has made studying for a Ph.D.~more bearable and enjoyable, and whose help I have always been able to rely on.  

I would also like to thank Prof.~M. Basil Altaie for introducing me to the exciting field of quantum time, and for sharing with me his great knowledge and enthusiasm for some of the more fundamental topics in theoretical physics; and Prof.~Jiannis Pachos whose kind help and support at various points during my Ph.D.~has always been appreciated and for which I am exceptionally grateful.  My sincerest thanks also to the EPSRC whose funding has made all of this research possible.

Finally I would like to thank my Mum and Dad who have always helped me in every way possible to make this Ph.D.~a success.  Although they can't help me much with the physics, they have done everything to make the other side of being a student as easy as possible, and for that I think I am most grateful of all. \\

\begin{singlespace}
\begin{center}
\textbf{Publications}
\end{center}

\begin{itemize}
	
\item Hodgson, D., Southall, J., Purdy, R. and Beige, A. Local Photons. Frontiers in Photonics, 2022, \textbf{3}: 978855.
	
\item Southall, J., Hodgson, D., Purdy, R. and Beige, A. Comparing Hermitian and Non-Hermitian Quantum Electrodynamics. Symmetry. 2022, \textbf{14} (9), 1816.
	
\item Altaie, M. B., Hodgson, D. and Beige, A. Time and Quantum Clocks: a review of recent developments. Frontiers in Physics. 2022, \textbf{10}: 897305.
	
\item Hodgson, D., Burgess, C., Altaie, M. B., Beige, A. and Purdy, R. An intuitive picture of the Casimir effect. arXiv:2203.14385, 2022.
	
\item Southall, J., Hodgson, D., Purdy, R. and Beige, A. Locally acting mirror Hamiltonians. J. Mod. Opt. 2021, \textbf{68} (12), pp.647-660.
	
\item Maybee, B., Hodgson, D., Beige, A. and Purdy, R. A physically-motivated quantisation of the electromagnetic field on curved spacetimes. Entropy. 2019, \textbf{21} (9), 844.
	
\end{itemize}   

\end{singlespace}

\newpage 
\begin{center}
\textbf{Abstract}
\end{center}
In quantum optics it is usual to describe the basic energy quanta of the electromagnetic (EM) field, photons, in terms of monochromatic waves which have a definite energy and momentum, and satisfy bosonic commutation relations.  Taking this approach, however, leads to several no-go theorems regarding the localisability and superluminal propagation of single photons.  Unfortunately, without a local quantum description of the EM field it becomes difficult to describe the specific dynamics of light in the presence of local interactions or local boundary conditions.

In this thesis we take an alternative approach and quantise the free EM field in both one and three dimensions in terms of quanta that are perfectly localised and propagate at the speed of light without dispersion.  Our approach has two characteristics that allow it to overcome earlier no-go theorems. Firstly, we make a clear distinction between particles, which can always be localised, and the electric and magnetic fields, which cannot; and secondly, we remove the lower bound on the Hamiltonian, thereby introducing negative-frequency photons from basic principles.

Afterwards we test our quantisation scheme by studying the propagation of light in a linear optics experiment analogous to that studied in Fermi's two-atom problem \cite{Fer}.  Here we show that, unlike standard quantisation schemes, our approach predicts the causal propagation of localised photonic wave packets. We also use our theory to provide a new perspective on the Casimir effect in both one and three dimensions.  In this part of the thesis we predict an attractive force between two highly-reflecting metallic plates without having to invoke regularisation procedures.

\end{onehalfspace}

\begin{singlespace}
\tableofcontents

\listoffigures
\end{singlespace}

\newpage

\setcounter{page}{8}

\chapter{Introduction}

\label{Chapter:introduction}

\section{Our perception of light}
Most of the things that we see about us are slow and heavy, and occupy very limited regions of space.  Light, on  the other hand, has a very different kind of existence: it travels extremely quickly, doesn't appear to weigh anything, and seemingly has no fixed size.  Although these unconventional properties are quite familiar to us, they rarely leave much of an impression on our thoughts; usually we are only interested in what we can see by the light rather than the light itself.  Nevertheless, light possesses certain unmistakable characteristics that have greatly influenced our theoretical ideas, from the discovery of Lorentz invariance to development of wave-particle duality.

We imagine, for instance, that when a light source is turned on the light propagates away from the source in the form of long narrow beams or rays.  We find ourselves inclined to this idea because we are used to seeing the way sunlight casts shadows on the ground when we obstruct its path.  We also know that light can be reflected or refracted when passing from one material to another of a different kind.  The study of the geometric relations between the trajectories of these rays is called geometrical optics \cite{Jen}, and captures well the dynamics of light through the simplest optical devices, such as mirrors, lenses and pin-hole cameras.

Sir Isaac Newton believed, based on the many experiments he conducted, that rays of light are composed of a stream of very light elastic particles, each travelling at large but finite speeds in a straight line \cite{New}.  This theory is a very instinctive one that likens the propagation, reflection and refraction of light to the mechanical interactions of small heavy particles.  The corpuscular theory, as it is called, is also one that Newton believed was necessary to explain many other properties of light, for example, the separation of white light into a spectrum of different colours by a prism.  

The Dutch mathematician and contemporary of Newton, Christiaan Huygens \cite{Huy}, was able to explain many of the known properties of light by means of a wave theory.  Huygens suggested, taking inspiration from the properties of sound, that light is not itself made up of small particles, but is rather a wave or ripple through some invisible corpuscular substance.  When a corpuscle was set in motion it would clatter against its neighbours, and they against theirs, causing the energy and momentum of the initial motion to spread throughout space.  Huygens demonstrated that the compounded motion of these corpuscles propagated in straight lines and was refracted at surface interfaces.  Despite some initial failings, the wave theory eventually became favoured over the corpuscular theory following further advancements by Augustin-Jean Fresnel and Thomas Young. 

The most important advancement in our understanding of light came in the early nineteenth century following the discoveries of Michael Faraday and Hans Christian Oersted \cite{Jen, Tip}, who demonstrated that a fluctuating magnetic field will generate an electric field and, conversely, that a current or fluctuating electric field will generate a magnetic field.  The inseparable dynamics of the electric and magnetic fields became collectively expressed through the dynamics of a combined electromagnetic (EM) field. 
There are four equations that govern the dynamics of the EM field known collectively as the Maxwell-Heaviside equations, or simply Maxwell's equations, after the two scientists who developed them: James Clerk-Maxwell and Oliver Heaviside \cite{Tip, Dobb}.  By studying these equations, Maxwell found that certain components of both the electric and magnetic fields propagate across space in the form of waves.  Crucially, he noticed that the calculated velocity of these electromagnetic waves was very close to the measured velocity of light.  The connection seemed obvious. Maxwell had discovered that light is an electromagnetic wave.

What is most interesting about these theories is that they all clearly exhibit the characteristics of locality.  In each of these theories, the properties of light can be independently described at each point in space and time.  In Newton's theory, for example, each light particle has a definite position in space whose motion can be predicted; in Huygens's theory, the position and motion of each corpuscle of the wave medium can similarly be defined; and, most importantly, in Maxwell's theory, the amplitudes of the electric and magnetic fields can be specified at each point in space and time. When described in this way we can understand the behaviour of light by looking at how the field changes from place to place.  This is an incredibly useful and insightful means of studying the EM field; not only because it is an idea congenial to our ordinary mode of thinking, but because the interactions between light and matter occur in a truly position-dependent way, even at the most fundamental level.

In modern electrodynamics, however, light is not described by the classical EM field, but is instead expressed in terms of a quantised set of electric and magnetic field observables.  Such field observables do not take a specific numerical value at each point in space and time as the classical fields do, but rather are a set of operators that act on a Hilbert space of quantum states.  The most natural way of characterising these states is in terms of the eigenstates of the conserved and commuting generators of the Poincar\'e algebra; namely the energy, momentum and angular momentum operators.  The position operator, however, not being part of the Poincar\'e algebra, is only a secondary construction defined more for our own convenience than anything else, if it can be defined at all.  The quantum states of the EM field, therefore, are not naturally expressed in a position-dependent way.  As an alternative, we are often satisfied to look only at the overall scattering dynamics of a particular interaction.  This method, however, does not give a full understanding of the intermediate evolution of a state.  Moreover, it is not always possible to construct a Hamiltonian for every system.  In such cases, other methods such as the triplet mode \cite{Zak, Car} and mirror image \cite{Fur, BD} approaches must be utilised, which often means introducing additional unphysical degrees of freedom.  A position-dependent description of photon wave packets is needed.  

\section{The search for a photon position wave function}    

One of the most well known attempts to derive a mathematically rigorous single-particle position operator was carried out by Newton and Wigner in 1948 \cite{NW, Fle3, Hal2, Jor1, Sch, Hof1, Ful} which built on the earlier work of Pryce \cite{Pry}.  The eigenstates of the Newton-Wigner (NW) position operator describe particles that are localised in every direction, are spherically symmetric and transform correctly under rotations.  Examples of alternative localisation criteria can be found, for example, in  Refs.~\cite{Shj, Wes}.  Although Newton and Wigner succeeded in defining such an operator for particles of finite mass and arbitrary spin, a position operator could not be defined for massless particles with a spin greater than one half (the photon is a spin-one particle).  A short proof of this can be found in Ref.~\cite{Jor2}.  In particular, position eigenstates cannot be defined that are both localised and spherically symmetric.  More recently, however, Hawton \cite{Haw4} has noticed that it would be more accurate to say that the photon position operator must have a cylindrical rather than spherical symmetry due to the divergence condition on the free EM fields.  Further research on NW localisation was carried out for example by Wightman \cite{Wig} and Fleming \cite{Fle1, Fle2} who similarly found the photon could not be localised.  Wightman, for instance, reformulated the work of Newton and Wigner in the framework of imprimitive representations of the Euclidean group \cite{Ros}.  Jauch and Piron \cite{Jau} have since generalised some of the axioms in Wightman's scheme developing the notion of weak localisation for photons.  See also Ref.~\cite{Amr}.  

Notwithstanding the results of Newton and Wigner, several different approaches for constructing photon position wave functions have been introduced \cite{Pik,Gla3, Chan, Cre, Hof2}.  For instance, Hawton was able to construct a photon position operator with commuting components and transversely polarised eigenstates by taking into consideration the longitudinally polarised components of the photon wave function \cite{Haw4,Haw10, Haw9, Haw2, Haw1}.  These components were previously discounted due to the divergence condition on the photon wave function which removed such contributions.  Hawton's position operator can be determined from the Pryce position operator by introducing a term closely related to Bia\l ynicki-Birula's phase invariant derivative \cite{Haw10, Haw2, BB3, Ska1}.  This position operator also has a cylindrical rather than a spherical symmetry \cite{Haw4}.  In close relation to the local photo-detection operators constructed by Glauber \cite{Gla2}, other authors have suggested that a meaningful single-photon wave function ought to be locally related to the field observables, which would then have a more direct physical interpretation.  For example, Bia\l ynicki-Birula \cite{BB1, BB2, BB4}, Sipe \cite{Sip}, and Smith and Raymer \cite{Ray2} constructed both first and second quantised solutions of a massless Dirac-like equation and obtained wave functions that are locally related to the Riemann-Silberstein vector \cite{BB7}, and, therefore, the electric and magnetic field observables.  On the other hand, in the view of Knight \cite{Kni} and Licht \cite{Lic1, Lic2} a photon can only be localised if the EM field expectation values are identical to the ground state expectation values everywhere but at the point of localisation.  From this point of view, however, when the field observables do not commute, single photons cannot be localised \cite{BB6}.

In the approach of Bia\l ynicki-Birula and others, when the wave function is locally related to the field observables, the typical Born rule now provides an energy rather than a probability causing further difficulties for the interpretation of the wave function.  There are two ways of circumventing this problem.  One method is to introduce a modified inner product that has the correct dimensions.  This can be done by either normalising the photon wave function with respect to the photon energy, as was done in Refs.~\cite{BB1,BB4, Ray2, Grs, Kai}, or by treating the system as a biorthogonal system \cite{Gla3, Haw9, Ray2, Haw11, Haw5, Jake2, Ray, Dob}.  For further reading on biorthogonal systems see, for example, Refs.~\cite{Ben, Ben2, Mos1, Mos2, Mos3, ElG}.  This latter approach introduces a non-standard inner product that normalises the wave functions by a term with units of energy.  The inner product between field states then has the typical units of probability density and may retain its usual probability interpretation.

Although useful \cite{Kun}, introducing a new inner product can often be impractical and some intuition for the wave function may be lost.  A second and simpler alternative is to consider the excitations with the correct units as physical regardless of their relation to the field observables.  This approach was adopted in the development of the Landau-Peierls wave function \cite{LP}.  The Landau-Peierls wave function has been criticised for its non-local transformation properties \cite{Fle3}; nevertheless, it has since been revived by Cook \cite{Ck1, Ck2} and Mandel \cite{Man1, MW} who have constructed second quantised position-dependent excitations for which the corresponding wave function represents the probability distribution of the state.  In spite of all this there is as yet no clear choice for the position wave function of the photon, and many different factors must be considered in each choice. 

\section{Difficulties with causality}

\begin{figure}[h]
\label{Fig:Lightcone1}
\centering
\includegraphics[width = 0.55\textwidth]{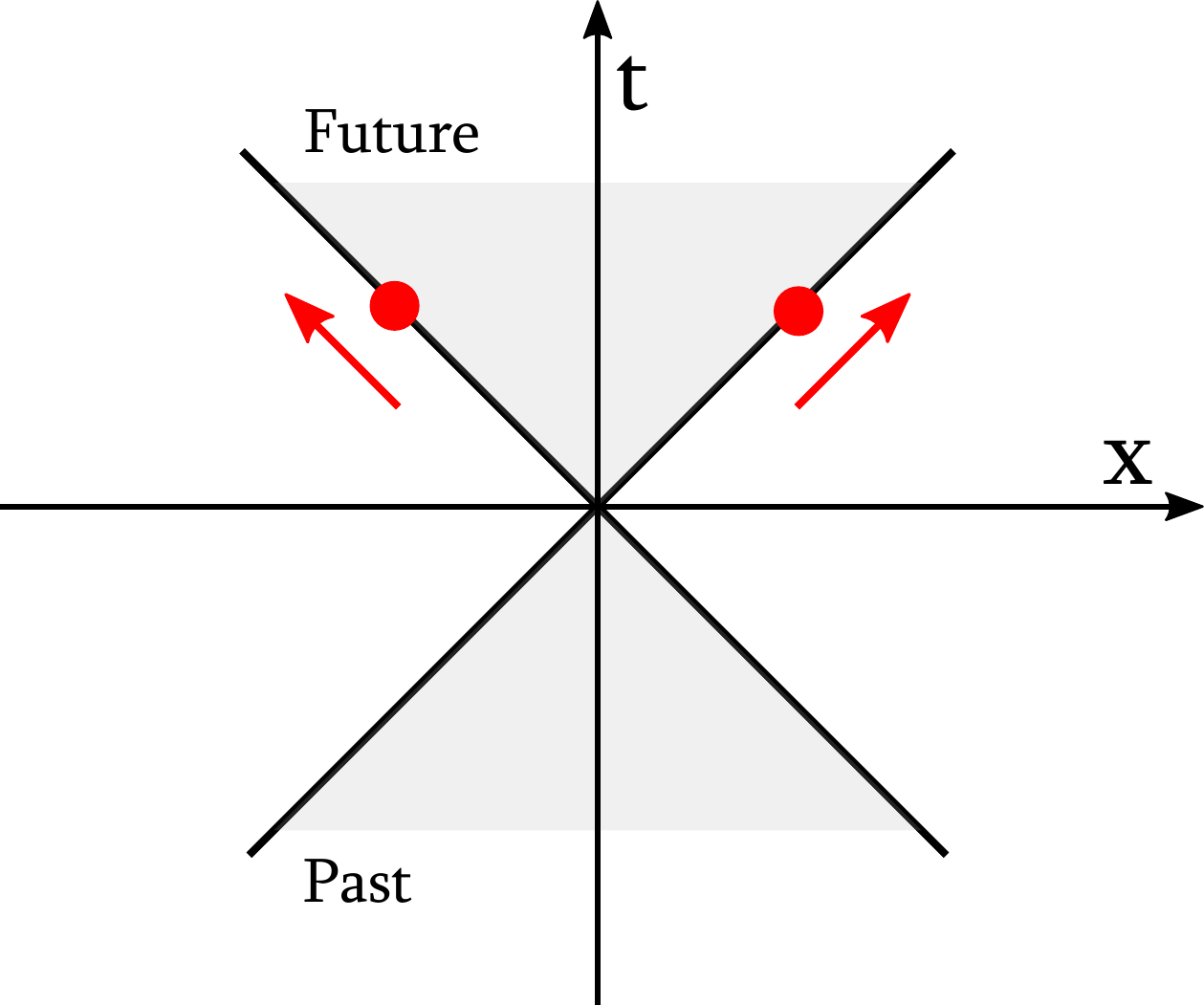}
\caption[A 1+1-dimensional light-cone]{A light-cone in 1+1-dimensional space-time. The localised solutions of Maxwell's equations, indicated by the red spots, propagate along the boundaries of the light-cones from the past into the future.  The light-cone boundaries are parametrised by $x = \pm ct$}.
\end{figure}

Another important aspect to consider when describing the local behaviour of light is the light principle \cite{Fren}.  The solutions of Maxwell's equations in a vacuum describe a set of waves that propagate away from the source at a constant and finite speed $c$, the speed of light, along the boundaries of the light-cone (see Fig.~\ref{Fig:Lightcone1}).  The speed of light is constant with respect to all observers, and naturally this places a lower bound on the time it takes for a signal to propagate from its sender to a receiver.  Needless to say, it is important that this lower bound be apparent in any quantum theory of light.  The localised quanta of light, therefore, should propagate from one place to another at a constant and finite speed, without dispersion, and take a finite time to make their journey.  The standard formulation of quantum field theory is inherently relativistic and some notion of causality arises naturally; in particular, the microcausality condition always applies. It is a result of this condition that there are no causal relationships between measurements made at points that are space-like separated or, in other words, do not lie in each other's light-cones \cite{Dic}.  This prohibits any form of signalling between two such points.  However, the localisation of single photons results in another problem \cite{Heg8, Mal, Heg1, Heg3, Heg5, Heg7, Hal1, Ska2, Per, Kim, Ra, Rui, Sor}.  In a paper published in 1974, Hegerfeldt \cite{Heg8} provided a short proof that, if the probability of detecting a particle in a certain region of space is given by the expectation value of some suitably chosen projection operator, then that same particle will either spread out superluminally or remain stationary \cite{Heg6}.  This has led some to think that particles cannot be localised at all \cite{Gul, Vaz, BB5}.  Similar proofs were also found by Malament \cite{Mal}, and Halvorsen and Clifton, \cite{Hal1}.  The only assumptions made in this derivation were that the particle Hamiltonian is bounded below and the system is translation invariant; however, more recently it has been shown that the sole cause of the spreading is a lower bound placed on the Hamiltonian \cite{Heg6, Heg2}.  

A particular problem that has been greatly studied in the context of causality violations is Fermi's two-atom problem.  The aim of the problem, originally posed by Fermi in 1932 \cite{Fer}, is to calculate the time taken by an excited atom to excite a distant second atom, initially in its ground state, by the transmission of a single photon.  Causality considerations would suggest that there is a minimum delay between the emission and absorption of the photon which would allow for the causal propagation of the light signal between the atoms.  Fermi originally concluded that the process occurs causally, but it was later shown that this result depends on an approximation.  Without this approximation it is predicted that the second atom will be excited immediately \cite{Shi}.  There is some disagreement on whether there is a causality violation in this experiment.  Most authors seem to believe that causality is preserved, but this is largely in the sense of no signalling rather than of no interatomic correlations, and that strict Einstein causality is lost \cite{Bis, Bor, Heg4, Pow, Rub, Val, Cr}.  It may also be possible to experimentally measure these correlations \cite{Set}.  Others believe there are no violations at all \cite{BenB1, BY, Mil1, Mil2, Sab1, Sab2, Tjo, For1, Pli, Kau}.  In many cases, the result is strongly tied up with the approximations used to define a coupling between the field and atoms \cite{Bis, Mil2}, the final state of the source \cite{Heg4, Pow} or the way the system evolves \cite{BenB1}, which makes it difficult to draw any clear conclusions.  Nevertheless, from the point of view of localised wave packets, instantaneous excitation makes no intuitive sense. 

\section{The Problem}

From our experience we know that light is something that ought to be thought of as having a definite position in space and time.  This is the case in classical electromagnetism where light is represented by a pair of electric and magnetic field vectors parametrised by a coordinate $\mathbf{r}$ and a time $t$. In quantum physics too, the position of a photon has a proper realisation in many experiments such as the measurement of position-momentum correlations.  In Ref.~\cite{Bhat}, for example, the authors used a two-photon position-dependent wave function in order to theoretically calculate the measurable position cross-spectral density function.  What is more, from a theoretical point of view, a position-dependent wave function approach to quantum optics is also necessary for modelling the exact dynamics of many locally-interacting systems, as in, for example, Ref.~\cite{Jake1}.

In spite of this, in standard quantum electrodynamics it has not been possible to construct a single-photon position operator $\hat{\mathbf{r}}$ with commuting components and spherically symmetric eigenvalues that also satisfies the usual Heisenberg relation $\big[\hat{\mathbf{r}}, \hat{\mathbf{p}}\big] = i\hbar$.  This prohibits us from constructing photon wave functions in a localised position basis for light polarised in any fixed direction.  It is possible to construct approximate wave functions that may represent the local behaviour of light in some sensible and useful way, but even so, when the Hamiltonian operator is bounded below, all initially localised wave functions will disperse, immediately filling all of space after any non-zero time.  This implies that an initially localised photon can be found at a position outside of its own light-cone.  If one wishes to construct a wave function whose square modulus represents the position of a photon, such a possibility cannot be allowed.  

The problem we face is that the current theory of the quantised EM field does not permit a description of localised photon wave packets that propagate causally as are experienced in experiments.  The purpose of this thesis is to construct a theory of local photon wave packets in the free EM field theory that overcomes the problems described above and brings new insight into the behaviour of single-photon wave packets. The outlook for this work is to apply our new theory to studies of light in systems whose properties differ from place to place, such as inhomogeneous media, gravitational fields or non-inertial reference frames \cite{May,Piw}, and to studies of wave packets with more complex structures, such as waves carrying orbital angular momentum \cite{Bnt1}.  Many interesting and strange effects arise when studying quantum systems in non-inertial frames \cite{Res,Bz,Bel, BenB2, Scu1, Svi, Scu2, Scu3, Svi2}, and we expect a local description of the quantised EM field to provide the tools necessary for describing the complete dynamics of light in these systems in a very intuitive way.

This thesis is divided into eight chapters.  In Chapters \ref{Chap:classical} and \ref{Chap:quantum} we shall review the classical and quantum theories of the free EM field.  In each case we shall pay special attention to the degrees of freedom used to characterise wave packets of light, noting both the similarities and differences.  In particular we notice that light in the classical description is characterised by a greater number of parameters than in the quantum theory.  In Chapters \ref{Chapter:1D} and \ref{Chapter:3D} we point out that, due to a deficiency in the available degrees of freedom, it is not possible to construct localised solutions of Maxwell's equations that propagate at the speed of light using the current quantum theory.  We therefore construct a more complete quantum theory of light in both one and three dimensions that permits such solutions. In this scheme the existence of localised particles is assumed and the field observables then derived by demanding consistency with Maxwell's equations.  A relevant Schr\"odinger equation is also derived for this system, and the relation of this new theory to the standard theory of the free EM field will be discussed.  In Chapter \ref{Chapter:Fermi} we shall study an experiment of my own design that provides an analogue to Fermi's two-atom problem without the need for atoms.  In this chapter, we shall investigate the apparent problems relating to causality inherent to the standard description of photons and show that our new description does not suffer from the same difficulties.  In Chapter \ref{Chapter:Casimir} we shall apply the new theory to a study of the Casimir effect between two perfectly reflecting parallel-plate conductors.  This will allow us not only to test the validity of the new theory against well known predictions, but also to uncover a new understanding of the underlying mechanisms of this effect in the more tangible position representation.  We shall conclude with a discussion of results in Chapter \ref{Chap:conclusions}.

\part{Preliminaries}

\chapter{Classical electromagnetism}

\label{Chap:classical}

In this chapter we look at some background material on the classical theory of the free EM field that will be relevant in the next chapter, where we discuss standard quantisations of the free EM field, and for new work in Chapters \ref{Chapter:1D} to \ref{Chapter:Casimir}.  In this section we define the free-space Maxwell's equations in both the position and momentum representations, which provides the fundamental equations of motion for light in a vacuum, and define the energy and Poynting vector of the field.  Later we shall calculate the solutions of Maxwell's equations in both one and three dimensions which will be useful in Chapters \ref{Chapter:1D} and \ref{Chapter:3D}.  We shall pay particular attention to the degrees of freedom that arise in the classical theory as these differ from those available in the quantum theory.  We shall also show that the Hamiltonian for the free EM field takes the form of a simple harmonic oscillator Hamiltonian in order to motivate discussions in the next chapter.

\section{The classical theory of the free radiation field}

\label{Sec:clasicalEM}

\subsection{Electric and magnetic fields}

\label{Sec:TheEandBfields}

\subsubsection{Electric and Magnetic fields in the position representation}
The theory of electromagnetism is concerned with the properties and dynamics of two fundamental quantities: the electric field $\mathbf{E}(\mathbf{r},t)$ and the magnetic field $\mathbf{B}(\mathbf{r},t)$. These two fields are vector valued, having components in all three space dimensions, and are parametrised by a position vector $\mathbf{r}$ and a time $t$, which represent the position and time at which the fields are measured.  Both of these fields are real; however, in the following we shall reserve the notation $\mathbf{E}(\mathbf{r},t)$ and $\mathbf{B}(\mathbf{r},t)$ for the complex electric and magnetic fields respectively unless we make specific mention otherwise.  The total real fields are given by the superposition $(1/2)(\mathbf{O} + \mathbf{O}^*)$ where $\mathbf{O} = \mathbf{E}, \mathbf{B}$.  Here and in the remainder of this thesis, * denotes complex conjugation.

\subsubsection{The Lorentz force}
When a charged material is placed in a non-vanishing electric or magnetic field, both fields will exert a force $\mathbf{f}(\mathbf{r},t)$ on the material at each point $\mathbf{r}$ and time $t$.  This force is the Lorentz force, and depends on both the charge density $\rho(\mathbf{r},t)$ and current density $\mathbf{j}(\mathbf{r},t)$ of the material, and the magnitude of the total electric and magnetic fields at the point $\mathbf{r}$ at a time $t$.  This force can be calculated using the following equation:
\begin{equation}
\label{LorentzForce1}
{\bf f}({\bf r},t) = \rho({\bf r},t){\bf E}({\bf r},t) + {\bf j}({\bf r},t) \times {\bf B}({\bf r},t).
\end{equation} 
Here $\mathbf{E}(\mathbf{r},t)$ and $\mathbf{B}(\mathbf{r},t)$ are the real fields. Like the electric and magnetic fields, the Lorentz force is a three-dimensional vector or 3-vector, and its direction indicates the direction of the applied force.

\subsubsection{Electric and magnetic fields in the momentum representation}

\label{Sec:momfields}

In this thesis it will often be convenient to express the position-dependent electric and magnetic fields in their Fourier representations.  The Fourier components of the electric and magnetic fields, like the fields themselves, are 3-vector valued, and parametrised by a time $t$.  In the momentum representation, the fields are also parametrised by a real 3-vector $\mathbf{k}$ that replaces the original space component $\mathbf{r}$.  We denote the Fourier components of the complex electric and magnetic fields $\widetilde{\mathbf{E}}(\mathbf{k},t)$ and $\widetilde{\mathbf{B}}(\mathbf{k},t)$ respectively.  Hence, the Fourier representations of the electric and magnetic fields are given by
\begin{eqnarray}
\label{FieldFT1}
\mathbf{O}(\mathbf{r},t) &=& \int_{\mathbb{R}^3}\frac{\text{d}^3\mathbf{k}}{({2\pi})^{3/2}}\;e^{\pm i\mathbf{k}\cdot\mathbf{r}}\widetilde{\mathbf{O}}(\mathbf{k},t)
\end{eqnarray} 
where $\mathbf{O} = \mathbf{E},\mathbf{B}$ and $\widetilde{\mathbf{O}} = \widetilde{\mathbf{E}}, \widetilde{\mathbf{B}}$ accordingly.  The inverse transformation, which expresses the Fourier components as a superposition of the original field amplitudes, is given by 
\begin{eqnarray}
\label{FieldFT2}
\widetilde{\mathbf{O}}(\mathbf{k},t) &=& \int_{\mathbb{R}^3}\frac{\text{d}^3\mathbf{r}}{(2\pi)^{3/2}}\;e^{\mp i\mathbf{k}\cdot\mathbf{r}}\mathbf{O}({\bf r},t).
\end{eqnarray} 

\subsection{Maxwell's equations and the wave equation}
\label{sectionMaxwellsequations}

\subsubsection{Maxwell's equations in a vacuum}
The dynamics of the electric and magnetic fields are governed by four local and first-order differential equations known as Maxwell's equations \cite{Dobb, Hei, Coh, Grif}.  Two of these four equations have an explicit dependence on the charge density and the charge current density, which influence the surrounding electric and magnetic fields.  The Maxwell equations for the real fields are
\begin{eqnarray}
\label{Maxwell1}
\mathbf{\nabla}\cdot \mathbf{E}(\mathbf{r},t) &=& \frac{\rho(\mathbf{r},t)}{\varepsilon_0}\\
\label{Maxwell2}
\mathbf{\nabla}\cdot \mathbf{B}(\mathbf{r},t) &=& 0\\
\label{Maxwell3}
\mathbf{\nabla}\times\mathbf{E}(\mathbf{r},t) &=& -\frac{\partial}{\partial t}\mathbf{B}(\mathbf{r},t)\\
\label{Maxwell4}
\mathbf{\nabla}\times\mathbf{B}(\mathbf{r},t) &=& \frac{1}{c^2}\frac{\partial}{\partial t}\mathbf{E}(\mathbf{r},t) + \mu_0\, \mathbf{j}(\mathbf{r},t). 
\end{eqnarray}
In the above $\varepsilon_0$ and $\mu_0$ are both constants denoting the permittivity and permeability of the vacuum respectively.   The factor $c$ is also a constant related to the permittivity and permeability of the vacuum according to the relation $c = 1/\sqrt{\varepsilon_0 \mu_0}$.  

In a system in which there are neither electrical currents nor charged matter, the charge density $\rho(\mathbf{r},t)$ and charge current density $\mathbf{j}(\mathbf{r},t)$ vanish everywhere: $\rho(\mathbf{r},t) = \mathbf{j}(\mathbf{r},t) = 0$ for all $\mathbf{r}$ and $t$.  A system of this kind is called free space, and we may determine for this system a simpler set of Maxwell's equations.  In free space, Maxwell's equations take the form
\begin{eqnarray}
\label{fMaxwell1}
\mathbf{\nabla}\cdot \mathbf{E}(\mathbf{r},t) &=& 0\\
\label{fMaxwell2}
\mathbf{\nabla}\cdot \mathbf{B}(\mathbf{r},t) &=& 0\\
\label{fMaxwell3}
\mathbf{\nabla}\times\mathbf{E}(\mathbf{r},t) &=& -\frac{\partial}{\partial t}\mathbf{B}(\mathbf{r})\\
\label{fMaxwell4}
\mathbf{\nabla}\times\mathbf{B}(\mathbf{r},t) &=& \frac{1}{c^2}\frac{\partial}{\partial t}\mathbf{E}(\mathbf{r},t) 
\end{eqnarray}
where the above fields are now complex.  The fields that propagate in free space are known as the free fields.

\subsubsection{Free field Maxwell's equations in the momentum representation}

In Section
\ref{Sec:momfields}, we expressed the electric and magnetic fields as Fourier transforms of the vector fields $\widetilde{\mathbf{E}}(\mathbf{k},t)$ and $\widetilde{\mathbf{B}}(\mathbf{k},t)$ respectively.  The free-space Maxwell's equations for the position-dependent fields therefore imply four equivalent equations for the components of the $\mathbf{k}$-dependent fields.  These four new equations are given by
\begin{eqnarray}
\mathbf{k}\cdot\widetilde{\mathbf{E}}(\mathbf{k},t) &=& 0 \label{Maxwellm1}\\
\mathbf{k}\cdot \widetilde{\mathbf{B}}(\mathbf{k},t) &=& 0\label{Maxwellm2}\\
\pm i\mathbf{k}\times \widetilde{\mathbf{E}}(\mathbf{k},t) &=& -\frac{\partial}{\partial t}\widetilde{\mathbf{B}}(\mathbf{k},t)\label{Maxwellm3}\\
\pm i\mathbf{k}\times\widetilde{\mathbf{B}}(\mathbf{k},t) &=& \frac{1}{c^2}\frac{\partial}{\partial t}\widetilde{\mathbf{E}}(\mathbf{k},t).\label{Maxwellm4}
\end{eqnarray}
Which sign to pick depends on the choice of sign made in Eq.~(\ref{FieldFT1}).

\subsubsection{The wave equation}

The dynamics of the electric and magnetic fields are specified by the two equations Eqs.~(\ref{fMaxwell3}) and (\ref{fMaxwell4}).  These equations are not independent of each other, but couple together different components of both the electric and magnetic field vectors by expressing the dynamics of one component in terms of various derivatives of the others.   If one considers, however, the curl of Eqs.~(\ref{fMaxwell3}) and (\ref{fMaxwell4}), then, by making careful substitutions of the other Maxwell equations, one can show that, in free space, the electric and magnetic field vectors each obey their own wave equation.  The wave equation mentioned is given by
\begin{eqnarray}
\label{waveequation1}
\left[\mathbf{\nabla}^2 - \frac{1}{c^2}\frac{\partial^2}{\partial t^2}\right]\mathbf{O}(\mathbf{r},t) = 0.
\end{eqnarray}
When the fields are expressed in a Cartesian basis, each component of $\mathbf{E}(\mathbf{r},t)$ and $\mathbf{B}(\mathbf{r},t)$ will independently satisfy the above wave equation.  This wave equation is a second order equation that describes the dynamics of a wave that propagates at a velocity $c$.  The constant $c = 1/\sqrt{\epsilon_0\mu_0}$ is the speed of light.

\subsection{The energy of the radiation field}
\label{Sec:classicalenergy}

The electromagnetic field will exert a force, determined by the Lorentz force law (\ref{LorentzForce1}), on any charged matter, and must therefore contain a certain amount of energy that enables it to do work on the charged material.  One of the most direct ways to determine the energy of the free field in a particular region of space is to consider the work done by the field on any charged matter that is placed in the vicinity \cite{Hei}.  The expression for the rate of work done on any charged matter propagating at a velocity $\mathbf{v}(\mathbf{r},t)$ in a region $V$ is 
\begin{eqnarray}
\label{classicalenergy1}
\frac{\text{d}}{\text{d}t}\int_V\text{d}^3\mathbf{r}\;W(\mathbf{r},t) &=& \int_V\text{d}^3\mathbf{r}\;\mathbf{f}(\mathbf{r},t)\cdot\mathbf{v}(\mathbf{r},t)\nonumber\\
&=&\int_V\text{d}^3\mathbf{r}\;\frac{1}{\rho(\mathbf{r},t)}\mathbf{f}(\mathbf{r},t)\cdot \mathbf{j}(\mathbf{r},t)\nonumber\\
&=& \int_V\text{d}^3\mathbf{r}\;\mathbf{E}(\mathbf{r},t)\cdot\mathbf{j}(\mathbf{r},t).
\end{eqnarray}
In the above expression, $\mathbf{f}(\mathbf{r},t)$ is the Lorentz force density exerted by the fields on the charged matter, which is substituted into the third line of Eq.~(\ref{classicalenergy1}).  The magnetic field does not appear in the final expression because it always exerts a force that is orthogonal to the velocity of the charged matter.  The vector field $\mathbf{v}(\mathbf{r},t)$, which represents the velocity density of the charged matter, is equivalent to the charge current density $\mathbf{j}(\mathbf{r},t)$ divided by the charge density $\rho(\mathbf{r},t)$ at the same location and time.

If Eq.~(\ref{Maxwell4}) is used to make a substitution for $\mathbf{j}(\mathbf{r},t)$ in the right-hand side of Eq.~(\ref{classicalenergy1}), an expression is found for the work done on charged matter in terms of the changes in the electromagnetic field in contact with that charged matter.  After carrying out this substitution, and by using Eq.~(\ref{Maxwell3}) in the manipulation of terms, one is able to show that Eq.~(\ref{classicalenergy1}) can be rewritten in the form
\begin{eqnarray}
\label{classicalenergy2}
\frac{\text{d}}{\text{d}t}\int_V\text{d}^3\mathbf{r}\;W = -\frac{\partial}{\partial t}\int_V\text{d}^3\mathbf{r}\; \varepsilon(\mathbf{r},t) - \int_{\partial V}\text{d}\mathbf{n}\cdot \mathbf{S}(\mathbf{r},t) 
\end{eqnarray}
where
\begin{equation}
\label{classicalenergy3}
\varepsilon(\mathbf{r},t) = \frac{\varepsilon_0}{2}\left[|\mathbf{E}(\mathbf{r},t)|^2 + c^2|\mathbf{B}(\mathbf{r},t)|^2\right]
\end{equation}
and
\begin{eqnarray}
\label{Poyntingvector}
\mathbf{S}(\mathbf{r},t) = \frac{1}{\mu_0}\left[\mathbf{E}(\mathbf{r},t)\times \mathbf{B}(\mathbf{r},t)\right].
\end{eqnarray}
In the above $\mathbf{E}(\mathbf{r},t)$ and $\mathbf{B}(\mathbf{r},t)$ refer to the total real fields.  One may determine from this expression that the rate of work done is equal to the difference between the rate at which the total $\varepsilon (\mathbf{r},t)$ decreases and the flux of $\mathbf{S}(\mathbf{r},t)$ out of the volume $V$.  If the total work done on some charged matter is attributed purely to an electromagnetic force, then, by considering the conservation of electromagnetic energy, one may naturally associate the $\varepsilon(\mathbf{r},t)$ in Eq.~(\ref{classicalenergy3}) with the energy density of the EM field, and $\mathbf{S}(\mathbf{r},t)$ with the change in the energy flux density of the EM field per unit time.  $\mathbf{S}(\mathbf{r},t)$ is also known as the Poynting vector.  

If we return now to the free-space system containing no charged particles, and consider the system in which $V$ encompasses all of space, then, assuming that the Poynting vector vanishes at the infinite boundaries of our system, the total energy of the free EM field is given by
\begin{equation}
\label{classicalenergy4}
H_{\text{energy}} = \int_{\mathbb{R}^3}\text{d}^3\mathbf{r}\frac{\varepsilon_0}{2}\left[|\mathbf{E}(\mathbf{r},t)|^2 + c^2|\mathbf{B}(\mathbf{r},t)|^2\right].
\end{equation}
The total energy of the system is constant.

\subsection{The solutions of Maxwell's equations in one dimension}
\label{Sec:1Dclassicalfields}

A particularly important class of solutions to the wave equation, Eq.~(\ref{waveequation1}), are the one-dimensional solutions.  The components of the one-dimensional EM field have a dependence on only one of the three space coordinates, in addition to the usual time coordinate, and remain constant in the two remaining dimensions.  Consequently,  propagation of electromagnetic waves takes place in this one dimension only.  It is important to note that, although the one-dimensional solutions propagate in one direction only, the electric and magnetic fields still inhabit three-dimensional space, and therefore the vector field amplitudes may be directed or polarised along any axis. 

Let us consider the electric and magnetic field vectors whose amplitudes depend only upon the $x$ Cartesian coordinate, measured along the $x$ axis, and a time $t$: $\mathbf{E}(x,t)$ and $\mathbf{B}(x,t)$.  With this simplification, the wave equation (\ref{waveequation1}) reduces to the following:
\begin{equation}
\label{1DMaxwell1}
\left[\frac{\partial^2}{\partial x^2} - \frac{1}{c^2}\frac{\partial^2}{\partial t^2} \right]O_i(x,t) = 0.
\end{equation}
By imposing the divergence conditions (\ref{fMaxwell1}) and (\ref{fMaxwell2}) on the one-dimensional fields, one finds that the $x$ components of the electric and magnetic fields are constant.  We assume here that the field vectors must vanish somewhere, and hence that this constant must be zero.  For a free field propagating in the $x$ direction only, the electric and magnetic fields are both polarised purely in the $y$ and $z$ directions.

The wave equation (\ref{1DMaxwell1}) can be factorised into a product of two, first-order differential operators:
\begin{equation}
\label{1DMaxwell2}
\left(\frac{\partial}{\partial x} - \frac{1}{c}\frac{\partial}{\partial t}\right)\left(\frac{\partial}{\partial x} + \frac{1}{c}\frac{\partial}{\partial t}\right)O_i\mathbf(x,t) = 0.
\end{equation}
D'Alembert's solution to the wave equation is the superposition of the solutions of these two, first-order differential operators.  After taking into account the allowed polarisations, one therefore finds that the general solution for the electric field vector propagating along the $x$ axis is given by
\begin{eqnarray}
\label{Efield1}
\mathbf{E}(x,t) = \sum_{s = \pm 1} \left[E_{ys}(x,t)\, \widehat{\boldsymbol{y}} + E_{zs}(x,t)\, \widehat{\boldsymbol{z}}\right]
\end{eqnarray}
where $E_{is}(x,t)$ satisfies the first-order differential equation
\begin{eqnarray}
\label{1stwaveequation}
\left(\frac{\partial}{\partial x} + \frac{s}{c}\frac{\partial}{\partial t}\right)E_{is}(x,t) = 0
\end{eqnarray}
for both $i = y,z$ and $s = \pm1$.  Here, $\widehat{\boldsymbol{y}}$ and $\widehat{\boldsymbol{z}}$ are unit vectors that lie parallel to the $y$ and $z$ axes respectively and are oriented in the direction of the increasing coordinate.  The solutions of this first-order equation have a dependence on the space-time distance $x-sct$ only: $E_{is}(x,t) = E_{is}(x-sct)$.  As previously mentioned, the electric and magnetic field vectors are not independent of each other, but are coupled through Maxwell's equations.  For this reason, the magnetic field vector may be calculated directly from the expression for the electric field in Eq.~(\ref{Efield1}), and is given by
\begin{eqnarray}
\label{Bfield1}
\mathbf{B}(x,t) = \sum_{s = \pm 1} \frac{s}{c}\left[E_{ys}(x,t)\widehat{\boldsymbol{z}} - E_{zs}(x,t)\widehat{\boldsymbol{y}}\right].
\end{eqnarray}
Eqs.~(\ref{Efield1}) and (\ref{Bfield1}) provide a complete set of solutions to Maxwell's equations in one-dimensional free space. In order to determine the particular choice of $E_{is}(x,t)$ satisfying Eq.~\ref{1stwaveequation}, one will need to impose a set of boundary conditions on the fields and their time derivatives at some chosen time.

\subsection{The solutions of Maxwell's equations in three dimensions}
\label{Sec:3Dclassicalfields}

The most general solutions of the free-space Maxwell equations are those that have a dependence on all spatial coordinates in addition to time.  Propagation is therefore also permitted in any direction.  When parametrised in this way, the three-dimensional solutions $\mathbf{E}(\mathbf{r},t)$ and $\mathbf{B}(\mathbf{r},t)$ satisfy the full wave equation (\ref{waveequation1}), which is no longer factorisable.  The simplest way to proceed is therefore to expand the field vectors into their Fourier representations and solve the appropriate set of equations on the Fourier components $\widetilde{\mathbf{E}}(\mathbf{k},t)$ and $\widetilde{\mathbf{B}}(\mathbf{k},t)$.

Consider again the wave equation (\ref{waveequation1}) for the electric and magnetic field vectors.  By expressing the field vectors in their Fourier representations, given in Eq.~(\ref{FieldFT1}), one finds that the wave equation is satisfied when the following relation holds:   
\begin{eqnarray}
\label{Fourierwaveequation1}
\left[|\mathbf{k}|^2 +\frac{1}{c^2}\frac{\partial^2}{\partial t^2}\right]\widetilde{\mathbf{E}}(\mathbf{k},t) = 0.
\end{eqnarray}
The most general solution of this equation is
\begin{eqnarray}
\label{frequencymode1}
\widetilde{\mathbf{E}}(\mathbf{k},t) = \boldsymbol{\alpha}_{\mathbf{k}}\,e^{-i\kappa ct}
\end{eqnarray}
where the Greek letter $\kappa$ is equal to $|\mathbf{k}|$, and where $\boldsymbol{\alpha}_{\mathbf{k}}$ can be any complex 3-vector-valued function of $\mathbf{k}$.  For future reference, the exponential term $\text{exp}[-i\kappa ct]$ for any positive $\kappa$ is known as a positive frequency mode, whereas the complex conjugate of this term, $\text{exp}[i\kappa ct]$, is known as a negative frequency mode.

Using Gauss's laws for the electric and magnetic fields (\ref{Maxwellm1}) and (\ref{Maxwellm2}) one will find that 
\begin{eqnarray}
\label{Fourierdivergence1}
\boldsymbol{\alpha}_{\mathbf{k}}\cdot \mathbf{k} = 0.
\end{eqnarray}
This equation implies that the mode coefficient $\boldsymbol{\alpha}_{\mathbf{k}}$ is a vector function that is polarised in the plane orthogonal to the direction of the wave vector $\mathbf{s} = \mathbf{k}/\kappa$.  Therefore, $\boldsymbol{\alpha}_{\mathbf{k}}$ may be expressed in terms of two real, three-dimensional basis vectors lying tangent to this plane, themselves orthogonal to $\mathbf{k}$.  We shall denote these two basis vectors $\boldsymbol{e}_{\mathbf{s}\lambda}$ where $\lambda \in \{\mathsf{H}, \mathsf{V}\}$ such that $\boldsymbol{\alpha}_{\mathbf{k}} = \sum_{\lambda = \mathsf{H}, \mathsf{V}}\, \alpha_{\mathbf{k}\lambda}\,\boldsymbol{e}_{\mathbf{s}\lambda}$.  Although other relations may be specified in order to define more explicitly the particular choice of basis vectors, the following relations shall always hold:
\begin{eqnarray}
\label{polarisationvectors1}
\boldsymbol{e}_{\mathbf{s}\mathsf{H}} \cdot \boldsymbol{e}_{\mathbf{s}\mathsf{V}} &=& 0 \nonumber\\
\boldsymbol{e}_{\mathbf{s}\mathsf{H}} \cdot \boldsymbol{e}_{\mathbf{s}\mathsf{H}} = \boldsymbol{e}_{\mathbf{s}\mathsf{V}} \cdot \boldsymbol{e}_{\mathbf{s}\mathsf{V}} &=& 1\nonumber\\
\boldsymbol{e}_{\mathbf{s}\mathsf{H}} \cdot\mathbf{k} = \boldsymbol{e}_{\mathbf{s}\mathsf{V}} \cdot \mathbf{k} &=& 0.
\end{eqnarray}

In order to determine the complete expression for the electric field vector in three dimensions we must piece together the original Fourier transform having found a general solution for the Fourier coefficients $\widetilde{\mathbf{E}}(\mathbf{k},t)$.  The general electric field solution is
\begin{eqnarray}
\label{3dEfield1}
\mathbf{E}(\mathbf{r},t) = \sum_{\lambda = \mathsf{H}, \mathsf{V}}\int_{-\infty}^{\infty}\frac{\text{d}^3 \mathbf{k}}{(2\pi)^{3/2}}\;e^{\pm i\mathbf{k}\cdot\mathbf{r}-i\kappa ct} \mathcal{\alpha}_{\mathbf{k}\lambda}\,\boldsymbol{e}_{\mathbf{s}\lambda} + \text{c.c.}
\end{eqnarray}
Here c.c denotes the complex conjugate. Note here that because the fields are real, both the positive- and negative-frequency modes contribute to the total field solution.  As in Section \ref{Sec:1Dclassicalfields}, the general solution for the magnetic field is calculated directly from Eq.~(\ref{3dEfield1}) by employing Faraday's law (\ref{fMaxwell3}).  The corresponding solution is
\begin{eqnarray}
\label{3dBfield1}
\mathbf{B}(\mathbf{r},t) &=& \pm\frac{1}{c}\sum_{\lambda = \mathsf{H}, \mathsf{V}}\int_{-\infty}^{\infty}\frac{\text{d}^3 \mathbf{k}}{(2\pi)^{3/2}}\;e^{\pm i\mathbf{k}\cdot\mathbf{r}-i\kappa ct} \mathcal{\alpha}_{\mathbf{k}\lambda}\left(\mathbf{s}\times\boldsymbol{e}_{\mathbf{s}\lambda}\right) + \text{c.c.}
\end{eqnarray}
The values for the electric and magnetic fields above provide a complete set of solutions to Maxwell's equations in three dimensions.  The remaining unknown coefficient $\alpha_{\mathbf{k}\lambda}$ is specific to the particular problem being investigated and can be fully determined by a set of initial conditions on the field vectors across all space at a given time.

We see in the derivation of these solutions that both the electric field and the magnetic field can be fully characterised by the mode coefficients $\alpha_{\mathbf{k}\lambda}$.  We must also be able, therefore, to express the energy of the free field in terms of these mode coefficients.  By substituting the electric and magnetic field vectors found in Eqs.~(\ref{3dEfield1}) and (\ref{3dBfield1}) into the expression for the electromagnetic energy in Eq.~(\ref{classicalenergy4}) one finds that
\begin{eqnarray}
\label{fclassicalenergy1}
H_{\text{energy}} &=& 2\varepsilon_0\sum_{\lambda = \mathsf{H},\mathsf{V}}\int_{-\infty}^{\infty}\text{d}^3\mathbf{k}\;|\alpha_{\mathbf{k}\lambda}(t)|^2.
\end{eqnarray} 
By expanding $\alpha_{\mathbf{k}\lambda}(t) = \text{exp}[-i\kappa ct]\alpha_{\mathbf{k}\lambda}$, one can check that the energy of the classical field is constant in time, as was expected.

\subsection{The free EM field as an harmonic oscillator}
\label{Sec:Harmonicoscillator}

The Hamiltonian formalism of classical mechanics is a methodology commonly used to determine the dynamics of a mechanical system.  In this formalism, a system is characterised entirely by a set of canonical position coordinates $q_i$ and an associated but independent set of canonical momentum coordinates $p_i$.  The indices $i$ may be either discrete or continuous.  The equations of motion for the canonical variables $q_i$ and $p_i$ that characterise the system are given by two first-order differential equations of motion known as Hamilton's equations.  These are \cite{Gold}
\begin{eqnarray}
\label{Hamiltonsequations}
\begin{matrix}
\dot{q}_{i} = \frac{\partial H}{\partial p_{i}} & \dot{p}_{i} = -\frac{\partial H}{\partial q_{i}}.
\end{matrix}
\end{eqnarray}
In these equations, the dynamics of $q_i$ and $p_i$ are first-order derivatives with respect to a function $H(q_i, p_i, t)$ known as the Hamiltonian which depends on the canonical variables and time.  The Hamiltonian is given by the expression for the energy of that system.     

In the previous two sections we determined the dynamics of the electric and magnetic fields by solving Maxwell's equations explicitly.  When a system is quantised, however, it is common to begin from a Hamiltonian description of the classical system.  In order to motivate certain results and aid discussions later on, it is convenient to demonstrate here that the free electromagnetic field can also be expressed as a Hamiltonian system, and that the expression for the energy of the free fields acts as the Hamiltonian for that system.

In Section \ref{Sec:3Dclassicalfields} it was shown by solving Maxwell's equations that the mode functions $\alpha_{\mathbf{k}\lambda}$ are the coefficients of the positive frequency modes, and therefore satisfy the equation of motion
\begin{eqnarray}
\label{normalvariableequation}
\dot{\alpha}_{\mathbf{k}\lambda}(t) + i\kappa c\,\alpha_{\mathbf{k}\lambda}(t) = 0.
\end{eqnarray}
The terms $\alpha_{\mathbf{k}\lambda}(t)$ are often known as normal variables as they oscillate at a single frequency and therefore describe the normal modes of vibration.  This equation describes the evolution of a simple harmonic oscillator when the normal mode is related to the position and momentum of the oscillator in the following way \cite{Hei}:
\begin{eqnarray}
\label{normalvariable1}
\alpha_{\mathbf{k}\lambda} = \sqrt{\frac{m\kappa^2c^2}{4\epsilon_0}}\left(q_{\mathbf{k}\lambda}+ \frac{i}{\kappa c m}p_{\mathbf{k}\lambda}\right).
\end{eqnarray}
Here $q_{\mathbf{k}\lambda}$ and $p_{\mathbf{k}\lambda}$ denote the position and momentum of the $\mathbf{k}\lambda$-th oscillator respectively.  As $\mathbf{k}$ is an infinite and continuous variable there is an infinite continuum of harmonic oscillators.

In terms of these new variables the energy observable (\ref{fclassicalenergy1}) is given by
\begin{eqnarray}
H_{\text{energy}}(t) &=& \sum_{\lambda = \mathsf{H},\mathsf{V}}\int_{\mathbb{R}^3}\text{d}^3\mathbf{k}\; \left\{\frac{p_{\mathbf{k}\lambda}^2}{2m} + \frac{m\kappa^2c^2}{2}q_{\mathbf{k}\lambda}^2\right\}.
\end{eqnarray}
This is the expression for the total energy of a continuous system of uncoupled harmonic oscillators that have positions $q_{\mathbf{k}\lambda}$ and momenta $p_{\mathbf{k}\lambda}$.  In order to demonstrate that the energy observable is also the Hamiltonian that generates the dynamics of the system, we must show that the dynamics of the position and momentum variables as predicted by Hamilton's equations (\ref{Hamiltonsequations}) are equivalent to Eq.~(\ref{normalvariableequation}).  We must remember at this stage that, as the canonical variables form a continuum in $\mathbf{k}$, the partial derivatives in Hamilton's equations must be replaced by functional derivatives.  Having taken this into account, one is able to verify that 
\begin{eqnarray}
\dot{p}_{\mathbf{k}\lambda} &=& -m\kappa^2c^2\,q_{\mathbf{k}\lambda}\nonumber\\
\dot{q}_{\mathbf{k}\lambda} &=& \frac{\,p_{\mathbf{k}\lambda}}{m}.
\end{eqnarray}
It is now possible, using Eq.~(\ref{normalvariable1}), to show that these equations are equivalent to Eq.~(\ref{normalvariableequation}).

\chapter{The quantum theory of light}

\label{Chap:quantum}

In this chapter we shall review a recent quantisation of the EM field that provides the standard result for the quantised free field observables.  This quantisation scheme will provide an important set of guidelines for quantising the EM field in position space, which will be the topic of Chapters \ref{Chapter:1D} and \ref{Chapter:3D}.  Before we discuss this quantisation scheme we review some of the basic principles of quantum mechanics that are preliminary to the remainder of this thesis.

\section{A short review of quantum mechanics}

\label{Sec:shortQM}

\subsection{The Hilbert space of quantum states}

\subsubsection{Quantum states}

In classical physics the state of a system is characterised by a complete set of variables such as the canonically conjugate pair $q_i$ and $p_i$.  In quantum physics, all information pertaining to a system is contained within the quantum mechanical state vector or wave-function of that system.  The collection of all possible state vectors spans a complex (and possibly infinite-dimensional) vector space $\mathscr{H}$ known as the Hilbert space.  A common notation that shall be adopted in this thesis will be the Dirac bra-ket notation.  In this notation, the state of a system $\Psi$ will be denoted by the ket-vector $\ket{\Psi}$ which exists in $\mathscr{H}$.

\subsubsection{The inner product}

\label{Sec:innerproduct}

There exists an inner product on Hilbert space such that, for any two state vectors $\ket{\Psi}$ and $\ket{\Phi}$ in $\mathscr{H}$, their inner product, denoted in the bra-ket notation as $\braket{\Phi|\Psi}$, is a complex number.  The inner product satisfies the following three conditions \cite{Wei}:
\begin{eqnarray}
\begin{matrix*}[l]
\text{1. Linearity:} & \text{For all } \ket{\Psi_1}, \ket{\Psi_2}, \ket{\Phi} \in \mathscr{H}\text{, and for all } \alpha, \beta \in \mathbb{C},\\
& \bra{\Phi}\left(\alpha\ket{\Psi_1} + \beta \ket{\Psi_2}\right) = \alpha\braket{\Phi|\Psi_1} + \beta\braket{\Phi|\Psi_2}.\\
\text{2. Conjugate symmetry:} & \text{For all }\ket{\Psi}, \ket{\Phi} \in \mathscr{H},\\
& \braket{\Phi|\Psi} = \braket{\Psi|\Phi}^*.\\
\text{3. Positivity:} & \text{For all }\ket{\Psi} \neq \mathbf{0} \in \mathscr{H},\\ 
& \braket{\Psi|\Psi} >0.
\end{matrix*}
\end{eqnarray}
Using this inner product it is possible for us to define a norm for state vectors in the Hilbert space.  The norm or magnitude of a state vector $\ket{\Psi}$ is given by $\|\braket{\Psi|\Psi}\|^{1/2}$ which is always real (condition 2) and positive (condition 3).  When we normalise a state vector we mean that we multiply that state vector by an overall constant such that the magnitude of the new vector is equal to unity.

By themselves, the backwards facing vectors in the inner product above are known as bra-vectors, and represent a single continuous mapping from points in $\mathscr{H}$ to a point in the complex plane.  Like the ket-vectors, the bra-vectors also span a vector space which in this latter case is known as the dual space $\mathscr{H}^*$.  For each ket-vector $\ket{\Psi}$ in the Hilbert space there is a unique bra-vector $\bra{\Psi}$ that exists in the dual space.

In quantum theory, the physical interpretation given to the inner product between two states $\ket{\Psi}$ and $\ket{\Phi}$ in $\mathscr{H}$, $\braket{\Phi|\Psi}$, is the probability amplitude for the state $\ket{\Psi}$ to transition into the state $\ket{\Phi}$.  The probability for this transition to take place is the real squared magnitude of this complex probability amplitude $\|\braket{\Phi|\psi}\|^2$.  This postulate is known as the Born rule.

\subsection{Operators, observables and expectation values}

\subsubsection{Quantum mechanical operators}
\label{Sec:qmoperators}

Any change of the physical state of a system will be accompanied by a corresponding change in it's state vector.  Such changes are brought about by acting on the state vector with a quantum mechanical operator that maps the Hilbert space onto itself, for example, $A:\mathscr{H} \to \mathscr{H}$, which will map a state $\ket{\Psi}$ to the new state A$\ket{\Psi}$ where $A$ is always understood to act to the right.  Changes of reference frame and the act of measurement both introduce changes to the system, and are therefore represented by operators acting on the Hilbert space. 

For every operator $A$ there will also exist an operator $A^\dagger$ known as the adjoint of $A$.  The adjoint of $A$ is defined such that
\begin{eqnarray}
	\label{hermiticity}
	\braket{\Phi|A|\Psi} = \braket{\Psi|A^\dagger|\Phi}^*.
\end{eqnarray}
A particularly interesting set of operators are those for which $A = A^\dagger$, which are known as self-adjoint.  In this thesis we shall use the expressions self-adjoint and Hermitian interchangeably, although in general not all Hermitian operators are self-adjoint.  The eigenvalues of an Hermitian operator are always real and eigenstates with distinct eigenvalues are orthogonal under the inner product that defines the adjoint operator.

\subsubsection{Observables}

When handed a system to play with, we often want to take measurements of some relevant quantity in that system: the electric field, for example.  When a measurement is made on a quantum system, the possible outcomes of that measurement are the eigenvalues of the operator that represents the particular observable being measured.  After that measurement is taken the state of the system is projected onto an eigenstate associated with the measured outcome with a probability determined by the Born rule. As the Hilbert space is a complex vector space, the eigenvalues of a general operator will be complex.  The measurable outcomes of an experiment, on the other hand, must always be real.  The eigenvalues of the observable must therefore also be orthogonal to ensure that a measurement repeated immediately will yield the same result as the first.  The class of operators we are describing are the Hermitian operators.  Consequently all observables are represented by Hermitian operators.

\subsubsection{Expectation values}

When an observable is measured, unless the state prior to the measurement is an eigenstate of the operator for that measurement, measurements repeated on separate identical systems will yield different results that occur with a known probability.  Consider repeating such an experiment an infinite number of times, each time beginning with the same initial state.  The average of all the measured values for that particular observable is known as the expectation value of that observable with respect to a given state.  If it is assumed that our state is initially given by the ket-vector $\ket{\Psi}$, then the expectation value of the observable $A$ is given by
\begin{eqnarray}
\label{expvalue1}
\braket{A}_\Psi = \braket{\Psi|A|\Psi}.
\end{eqnarray}
In the classical limit of quantum mechanics, the possible outcomes of the measurement of a particular observable all approach the expectation value of that observable.  The expectation value of a quantum observable represents the classical value of that observable.

\subsection{The dynamics of a quantum system}
\label{Sec:Dynamics1}

\subsubsection{The Schr\"odinger equation}

Between measurements a system and its corresponding state vector will naturally evolve of their own accord.  The equation of motion that determines this evolution is the Schr\"odinger equation:
\begin{eqnarray}
\label{Schrodingerequation}
i\hbar\frac{\text{d}}{\text{d}t}\ket{\Psi(t)} = H(t)\ket{\Psi(t)},
\end{eqnarray}
where the constant $\hbar = h/2\pi$ is the reduced Planck's constant.  In the above equation, the operator $H(t)$ is the Hamiltonian of the system.  It acts as the generator of time translations in a quantum system and, similar to as in the classical formalism, is equal to the energy observable.  In the particular case that the Hamiltonian is independent of time, the solution of the Schr\"odinger equation is  
\begin{eqnarray}
\label{Schrodingersolution1}
\ket{\Psi(t)} = U(t,t_0)\ket{\Psi(t_0)}
\end{eqnarray}
where $U(t,t_0) = \text{exp}[-iH(t-t_0)/\hbar]$ and $\ket{\Psi(t_0)}$ is the state at an initial time $t_0$.  The description of a system in which the dynamics of the system is encoded within the time dependence of state vectors is known as the Schr\"odinger picture.

\subsubsection{The Heisenberg equation}

When investigating the dynamics of a quantum system, our primary task will be to calculate expectation values of different observables at particular times.  The expectation of an observable $A$ calculated with respect to a state specified at a time $t$, which is calculated using Eq.~(\ref{expvalue1}), may be expressed as 
\begin{eqnarray}
\braket{\Psi(t_0)|U^\dagger(t,t_0)AU(t,t_0)|\Psi(t_0)} = \braket{\Psi(t_0)|A(t)|\Psi(t_0)}
\end{eqnarray}
where $t_0$ is some fixed reference time and $A = A(t_0)$.  When calculating a time-dependent expectation value, it does not matter whether the calculation is carried out using a time-independent operator with respect to a state evolving according to the Schr\"odinger equation or whether it is calculated for a time-dependent operator $A(t) = U^\dagger(t,t_0)AU(t,t_0)$ with respect to a time-independent state $\ket{\Psi(t_0)}$.  

The description of a system in which its dynamics are encoded in the operators is known as the Heisenberg picture of quantum mechanics.  In the Heisenberg picture, as we have just seen, the time dependence of an operator is already specified by the time-evolution operators $U(t,t_0)$.  An equation of motion for the time-dependent operators is therefore found by taking the time derivative of this operator.  The equation of motion for the time-dependent operator is known as Heisenberg's equation of motion and is given by
\begin{eqnarray}
\label{Heisenbergequation1}
\frac{\text{d}}{\text{d}t}A(t) = -\frac{i}{\hbar}\left[A(t), H(t)\right]
\end{eqnarray}
where the commutator of two operators $\left[A,B\right]$ is equal to $AB - BA$.

\section{Quantisation of the free EM field}

\label{Sec:Quantisation1}

\subsection{The Hamiltonian operator}

\label{Sec:backgroundHamiltonian}

\subsubsection{The Photon}

The process of quantising the classical electromagnetic theory aims to achieve two things.  It must determine a Hilbert space that is appropriate for describing the possible  quantum states of the free radiation field, and it must also provide us with a set of observables that represent the electric and magnetic field vectors that act on this Hilbert space.  There are several ways in which one may go about finding the quantised observables of the free EM field.  Using a canonical quantisation prescription, for example, textbooks usually obtain expressions for the basic field observables by expanding the vector potential $\boldsymbol{A}$ of the classical EM field into its Fourier components. The Fourier coefficients are then replaced by photon creation and annihilation operators with bosonic commutation relations \cite{Mil3,Sak,Lou}.  In some cases authors only consider standing waves inside a box (see for example Refs.~\cite{MW, Coh,Lou}).  The particular method that we shall follow in this section is the one presented by Bennett \textit{et al.}~\cite{RB2} that considers the running wave solutions of Maxwell's equations which are characterised by a wave number $\mathbf{k} \in \mathbb{R}^3$, a frequency $\omega = |\mathbf{k}|c$ and a polarisation $\lambda \in \{\mathsf{H}, \mathsf{V}\}$.  This characterisation can be traced back to Planck's 1901 modelling of black body radiation \cite{Pla} and Einstein's 1917 analysis of the photoelectric effect \cite{Ein}. 

It was noted in Section \ref{Sec:Harmonicoscillator} that the Hamiltonian for the free EM field can be expressed as an infinite collection of uncoupled harmonic oscillators with a frequency $\omega = \kappa c$.  As is well known from quantum mechanics, the energy spectrum of the quantum harmonic oscillator consists of an infinite ladder of energy eigenvalues.  Each rung on the ladder is separated from the one above and the one below by an energy spacing of $\hbar \omega$ where $\omega$ is the frequency of the oscillator.  Such a spectrum is equivalent to that for a system of many identical particles: as a particle is added or removed from the system, we move up or down a rung on the ladder, adding or removing a fixed amount of energy.  In this scheme, therefore, the initial and only assumption shall be that the Hilbert space is spanned by a set of countable energy quanta characterised by a wave vector $\mathbf{k}$ and a polarisation $\lambda$.  Excitations of this type are known as photons and are the basic excitations of the free EM field.

\subsubsection{Annihilation and creation operators}

A useful notation for a state containing $n$ photons, each characterised by a wave vector $\mathbf{k}_i$ with a corresponding polarisation $\lambda_i$, is
\begin{equation}
\label{photonstate1}
\ket{\mathbf{k}_1\, ...\, \mathbf{k}_n}.
\end{equation}
Here $i$ takes all integer values between $1$ and $n$.  As photons are bosons, this state is symmetric under a reordering of any of the $\mathbf{k}_i$.  As states characterised by a definite $\mathbf{k}$ and $\lambda$ are eigenstates of the Hermitian energy observable, they are all orthogonal to one another.  Consequently \cite{Lan}
\begin{eqnarray}
\label{innerproduct1}
\braket{\mathbf{k}_1\, ...\, \mathbf{k}_n|\mathbf{k}'_1\,...\,\mathbf{k}'_m} = \delta_{nm}\sum_\mathcal{P}\prod_{i = 1}^n\; \delta(\mathbf{k}_i - \mathbf{k}'_{\mathcal{P}(i)})\delta_{\lambda_i, \lambda_{\mathcal{P}(i)}}
\end{eqnarray}
where $\mathcal{P}$ represents a permutation of the set $\{1, ..., m\}$ and $\mathcal{P}(i)$ is the $i$-th member of that permutation. All permutations must be considered in this inner product to ensure that the inner product is also unchanged under any reordering of the $\mathbf{k}_i$.  Here $\delta(\mathbf{k})$ refers to the Dirac delta function and $\delta_{nm}$ is the Kronecker delta.

It is convenient at this stage to introduce an operator that will raise an energy eigenstate one rung higher on the ladder by creating a photon with a wave vector $\mathbf{k}$ and polarisation $\lambda$.  Such an operator is aptly named the creation operator and denoted $a^\dagger_{\mathbf{k}\lambda}$.  Consider the earlier $n$-photon state given in Eq.~(\ref{photonstate1}).  By applying the creation operator $a^\dagger_{\mathbf{k}\lambda}$ to this state we generate the following $(n+1)$-photon state:
\begin{equation}
\label{creation1}
a^\dagger_{\mathbf{k}\lambda}\ket{\mathbf{k}_1\,...\,\mathbf{k}_n} = \ket{\mathbf{k},\mathbf{k}_1\,...\,\mathbf{k}_n}.
\end{equation}
Since the state is unchanged by the reordering of any $\mathbf{k}_i$, all creation operators commute with one another.

The Hermitian conjugate operator of the creation operator $a^\dagger_{\mathbf{k}\lambda}$ is denoted $a_{\mathbf{k}\lambda}$.  Using the definition of the adjoint operator given in Eq.~(\ref{hermiticity}) and the inner product given in Eq.~(\ref{innerproduct1}), one may show that when $a_{\mathbf{k}\lambda}$ is applied to the $n$-photon state given in Eq.~(\ref{photonstate1}), it generates the following state:
\begin{equation}
\label{annihilation1}
a_{\mathbf{k}\lambda}\ket{\mathbf{k}_1\, ...\, \mathbf{k}_n} = \sum_{i=1}^n \delta(\mathbf{k}-\mathbf{k}_i)\,\delta_{\lambda, \lambda_i}\ket{\mathbf{k}_1\, ...\, \mathbf{k}_{i-1}, \mathbf{k}_{i+1}\, ...\, \mathbf{k}_n}.
\end{equation}
The resulting state contains exactly $n-1$ photons.  The operator $a_{\mathbf{k}\lambda}$ has moved the state down one rung of the ladder by removing one photon creating an $(n-1)$-photon energy eigenstate. For this reason, they are referred to as annihilation operators.  As the final state on the right-hand side of Eq.~(\ref{annihilation1}) is unchanged by the reordering of $\mathbf{k}_i$, all annihilation operators commute with one another.

Since every time a photon is removed the total energy of the system drops, to ensure that the energy is bounded below we must assume that there comes a point when we reach a lowest energy state that is destroyed by all annihilation operators.  Such a state is known as the vacuum state and denoted $\ket{0}$: the state containing no photons.  In order to satisfy the conditions of a vacuum state we insist that $a_{\mathbf{k}\lambda}\ket{0} = 0$ for all $\mathbf{k}$ and $\lambda$.  It is assumed that the vacuum state is normalised to unity: $\braket{0|0} = 1$.

In Eq.~(\ref{creation1}), we defined a set of creation operators that add a single photon to a state.  As we now have a vacuum state $\ket{0}$ that has been shown to contain exactly no photons, we can construct states containing any number of photons we like by applying the creation operators repeatedly to the vacuum state.  We refer to states containing a specific number of photons as number states.  A state containing $n$ identical photons is denoted
\begin{eqnarray}
	\label{numberstates1}
\ket{n_{\mathbf{k}\lambda}} = \frac{1}{\sqrt{n!}}\,{a^\dagger_{\mathbf{k}\lambda}}^n\ket{0}.
\end{eqnarray}
The $\ket{n_{\mathbf{k}\lambda}}$ state contains $n$ photons that have a wave vector $\mathbf{k}$ and are polarised in the $\lambda$ direction.  We refer to this state as an $n$-photon state.

Using Eqs.~(\ref{creation1}) and (\ref{annihilation1}), one is able to show by direct calculation that the commutator between an annihilation and creation operator satisfies the relation
\begin{equation}
\left[a_{\mathbf{k}\lambda}, a^\dagger_{\mathbf{k}'\lambda'}\right]\ket{\mathbf{k}_1\,...\,\mathbf{k}_n} = \delta(\mathbf{k}-\mathbf{k}')\,\delta_{\lambda,\lambda'}\ket{\mathbf{k}_1\,...\,\mathbf{k}_n}.
\end{equation}
Consequently, we find that 
\begin{equation}
\label{standardcommutator1}
\left[a_{\mathbf{k}\lambda}, a^\dagger_{\mathbf{k}\lambda'}\right] = \delta(\mathbf{k}-\mathbf{k}')\,\delta_{\lambda,\lambda'}\, \text{id}
\end{equation}
where id is the identity operator.  Since this commutator is proportional to the identity operator, I shall drop the $\text{id}$ and treat the commutator as simply a number, in which case it may be shown that 
\begin{equation}
\label{commutatorfromproduct}
\braket{1_{\mathbf{k}\lambda}|1_{\mathbf{k}'\lambda'}} = \left[a_{\mathbf{k}\lambda}, a^\dagger_{\mathbf{k}'\lambda'}\right].
\end{equation}
In future we shall use this definition to define the commutator between creation and annihilation operators.

\subsubsection{The Hamiltonian observable}

In standard descriptions of the quantised EM field \cite{MW, Hei, Coh, Lou, RB2}, the energy observable is given by the harmonic oscillator Hamiltonian for a collection of uncoupled oscillators.  We can understand this result in terms of the discussion on the photon at the beginning of this section.  This operator can be expressed in terms of the photon creation and annihilation operators in the following way:  
\begin{eqnarray}
	\label{energyobs1}
H(t) = \sum_{\lambda = 1,2}\int_{\mathbb{R}^3}\text{d}^3\mathbf{k}\;\hbar|\mathbf{k}| c\, a^\dagger_{\mathbf{k}\lambda}a_{\mathbf{k}\lambda} + H_{\text{ZPE}}.
\end{eqnarray}
The $H_{\text{ZPE}}$ term is a further numerical term that is, as yet, unknown to us.  It represents the zero-point energy of the system, which is the energy of the vacuum state.  It acts to simultaneously shift the energy of all $(\mathbf{k},\lambda)$ states by a fixed amount.  The zero-point energy does not play a role in determining the dynamics of a state or any observables, but it does still contribute to many important and observable effects, as shall be seen in Chapter \ref{Chapter:Casimir} of this thesis.  One may check, using Eqs.~(\ref{numberstates1}) and (\ref{standardcommutator1}), that the photon number states are eigenstates of the Hamiltonian observable and have eigenvalues $n\hbar |\mathbf{k}| c + H_{\text{ZPE}}$, as expected.

\subsection{Photon wave packets}

\label{Sec:wavepackets}

\subsubsection{Single-photon wave packets}

In the context of linear optics experiments \cite{Sho, Zuk,Lim,Kok1, Br}, it is usual to talk about single photons when referring to particles whose state vectors $\ket{1}$ can be expressed as $\ket{1} = a^\dagger \ket{0}$ where $a$ is an annihilation operator that satisfies the commutation relation
\begin{eqnarray}
	\label{wavepacketcommutator}
	\left[a,a^\dagger\right] = 1.
\end{eqnarray}
The single-photon states $\ket{1_{\mathbf{k}\lambda}}$ defined in the previous section, however, are not normalisable.  The annihilation operator $a$ can be constructed by linearly superposing a number of the $\mathbf{k}$-dependent annihilation operators $a_{\mathbf{k}\lambda}$ in the following way
\begin{eqnarray}
	\label{wavepacket1}
	a = \sum_{\lambda = \mathsf{H}, \mathsf{V}}\int_{\mathbb{R}^3}\text{d}^3\mathbf{k}\;\psi^*_{\mathbf{k}\lambda}\,a_{\mathbf{k}\lambda}
\end{eqnarray}
with complex coefficients $\psi^*_{\mathbf{k}\lambda}$.  When the wave packet coefficients satisfy the normalisation condition  
\begin{eqnarray}
	\label{wavepacketnormalisation1}
	\sum_{\lambda =\mathsf{H},\mathsf{V}}\int_{\mathbb{R}^3}\text{d}^3\mathbf{k}\;|\psi_{\mathbf{k}\lambda}|^2 = 1
\end{eqnarray}
the state $\ket{1}$ is normalised and Eq.~(\ref{wavepacketcommutator}) is satisfied.

\subsubsection{Many-photon wave packets}

Quantum states of light, such as the single-photon state above, are not restricted to contain only one photon, but may contain any number of photons we like.  A many-photon state  may be constructed in the following way: 
\begin{eqnarray}
\label{manyphoton}
\ket{\psi} &=& \sum_{n=0}^\infty {c_n \over \sqrt{n!}}\, {a^\dagger}^n \ket{0} 
\end{eqnarray}
where $a^\dagger$ is the Hermitian conjugate of $a$ defined in Eq.~(\ref{wavepacket1}).  This state is a superposition of states containing precisely $n$ $a$-photons for all positive integers $n$.  The contribution of the $n$ number state to the total state $\ket{\psi}$ is determined by the coefficient $c_n$.  Using the commutation relation given above in Eq.~(\ref{wavepacketcommutator}), one can verify that the state $\ket{\psi}$ is normalised when the coefficients $c_n$ are chosen such that $\sum_{n=0}^\infty |c_n|^2 = 1$.  The state $\ket{\psi}$ is not the most general many-photon state that may be constructed, but there will be no need to construct more general states in this thesis.

\subsubsection{Coherent wave packets}

\label{Sec:coherentwavepackets}

One possible choice for the normalised coefficients $c_n$ is 
\begin{eqnarray}
	\label{coherent1}
	c_n &=& {\rm e}^{- |\alpha|^2/2} \, \alpha^n  \, 
\end{eqnarray}
for a complex number $\alpha$.  For this particular choice of $c_n$, the state $\ket{\psi}$ corresponds to the coherent state $|\alpha \rangle$ of a short laser pulse.  Coherent states are minimal uncertainty quantum states that do not exhibit correlations of any order \cite{Gla1}.  They are also the eigenstates of the annihilation operators.  Consider the annihilation operator $a$ for a normalised single-photon wave packet as defined in Eq.~(\ref{wavepacket1}).  A coherent wave packet $\ket{\alpha}$ satisfies the eigenstate equation
\begin{equation}
a\ket{\alpha} = \alpha\ket{\alpha}.
\end{equation}
This equation may be verified using Eqs.~(\ref{wavepacketcommutator}), (\ref{manyphoton}) and (\ref{coherent1}).  Hence, a coherent state is an infinite superposition of all possible number states.

It is convenient to express the state $\ket{\alpha}$ in terms of an operator $D(\alpha)$ that generates $\ket{\alpha}$ from the vacuum state:
\begin{equation}
\label{Displacement1}
\ket{\alpha} = D(\alpha)\ket{0} = e^{(\alpha a^\dagger - \alpha^*a)}\ket{0}=  e^{-\frac{1}{2}|\alpha|^2}\sum_{n=1}^{\infty}\frac{\alpha^n}{n!}{a^\dagger}^n\ket{0}.
\end{equation}
The operator $D(\alpha)$ is known as the displacement operator and is unitary: $D(\alpha)D^\dagger(\alpha) = \text{id}$.

 \subsection{The electric and magnetic field observables}
 
 \label{Sec:Bennetquantisation}

\subsubsection{An ansatz for the field observables}

In Section \ref{Sec:classicalenergy} we demonstrated that the energy of the EM field is quadratic in both the electric and magnetic fields.  In Section \ref{Sec:backgroundHamiltonian} we also found that the energy observable is quadratic in the creation and annihilation operators $a^\dagger_{\mathbf{k}\lambda}$ and $a_{\mathbf{k}\lambda}$.  A suitable ansatz for the electric and magnetic fields therefore is one that is linear in both the annihilation and creation operators.  For simplicity we shall only follow the quantisation procedure for one-dimensional fields, and state the three-dimensional result towards the end.  In the one-dimensional system, the wave vector $\mathbf{k} = k\,\widehat{\boldsymbol{x}}$ is characterised by its magnitude $\kappa$ and its sign $s$ which takes  the value $s =\mathsf{L}$ for left-propagating light and $s=\mathsf{R}$ for right-propagating light.  The one-dimensional electric and magnetic field observables in the Heisenberg picture are given by
\begin{eqnarray}
\label{Efieldansatz}
\mathbf{E}(x,t) &=& \sum_{\lambda = \mathsf{H}, \mathsf{V}}\sum_{s = \mathsf{L},\mathsf{R}}\int_{0}^{\infty}\frac{\text{d}\kappa}{\sqrt{2\pi}}\; f_{s\lambda}(x,t;\kappa)\,a^\dagger_{s\lambda}(\kappa)\,\boldsymbol{e}_{s\lambda} + \text{H.c.}\\
\label{Bfieldansatz}
\mathbf{B}(x,t) &=& \sum_{\lambda = \mathsf{H}, \mathsf{V}}\sum_{s = \mathsf{L},\mathsf{R}}\int_{0}^{\infty}\frac{\text{d}\kappa}{\sqrt{2\pi}}\; g_{s\lambda}(x,t;\kappa)\,a^\dagger_{s\lambda}(\kappa)\,\boldsymbol{e}_{s\lambda} + \text{H.c.}
\end{eqnarray}
Here the coefficients $f_{s\lambda}(x,t;\kappa)$ and $g_{s\lambda}(x,t;\kappa)$ are complex and H.c.~denotes the Hermitian conjugate.  As this is a one-dimensional solution we shall assume that $\boldsymbol{e}_{s \mathsf{H}} = \widehat{\boldsymbol{y}}$ and $\boldsymbol{e}_{s \mathsf{V}} = \widehat{\boldsymbol{z}}$.

\subsubsection{Field dynamics}

In the Heisenberg picture, quantum mechanical operators evolve according to the Heisenberg equation (\ref{Heisenbergequation1}).  As was the case in the classical system, the energy observable (\ref{energyobs1}) is the generator of time translations, which now appears in the Heisenberg equation.  By substituting the expression for the electric field observable (\ref{Efieldansatz}) and the Hamiltonian observable into the Heisenberg equation one finds that
\begin{eqnarray}
\frac{\partial f_{s\lambda}(x,t;\kappa)}{\partial t} = i\kappa c f_{s\lambda}(x,t;\kappa).
\end{eqnarray}
By following the same process for the magnetic field one will find an identical equation for the $g_{s\lambda}(x,t;\kappa)$ coefficients.  The solutions of these equations are
\begin{eqnarray}
f_{s\lambda}(x,t;\kappa) &=& e^{i\kappa c t}f_{s\lambda}(x;\kappa)\nonumber\\
g_{s\lambda}(x,t;\kappa) &=& e^{i\kappa c t}g_{s\lambda}(x;\kappa)
\end{eqnarray}
where $f_{s\lambda}(x;\kappa)$ and $g_{s\lambda}(x;\kappa)$ are two time-independent functions left to be determined.  One may notice that the annihilation and creation operators evolve just like the normal variables in the classical theory.  In the quantised theory, in which the classical electric and magnetic fields become operators acting on the Hilbert space, one could say that the normal variables $\alpha_{\mathbf{k}\lambda}$ and their complex conjugates are promoted to annihilation and creation operators respectively.  Note in particular, however, that in contrast to the real classical solutions of Maxwell's equations, the evolution of the photon creation operators is determined by the negative-frequency modes only.

\subsubsection{Maxwell's equations}

In the quantised theory, the expectation values of the electric and magnetic field observables calculated with respect to any normalised state are assumed to evolve according to Maxwell's equations.  As all space and time dependence is contained within the observables, we may equivalently demand that the field observables themselves obey Maxwell's equations.  By substituting the field observables (\ref{Efieldansatz}) and (\ref{Bfieldansatz}) into the free-space Maxwell's equations and solving these equations, taking into account the time dependence of the $f_{s\lambda}(x,t;\kappa)$ and $g_{s\lambda}(x,t;\kappa)$ coefficients, one will find that
\begin{eqnarray}
f_{s\lambda}(x;\kappa) &=& e^{i\kappa x}f^{(1)}_{\mathsf{L}\lambda}(\kappa) + e^{-i\kappa x}f^{(2)}_{\mathsf{R}\lambda}(\kappa)\\
g_{s\lambda}(x;\kappa) &=& \pm\frac{1}{c}\left\{e^{i\kappa x}f^{(1)}_{\mathsf{L}p}(\kappa) - e^{-i\kappa x}f^{(2)}_{\mathsf{R}p}(\kappa)\right\}
\end{eqnarray}
where $\lambda \neq p$.  Here the positive sign applies when $\lambda =\mathsf{H}$ and the negative sign when $\lambda = \mathsf{V}$.  As the field associated with a right-propagating photon must be a function of $x-ct$ only, and $x+ct$ for a left-propagating photon, we have already established that $f^{(1)}_{\mathsf{R}\lambda}$ and $f^{(2)}_{\mathsf{L}\lambda}$ equal zero.

\subsubsection{The electric and magnetic field observables}

By solving Heisenberg's equation and Maxwell's equations we have determined the complete expressions for the electric and magnetic field observables up to the two remaining $\kappa$- and $\lambda$-dependent terms $f^{(1)}_{\mathsf{L}\lambda}(\kappa)$ and $f^{(2)}_{\mathsf{R}\lambda}(\kappa)$.  To determine these factors, it is necessary only to substitute the field observables into the expression for the field energy (\ref{classicalenergy4}) and equate the resulting operator with the Hamiltonian observable (\ref{energyobs1}).  Having performed this calculation one finds that the field energy and the Hamiltonian observable exactly coincide when 
\begin{eqnarray}
|f^{(1)}_{\mathsf{L}\lambda}(\kappa)|^2 = |f^{(2)}_{\mathsf{R}\lambda}(\kappa)|^2 = \frac{\hbar \kappa c}{2A\varepsilon_0}
\end{eqnarray}
where $A$ is the cross-sectional area inhabited by the field in the plane perpendicular to the direction of propagation.  In this way, the field observables are determined up to an overall phase.

Putting the results of this section together, we have, up to an overall phase, a set of observables for the electric and magnetic field vectors.  These are
\begin{eqnarray}
\label{Efield2}
\mathbf{E}(x,t) &=& \sum_{\lambda = \mathsf{H},\mathsf{V}}\int_{0}^{\infty}\text{d}\kappa\;\sqrt{\frac{\hbar \kappa c}{4\pi \varepsilon_0A}} \,\left[e^{i\kappa(x+ct)}a^\dagger_{\mathsf{L}\lambda}(\kappa) + e^{-i\kappa(x-ct)}a^\dagger_{\mathsf{R}\lambda}(\kappa)\right]\boldsymbol{e}_{s\lambda}\nonumber\\
&& \hspace*{5cm} + \text{H.c}\\
\label{Bfield2}
\mathbf{B}(x,t) &=& \pm \sum_{\lambda = \mathsf{H},\mathsf{V}}\int_{0}^{\infty}\text{d}\kappa\;\frac{1}{c}\sqrt{\frac{\hbar \kappa c}{4\pi \varepsilon_0A}} \,\left[e^{i\kappa(x+ct)}a^\dagger_{\mathsf{L}p}(\kappa) - e^{-i\kappa(x-ct)}a^\dagger_{\mathsf{R}p}(\kappa)\right]\boldsymbol{e}_{s\lambda}\nonumber\\
&& \hspace*{5cm} + \text{H.c}
\end{eqnarray}
where $p \neq \lambda$ and the positive sign applies when $\lambda = \mathsf{H}$ and the negative sign when $\lambda = \mathsf{V}$.  Now that we have a set of observables for the electric and magnetic fields, and therefore are able to calculate the energy observable exactly using Eq.~(\ref{classicalenergy4}), we may also determine the value of the $H_{\text{ZPE}}$ left unspecified earlier.  We find that this zero-point energy is
\begin{eqnarray}
H_{\text{ZPE}} = \int_{0}^{\infty}\text{d}\kappa\; 2\hbar \kappa c
\end{eqnarray} 
which is infinite.

If one proceeds in a similar manner, making first the assumption that the Hamiltonian is given by Eq.~(\ref{energyobs1}), one is able to show that the three-dimensional electric and magnetic field observables are given by the expressions
\begin{eqnarray}
\label{Efield3}
\mathbf{E}(\mathbf{r},t) &=& \sum_{\lambda = \mathsf{H}, \mathsf{V}}\int_{\mathbb{R}^3}\frac{\text{d}^3\mathbf{k}}{(2\pi)^{3/2}}\;\sqrt{\frac{\hbar \kappa c}{2\varepsilon_0}} \,e^{i(\mathbf{k}\cdot \mathbf{r}+\kappa ct)}a^\dagger_{\mathbf{k}\lambda}\,\boldsymbol{e}_{\mathbf{s}\lambda} + \text{H.c}\\
\label{Bfield3}
\mathbf{B}(\mathbf{r},t) &=& \sum_{\lambda = \mathsf{H},\mathsf{V}}\int_{\mathbb{R}^3}\frac{\text{d}^3\mathbf{k}}{(2\pi)^{3/2}}\;\frac{1}{c}\sqrt{\frac{\hbar \kappa c}{2\varepsilon_0}} \,e^{i(\mathbf{k}\cdot\mathbf{r}+\kappa ct)}a^\dagger_{\mathbf{k}\lambda}\left(\mathbf{s}\times\boldsymbol{e}_{\mathbf{s}\lambda}\right) + \text{H.c}
\end{eqnarray}
where again $\mathbf{k} = \kappa \mathbf{s}$.  By substituting the expressions for the field observables (\ref{Efield3}) and (\ref{Bfield3}) into Eq.~(\ref{classicalenergy4}), one will arrive at the expression for the energy observable (\ref{energyobs1}) with a zero-point energy given by
\begin{eqnarray}
H_{\text{ZPE}} = \int_{\mathbb{R}^3}\text{d}^3\mathbf{k}\;\hbar\kappa c.
\end{eqnarray}

\part{A local quantisation of the free radiation field}

\label{Part:theory}

\chapter{Local photons in one dimension}

\label{Chapter:1D}

In this chapter we construct an alternative quantisation for the free EM field in one dimension in both the position and momentum representations.  Analogous to the classical solutions of the free-space Maxwell's equations in one dimension, in this chapter the basic quanta of the EM field are localised particles that propagate along a single axis at the speed of light in a fixed direction without dispersion.  We motivate these excitations in Section \ref{Sec:1Dintroduction} by considering the evolution of localised wave packets that are propagating in opposite directions. Later, in Section \ref{Sec:1Dposition}, we describe their basic properties, develop an appropriate equation of motion for these localised particles, and determine expressions for the related electric and magnetic field observables.  We also derive a Hamiltonian for these particles that evolves the system according to the Schr\"odinger equation.  As in the standard theory, by writing the equation of motion of the system in this way we can more easily introduce interactions in future if desired.  Moreover, by having a Hamiltonian for the new system we can more easily compare our approach with standard quantum optical approaches.  In Section \ref{Sec:1Dmomentum} we examine these same particles and the associated field observables in their momentum representation,  which we then compare in Section \ref{Sec:comparison1} with the standard quantisation of the free EM field.  We end this chapter with a discussion in Section \ref{Sec:photonsdiscussion}.

\section{Introduction: The importance of a complete Hilbert space}

\label{Sec:1Dintroduction}

In Chapter \ref{Chap:quantum} we summarised a recent quantisation scheme for the free EM field which describes a system of energy quanta characterised by a wave vector $\mathbf{k}$ and a polarisation $\lambda$ \cite{RB2}.  This approach assumed that the electric and magnetic field observables were linear in the photon creation and annihilation operators, and then found the complete solutions by first solving Maxwell's equations and afterwards making sure that the photons had the correct energy.  This quantisation scheme has some similarities with that of Ornigotti \textit{et al}.~\cite{Orn} which quantises a paraxial electromagnetic field in a dispersive medium.  Here the field is quantised in a basis of orthogonal, orbital angular momentum carrying X waves \cite{Her} which, like the monochromatic plane waves, have an infinite norm.  Aiello \cite{Ai1, Ai2} recently also obtained a non-standard description of the EM field by quantising the monochromatic solutions of the paraxial wave equation for light in free space.  

The purpose of this section is, by using the approach of Bennett \textit{et al}., to quantise the one-dimensional free electromagnetic field in terms of localised and causally propagating particles.  From our discussion in the introduction, we have seen that there are different ways in which one may go about constructing locally orthogonal single-photon wave packets.  In this chapter, however, we quantise a set of locally bosonic particles that are distinct from the fields, and may instead be viewed as carriers of the fields.  The corresponding electric and magnetic field observables in this system will also be derived.  The theorems of Hegerfeldt \cite{Heg8} and Malament \cite{Mal} have also proven that localised states do not propagate in a way that would be expected for a short light pulse.  To avoid the consequences of Hegerfeldt's theorem, in this chapter we find that we must quantise both the positive- and negative-frequency solutions of Maxwell's equations, thus doubling the usual Hilbert space.

\begin{figure}
\centering
\includegraphics[width = \textwidth]{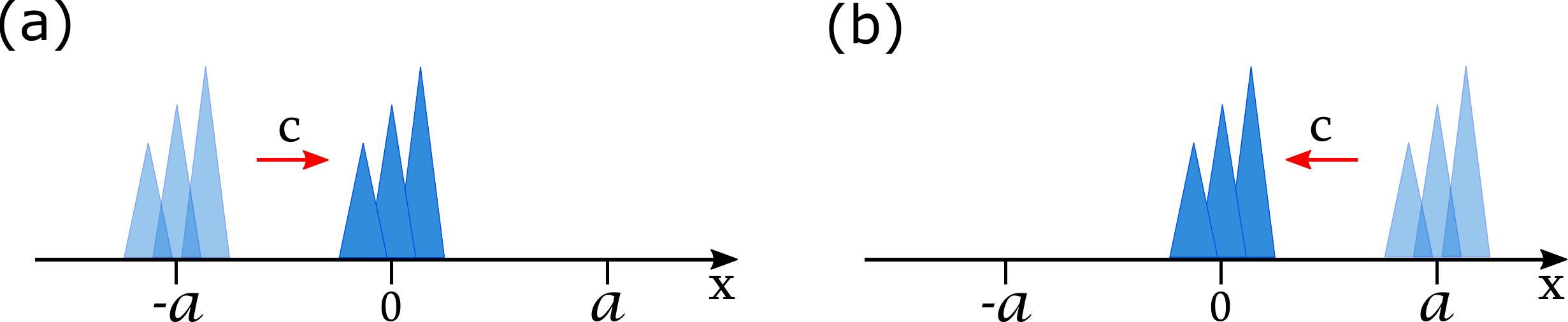}
\caption[Short light pulses in one dimension]{The figure shows two different scenarios in which a single-photon wave packet travels at the speed of light in a well-defined direction.  (a) Here a wave packet initially localised around the point $x = -a$ propagates to the right. (b) Here a wave packet initially localised around the point $x = a$ propagates to the left.  At $t=0$ both wave packets are easily distinguishable and occupy distinct sections of the real line. At $t=a/c$ both wave packets reach the point $x = 0$ and in both scenarios the wave packets appear to be the same.}
\label{Fig:photons1}
\end{figure}

It is possible to demonstrate this result with a simple thought experiment.  Consider the two scenarios depicted in Fig.~\ref{Fig:photons1}.  In the first scenario (part (a) of Fig.~\ref{Fig:photons1}), a right-moving single-photon wave packet with an initial wave function $\psi_1(x,0)$ is placed near the point $x = -a$.  In the second scenario (part (b) of Fig.~\ref{Fig:photons1}), a similar left-moving single-photon wave packet with an initial wave function $\psi_2(x,0)$ is placed near the point $x = a$.  If the two wave packets have the same shape, then it must follow that
\begin{equation}
\label{example1}
\psi_1(x,0) = e^{i\varphi}\,\psi_2(x+2a,0)
\end{equation}
where $\varphi$ denotes a general phase.  This statement ensures, using the Born rule, that the probability of finding the first photon at $x$ is the same as finding the second photon at $x + 2a$.  Furthermore, if the two wave packets do not overlap each other then they are easily distinguishable and their state vectors are pairwise orthogonal:
\begin{equation}
\label{example2}
\braket{\psi_1(0)|\psi_2(0)} = 0.
\end{equation} 

Let us assume that a quantum state evolves unitarily with a time-evolution operator $U(t,0)$ where $U^\dagger(t,0)U(t,0) = \text{id}$.  From Eq.~(\ref{example2}) it follows that the overlap of the two state vectors $\ket{\psi_1(t)}$ and $\ket{\psi_2(t)}$ remains zero at all times:
\begin{equation}
\label{example3}
\braket{\psi_1(t)|\psi_2(t)} = \braket{\psi_1(0)|U^\dagger(t,0)U(t,0)|\psi_2(0)} = \braket{\psi_1(0)|\psi_2(0)} = 0.
\end{equation}
This cannot be the case, however, as at a time $t = a/c$ the two wave packets overlap and are no longer distinguishable.  At this time, the two wave functions differ by at most a phase factor since
\begin{equation}
\label{example4}
\psi_1(x,t) = \psi_1(x-ct,0) = e^{i\varphi}\,\psi_2(x-ct + 2a,0) = e^{i\varphi}\,\psi_2(x,t).
\end{equation}
If Eq.~(\ref{example3}) holds at all times, then the quantum states of wave packets travelling to the left and to the right must belong to separate Hilbert spaces.  This means that particle wave packets must be characterised by an additional degree of freedom: their direction of propagation.  

In the following, we therefore distinguish two different types of photon by introducing a parameter $s = \pm 1$ to signify a direction of propagation where $ s= 1$ indicates right-moving wave packets and $s = -1$ indicates left-moving wave packets.  Now the states for wave packets travelling in opposite directions remain orthogonal at all times.  Although in this thesis we are specifically concerned with photons, in principle the above argument, which justifies the introduction of an additional parameter s, would also apply to localised wave packets of any other particle type.  The problem raised by the argument above, however, is much more obvious in the case of photons.  Since photons are massless, they must always propagate at the speed of light.  It is not possible, therefore, by means of a Lorentz transformation, to slow down and reverse the direction of a photon in one-dimension.  On the other hand, for massive particles, e.g, the electron, a change of reference frame can always be made that slows the electron to a stop and then sends it in the opposite direction.  In this way a connection can be established between the right-propagating and left-propagating states.  For photons there is no such connection and right-propagating and left-propagating wave packets must therefore inhabit distinct regions of the Hilbert space. 

One should also notice that photons with a well-defined direction of propagation travel at the speed of light.  It follows, therefore, that
\begin{equation}
\label{example5}
\psi_{s\lambda}(x,t) = \psi_{s\lambda}(x-sct,0).
\end{equation}
By expressing the above photon wave function as a Fourier transform of the momentum-dependent wave function $\widetilde{\psi}_{s\lambda}(k,t)$ one can show that
\begin{equation}
\label{example6}
\int_{-\infty}^{\infty}\text{d}k\;\widetilde{\psi}_{s\lambda}(k,t)  \, e^{ikx} = \int_{-\infty}^{\infty}\text{d}k\; \widetilde{\psi}_{s\lambda}(k,0)\,e^{ik(x-sct)}.
\end{equation} 
Since this relation holds for wave packets of any shape it is implied that
\begin{equation}
\label{example7}
\widetilde{\psi}_{s\lambda}(k,t) = \widetilde{\psi}_{s\lambda}(k,0)\,e^{iskct}
\end{equation}
for all $s$, $k$ and $\lambda$.  As we have seen in Section \ref{Sec:Bennetquantisation}, these dynamics can be generated by the Schr\"odinger equation for a collection of uncoupled harmonic oscillators, each oscillator characterised by a set of parameters $k$, $\lambda$ and $s$.  

We can see from Eq.~(\ref{example7}) that, as was also observed in Ref.~\cite{Jake1}, when the dynamics of the state $\psi_{s\lambda}(x,t)$ evolves according to the Schr\"odinger equation defined in Eq.~(\ref{Schrodingerequation}), the corresponding Hamiltonian must have eigenvalues $skc$.  We also know from Eq.~(\ref{example6}) that the wave number $k$ must take all real values in order to localise the initial wave packet illustrated in Fig.~\ref{Fig:photons1}.  In addition, $k$ must be independent of $s$ to ensure that the direction of propagation can be chosen freely for any wave packet.  The eigenstates of this Hamiltonian, therefore, are not bounded either from above or below, and they must take all real values.  In standard quantum mechanics, the Hamiltonian appearing in the Schr\"odinger equation (\ref{Schrodingerequation}) is given by the energy observable of the system being described.  It is strictly necessary that the eigenvalues of the energy observable have a lower bound in order to ensure that the system cannot be used as a source of infinite energy.  This can only be guaranteed when the direction of propagation $s$ and the orientation of $k$ coincide.  This would mean that the two are no longer independent.

The argument presented above demonstrates that there is a contradiction between the current theory of the quantised EM field (see for example Refs.~\cite{Hei, Coh, Lou}) and a theory that would allow us to construct quantised descriptions of all possible solutions to Maxwell's equations.  As we have seen in Section \ref{Sec:1Dclassicalfields}, these include localised wave packets that propagate at the speed of light without dispersion.  More precisely, the propagation of localised photon wave packets in a given direction, like those depicted in Fig.~\ref{Fig:photons1}, cannot be described by the unitary evolution of photons characterised by a wave number $k$, a polarisation $\lambda$ and a positive frequency $\omega$, as is the case in the current theory.  Hence we find that a more complete description of the quantised EM field is required that considers states that evolve with negative frequencies, in addition to the usual positive-frequency states.  Within this new description, the dynamical Hamiltonian of the system, which always has both positive and negative eigenvalues, can no longer coincide with the energy observable of the EM field, which must always be positive.

From the discussion above we see that when a source emits a localised wave packet it must emit a series of monochromatic waves oscillating with both positive and negative frequencies, which are later absorbed by the receiver.  This model bears a resemblance to Cramer's transactional interpretation of quantum mechanics \cite{Cra}.  In Cramer's description, the emitter of a particle emits positive-frequency waves forwards in time towards the absorber, but also negative-frequency excitations backwards in time.  The total process is time symmetric.  The absorber of the particle also emits positive-frequency waves forwards in time and negative-frequency waves backwards in time towards the emitter, again in an overall time-symmetric process.  These waves interfere leaving only a single particle propagating from the emitter towards the absorber.  In this thesis, both positive- and negative-frequency photons are required to generate the correct interference to produce a localised, non-dispersive wave packet with a fixed direction of propagation.  Although here we do not think of the negative-frequency modes as propagating backwards in time from the emitter, this interpretation is adopted in, for instance, Ref.~\cite{Haw6}.

In the remainder of this chapter these considerations are taken into account as we investigate an alternative approach to quantising the free one-dimensional EM field.  Our starting point for this approach will be the assumption that, like the classical solutions of Maxwell's equations given in Eqs.~(\ref{Efield1}) and (\ref{Bfield1}), single-photon states can be localised and travel at the speed of light without dispersion.  Like Mandel \cite{Man1, MW} and Cook \cite{Ck1, Ck2}, in this chapter and the next we shall assume that particles are localised when they are orthogonal to one another under the usual inner product of quantum mechanics.  This assumption allows us to avoid the need for biorthogonal quantum physics \cite{Jake2}.  Insisting that localised states are orthogonal to one another also implies that the annihilation and creation operators obey bosonic commutation relations.  This is in good agreement with linear optics experiments with ultra-broadband photons, which confirms the bosonic nature of these localised particles \cite{Nasr, Ok, Jav, Tak}.

\section{Quantisation in position space}

In this section we introduce the annihilation and creation operators for a set of locally bosonic particles that are characterised by the additional degree of freedom $s$. We determine the commutation properties of these particles and their fundamental equation of motion in addition to defining an appropriate set of field observables in terms of these operators.  We find in this section that the field observables are non-locally related to the particle states and that the transformation between the two must be implemented by means of a regularisation operator.  We discuss this operator more in Section \ref{Sec:1Dmomentum}.  Towards the end of this section we determine the energy observable of the system and construct the generator for time translations that appears in Schr\"odinger's equation.  As mentioned in Section \ref{Sec:1Dintroduction}, we find that the two are not the same. 

\label{Sec:1Dposition}

\subsection{A new Hilbert Space}

\label{Sec:bliphilbertspace}

\subsubsection{Blip states}

In the past we have assumed that the basic building blocks of the EM field are the monochromatic photon states.  Let us now take a different approach and assume that the fundamental excitations are a set of spatially localised particle states that propagate along the $x$ axis of a Cartesian coordinate system.  We shall call these localised excitations blips, which is an abbreviation of the name ``bosons localised in position."  At any given time $t$, a blip state can be fully characterised by its position $x$ along the $x$ axis, a polarisation $\lambda$ and a direction of propagation $s$.  Here $s$ takes values $\pm 1$ with $s = +1$ indicating propagation in the direction of increasing $x$ and $s = -1$ indicating propagation in the direction of decreasing $x$.  As is usual in one dimension, $\lambda = \mathsf{H},\mathsf{V}$.

\subsubsection{Annihilation and creation operators}

As in Chapter \ref{Chap:quantum}, we may define a set of annihilation operators for the blip states.  We shall denote these operators $a_{s\lambda} (x,t)$ in the Heisenberg picture and $a_{s\lambda} (x,0)$ in the Schr\"odinger picture.  To identify a Hilbert space, we proceed as before and first define a vacuum state $\ket{0}$ for this system.  The vacuum state is the lowest energy state and is annihilated by all annihilation operators $a_{s\lambda}(x,t)$ for any $x$, $t$, $s$ and $\lambda$: 
\begin{eqnarray} 
\label{annihilate1}
	a_{s\lambda}(x,t) \, \ket{0} &=& 0.
\end{eqnarray}
The vacuum state is normalised such that $\braket{0|0} = 1$.

The Hermitian conjugate of $a_{s\lambda}(x,t)$, $a^\dagger_{s\lambda}(x,t)$, is the creation operator that generates a single blip excitation when applied to the vacuum state:
\begin{eqnarray}
\label{blipstate1}
\ket{1_{s\lambda}(x,t)} = a^\dagger_{s\lambda}(x,t)\ket{0}.
\end{eqnarray}
These states span the single-particle Hilbert space.  The total Hilbert space is a symmetric Fock space of such Hilbert spaces.  Accordingly, by applying blip creation operators to the vacuum state repeatedly we are able to generate a complete set of multi-particle states.  In general, a state containing $n$ identical blips localised at a position $x$, propagating in the $s$ direction and carrying a polarisation $\lambda$ is denoted
\begin{eqnarray}
	\label{blipstate2}
\ket{n_{s\lambda}(x,t)} = \frac{1}{\sqrt{n!}}\,a^\dagger_{s\lambda}(x,t)^n\ket{0}.
\end{eqnarray}

\subsubsection{A fundamental equation of motion} 

In Chapter \ref{Chap:quantum}, the dynamics of the quantised field observables were calculated using Heisenberg's equation of motion.  This was a very straightforward procedure because the monochromatic photon states are the eigenstates of the energy observable that generates the dynamics of the system.  The blip states, on the other hand, have a well defined position in space and time, and Heisenberg's uncertainty relation would tell us that their momenta and energies are completely unknown.  Consequently, at this stage we cannot carry out a similar set of calculations to determine the time dependence of the blip states.  Fortunately, we may determine their dynamics by another method.

Blip states represent the localised excitations of the EM field that propagate at the speed of light.  This assumption places a constraint on the expectation values of the EM fields at different times to ensure propagation at a constant speed, which is given by $\braket{a_{s\lambda}(x,t)} = \braket{a_{s\lambda}(x-sct,0)}$.  Since this relation holds for any time-independent state we can deduce the following relation:
\begin{eqnarray}
\label{blipconstraint1}
a_{s\lambda}(x,t) = a_{s\lambda}(x-sct,0).
\end{eqnarray} 
This equality asserts that, when allowed to propagate freely, a blip state placed at a position $x$ at a time $t=0$ will be found at a position $x+sct$ at the later time $t$.  Rather than invoking Heisenberg's equation, we are able to determine the equation of motion for a blip state using the above condition. By taking the time derivative of the blip operator in Eq.~(\ref{blipconstraint1}) one may show that
\begin{eqnarray}
\label{blipmotion1}
\frac{\text{d}}{\text{d}t}a_{s\lambda}(x,t) = -sc\frac{\text{d}}{\text{d}x}a_{s\lambda}(x,t).
\end{eqnarray}
This is the fundamental equation of motion for the blip operators.  This equation tells us that the blip operators are constant along the light-cones defined by the coordinates $x = sct$, which is in agreement with the expected behaviour illustrated in Fig.~\ref{Fig:Lightcone1}.

\subsection{The blip commutation relations}

\subsubsection{Bosonic commutation relations}

Blips are bosons, and the states that represent them, therefore, are unchanged when an exchange of blips takes place.  Hence, we assume in the following that
\begin{eqnarray}
	\label{blipcommutator1}
\Big[a_{s\lambda}(x,t),a_{s'\lambda'}(x',t')\Big] = \Big[a^\dagger_{s\lambda}(x,t), a^\dagger_{s'\lambda'}(x',t')\Big] = 0.
\end{eqnarray}
Using the definition of a blip state in Eq.~(\ref{blipstate1}), it is possible to show that the commutation relation between a blip creation and annihilation operator is identical to the inner product between two single-blip states:
\begin{eqnarray}
\braket{1_{s\lambda}(x,t)|1_{s'\lambda'}(x',t')} &=& \bra{0}a_{s\lambda}(x,t) a^\dagger_{s'\lambda'}(x',t')\ket{0}\nonumber\\
&=& \Big[a_{s\lambda}(x,t), a^\dagger_{s'\lambda'}(x',t')\Big].
\end{eqnarray}

\subsubsection{Orthogonal states}

Single-blip states defined at different positions at a fixed time are orthogonal to one another.  This is a necessary condition for a correct interpretation of a quantum state as it means a localised state found at one position has no chance of being found elsewhere.  In the following we shall assume that such states are orthogonal under the usual inner product of quantum mechanics in order to avoid the unnecessary complexities of biorthogonal physics \cite{Jake2}.  Polarisation, we know, is a measurable quantity, but, as was shown before, to effectively model the dynamics of a state we must also consider $s$ as an additional parameter.  Blip states characterised by different $s$ and $\lambda$ must therefore be orthogonal.  Hence, in the following we demand that 
\begin{eqnarray} 
\label{blipproduct1}
&& \hspace*{-1cm} \langle 1_{s\lambda} (x,t) |1_{s'\lambda'} (x',t') \rangle \nonumber \\
&=& \langle 1_{s\lambda} (x-sct,0) |1_{s'\lambda'} (x'-s'ct',0) \rangle \nonumber \\
&=&\delta_s\big((x-sct)-(x'-s'ct')\big) \, \delta_{s,s'} \, \delta_{\lambda,\lambda'} 
\end{eqnarray}
where the $\delta_s(x-x')$ is given by
\begin{eqnarray} 
\label{commutator1}
	\delta_s(x-x') &=& \frac{1}{2\pi}\int_{-\infty}^{\infty}\text{d}k \, {\rm e}^{{\rm i}sk(x-x')}.
\end{eqnarray}
Although not strictly necessary, in the above we choose to add a factor of $s$ into the exponent so that a change in the direction of propagation $(s \mapsto -s)$ leads to the reversal of the exponent, as was the case for the Fourier transforms in Section \ref{Sec:Bennetquantisation}. 

The above inner product is a good choice because it is strictly positive, translation-invariant, symmetric with respect to the position of the blips and real valued.  There is also a unit probability of finding the blip within $(-\infty,\infty)$.  Furthermore, $\delta_{s}(x-x')$ depends on $x$ and $x'$, but not on $s$ or $\lambda$, as one would expect.  The reason for the additional $s$ label in Eq.~(\ref{commutator1}) is to ensure that the $s=1$ and the $s=-1$ cases do not become formally the same.  Hence, we find that at equal times
\begin{eqnarray} 
	\label{blipcommutator2}
	\big[ a_{s\lambda} (x), a^\dagger_{s'\lambda'} (x') \big] &=& \delta_s(x-x')  \, \delta_{s,s'} \, \delta_{\lambda,\lambda'} \, , ~~~
\end{eqnarray}
which is the expected bosonic commutation relation.

\subsection{Field observables in the position representation}

\label{Sec:1Dposfields}

\subsubsection{Field observables}

In Section \ref{Sec:bliphilbertspace}, we constructed a new Hilbert space spanned by the blip number states defined in Eq.~(\ref{blipstate2}). Next we shall obtain a set of expressions for the (complex) field observables ${\bf E}(x,t)$ and ${\bf B}(x,t)$, and the energy observable $H_{\rm energy}(t)$ in the position representation.  As was shown in Section \ref{Sec:1Dclassicalfields}, the classical solutions of Maxwell's equations in one dimension obey the blip equation of motion (\ref{blipmotion1}).  Consequently, like the blips, the solutions of Maxwell's equations in a vacuum are wave packets which travel at the speed of light along the $x$ axis.  Hence,  in the following we postulate that the observables of the complex vectors ${\bf E}(x,t)$ and ${\bf B}(x,t)$ are given by
\begin{eqnarray} 
\label{fieldobservables1}
\textbf{E}(x,t) &=& \sum_{s= \pm 1} c \, \left\{\mathcal{R} \left[a_{s\mathsf{H}}\right](x,t)\,\widehat{\boldsymbol{y}} + \mathcal{R}\left[a_{s\mathsf{V}}\right](x,t)\,\widehat{\boldsymbol{z}}\right\} , \nonumber \\
\textbf{B}(x,t) &=& \sum_{s= \pm 1} s \, \left\{\mathcal{R} \left[a_{s\mathsf{H}}\right](x,t)\,\widehat{\boldsymbol{z}} - \mathcal{R}\left[a_{s\mathsf{V}}\right](x,t)\,\widehat{\boldsymbol{y}}\right\}.
\end{eqnarray}
The role of $\mathcal{R}$ will be discussed below and is included here to ensure that the Lorentz covariance of the field observables can be preserved.  The above operators are non-Hermitian, and their expectation values are complex by construction.  Here the actual field observables are given by the real combination $(1/2)\left(\mathbf{O}+\mathbf{O}^\dagger\right)$ where $\mathbf{O} = \mathbf{E}, \mathbf{B}$.

\subsubsection{The regularisation operator} 

In Eq.~(\ref{fieldobservables1}), $\mathcal{R}$ is a symmetric and translation-invariant superoperator that we shall refer to as the regularisation operator.  We shall determine the explicit form of $\mathcal{R}$ later in this section. $\mathcal{R}$ can be understood as a distribution that acts on the annihilation operator $a_{s\lambda}(x,t)$ in such a way that
\begin{eqnarray}
\label{Rsuperposition1}
\mathcal{R}\Big[a_{s\lambda}\Big](x,t) = \int_{-\infty}^{\infty}\text{d}x'\;\mathcal{R}_{s\lambda}(x-x')\,a_{s\lambda}(x',t).
\end{eqnarray}
Similarly, $\mathcal{R}\Big[a^\dagger_{s\lambda}\Big](x,t)$ is defined such that
\begin{eqnarray}
\label{Rsuperposition2}
\mathcal{R}\left[a^\dagger_{s\lambda}\right](x,t) = \int_{-\infty}^{\infty}\text{d}x'\;\mathcal{R}^*_{s\lambda}(x-x')\,a^\dagger_{s\lambda}(x',t).
\end{eqnarray}
The purpose of this distribution is to relate the measurable field observables to a local particle in a possibly non-local way.  Without this regularisation it would not be possible later on to assign an energy $\hbar \omega$ to each photon.  In this way, we distinguish between localised blip excitations and the excitations of the field observables.  This is in contrast to Refs.~\cite{BB4, Sip, BB6, Jake2, Haw3}.  As the $\mathcal{R}$ superoperator is translation invariant, both the field observables and the blip states that propagate in a fixed direction satisfy the equation of motion given in Eq.~(\ref{blipmotion1}).  It is then possible to show that the observables (\ref{fieldobservables1}), and their expectation values, obey the free-space Maxwell equations.  Symmetry of $\mathcal{R}(x-x')$ is assumed because, although their direction of propagation will be reversed, a parity transformation will only displace a blip and not change its shape.  It is therefore important that the electric and magnetic fields associated with a single-blip state also only undergo a change of direction and a translation under a parity transformation.  This is ensured by the symmetry of $\mathcal{R}(x-x')$.

\subsubsection{Field commutation relations}

Using the expressions for the field observables in Eq.~(\ref{fieldobservables1}) it is possible to show that all components of the field observables commute with one another:
\begin{equation}
\label{fieldcommutator1}
\left[O_i(x,t), P_j(x',t)\right] = 0,
\end{equation}
where $O_i(x,t)$ and $P_j(x,t)$ are any two components of the electric and magnetic field vectors.  When the direction of propagation is known, the electric and magnetic fields differ by only a rotation and a constant $c$.  It follows therefore that whenever the electric field is known, so is the magnetic field.  Many quantisation schemes produce field observables that do not commute \cite{Hei, Bohr}.  By introducing the parameter $s$, in our scheme the field observables are guaranteed to commute.  Notice also that this means that single-photon states can now be localised in the sense of Knight \cite{Kni} and Licht \cite{Lic1, Lic2}, \cite{BB6}.

\subsubsection{The energy observable}

To determine the energy observable in terms of the blip creation and annihilation operators we proceed as in Section \ref{Sec:Bennetquantisation} and substitute the field observables (\ref{fieldobservables1}) into the expression for the classical energy (\ref{classicalenergy4}).  The resulting expression is
\begin{eqnarray} 
	\label{Heng1}
	&& \hspace*{-0.7cm} H_{\text{energy}}(t) \nonumber \\
	&=& \sum_{s= \pm 1} \sum_{\lambda = {\sf H},{\sf V}} {\varepsilon_0 A c^2 \over 4} \int_{-\infty}^{\infty}\text{d}x \, \big\{ \,\mathcal{R}\left[a_{s\lambda}\right](x,t) + {\rm H.c.} \, \big\}^2 . ~~~~~
\end{eqnarray}
Thus $\mathcal{R}$ determines the energy of a single-blip state.  Most importantly, notice that because of the quadratic form of this observable, energy expectation values are always positive, just as they are in classical electrodynamics.

\subsection{The dynamical Hamiltonian}

\label{Sec:dynHamiltonian1}

In this final subsection we would like to show that the equation of motion for a blip operator, given in Eq.~(\ref{blipmotion1}), can be written as a Schr{\"o}dinger equation.  In particular we would like to show that the field observables ${\bf O}(x,t)$ evolve in the Heisenberg picture according to Heisenberg's equation of motion,
\begin{eqnarray}
	\label{Heisenberg's equation}
	{{\rm d} \over {\rm d}t} \, {\bf O}(x,t) &=& -\frac{\rm i}{\hbar}\left[ {\bf O}(x,t), H_{\rm dyn}(t) \right],
\end{eqnarray}
for some dynamical Hamiltonian $H_{\text{dyn}}$.  Heisenberg's equation of motion is the most important tool we have for determining the dynamics of more complex interacting systems, and it is important therefore to establish a Hamiltonian formalism for the free field dynamics so that interactions can be added if needed.

In the following we shall deduce the dynamical Hamiltonian $H_{\rm dyn}(t)$ using the blip equation of motion (\ref{blipmotion1}).  Consider initially Heisenberg's equation of motion for the operator $a_{s\lambda} (x,t)$. In this case, Eq.~(\ref{Heisenberg's equation}) leads us to the relation
\begin{eqnarray} 
\label{blipmotion2}
{{\rm d} \over {\rm d}t} \, a_{s \lambda}(x,t) &=& - {{\rm i} \over \hbar} \left[ a_{s\lambda}(x,t), H_{\rm dyn}(t) \right] \, .
\end{eqnarray}
What is interesting about the blip annihilation operators is that their equation of motion is already known to us from Eq.~(\ref{blipmotion1}).  We may therefore replace the time derivative on the left-hand side of this equation with a space derivative.  Heisenberg's equation is rewritten
\begin{eqnarray}
\label{blipmotion3}
\frac{\text{d}}{\text{d}x} \, a_{s\lambda}(x,t) &=& {{\rm i}s \over \hbar c} \left[a_{s\lambda}(x,t), H_{\text{dyn}}(t) \right] \, .
\end{eqnarray}
The above equation of motion suggests that the dynamical Hamiltonian affects the position, but not the time coordinate, of $a_{s\lambda}(x,t)$. This is not surprising: the purpose of $H_{\rm dyn}(t)$ is to propagate wave packets at the speed of light along the $x$ axis. As the generator of such dynamics, the Hamiltonian must continuously annihilate blips while simultaneously replacing them with excitations of equal amplitudes at nearby positions.

Taking this into account, we may construct an exchange Hamiltonian for blips at different locations.  A Hamiltonian of this type is given by
\begin{eqnarray} 
\label{Hdyn1}
H_{\text{dyn}}(t) &=& \sum_{s=\pm 1} \sum_{\lambda = {\sf H}, {\sf V}} \int_{-\infty}^{\infty}\text{d}x' \int_{-\infty}^{\infty}\text{d}x'' \, \hbar sc \, f_{s \lambda}(x'',x') \nonumber \\
&& \hspace*{3cm}\times\, a^\dagger_{s\lambda}(x'',t)\, a_{s\lambda}(x',t) \, ,
\end{eqnarray}
where $f_{s\lambda}(x'',x')$ is a complex term left to be determined.  By substituting the Hamiltonian (\ref{Hdyn1}) into our modified Heisenberg's equation (\ref{blipmotion3}) one will find that
\begin{eqnarray} 
\label{blipmotion4}
\frac{\text{d}}{\text{d}x} \, a_{s\lambda}(x,t) &=& {\rm i} \int_{-\infty}^{\infty}\text{d}x' \, f_{s \lambda}(x,x') \, a_{s\lambda}(x',t) \, .
\end{eqnarray} 
One is then able to verify that
\begin{eqnarray}
\label{deltaderivative}
f_{s \lambda}(x,x') &=& {1 \over 2\pi} \int_{-\infty}^{\infty}\text{d}k \, sk \, {\rm e}^{{\rm i}sk(x-x')} \nonumber \\
&=& - {\rm i} \, {{\rm d} \over {\rm d}x} \left[ {1 \over 2\pi} \int_{-\infty}^{\infty}\text{d}k \, {\rm e}^{{\rm i}sk(x-x')} \right] \nonumber \\
&=& - {\rm i} \, \delta_s'(x-x') 
\end{eqnarray}
where $\delta_{s}'(x-x')$ denotes the derivative of the delta function $\delta_s(x-x')$ with respect to $x$.  Overall, the dynamical Hamiltonian in Eq.~(\ref{Hdyn1}) equals
\begin{eqnarray}
\label{Hdyn2}
H_{\text{dyn}}(t) &=& \sum_{s=\pm 1} \sum_{\lambda = {\sf H}, {\sf V}} \int_{-\infty}^{\infty}\text{d}x' \int_{-\infty}^{\infty}\text{d}x''  \int_{-\infty}^\infty {\rm d} k \, {\hbar c k \over 2 \pi} \nonumber \\
&& \times\, {\rm e}^{{\rm i}sk (x''-x')} \, a^\dagger_{s\lambda}(x'',t) \,a_{s\lambda}(x',t).
\end{eqnarray}
This Hamiltonian is Hermitian and therefore a generator of unitary dynamics.  It also satisfies a number of relevant properties.  For example, $f_{s\lambda}(x-x')$ is antisymmetric under an exchange of $x$ and $x'$.  This means that a state that propagates from $x$ to $x'$ will only propagate from $x'$ to $x$ if either $s$ is reversed or time is reversed.  It also means that one blip cannot be replaced by another at the same position or, in other words, that stationary photons do not exist.  The function $f_{s\lambda}(x-x')$ is also translation invariant, which means that blips propagate identically at all positions, as would be expected in free space.

Unlike the energy observable, the dynamical Hamiltonian (\ref{Hdyn2}) has both positive and negative eigenvalues as was shown in the introduction to this chapter.  Usually the dynamical Hamiltonian and energy observable are equal; this is no longer the case.  More will be said on this in Section \ref{Sec:photonsdiscussion}, but at this moment it is necessary for us only to check that the energy of the system is conserved.  According to Heisenberg's equation, the observable for a conserved quantity commutes with the dynamical Hamiltonian.  When the dynamical Hamiltonian and energy observable coincide, this property is ensured automatically because all observables commute with themselves.  We must now check that  
\begin{eqnarray}
\label{energyconservation}
\left[ H_{\rm energy}(t), H_{\rm dyn}(t) \right] &=& 0 . ~~~~
\end{eqnarray} 
Owing to the symmetry of the distribution $\mathcal{R}(x-x')$ and the antisymmetry of $f_{s\lambda}(x,x')$, one can show that the commutator indeed vanishes.

In the next section, where we look at this new quantisation in the momentum representation, we will see that the dynamical Hamiltonian (\ref{Hdyn2}) takes a simpler and more recognisable form.  Moreover we shall see that when restricted to a subset of parameters, the dynamical Hamiltonian (\ref{Hdyn2}) and the energy observable (\ref{Heng1}) coincide both with each other and the standard energy observable (\ref{energyobs1}).

\section{Quantisation in momentum space}

\label{Sec:1Dmomentum}

In classical electrodynamics, the position and momentum representations of the EM field complement each other well, and we may use either representation interchangeably for our convenience.  For example, we often describe light scattering experiments using Maxwell's equations, which involve only local field amplitudes.  In such situations, it might be best to use the position representation.  In other situations, classical electrodynamics introduces optical Green's functions and decomposes the EM field into monochromatic waves to predict general optical properties \cite{RB,Phil}.  Here a momentum space description might be more convenient.

The purpose of this section is to construct a momentum-dependent representation of the free quantised EM field described in Section \ref{Sec:1Dposition}.  We shall approach this problem analogously by first defining the set of monochromatic states in the blip Hilbert space, and then afterwards construct the field and energy observables and determine a dynamical Hamiltonian.  In addition to providing us with a more complete formulation of the quantised EM field, by investigating the momentum representation, in Section \ref{Sec:comparison1} we shall be able to examine more closely the relationship between standard quantum optics descriptions of the quantised EM field \cite{RB2} and that given in Section \ref{Sec:1Dposition}.

\subsection{A photon Hilbert space}

\label{Sec:photonHilbertspace}

\subsubsection{Annihilation and creation operators}

From classical electrodynamics we know that the set of travelling waves with wave numbers $k \in (-\infty, \infty)$ and two different polarisations $\lambda \in \{\mathsf{H},\mathsf{V}\}$ provide a complete set of solutions to Maxwell's equations in one dimension.  Therefore, in the momentum representation, we usually express the quantised EM field in terms of the single-photon annihilation operators $a_{k\lambda}(t)$ where $k$ and $\lambda$ refer to the wave number and polarisation of the photon respectively.  By applying a Fourier transform to the blip operators $a_{s\lambda}(x,t)$, defined in Section \ref{Sec:bliphilbertspace}, however, we obtain the operators $\widetilde{a}_{s\lambda}(k,t)$.  In the following we assume that the operators
\begin{equation}
\label{1Dphoton1}
\widetilde{a}_{s\lambda}(k,t) =  \int_{-\infty}^{\infty}\frac{\text{d}x}{\sqrt{2\pi}}\;e^{-iskx}\,a_{s\lambda}(x,t)
\end{equation}
represent the annihilation operators of monochromatic photons.  The inverse transformation is given by
\begin{equation}
\label{1Dphoton2}
a_{s\lambda}(x,t) = \int_{-\infty}^{\infty}\frac{\text{d}k}{\sqrt{2\pi}}\;e^{iskx}\,\widetilde{a}_{s\lambda}(k,t).
\end{equation} 
In the above the operators $\widetilde{a}_{s\lambda}(k,t)$ are characterised by $k \in (-\infty,\infty)$, $\lambda \in \{\mathsf{H},\mathsf{V}\}$, but now also, like the blip states, an additional parameter $s \in \{+1, -1\}$.  As pointed out in Section \ref{Sec:1Dposition}, the EM field introduced in this thesis requires a Hilbert space that is doubled when compared to the standard description of the quantised EM field.

Usually for light propagating along a single axis, the wave number $k$ is the $x$ component of the wave vector oriented in the direction of propagation.  In Eqs.~(\ref{1Dphoton1}) and (\ref{1Dphoton2}) we adopt the alternative convention that the wave number is given by the factor $ks$.  As before, this means that a change in the direction of propagation, that is, replacing $s$ with $-s$, results in a change in the sign of the wave number.  The parameter $k$, however, then loses its traditional interpretation as being oriented in the direction of propagation.  Nevertheless, to ensure that Eq.~(\ref{1Dphoton1}) is invertible, $k$ must take all values along the real line \cite{How}.  Another consequence of this change is that left- and right-moving wave packets have different Fourier decompositions in momentum space, even if the wave packets are the same shape.  This later enables us to predict different dynamics for left- and right-moving wave packets without having to introduce more than one equation of motion.

In the following we refer to the states obtained by applying the operator $\widetilde{a}^\dagger_{s\lambda}(k,t)$ to the vacuum state as photon states.  The single-photon Hilbert space is spanned by the states $a^\dagger_{s\lambda}(k,t)\ket{0}$ for all $s$, $k$ and $\lambda$.  These photon states are not exactly the same as those defined in standard theories however.  For example, the operators $\widetilde{a}_{s\lambda}(k,t)$ and $\widetilde{a}_{-s\lambda}(-k,t)$ both describe photons whose coherent states have exactly the same electric field expectation values.  The two operators evolve with different dynamics, however, and whereas the $\widetilde{a}_{s\lambda}(k,t)$ operators describes photons propagating in the $s$ direction, the $\widetilde{a}_{-s\lambda}(-k,t)$ operators describe photons propagating in the $-s$ direction.  Although the standard $a_{k\lambda}$ operators can be used to construct wave packets with electric field expectation values of any shape, they cannot be used to construct localised wave packets of an arbitrary shape that propagate at the speed of light in a fixed direction.

\subsubsection{Photon states}

To construct the relevant Hilbert space of the quantised EM field, we start again by defining a vacuum state $\ket{0}$. Applying any annihilation operator to the vacuum state destroys the state:  
\begin{eqnarray}
\label{annihilate2}
\widetilde{a}_{s \lambda}(k,t) \, \ket{0} &=& 0 \, .
\end{eqnarray}
By applying both the left- and right-hand sides of Eq.~(\ref{1Dphoton2}) to the vacuum state, one can see that the vacuum states in the position and momentum representations coincide.

By applying a creation operator $\widetilde{a}^\dagger_{s\lambda}(k,t)$ to the vacuum state an arbitrary number of times we may generate the full symmetric Fock space of states containing any number of photons.  A state containing $n$ identical photons is denoted
\begin{eqnarray} 
\label{Photonstate1}
\ket{n_{s\lambda} (k,t)} &=& \frac{1}{\sqrt{n!}}\,{\widetilde{a}^{\dagger}_{s\lambda}(k, t)}^n \ket{0}.
\end{eqnarray}
When $n=1$ we shall refer to these states as single-photon states.

\subsection{The photon commutation relations}

\label{Sec:momcomms1}

\subsubsection{Commutation relations}

In this section we look at the commutation relations for the photon creation and annihilation operators $\widetilde{a}^\dagger_{s\lambda}(k,t)$ and $\widetilde{a}_{s\lambda}(k,t)$ defined in the previous section. As photons are bosons, many-photon states are unchanged when two or more photons are exchanged.  This leads us to the typical commutation relation for bosonic particles:
\begin{eqnarray}
\label{photoncommutator1}
\Big[ \widetilde{a}_{s\lambda}(k,t), \widetilde{a}_{s'\lambda'}(k',t')\Big] = \Big[ \widetilde{a}^\dagger_{s\lambda}(k,t), \widetilde{a}^\dagger_{s'\lambda'}(k',t')\Big] =0 \,  \nonumber \\ 
\end{eqnarray}
for any $s$, $s'$, $\lambda$, $\lambda'$, $k$, $k'$, $t$ and $t'$.  These relations are consistent with Eq.~(\ref{blipcommutator1}).  In order to determine the commutation relation between photon creation and annihilation operators we use the Fourier transformations of the photon operators given in Eq.~(\ref{1Dphoton1}).  Using the blip commutation relation given in Eq.~(\ref{blipcommutator2}) we find that 
\begin{eqnarray}
\label{photoncommutator2}
\big[\widetilde{a}_{s\lambda}(k,t), \widetilde{a}^\dagger_{s'\lambda'}(k',t)\big] = \delta(k-k') \, \delta_{s,s'} \, \delta_{\lambda,\lambda'} \, .
\end{eqnarray}
Using this commutator one may also show that single-photon states are orthogonal to one another as
\begin{eqnarray} 
\label{photoncommutator3}
\hspace*{-1.2cm} \braket{1_{s\lambda}(k,t)|1_{s'\lambda'}(k',t)} &=&\braket{0|\big[ \widetilde{a}_{s\lambda} (k,t), \widetilde{a}^\dagger_{s'\lambda'} (k',t) \big]|0} \nonumber \\
&=& \delta(k-k') \, \delta_{s,s'} \, \delta_{\lambda,\lambda'} \, .
\end{eqnarray}
As one would expect, the photons that we consider in this section are bosonic particles.

\subsubsection{The fundamental equation of motion}

By using the Fourier expressions of the blip operators we may also determine a fundamental equation of motion for the photon operators.  By substituting the right-hand side of Eq.~(\ref{1Dphoton2}) into Eq.~(\ref{blipmotion1}) one finds that the photon operators satisfy the equation
\begin{equation}
\label{photonmotion1}
\frac{\text{d}}{\text{d} t} \widetilde{a}_{s\lambda}(k,t)  = - ikc\, \widetilde{a}_{s\lambda}(k,t).
\end{equation}
The solution of this equation is
\begin{equation}
\label{photonmotion2}
\widetilde{a}_{s\lambda}(k,t) = e^{-ikct}\,\widetilde{a}_{s\lambda}(k,0).
\end{equation}
This equation is the usual equation of motion of the quantised EM field in momentum space and shows that photons oscillate with an angular frequency $\omega = kc$.  Since $k$ now takes all values along the real line, $\omega$ can now take all real values.  As we have seen already and shall see again, however, this is not a problem because the energy observable is always positive.  This is possible because the energy observable and the generator of the system's dynamics are no longer the same.

\subsection{Field observables in the momentum representation}

\subsubsection{The field observables}

\label{Sec:1Dmomfields}

In this section we consider how to express the complex electric and magnetic field observables in their momentum representations.  As in Section \ref{Sec:clasicalEM}, we denote these fields $\widetilde{\mathbf{E}}(k,t)$ and $\widetilde{\mathbf{B}}(k,t)$.  These fields are three-vector valued and parametrised by a time $t$ and a wave number $k$.  As we have seen in Section \ref{Sec:1Dposfields}, the electric and magnetic field vectors are a linear superposition of travelling wave solutions with a fixed direction of propagation and polarisation.  As the Fourier transform is linear, the same applies in the momentum representation.  Thus
\begin{equation}
\widetilde{\mathbf{O}}(k,t) = \sum_{s = \pm 1}\sum_{\lambda = {\mathsf{H},\mathsf{V}}}\, \widetilde{\mathbf{O}}_{s\lambda}(k,t) 
\end{equation}
where $\widetilde{\mathbf{O}} = \widetilde{\mathbf{E}}, \widetilde{\mathbf{B}}$.
As in Eqs.~(\ref{1Dphoton1}) and (\ref{1Dphoton2}), we include a factor of $s$ in the definition of the wave vector.  Hence
\begin{eqnarray}
\label{fieldtransforms}
\mathbf{O}_{s\lambda}(x,t) = \int_{-\infty}^{\infty}\frac{\text{d}k}{\sqrt{2\pi}}\;e^{iskx}\,\widetilde{\mathbf{O}}_{s\lambda}(k,t)\nonumber\\
\widetilde{\mathbf{O}}_{s\lambda}(k,t) = \int_{-\infty}^{\infty}\frac{\text{d}x}{\sqrt{2\pi}}\;e^{-iskx}\,\mathbf{O}_{s\lambda}(x,t).
\end{eqnarray}
Like the photon operators themselves, the field observables in the momentum representation satisfy the equation
\begin{equation}
\label{photonmotion3}
\frac{\text{d}}{\text{d}t}\widetilde{\mathbf{O}}_{s\lambda}(k,t) = -ikc\,\widetilde{\mathbf{O}}_{s\lambda}(k,t).
\end{equation}

In analogy to standard quantisations of the EM field, in which the system can be described as a collection of simple harmonic oscillators (see Section \ref{Sec:Harmonicoscillator} and Ref.~\cite{Hei}), and in agreement with the transformations given in Eqs.~(\ref{1Dphoton1}) and (\ref{1Dphoton2}), we assume that the complex electric and magnetic field observables can be expressed as a linear combination of the photon annihilation operators:
\begin{eqnarray} 
\label{FTfieldcomponents1}
\widetilde{\mathbf{E}}(k,t) &=& \sum_{s=\pm 1} c\, \Omega(k)\, e^{i\varphi(k)} \left[ \widetilde{a}_{s {\sf H}}(k,t) \, \widehat{\boldsymbol{y}} +  \widetilde{a}_{s {\sf V}}(k,t) \, \widehat{\boldsymbol{z}} \right]  \, , \nonumber\\
\widetilde{\mathbf{B}}(k,t) &=& \sum_{s=\pm 1} s\, \Omega(k) \,e^{i\varphi(k)}\left[ \widetilde{a}_{s {\sf H}}(k,t) \, \widehat{\boldsymbol{z}} - \widetilde{a}_{s {\sf V}}(k,t) \, \widehat{\boldsymbol{y}} \right].
\end{eqnarray}
Here $\Omega(k)$ is a $k$-dependent numerical factor. By introducing the $k$-dependent phase $\varphi(k)$ we may assume that $\Omega(k)$ is real.  By substituting Eq.~(\ref{photonmotion2}) into Eq.~(\ref{FTfieldcomponents1}) one can show that Eq.~(\ref{photonmotion3}) is satisfied.

\subsubsection{The field amplitudes}
\label{Sec:fieldamplitudes}

The equation of motion (\ref{photonmotion3}), however, does not determine the particular value of $\Omega(k)\,e^{i\varphi(k)}$.  The factor $\Omega(k)$ is a function that determines the amplitude of the field observables and, therefore, the energy of a single photon in the $(k,s,\lambda)$ mode.  Due to the isotropy of free space, we can safely assume that $\Omega(k)$ does not have any dependence on either $s$ or $\lambda$.  Furthermore, to ensure the symmetry of the field excitations, $\Omega(k)$ must be an even function.  For the time being we shall not specify $\Omega(k)$ any further, but shall return to this function in Section \ref{Sec:Lorentzcovariance}.

\subsection{Energy in the momentum representation}

\label{Sec:momenergy}

\subsubsection{The energy observable}

\label{Sec:momenergyobservable}

For completeness we now also derive the energy observable of the quantised EM field $H_{\text{energy}}(t)$ in the momentum representation.  Taking as our starting point the expression for the classical field energy in Eq.~(\ref{classicalenergy4}), and substituting in for the classical fields the position-dependent field observables in their Fourier representations, one finds that
\begin{eqnarray}
\label{Heng3}
H_{\text{energy}}(t) &=& {\varepsilon_0 A c^2 \over 4} \sum_{s= \pm 1} \sum_{\lambda = {\sf H},{\sf V}} \int_{-\infty}^{\infty}\text{d}k \, \nonumber \\
&& \hspace*{-1.4cm} \times \big\| \Omega(k)\,e^{i\varphi(k)} \, \widetilde{a}_{s\lambda}(k,t) + \Omega^*(-k)\, e^{-i\varphi(-k)} \, \widetilde{a}^\dagger_{s\lambda}(-k,t) \big\|^2 \, . ~~
\end{eqnarray}
The expectation value of the energy observable above is again positive which is guaranteed by the squared expression in the integrand.  

In Eq.~(\ref{Heng3}), the energy observable has an explicit dependence on the choice of phase $\varphi(k)$.  When we restrict this theory to positive wave numbers only, this dependence disappears, but we cannot make that assumption here.  The absolute phase of a field is not observable, and therefore should not have an explicit dependence in the energy of the field.  Therefore, we must impose the following condition:
\begin{equation} 
\varphi(k) = -\varphi(-k).
\end{equation}
One can see that the phase gained by evolving the system in time is of this form.  By substituting Eq.~(\ref{photonmotion2}) into Eq.~(\ref{Heng3}) one may verify that the energy observable is time-independent.

\subsubsection{Interference effects}  
Unlike the usual Hamiltonian for a system of uncoupled harmonic oscillators, the above energy observable contains additional terms consisting of pairs of annihilation or creation operators only.  These terms have a natural explanation as they are necessary in order to accommodate interference effects that may arise in this new model.  In particular, the real field expectation values associated with photon states occupying the positive-frequency or negative-frequency part of the Fourier spectrum of a blip can interfere with, and even cancel, each other when characterised by the same values of $s$ and $\lambda$.    

To see this consider the operator
\begin{equation}
\label{Displacement2}
D = D_k(\alpha)\,D_{-k}(-\alpha^*).
\end{equation}
Here $D_k(\alpha)$ is the displacement operator defined in Eq.~(\ref{Displacement1}) for states with a single polarisation and direction of propagation.  The subscript $k$ is to remind us that this is a coherent superposition of states with a single frequency $k$ rather than of normalised wave packets.  By again using Eq.~(\ref{Displacement1}) and the expression for the field observables in Eq.~(\ref{FTfieldcomponents1}), one may show that the operator $D$ defined in Eq.~(\ref{Displacement2}) commutes with both the electric and magnetic field observables.  It follows therefore that  
\begin{eqnarray}
\label{Displacement3}
\braket{\alpha_k,-\alpha^*_{-k}|\textbf{O}(x,t)|\alpha_k,-\alpha^*_{-k}}
&=& \braket{0|D^\dagger \mathbf{O}(x,t)D|0}\nonumber\\
&=& \braket{0|D^\dagger D\, \mathbf{O}(x,t)\ket{0}}\nonumber\\
&=&\braket{0|\mathbf{O}(x,t)|0}\nonumber\\
&=& 0
\end{eqnarray}
where $\mathbf{O} = \mathbf{E},\mathbf{B}$.   The fourth line follows from the third by the unitarity of displacement operators.  In the above we see that, whereas a single coherent state $\ket{\alpha_k}$, for example, has a non-zero field expectation value, by introducing to the system the additional state $\ket{-\alpha^*_{-k}}$, the field expectation values will vanish everywhere.

When we consider wave packets containing an integer number of photons characterised by a single polarisation and value of $s$, the positive-frequency and negative-frequency components of the wave packet cannot interfere with each other either totally constructively or totally destructively because their wave fronts propagate in opposite directions.  On the other hand, when, as in Eq.~(\ref{Displacement3}), we calculate the expectation value of the classical field observables with respect to a coherent state, a state containing contributions from the positive part of the Fourier spectrum of a blip will contribute to both the positive-frequency part of the field expectation value and its negative-frequency complex conjugate.  Similarly, a state containing photons belonging to the negative-frequency part of a blip will contribute to both the negative-frequency part of the expectation value and its positive-frequency complex conjugate.

As a result of this, by introducing negative-frequency excitations we are able to construct distinct states that have opposing field expectation values and therefore interfere destructively.  Similarly the state $\ket{\alpha_k, \alpha^*_{-k}}$ has an energy four times that of either coherent state independently, which is due to constructive interference.  Without the additional terms, this would not be correctly represented in the energy expectation value of the state.  This means, however, that in the classical limit of this theory there is a redundancy between the positive- and negative-frequency photons characterised by $\omega > 0$ and $\omega < 0$ respectively.  In both cases it is possible to construct identical field expectation values that evolve identically, but both are needed to construct localised single-photon wave packets that propagate without dispersion.  In future, in order to avoid any problems that may arise due to this redundancy, it may be required that the classical observables are projected onto a subspace of the total Hilbert space containing the usual positive-frequency excitations only.

\subsection{The dynamical Hamiltonian}

\label{Sec:dynHamiltonian2}

Eq.~(\ref{photonmotion1}) shows that the n-photon states $\ket{n_{s\lambda}(k,t)}$ are, up to the accumulation of a time-dependent phase factor, invariant under the dynamical evolution of the EM field.  They are therefore eigenstates of the dynamical Hamiltonian.  Using the photon commutation relations in Eq.~(\ref{photoncommutator2}) it can be shown that
\begin{eqnarray}
\label{Hdyn3}
H_{\text{dyn}} (t) &=& \sum_{s= \pm 1} \sum_{\lambda = {\sf H},{\sf V}} \int_{-\infty}^{\infty}\text{d}k \, \hbar ck \, \widetilde{a}^\dagger_{s\lambda}(k,t) \widetilde{a}_{s\lambda}(k,t) \, .
\end{eqnarray}
Using the transformations given in Eqs.~(\ref{1Dphoton1}) and (\ref{1Dphoton2}), the above Hamiltonian can be shown to be identical to that given in Eq.~(\ref{Hdyn2}).

In Section \ref{Sec:dynHamiltonian1} we pointed out that the dynamical Hamiltonian for blip states had both positive and negative eigenvalues.  This ensured that the dynamics of localised light pulses was reversible: light moving to the left is indistinguishable from light moving to the right when time is reversed.  Here we have found that, if our system contains photons of negative $k$, then the dynamical Hamiltonian in this representation also possesses positive and negative eigenvalues.  In the momentum representation, the dynamical Hamiltonian takes a much simpler form than the equivalent expression in the position representation (cf. Eqs.~(\ref{Hdyn2}) and (\ref{Hdyn3})).

\section{The importance of the position and momentum representations, and the role of the regularisation operator}

\label{Sec:comparison1}

In this section let us emphasize that both the position and momentum space quantisation approaches are important in obtaining a complete picture of the quantised EM field.  For example, studying the EM field in position space has helped us to identify an otherwise hidden degree of freedom: the parameter $s$ which characterises the direction of propagation of wave packets of light.  In Section \ref{Sec:1Dposition} we determined the Hilbert space for modelling localised wave packets in one dimension.  By solving Maxwell's equations we were also able to determine a set of field observables in addition to deriving a dynamical Hamiltonian that describes the evolution of the system.  However, we were not able to determine the particular value of the regularisation operator $\mathcal{R}$ which establishes a relationship between the local blip operators and the electric and magnetic field observables.

Although both the position and momentum representations are essentially equivalent, it is easier to determine these regularisation factors in momentum space.  In the following we shall determine this factor.  The Fourier transforms in Eqs.~(\ref{1Dphoton1}) and (\ref{1Dphoton2}) allow us to alternate freely between the position and momentum representations.  Once we have identified $\Omega(k)$ we will then be able to determine the effect of the regularisation $\mathcal{R}$ on the blip operators.  Later in this section we shall look at the relationship between this and the standard description of the quantised EM field in momentum space.  It is shown that the standard description emerges when we restrict the photon annihilation operators that we consider to a certain subset of operators.

\subsection{Lorentz covariance}
\label{Sec:Lorentzcovariance}

In Section \ref{Sec:1Dmomfields}, we were only able to define the electric and magnetic field observables up to a $k$-dependent factor $\Omega(k)$ which was shown to be directly related to the energy of a photon.  One way of determining this factor is to take into account that spontaneous emission from an individual atom or ion results in exactly one photon.  This assumption is in good agreement with quantum optics experiments which generate single energy quanta on demand \cite{Cir, Kuh, Moe, Ste}. These behave as monochromatic waves with energies and frequencies determined by the atomic transition frequency $\omega_0$.  A different but equivalent method is to ensure that the electric and magnetic field observables transform correctly under the orthochronous Lorentz transformations.  In this section we shall follow this latter approach.

For light propagating along the $x$ axis, the possible transformations are translations in $x$ and $t$, and rotations about and boosts along the $x$ axis. A rotation changes the polarisation of the $\mathbf{E}$ and $\mathbf{B}$ fields, which are oriented in the $y$-$z$ plane, but leaves the direction of propagation unchanged.  The inner product between two states is a Lorentz scalar, and is therefore unchanged by a Lorentz transformation.  We shall denote a single such transformation by the Greek letter $\Lambda$.  Naturally, these changes of reference frame induce a unitary operation $U(\Lambda)$.  Under a translation or a rotation, the photon creation operator gains only a phase factor.  The Lorentz boosts along the $x$ axis, however, involve a more interesting transformation.  Let us choose the particular transformation $\Lambda$ that causes a Doppler shift of the wave number $k$ to the new wave number $p$ without changing the polarisation or direction of the photon.  By taking into account Eq.~(\ref{photoncommutator2}) and using the Lorentz invariant measure $\text{d}k/|k|$ one may show that the single-photon inner product given in Eq.~(\ref{photoncommutator3}) is both Lorentz- and form-invariant only when
\begin{eqnarray}
\label{photontransformation1}
U(\Lambda)\,\widetilde{a}_{s\lambda}(k)\,U^\dagger(\Lambda) =  \sqrt{\left|\frac{p}{k}\right|}\,\widetilde{a}_{s\lambda}(p).
\end{eqnarray}

When a change of reference frame is made, either by moving the field or moving ourselves, the new field is some transformed value of the original field strength measured at the untransformed position.  In classical electromagnetism, the electric and magnetic field vectors transform as the components of an antisymmetric rank-2 tensor.  We should expect the expectation values of the quantised field observables to transform in just the same way.  By unitarily transforming the position-dependent field observables in their momentum representations, one finds that the correct transformation occurs when
\begin{eqnarray}
\label{Omega1}
\Omega(k) = \sqrt{|k|}\,\Omega_0.
\end{eqnarray}  
This result is calculated in three dimensions in Appendix (\ref{App:Omegaproof}), but the one-dimensional result is analogous.  Furthermore, if we want the expectation values of the energy observable $H_{\text{energy}}(t) $ in Eq.~(\ref{Heng3}) and of the dynamical Hamiltonian $H_{\text{dyn}} (t)$ in Eq.~(\ref{Hdyn3}) to be the same, at least in some cases, we must also choose 
\begin{eqnarray}
\label{Omega2}
\left| \Omega_0 \right|^2 &=& {2 \hbar \over \varepsilon_0 A c} \, .
\end{eqnarray}
The latter equivalence only holds for states with positive values of $k$. In general, the above choice of $\Omega_0$ implies that the energy expectation value of a single photon in the $(s,k,\lambda)$ mode equals $\hbar c|k|$.

\subsection{The regularisation operator revisited}

\label{Sec:fieldcomparison}

In Section \ref{Sec:photonHilbertspace} we demonstrated how to transform between the blip and photon operators by means of a Fourier transform.  By combining Eqs.~(\ref{1Dphoton1}) and (\ref{1Dphoton2}) with Eq.~(\ref{fieldobservables1}) one will arrive at the expressions
\begin{eqnarray} 
\label{fieldobservables2}
\textbf{E}(x,t) &=&  \sum_{s= \pm 1} \int^{\infty}_{-\infty} \frac{\text{d}k}{\sqrt{2\pi}}\;c \, {\rm e}^{{\rm i}(skx + \varphi(k))}
\left( {\cal R} \left[ \widetilde{a}_{s{\sf H}}\right](k,t) \,\widehat{\boldsymbol{y}} + \mathcal{R}\left[\widetilde{a}_{s{\sf V}}\right](k,t)\,\widehat{\boldsymbol{z}}\right) , ~~~ \nonumber \\
\textbf{B}(x,t) &=& \sum_{s= \pm 1} \int^{\infty}_{-\infty} \frac{\text{d}k}{\sqrt{2\pi}}\; s \, {\rm e}^{{\rm i}(skx + \varphi(k))}
\left( {\cal R} \left[\widetilde{a}_{s{\sf H}}\right](k,t) \,\widehat{\boldsymbol{z}} - \mathcal{R}\left[\widetilde{a}_{s{\sf V}}\right](k,t)\,\widehat{\boldsymbol{y}} \right)\,.
\end{eqnarray}
Moreover, by substituting Eq.~(\ref{Omega1}) into the Fourier transforms (see. Eq.~(\ref{fieldtransforms})) of the electric and magnetic field observables given in Eq.~(\ref{FTfieldcomponents1}), one finds that
\begin{eqnarray}
\label{fieldobservables3}
\textbf{E}(x,t) &=&  \sum_{s= \pm 1} \int^{\infty}_{-\infty} \frac{\text{d}k}{\sqrt{2\pi}} \; c \,\sqrt{\frac{2\hbar|k|}{\varepsilon_0 Ac}} \, {\rm e}^{{\rm i}(skx+ \varphi(k))}  \left(\widetilde{a}_{s {\sf H}}(k,t) \, \widehat{\boldsymbol{y}} +  \widetilde{a}_{s {\sf V}}(k,t) \, \widehat{\boldsymbol{z}} \right) \, ,  \nonumber \\
\textbf{B}(x,t) &=& \sum_{s= \pm 1} \int_{-\infty}^{\infty}\frac{\text{d}k}{\sqrt{2\pi}} \; s \,\sqrt{\frac{2 \hbar|k|}{\varepsilon_0 A c}} \, {\rm e}^{{\rm i}(skx+\varphi(k))}
\left( \widetilde{a}_{s {\sf H}}(k,t) \, \widehat{\boldsymbol{z}} - \widetilde{a}_{s {\sf V}}(k,t) \, \widehat{\boldsymbol{y}}  \right). 
\end{eqnarray}
This equivalence between the complex field observables above allows us to determine the action of the superoperator $\mathcal{R}$ on a photon annihilation operator.  We find that
\begin{eqnarray} 
\label{regularise1}
{\cal R} \left[ \widetilde{a}_{s\lambda}\right](k,t) &=& \sqrt{2 \hbar |k| \over \varepsilon_0 A c} \, \widetilde{a}_{s\lambda}(k,t) \, .
\end{eqnarray}
In the momentum representation, in order to regularise the photon operators we multiply them by $(2\hbar|k|/\varepsilon_0 A c)^{1/2}$.  As expected from the symmetries of the quantised EM field, the superoperator ${\cal R}$ depends on neither $x$, $t$, $s$ nor $\lambda$.

In Eq.~(\ref{fieldobservables2}) we assumed that the regularised blip operators, which before regularisation are the Fourier transforms of the photon operators, are the Fourier transforms of the regularised photon operators.  The action of the regularisation superoperator $\mathcal{R}$ on a blip annihilation operator can therefore be determined by calculating the Fourier transform of Eq.~(\ref{regularise1}).  One will find that the action of the regularisation operator on a blip annihilation operator is of the type given in Eq.~(\ref{Rsuperposition1}) where the $R(x-x')$ is given by
\begin{eqnarray}
\mathcal{R}(x-x') = \int_{-\infty}^{\infty}\frac{\text{d}k}{2\pi}\; e^{ik(x-x')}\sqrt{\frac{2\hbar|k|}{\varepsilon_0 A c}}.
\end{eqnarray}

The above trick of introducing the superoperator ${\cal R}$ allows us to describe the quantised EM field in position space in terms of local bosonic blip operators without having to sacrifice the Lorentz covariance of the electric and magnetic field observables $\textbf{E}(x,t)$ an $\textbf{B}(x,t)$. Importantly, however, this distribution is not local in the same way that the Dirac delta function is.  We can see this by evaluating $\mathcal{R}(x-x')$ at values $x \neq x'$:
\begin{eqnarray}
\label{explicitR}
	\mathcal{R}(x-x') = -\sqrt{\frac{\hbar}{4\pi\varepsilon_0 Ac}}\frac{1}{|x-x'|^{3/2}}.
\end{eqnarray}
This means that the field observables are not a simple linear superposition of blip operators defined at the same point (see Fig.~\ref{Fig:regularisation} (c)).  However, since it is easier to work with bosonic annihilation and creation operators, we can perform all calculations in the Hilbert space created by the local bosonic operators.  As a final point we may mention that, due to the equivalence of the field observables, the two representations of the energy observable are equal.

\subsection{Comments on field and blip localisation}
\label{Sec:biorthogonal}

As was mentioned in the Introduction, some authors will prefer to work with non-locally acting operators that have a closer link to the field observables \cite{Haw9, Haw11, Haw5, Jake2}.  Such operators are given by
\begin{eqnarray} 
\label{fieldexcitation1}
A_{s \lambda}(x,t) &=& {\cal R}[a_{s \lambda}](x,t)
\end{eqnarray}
and describe excitations that share a vacuum with the blip operators.  The reason we differentiate between blip operators $a_{s\lambda}(x,t)$ and the field excitations $A_{s\lambda}(x,t)$ is that the blip operators possess a set of bosonic commutation relations (Eq.~(\ref{blipcommutator2})) with respect to the conventional inner product.  In contrast to this, the commutation relation for the $A_{s\lambda}(x,t)$ operators is 
\begin{eqnarray}
\label{fieldexcitationscommutator1}
\big[ A_{s\lambda}(x,t), A^\dagger_{s'\lambda'}(x',t) \big] &=& \int^\infty_{-\infty} \frac{\text{d}k}{2\pi} \, \frac{2\hbar|k|}{\varepsilon_0Ac} \, {\rm e}^{{\rm i}sk(x-x')} \, \delta_{ss'} \,\delta_{\lambda\lambda'} \, . \nonumber \\
\end{eqnarray}

Nevertheless, one can see from Eq.~(\ref{fieldobservables1}), for example, that the energy quanta associated with the $A_{s\lambda}(x,t)$ operators can be more easily associated with the electric and magnetic field observables.

With respect to the usual inner product, the single-excitation states $\ket{1^A_{s\lambda}(x,t)} = A^\dagger_{s\lambda}(x,t)\ket{0}$ are not orthogonal to one another.  Using Eq.~(\ref{fieldexcitationscommutator1}), one can show that 
\begin{equation}
\label{nonorthogonal1}
\braket{1^A_{s\lambda}(x,t)|1^A_{s'\lambda'}(x',t)}  = \braket{0|\big[ A_{s\lambda}(x,t), A^\dagger_{s'\lambda'}(x',t) \big]|0} \neq 0.
\end{equation}   
This is related to the well-known problem that the measurable field observables have the units of $(\text{energy density})^{1/2}$ and not $(\text{probability density})^{1/2}$ as would be expected for a wave function compatible with the Born rule.

\begin{figure}[t]
\centering
\includegraphics[width = 0.9\textwidth]{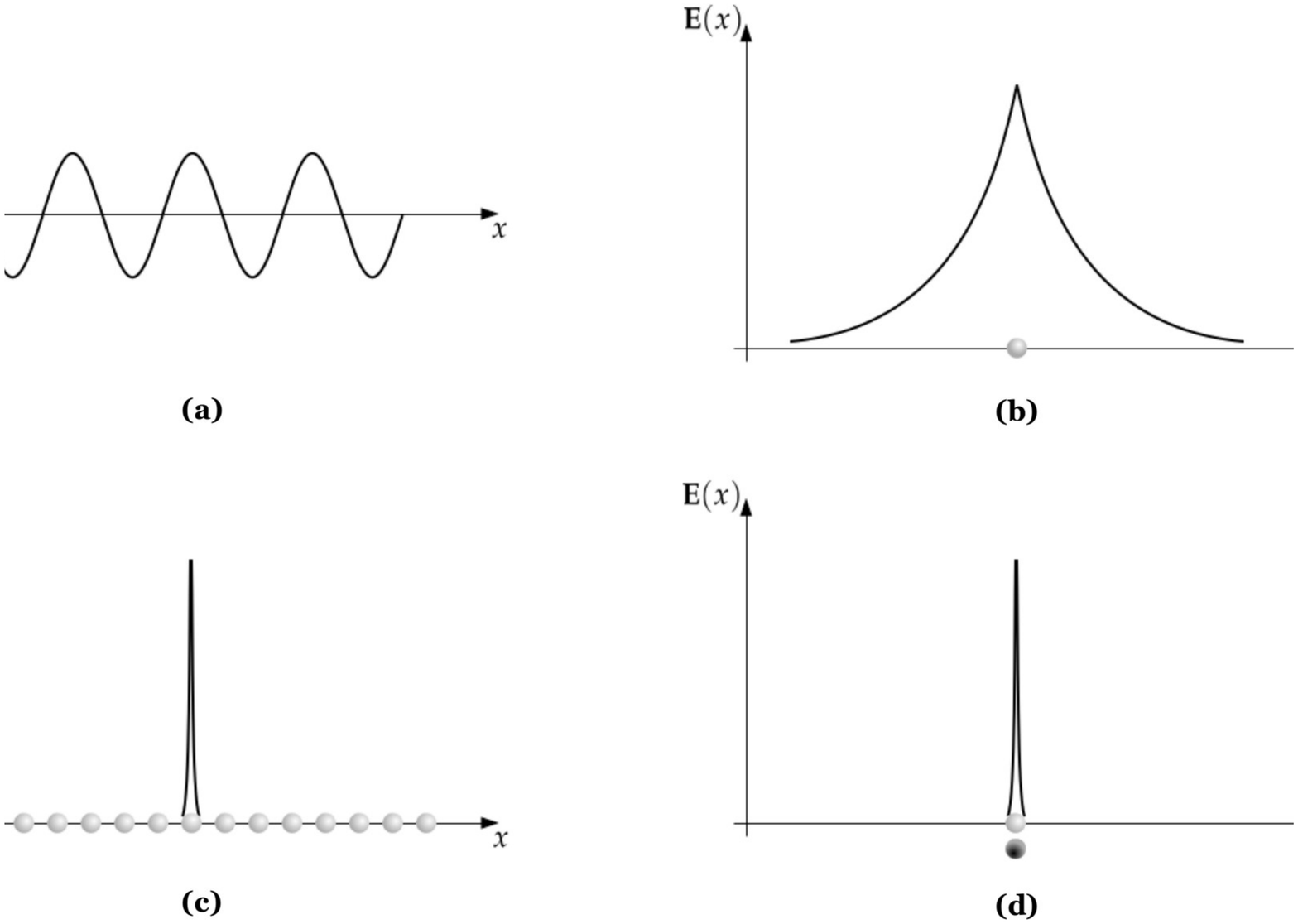}
\caption[The relation between particles and fields]{\cite{Jake2} This figure compares the locality of different excitations under the standard inner product and the inner product denoted A. (a) A monochromatic photon is equally likely to be found anywhere along the $x$ axis using either inner product. (b) Under the standard inner product a localised blip, illustrated by the white particle, carries with it a non-local field whose relation to the position of the blip is given by Eq.~(\ref{explicitR}). (c) Under the standard inner product, a localised field excitation, which is generated by the operators $A^\dagger_{s\lambda}(x,t)$ operator, is constructed from a non-local superposition of blips.  Under the biorthogonal inner product, the blips and fields can be localised simultaneously.}
\label{Fig:regularisation}
\end{figure}

Although we shall not make use of it here, for completeness it is worth describing a method by which we may impose bosonic commutation relations on the $A_{s\lambda}(x,t)$ operators.  This is achieved by introducing a new inner product -- labelled by superscript $A$ --  such that:
\begin{eqnarray} 
\label{bioproduct1}
\braket{ 1^{A}_{s\lambda}(x,t) | 1^{A}_{s'\lambda'}(x',t) }^{A} &=& \braket{ 1_{s\lambda}(x,t) | 1_{s'\lambda'}(x',t)} \nonumber \\
&=& \delta(x-x'). 
\end{eqnarray}
Under this new inner product the single-photon field excitations form an orthogonal basis of states in the position representation. This is shown in Fig.~\ref{Fig:regularisation} (d). However, this new inner product drastically alters the Hilbert space of the quantised EM field.  One consequence of this change, for instance, is that some previously Hermitian operators become non-Hermitian, whereas other previously non-Hermitian operators may become Hermitian \cite{Mos1, Mos2, Mos3}.  For this reason, this aspect of quantum physics is known as ``biorthogonal" or ``pseudo-Hermitian" quantum mechanics. It is an interesting area of physics that has attracted a lot of attention in the field of local quantum electrodynamics (QED) \cite{Haw9, Ray2, Haw11, Haw5, Haw3, Bro}.  Although this is a very elegant way of restoring orthogonality, as we have investigated in Ref.~\cite{Jake2}, introducing a biorthogonal system introduces complexities that are by no means necessary to our understanding of the local dynamics of photons.

\subsection{The relation to standard descriptions}

In this final subsection, we compare the description of the EM field in Section \ref{Sec:1Dmomentum} with the standard description of the quantised EM field in momentum space, and ask which additional assumptions have to be made for the latter to emerge. Looking at Eqs.~(\ref{Heng3}) and (\ref{Hdyn3}), and using the value of $\Omega(k)$ given by Eqs.~(\ref{Omega1}) and (\ref{Omega2}), we can see that the energy observable and dynamical Hamiltonian coincide when the negative-frequency modes are excluded and we restrict ourself to positive values of $k$ only.  As one can check, in this case the field observables in Eqs.~(\ref{fieldobservables2}) and (\ref{fieldobservables3}) become the same as the local field observables in the standard description of the quantised EM field \cite{RB2}. As we know, the positive-frequency photon states provide a complete description of the quantised EM field in the sense that they can be superposed to reproduce the right electric and magnetic field expectation values of wave packets of any shape. However, as we have seen in the introduction to this chapter, they are not sufficient to generate the quantum versions of all possible solutions of Maxwell's equations, like highly-localised wave packets that remain localised \cite{Heg8, Mal}.

\section{Discussion}

\label{Sec:photonsdiscussion}

In standard quantisations of the free EM field in one dimension, the basic excitations of the field are characterised by a wave number $k$, a frequency $\omega = |k|c$ and a polarisation $\lambda$.  These degrees of freedom, however, are not sufficient for constructing single-photon wave packets that are both localised and propagate at the speed of light without dispersion.  In this chapter we have constructed a complete description of the quantised EM field in one dimension by assuming that the basic quanta of the field are blips, which have a well-defined position, polarisation and direction of propagation. Using these conditions, we then identified the Schr\"odinger equation for this system and constructed electric and magnetic field observables $\mathbf{E}(x,t)$ and $\mathbf{B}(x,t)$ that are consistent with Maxwell's equations. The Lorentz covariance of these field operators is achieved with the help of a regularisation superoperator ${\cal R}$. Although the blips themselves are local, they generate non-local fields which can be felt in a region of the space surrounding them (see Fig.~\ref{Fig:regularisation} (b)).

By writing the blip excitations as superpositions of monochromatic photons, we have also shown that our approach is consistent with the standard theory of the quantised EM field with the addition of countable negative-frequency states.  Previously, these states have been widely overlooked but the concept of adopting them to negate the consequences of Hegerfeldt's theorem \cite{Heg8} is not new.  For example, Allcock pointed out that negative-frequency modes are necessary for defining states that have a well-defined and measurable time of arrival \cite{Al1, Al2, Al3}.  It is also possible to define a time of arrival for blip states.  Negative-frequency solutions have also been considered in Refs.~\cite{Chan,Kai, Mil1, For1, For2, Fac, Bos, Con}. In Refs.~\cite{Haw11, Haw6, Haw8, Bab} negative frequencies were also considered in the context of constructing real field excitations.  In this chapter, by introducing local particles of light with a given direction of propagation we have clarified how these solutions arise naturally in a covariant quantised theory. In addition, we have exposed some consequences of a theory containing these states such as the difference between the energy observable $H_\text{energy}(t)$ and the generator for time translations, that is, the dynamical Hamiltonian. 

In classical electrodynamics, a local description of the EM field is often preferable to a non-local description. Similarly, we expect that the modelling of the quantised EM field in terms of blip states will sometimes be preferable to the standard description in terms of monochromatic photon states.  For example, this theory should provide an extremely useful tool for the modelling of the quantised EM field in inhomogeneous dielectric media and on curved space-times \cite{May}. Other potential applications include modelling linear optics experiments with ultra-broadband photons \cite{Nasr, Ok, Jav, Tak} or local light-matter interactions. For example, in Ref.~\cite{Jake1}, we used the blip annihilation operators $a_{s \lambda}(x,t)$ to construct locally-acting mirror Hamiltonians.  This work might also inspire other areas of quantum physics where there is a need for a local description of quantum fields, for instance, the modelling of electrons in quantum transport problems.  Later in this thesis we shall show how the results of this paper also provide novel insight into the Fermi problem and the Casimir effect.

\chapter{Local photons in three dimensions}

\label{Chapter:3D}

This chapter provides a complete quantisation of the free EM field in three dimensions in terms of particles that are, like in one dimension, perfectly localised and propagate at the speed of light. In Section \ref{Sec:raypicture} we initially take a na\"ive approach to quantisation by describing light as rays, but later find that these excitations form an over-complete basis of the Hilbert space.  A complete basis is constructed in terms of plane wave excitations. In Section \ref{Sec:3Dlocalstates} we show that a reduced and complete Hilbert space can still be constructed in terms of perfectly localised excitations, and in Section \ref{Sec:3Dcomplete} we use these excitations to provide a new quantisation of the free EM field in three dimensions that is analogous to the results of the previous chapter.  Here, as before, we construct a set of field observables and determine a dynamical Hamiltonian for the system which differes from the usual energy observable. 

\section{Introduction: Lessons from one dimension}

In the previous chapter we quantised the free EM field in one dimension in terms of localised energy quanta, so-called blips, that have a well-defined direction of propagation and move without dispersion.  The purpose of this chapter is to extend the work of the previous chapter and determine a more complete quantisation of the free EM field that describes all possible solutions of Maxwell's equations in three dimensions.  A three-dimensional description of the EM field will have the benefit of providing a general formalism for describing light in a much more expansive array of problems, for example, predicting experimental verifications of the Casimir effect.  This theory will hopefully provide a greater set of tools that will enable new investigations into many different topics. 

In Section \ref{Sec:1Dclassicalfields}, we showed that the one-dimensional wave equation, satisfied by both $\mathbf{E}(x,t)$ and $\mathbf{B}(x,t)$, can be factorised into two first-order equations which solve for fields propagating either to the left or the right at the speed of light.  This enabled us to immediately construct a first-order differential equation of motion for the basic excitations of the EM field.  In three dimensions this is not possible as solutions of the wave equation (\ref{waveequation1}) cannot all be said to propagate in a given direction, such as left or right, but may propagate in all sorts of directions.  This makes it difficult to identify the basic solutions of Maxwell's equations in the position representation.  In the past, some authors have used a massless Dirac-like equation as the fundamental equation of motion for photons in the position representation.  In this chapter, however, we shall instead follow the one-dimensional approach as closely as possible by identifying the solutions of Maxwell's equations that have a fixed direction of propagation.

In Section \ref{Sec:1Dclassicalfields}, we considered waves that propagate along the $x$ axis only.  As the three-dimensional wave equation is rotationally invariant, any rotation of these solutions will provide a further set of solutions to Maxwell's equations. Taking this into account and letting $\mathbf{s}$ be a constant unit vector, any one-dimensional solution of the wave equation propagating in the $\mathbf{s}$ direction will satisfy the equation
\begin{equation}
\label{3Deom1}
\left[\mathbf{s}\cdot\nabla + \frac{1}{c}\frac{\partial}{\partial t}\right]\mathbf{O}_{\mathbf{s}}(\mathbf{r},t) = 0
\end{equation}
where $\mathbf{O}_{\mathbf{s}} = \mathbf{E}_{\mathbf{s}}, \mathbf{B}_{\mathbf{s}}$ are the solutions of the wave equation that propagate in the $\mathbf{s}$ direction only.  Here $\mathbf{s}$ is assumed to be constant with respect to all coordinates.

Using the equation of motion (\ref{3Deom1}), it is also possible to show that the three-dimensional electric and magnetic field vectors $\mathbf{E}_{\mathbf{s}}(\mathbf{r},t)$ and $\mathbf{B}_{\mathbf{s}}(\mathbf{r},t)$ are orthogonal to both $\mathbf{s}$ and each other. To show that this is the case, one must first note that, for a divergence-less vector field $\textbf{O}_{\mathbf{s}}$, the following relation holds:
\begin{equation}
\nabla \times (\mathbf{s} \times \mathbf O) = \mathbf{s}\cdot \nabla \mathbf{O}_{\mathbf{s}}.
\end{equation}
Using the above relation in conjunction with Faraday's and the Maxwell-Amp\`ere law given by Eqs.~(\ref{fMaxwell3}) and (\ref{fMaxwell4}), one may show that Eq.~(\ref{3Deom1}) is satisfied when
\begin{equation}
\label{3Dfieldrelations1}
\textbf{E}_{\mathbf{s}} = c\left(\mathbf{s}\times \textbf{B}_{\mathbf{s}}\right)
\end{equation}
and
\begin{equation}
\textbf{B}_{\mathbf{s}} = -\frac{1}{c}\left(\mathbf{s} \times \textbf{E}_\mathbf{s}\right).
\end{equation}
Thus $\textbf{E}_{\mathbf{s}}$, $\textbf{B}_\mathbf{s}$ and $\mathbf{s}$ form an orthogonal triad.  Moreover, since $\mathbf{E}_\mathbf{s}$ and $\mathbf{B}_\mathbf{s}$ can be rotated about $\mathbf{s}$ without changing their relative orientations, the electric and magnetic fields propagating in the direction of $\mathbf{s}$ may be further characterised by two linear polarisations orthogonal to $\mathbf{s}$.  Note also that $\mathbf{s}$ is parallel to the Poynting vector defined in Eq.~(\ref{Poyntingvector}).

The considerations above provide a good starting point for quantising the free, three-dimensional EM field in position space.  We have seen, for instance, that there are solutions of Maxwell's equations that propagate in a fixed direction $\mathbf{s}$, have two distinct polarisations, and satisfy a first-order equation of motion.  In the remainder of this chapter, we shall build upon these observations to develop a description of the three-dimensional quantised EM field in the position and momentum representations.

\section{A na\"ive approach to quantisation in the position representation}

\label{Sec:raypicture}

In this section we describe the free EM field in terms of localised excitations that can each travel in any single direction, and we define some of their basic properties.  We also define an appropriate set of field observables. Towards the end of this section we shall demonstrate that these excitations provide an over-complete basis of the Hilbert space and that plane wave excitations present a more suitable set of principal excitations. 

\subsection{The Hilbert space}

\label{Sec:3DHilbertspace}

\subsubsection{Blip states}

As in the one-dimensional theory, we assume here that the fundamental building blocks of the electric and magnetic field observables are a set of spatially localised particles.  We shall again refer to these excitation as blips.  In three dimensions, blips are localised to a single point characterised by the three-dimensional position vector $\mathbf{r} \in \mathbb{R}^3$.  As discussed in the preceding introduction, blips have a unique direction of propagation which is no longer fixed along a specific axis but may be oriented in any constant direction of our choosing.  The direction of propagation is characterised by the three-dimensional unit vector $\mathbf{s} \in S^2$ which is oriented in the direction of propagation.  Here $S^2$ is the unit two-sphere which consists of all possible unit vectors.  Blip states are also characterised by a polarisation $\lambda \in \{\mathsf{H},\mathsf{V}\}$ that specifies the orientation of the associated electric and magnetic field observables.

\begin{figure}[h]
\centering
\includegraphics[width = 0.55\textwidth]{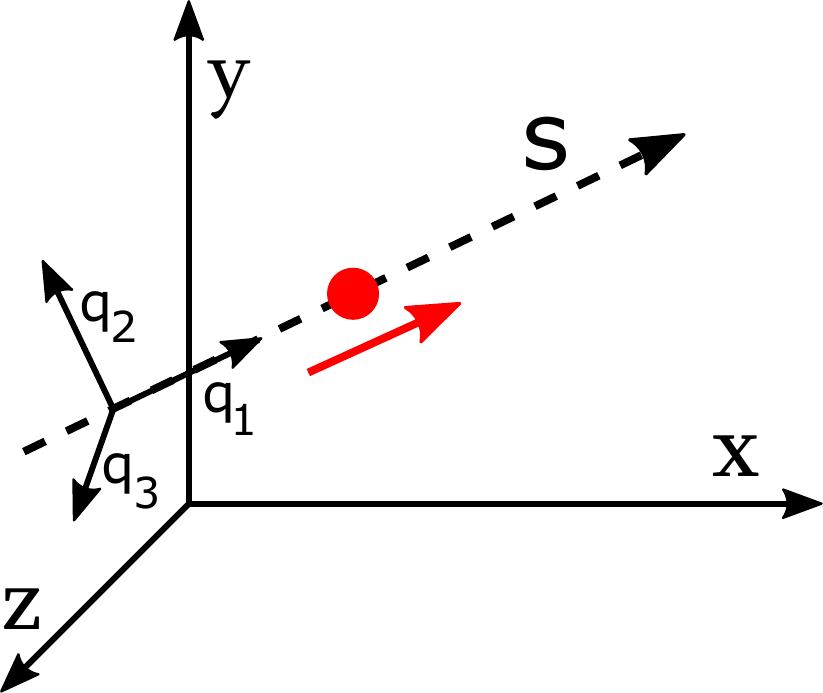}
\caption[A localised blip propagating in three dimensions]{The figure shows a single blip propagating in the direction of $\mathbf{s}$.  The blip remains localised as it propagates along the trajectory characterised by $\mathbf{s}$ and indicated by the dotted line.  The parameters $\mathbf{r}$ and $\mathbf{s}$ determine this trajectory uniquely. An identical blip characterised by a vector $-\mathbf{s}$ would propagate along the same path in the opposite direction.  The coordinates $(q_1,q_2,q_3)$ form a right-handed coordinate system where the $q_1$ axis is aligned with $\mathbf{s}$.}
\label{Fig:3Dblip1}  
\end{figure}

In this description, the basic elements of the Hilbert space are localised energy quanta that move with a fixed direction tracing out rays as illustrated in Fig.~\ref{Fig:3Dblip1}.  The reader may like to argue that light does not form perfectly narrow rays in reality.  This, however, would only be a consequence of having an over-complete Hilbert space rather than an under-complete one.  For now we shall proceed by assuming that the basic excitations of the field propagate along these rays and later determine any constraints on the arrangement of these rays that would effectively reduce the size of our Hilbert space.

\subsubsection{Annihilation and creation operators}

As in previous sections we define a set of creation and annihilation operators that add and remove blips from the system.  We denote the blip annihilation operators $a_{\mathbf{s}\lambda}(\mathbf{r},t)$ in the Heisenberg picture and $a_{\mathbf{s}\lambda}(\mathbf{r},0)$ in the Schr\"odinger picture.  The vacuum state $\ket{0}$ of the system, which contains exactly zero blip excitations, is defined such that
\begin{equation}
\label{3Dvacuum1}
a_{\mathbf{s}\lambda}(\mathbf{r},t)\ket{0} = 0
\end{equation}
for all $\mathbf{r},t,\mathbf{s}$ and $\lambda$.  The vacuum state is normalised to unity: $\braket{0|0} = 1$.

As before, a single-blip state is generated by applying the creation operator $a^\dagger_{\mathbf{s}\lambda}(\mathbf{r},t)$ to the vacuum state:
\begin{equation}
\label{3Dstate1}
\ket{1_{\mathbf{s}\lambda}(\mathbf{r},t)} = a^\dagger_{\mathbf{s}\lambda}(\mathbf{r},t)\ket{0}.
\end{equation}
The single-blip Hilbert space is therefore spanned by such states for all $\mathbf{r}$, $\mathbf{s}$ and $\lambda$ at a fixed time.  By applying a creation operator to the vacuum state repeatedly we can generate states that contain any number of blips.  In analogy to Eqs.~(\ref{numberstates1}) and (\ref{blipstate2}), a state that contains $n$ identical blips at a time $t$ is denoted
\begin{equation}
\label{3Dstate2}
\ket{n_{\mathbf{s}\lambda}(\mathbf{r},t)} = \frac{1}{\sqrt{n!}} a^\dagger_{\mathbf{s}\lambda}(\mathbf{r},t)^n\,\ket{0}.
\end{equation}
The total Hilbert space is the symmetric Fock space of the single-blip Hilbert space.

\subsubsection{A fundamental equation of motion}

\label{Sec:blipmotion1}

In the introduction to this chapter we pointed out that components of the electric and magnetic field vectors that propagate in a space-fixed direction $\mathbf{s}$ will satisfy the first-order equation of motion (\ref{3Deom1}).  The solutions to Eq.~(\ref{3Deom1}) describe waves that propagate in the direction of $\mathbf{s}$ at the speed of light.  As blips represent the basic excitations of the field, we propose here that blips characterised by a vector $\mathbf{s}$ will also propagate in this direction at the speed of light.  This condition places a constraint on the expectation values of the blip operators, and therefore the operators themselves, which leads us to the following equation of motion:
\begin{equation}
\label{3Deom2}
\left[ \mathbf{s}\cdot \nabla + \frac{1}{c}\frac{\partial}{\partial t}\right]\,a_{\mathbf{s}\lambda}(\mathbf{r},t) = 0.
\end{equation}

Although a blip state is characterised by three spatial coordinates that specify its position in space, as blips only move in a single direction (indicated by $\mathbf{s}$), the dynamics of a blip will only depend on a single coordinate.  The remaining two coordinates define the position of this one-dimensional trajectory.  To help us describe the solutions to Eq.~(\ref{3Deom2}), let us introduce a new Cartesian coordinate system $\mathbf{r} = (q_1, q_2, q_3)$ where the $q_1$ axis is oriented in the direction of $\mathbf{s}$ as illustrated in Fig.~\ref{Fig:3Dblip1}. Although this coordinate system depends on the orientation of $\mathbf{s}$, there is still a well-defined way of constructing this coordinate system for each choice of $\mathbf{s}$.  We shall assume that all three coordinate axes are orthogonal.  The solutions to Eq.~(\ref{3Deom2}) are then given by
\begin{equation}
\label{blipconstraint2}
a_{\mathbf{s}\lambda}(q_1,q_2,q_3;t) = a_{\mathbf{s}\lambda}(q_1-ct,q_2,q_3;0). 
\end{equation}
This equation of motion is analogous to Eq.~(\ref{blipconstraint1}).

\subsection{Blip commutation relations}
\label{Sec:3Dcommutators}

\subsubsection{Commutation relations}

As blips are bosons, a state containing many particles is unchanged by the exchange of any two blips within that state.  Consequently, all blip creation and annihilation operators commute with one another:
\begin{equation}
\label{3Dcommutator1}
	\left[a^{\dagger}_{\mathbf{s}\lambda}(\mathbf{r},t), a^{\dagger}_{\mathbf{s}'\lambda'}(\mathbf{r}',t')\right] = \Big[a_{\mathbf{s}\lambda}(\mathbf{r},t), a_{\mathbf{s}'\lambda'}(\mathbf{r}',t')\Big] = 0.
\end{equation}
for all $\mathbf{r}$, $\mathbf{r}'$, $t$ , $t'$, $\mathbf{s}$, $\mathbf{s}'$, $\lambda$ and $\lambda'$.  As in Eq.~(\ref{commutatorfromproduct}), the commutation relation between the blip creation and annihilation operators is determined by the inner product between two single-blip states:
\begin{equation}
\label{3Dcommutator3}
\braket{1_{\mathbf{s}\lambda}(\mathbf{r},t)|1_{\mathbf{s}'\lambda'}(\mathbf{r}',t')} = \Big[a_{\mathbf{s}\lambda}(\mathbf{r},t),a^\dagger_{\mathbf{s}'\lambda'}(\mathbf{r}',t')\Big].
\end{equation}
This inner product must now be determined.

\subsubsection{Orthogonal states}

As blips are entirely localised at a point specified by the vector $\mathbf{r}$, blips defined at different positions must all be orthogonal to one another at a given fixed time.  Moreover, as polarisation is a measurable quantity and blips characterised by different values of $\mathbf{s}$ have different dynamics, they must all be distinguishable and therefore belong to different Hilbert spaces.  Hence, in the following we find that
\begin{eqnarray}
\label{3Dcommutator4}
\braket{1_{\mathbf{s}\lambda}(\mathbf{r},t)|1_{\mathbf{s}'\lambda'}(\mathbf{r}',t')} &=& \braket{0|\Big[a_{\mathbf{s}\lambda}(\mathbf{r},t),a^\dagger_{\mathbf{s}'\lambda'}(\mathbf{r}',t')\Big]|0}\nonumber\\
&=& \delta^3(\mathbf{r}-\mathbf{r}')\,\delta^2(\mathbf{s}-\mathbf{s}')\,\delta_{\lambda,\lambda'}.
\end{eqnarray}

\subsubsection{Single-blip wave packets}

\label{Sec:3Dwavepackets}

As we have seen above, the single-blip states characterised by a single value of the continuous variables $\mathbf{s}$ and $\mathbf{r}$ are therefore non-normalisable.  Physical states of the system, on the other hand, are given by finitely normalised wave packets as were described in Section \ref{Sec:wavepackets}.  Hence, a single-photon wave packet is given by the superposition
\begin{equation}
\label{3Dwavepacket1}
\ket{\psi(t)} = a^\dagger(t)\,\ket{0} = \sum_{\lambda = \mathsf{H},\mathsf{V}}\int_{S^2}\text{d}^2\mathbf{s}\int_{-\infty}^{\infty}\text{d}^3\mathbf{r}\;\psi_{\mathbf{s}\lambda}(\mathbf{r})\ket{1_{\mathbf{s}\lambda}(\mathbf{r},t)}
\end{equation}
where $a^\dagger(t)$ is the single-photon creation operator for the wave packet with position wave function $\psi_{\mathbf{s}\lambda}(\mathbf{r})$. 

To ensure that this state is normalised and that $\left[a(t), a^\dagger(t)\right] = 1$, the wave function $\psi_{\mathbf{s}\lambda}(\mathbf{r})$ must satisfy the condition
\begin{equation}
	\label{3Dnormalise1}
	\sum_{\lambda=\mathsf{H},\mathsf{V}}\int_{S^2}\text{d}^2\mathbf{s}\int_{-\infty}^{\infty}\text{d}^3\mathbf{r}\;|\psi_{\mathbf{s}\lambda}(\mathbf{r})|^2 = 1.
\end{equation}  
Eq.~(\ref{3Dwavepacket1}) above will help us determine the complete physical Hilbert space of states later in this chapter.

\subsection{Field observables in the position representation}

\label{Sec:3Dposfields}

\subsubsection{Field observables}
In this section we shall determine a set of expressions for the complex three-dimensional field observables $\mathbf{E}(\mathbf{r},t)$ and $\mathbf{B}(\mathbf{r},t)$ in terms of the blip operators $a_{\mathbf{s}\lambda}(\mathbf{r},t)$ and $a^\dagger_{\mathbf{s}\lambda}(\mathbf{r},t)$.  Analogous to the results found in Sections \ref{Sec:Quantisation1} and \ref{Sec:1Dposfields}, we shall assume that the complex field observables are given by a linear superposition of blip annihilation operators.  Since both the components of the electric and magnetic field observables and the blip operators obey Eq.~(\ref{waveequation1}), we assume that
\begin{eqnarray}
\label{3Dfieldobservables1}
\mathbf{E}(\mathbf{r},t) &=& \int_{S^2}\text{d}^2\mathbf{s}\; c\,\left\{\boldsymbol{\cal{R}}_{\mathbf{s}\mathsf{H}}[a_{\mathbf{s}\mathsf{H}}](\mathbf{r},t) + \boldsymbol{\cal{R}}_{\mathbf{s}\mathsf{V}}[a_{\mathbf{s}\mathsf{V}}](\mathbf{r},t)\right\}\nonumber\\
\mathbf{B}(\mathbf{r},t) &=& \int_{S^2}\text{d}^2\mathbf{s}\; \left\{\boldsymbol{\cal{R}}_{\mathbf{s}\mathsf{H}} [a_{\mathbf{s}\mathsf{H}}](\mathbf{r},t) + \boldsymbol{\cal{R}}_{\mathbf{s}\mathsf{V}}[a_{\mathbf{s}\mathsf{V}}](\mathbf{r},t)\right\} \times \mathbf{s}.
\end{eqnarray}
The above operator $\boldsymbol{\cal{R}}_{\mathbf{s}\lambda}$ denotes a translation-invariant superoperator that shall again be referred to as the regularisation operator.  These field observables have been chosen such that the relations given in Eq.~(\ref{3Dfieldrelations1}) are satisfied.

\subsubsection{The regularisation operator}

As in Section \ref{Sec:1Dposfields}, we have introduced a position-independent regularisation operator in order to relate the electric and magnetic field observables to the local blip operators. This regularisation operator will again allow us to ensure that the field observables have the correct dimensionality and transformation properties without having to sacrifice the orthogonality of the blip operators.  Unlike the regularisation operator defined in Section \ref{Sec:1Dposfields}, the regularisation operator is now vector valued.  In one dimension, light is always polarised in the $y$-$z$ plane, and the constant polarisation vectors can therefore be forgotten.  In three dimensions, however, the polarisation vectors, like $\mathbf{s}$, can be oriented in any direction, so must be included in the definition of $\boldsymbol{\mathcal{R}}_{\mathbf{s}\lambda}$. 

The regularisation operator can again be understood in a distributional sense in the following way\;
\begin{eqnarray}
\boldsymbol{\mathcal{R}}_{\mathbf{s}\lambda}\Big[a_{\mathbf{s}\lambda}\Big](\mathbf{r},t) &=& \int_{\mathbb{R}^3}\text{d}^3\mathbf{r}'\;\boldsymbol{\cal{R}}_{\mathbf{s}\lambda}(\mathbf{r}-\mathbf{r}')\,a_{\mathbf{s}\lambda}(\mathbf{r}',t)\nonumber\\
\boldsymbol{\mathcal{R}}^*_{\mathbf{s}\lambda}\Big[a^\dagger_{\mathbf{s}\lambda}\Big](\mathbf{r},t) &=& \int_{\mathbb{R}^3}\text{d}^3\mathbf{r}'\;\boldsymbol{\cal{R}}_{\mathbf{s}\lambda}^*(\mathbf{r}-\mathbf{r}')\,a^\dagger_{\mathbf{s}\lambda}(\mathbf{r}',t)
\end{eqnarray}
In order to satisfy Eq.~(\ref{3Dfieldrelations1}), the distributions $\boldsymbol{\mathcal{R}}_{\mathbf{s}\lambda}(\mathbf{r}-\mathbf{r}')$ must be oriented in a direction orthogonal to $\mathbf{s}$.  Furthermore, a change in the polarisation will rotate $\boldsymbol{\mathcal{R}}_{\mathbf{s}\lambda}(\mathbf{r}-\mathbf{r}')$ about the direction of propagation.  The orientation of $\boldsymbol{\mathcal{R}}_{\mathbf{s}\lambda}(\mathbf{r}-\mathbf{r}')$ therefore depends on both $\mathbf{s}$ and $\lambda$.

\subsection{Physical particle states}

\label{Sec:Physicalstates}

\subsubsection{Field expectation values}

In Section \ref{Sec:3Dposfields} we constructed a set of field observables that satisfy both the equation of motion \ref{3Deom1} for light propagating in one direction and the field relations specified in Eq.~(\ref{3Dfieldrelations1}).  We must now ensure that the expectation values of these field observables always satisfy Maxwell's equations (\ref{fMaxwell1})-(\ref{fMaxwell4}).  The main outcome of this section will be to determine which wave functions $\psi_{\mathbf{s}\lambda}(\mathbf{r})$ correspond to physical solutions of Maxwell's equations. This will enable us later on to reduce a possibly over-complete Hilbert space to a complete set of states describing all possible physical solutions of Maxwell's equations.

Since field expectation values calculated with respect to a state containing an integer number of blips are always zero, we shall first construct a coherent state $\ket{\alpha(t)}$ from the normalised wave packets defined in Eq.~(\ref{3Dwavepacket1}) in the manner described in Section \ref{Sec:wavepackets}.  With the help of Eqs.~(\ref{manyphoton}), (\ref{coherent1}), (\ref{3Dcommutator4}) and (\ref{3Dwavepacket1}), one may then show that
\begin{equation}
\label{3Dcoherent1}
a_{\mathbf{s}\lambda}(\mathbf{r},t)\ket{\alpha(t)} = \alpha\, \psi_{\mathbf{s}\lambda}(\mathbf{r})\ket{\alpha(t)}.
\end{equation}
Hence, the state $\ket{\alpha(t)}$ is an eigenstate of the blip annihilation operator.  It follows from the equation above that the expectation values of the (complex) electric and magnetic field observables, $\mathbf{E}(\mathbf{r},t)$ and $\mathbf{B}(\mathbf{r},t)$, calculated with respect to $\ket{\alpha(t)}$ are given by 
\begin{eqnarray}
\label{fieldexpectationvalues1}
\braket{\mathbf{E}(\mathbf{r},t)} &=& \alpha \sum_{\lambda = {\mathsf{H},\mathsf{V}}}\int_{S^2}\text{d}^2\mathbf{s}\;c\,\boldsymbol{\cal{R}}_{\mathbf{s}\lambda}\left[\psi_{\mathbf{s}\lambda}\right](\mathbf{r})\nonumber\\
\braket{\mathbf{B}(\mathbf{r},t)} &=& \alpha\sum_{\lambda = {\mathsf{H},\mathsf{V}}}\int_{S^2}\text{d}^2\mathbf{s}\;\boldsymbol{\cal{R}}_{\mathbf{s}\lambda}\left[\psi_{\mathbf{s}\lambda}\right](\mathbf{r}) \times \mathbf{s}.
\end{eqnarray}
In the above expectation values the notation $\boldsymbol{\cal{R}}_{\mathbf{s}\lambda}\left[\psi_{\mathbf{s}\lambda}\right](\mathbf{r})$ is shorthand for the quantity
\begin{equation}
\label{regularisedwavefunction1}
\boldsymbol{\cal{R}}_{\mathbf{s}\lambda}\left[\psi_{\mathbf{s}\lambda}\right](\mathbf{r}) = \int_{\mathbb{R}^3}\text{d}^3\mathbf{r}'\;\boldsymbol{\cal{R}}_{\mathbf{s}\lambda}(\mathbf{r}-\mathbf{r}')\,\psi_{\mathbf{s}\lambda}(\mathbf{r}').
\end{equation}

\subsubsection{Characteristics of physical wave packets}
\label{Sec:physicalstates}

The field observables (\ref{3Dfieldobservables1}) have been oriented such that the relations given in Eq.~(\ref{3Dfieldrelations1}) are satisfied.  Therefore, as all field and blip operators satisfy Eq.~(\ref{3Deom1}) (see Eq.~(\ref{3Deom2})), all field expectation values are guaranteed to satisfy Faraday's law and the Maxwell-Amp\`ere law provided that the field expectation values are divergence-less.  Hence, in order to ensure that the field expectation values obey Maxwell's equations, we must only determine for which wave functions Gauss's laws for electric and magnetic fields, Eqs.~(\ref{fMaxwell1}) and (\ref{fMaxwell2}), are satisfied.

With this in mind let us first calculate the divergence of the field expectation values given in Eq.~(\ref{fieldexpectationvalues1}).  Using the spatial independence of $\boldsymbol{\mathcal{R}}_{\mathbf{s}\lambda}$, one may show that Gauss's laws are satisfied only when the gradient of the wave function $\psi_{\mathbf{s}\lambda}(\mathbf{r})$ is oriented in a direction orthogonal to both $\boldsymbol{\mathcal{R}}_{\mathbf{s}\lambda}$ and $\boldsymbol{\mathcal{R}}_{\mathbf{s}\lambda}\times \mathbf{s}$.  This result is determined in App.~(\ref{Rorthogonaltopsi}).  Since $\boldsymbol{\mathcal{R}}_{\mathbf{s}\lambda}$ is orthogonal to $\mathbf{s}$, the gradient of the wave function must lie either parallel or anti-parallel to $\mathbf{s}$.
Referring to the coordinate system $(q_1, q_2, q_3)$ defined in Section \ref{Sec:blipmotion1},  the wave functions that correspond to solutions of Maxwell's equations depend only on the coordinate $q_1$, but not on either $q_2$ or $q_3$.  This result implies that not all choices of wave function represent physical solutions to Maxwell's equations and, therefore, that our blip Hilbert space is over-complete.  In the next section we shall determine a set of annihilation and creation operators for the physical single-photon states that we have found in this section.

\subsection{A complete Hilbert space}
\label{Sec:planewaves}

\subsubsection{Plane wave excitations}

\begin{figure}
\centering
\includegraphics[width = 0.55\textwidth]{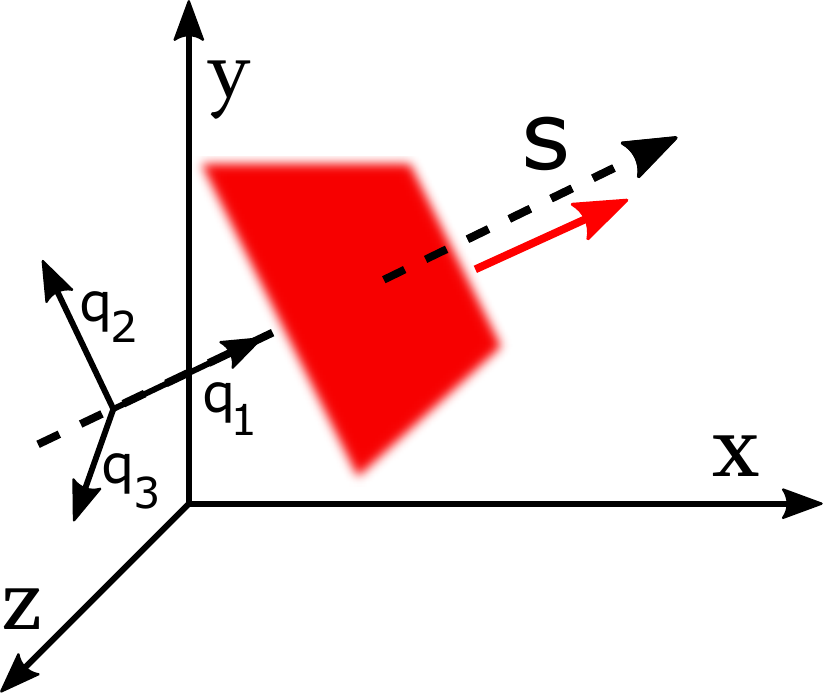}
\caption[A plane wave excitation]{The figure shows a single plane wave excitation propagating in the direction of $\mathbf{s}$.  The excitation is completely localised in the $q_1$ direction and completely delocalised in the $q_2$ and $q_3$ directions.  The plane propagates along the dotted trajectory at the speed of light in the direction of the red arrow without dispersion or rotation.}
\label{Fig:3Dblip2}
\end{figure}

Let us look again at the single-photon wave packet defined in Eq.~(\ref{3Dwavepacket1}). We now only consider the physical wave functions, which do not depend on either $q_2$ or $q_3$.  The single-photon wave function then takes the form
\begin{equation}
\label{3Dwavepacket2}
\ket{\psi(t)} = \sum_{\lambda = \mathsf{H},\mathsf{V}}\int_{S^2}\text{d}^2\mathbf{s}\int_{-\infty}^{\infty}\text{d}q_1\;\psi_{\mathbf{s}\lambda}(q_1)\ket{1_{\mathbf{s}\lambda}(q_1,t)}
\end{equation}
where
\begin{equation}
\label{3Dwavepacket3}
\ket{1_{\mathbf{s}\lambda}(q_1,t)} = a^\dagger_{\mathbf{s}\lambda}(q_1,t)\ket{0} =  \frac{1}{\sqrt{L_2L_3}}\int_{L_2}\text{d}q_2\int_{L_3}\text{d}q_3\;\,a^\dagger_{\mathbf{s}\lambda}(\mathbf{r},t)\ket{0}.
\end{equation}
In the above we must normalise the state $\ket{1_{\mathbf{s}\lambda}(q_1,t)}$ by the lengths $L_2$ and $L_3$ in the $q_2$ and $q_3$ directions in order to ensure that Eq.~(\ref{3Dnormalise1}) is still satisfied.  $L_2$ and $L_3$ are assumed to be very large.  The single-blip states defined in Eq.~(\ref{3Dwavepacket3}) are a superposition of blips characterised by a single $q_1$ for all values of $q_2$ and $q_3$.  Hence, such states represent plane waves that propagate at the speed of light along the $q_1$ axis, normal to the plane itself, in the direction of $\mathbf{s}$.  These states are illustrated in Fig.~\ref{Fig:3Dblip2}.  Plane wave states generated from the vacuum by the $a^\dagger_{\mathbf{s}\lambda}(q_1,t)$ operators are characterised by a polarisation $\lambda$, a direction of propagation $\mathbf{s}$ and a position along the $q_1$ axis. 

We have seen above that all physical single-photon states in our description can be constructed by superposing the plane wave excitations given in Eq.~(\ref{3Dwavepacket3}).  Although the localised blip excitations can also be used to construct all physical single-photon wave packets, they can be used to construct non-physical ones too.  For this reason our original Hilbert space was over-complete.  A better way of quantising the free EM field is to treat the plane wave excitations generated by the $a_{\mathbf{s}\lambda}(q_1,t)$ operators as the basic excitations of the free EM field.  This is in close analogy to the results in Chapter \ref{Chapter:1D}, where the basic excitations are plane waves propagating along the $x$ axis only.  Furthermore, in the classical theory, plane waves form a complete basis for the solutions of the vector wave equation \cite{Leo3}, which is satisfied by both the electric and magnetic fields. Later, in Section \ref{Sec:3Dlocalstates}, we shall see that there is an equivalent description of the reduced and complete Hilbert space in terms of localised excitations rather than plane waves.  Before we get to this, in the remainder of this section we shall examine some important properties of the plane wave excitations.

\subsubsection{A fundamental equation of motion}    

As both $q_2$ and $q_3$ are orthogonal to the direction of propagation, the blip equation of motion given in Eq.~(\ref{3Deom2}), which contains a derivative in the $q_1$ direction only, acts identically upon the plane wave excitations as it does upon the blip excitations.  We find, therefore, that 
\begin{equation}
\label{3Deom3}
\left[ c\mathbf{s}\cdot \nabla + \frac{\partial}{\partial t}\right]\,a_{\mathbf{s}\lambda}(q_1,t) = 0.
\end{equation} 
Analogous to the dynamical solutions of the one-dimensional excitations given in Eq.~(\ref{blipconstraint1}), the solutions of this equation are
\begin{equation}
\label{blipconstraint3}
a_{\mathbf{s}\lambda}(q_1,t) = a_{\mathbf{s}\lambda}(q_1-ct,0)
\end{equation}
In the above we must remember that the $q_1$ axis is oriented in the direction of $\mathbf{s}$ and is therefore reversed when the direction of propagation is reversed.  Both left- and right- propagating plane wave solutions are therefore present in Eq.~(\ref{blipconstraint3}).

\subsubsection{Commutation relations}

As we have done for other excitations, the commutation relation for plane wave excitations can be determined by calculating the inner product between two single-particle states.  Using Eqs.~(\ref{3Dcommutator4}) and (\ref{3Dwavepacket3}) one may show that the inner product between two plane wave excitations is given by
\begin{eqnarray}
\label{planewavecommutator1}
\braket{1_{\mathbf{s}\lambda}(q_1,t)|1_{\mathbf{s}'\lambda'}(q_1',t)} &=& \left[a_{\mathbf{s}\lambda}(q_1,t),a^\dagger_{\mathbf{s}'\lambda'}(q_1',t)\right]\nonumber\\
&=& \delta(q_1-q_1')\,\delta^2(\mathbf{s}-\mathbf{s}')\,\delta_{\lambda,\lambda'}.
\end{eqnarray}
This determines the plane wave commutation relation.  One may also show that, as usual, any creation or annihilation operator will commute with any other creation or annihilation operator respectively:
\begin{equation}
\label{planewavecommutator2}
\Big[a_{\mathbf{s}\lambda}(q_1,t),a_{\mathbf{s}'\lambda'}(q_1',t')\Big] = \Big[a^{\dagger}_{\mathbf{s}\lambda}(q_1,t),a^{\dagger}_{\mathbf{s}'\lambda'}(q_1',t')\Big] = 0. 
\end{equation}
for all $q_1$, $q'_1$, $\mathbf{s}$, $\mathbf{s}'$, $\lambda$, $\lambda'$, $t$ and $t'$.

\section{Localised states in three dimensions}

\label{Sec:3Dlocalstates}

In this section we show that there is an alternative and complete set of basis states for the free EM field that are localised, and can therefore replace plane waves as the principal excitations of the field.  We also introduce a set of vector-valued excitations that have the same dimensions as blips, but relate much more easily to the field observables which are distinct from the blips.

\subsection{Blips in the position representation}

\label{Sec:3Dlocalblips}

\subsubsection{Blip states}

\label{Sec:blipsin3D}

In the previous section we have shown that when light propagates in a fixed and constant direction, the basic excitations of the free EM field are plane waves.  Such excitations are not localised to a point, but, as the name suggests, localised to a plane and therefore completely delocalised in two out of three dimensions.  In the position representation of the EM field, we must be able to describe the fields at all points in space and time.  In order to make such a construction possible, however, it is necessary that the basic energy quanta of the field are point-like, and not stretched over a plane as the $a_{\mathbf{s}\lambda}(q_1,t)$ energy quanta are.  The purpose of this section is to show that there is an alternative complete basis of states in three dimensions that are completely localised in space.

Let us postulate the existence of localised energy quanta characterised by a position $\mathbf{r}$ and time $t$, a polarisation $\lambda$ and an additional discrete parameter $s = \pm 1$. Here we reduce the Hilbert space that was considered earlier on in this chapter by making the transition from $\mathbf{s} \in S^2$ to $s \in \{-1, +1\}$.  From the last section we saw that the plane wave excitations can still propagate in two directions along the $q_1$ axis.  In analogy to the one-dimensional blips, the parameter $s$ introduced here is intended to play a similar role and characterise the forwards and backwards motion of the plane.  In other words, a state characterised by a particular value $s$ behaves just like an identical state characterised by $-s$ if time is reversed.  As these particles are localised we shall refer to them as blips, and we will see later on that they provide a complete basis of the single-photon Hilbert space. We denote the annihilation operators for these excitations $a_{s\lambda}(\mathbf{r},t)$ and demand that they share a vacuum state with the blip operators defined in Section \ref{Sec:3DHilbertspace}.  Finally, we also assume that these excitations obey the wave equation (\ref{waveequation1}):
\begin{equation}
	\label{waveequation2}
	\left[\mathbf{\nabla}^2 - \frac{1}{c^2}\frac{\partial^2}{\partial t^2}\right] a_{s\lambda}(\mathbf{r},t) = 0.
\end{equation}

\subsubsection{Blip commutation relations}

\label{Sec:3Dblipcommutator1}

By applying the blip creation operators $a^\dagger_{s\lambda}(\mathbf{r},t)$ repeatedly to the vacuum state we can generate states containing any number of identical blips.  We denote the $n$-blip states 
\begin{equation}
\label{3Dstate3}
\ket{n_{s\lambda}(\mathbf{r},t)} = \frac{1}{\sqrt{n!}}\,a^\dagger_{s\lambda}(\mathbf{r},t)^n \,\ket{0}.
\end{equation}
When $n=1$ we refer to these states as single-blip states.  Since blips are localised, single-blip states defined at different positions at a fixed time are orthogonal to one another.  We also assume here that states characterised by different values of $s$ and $\lambda$ are distinguishable.  This implies that such states are also orthogonal to one another.  Hence, the single-blip inner product is given by
\begin{eqnarray}
\label{3Dcommutator5}
\braket{1_{s\lambda}(\mathbf{r},t)|1_{s'\lambda'}(\mathbf{r}',t)} = \delta^3(\mathbf{r}-\mathbf{r}')\,\delta_{s,s'}\,\delta_{\lambda,\lambda'}.
\end{eqnarray} 
Taking into account that the bosonic creation and annihilation operators commute with all other creation and annihilation operators respectively, the inner product defined above in Eq.~(\ref{3Dcommutator5}) implies the following commutation relations for the blip operators:
\begin{eqnarray}
\label{3Dcommutator6}
\Big[a_{s\lambda}(\mathbf{r},t),a^\dagger_{s'\lambda'}(\mathbf{r}',t)\Big] &=& \delta^3(\mathbf{r}-\mathbf{r}')\,\delta_{s,s'}\,\delta_{\lambda,\lambda'},\nonumber\\
\Big[a^{\dagger}_{s\lambda}(\mathbf{r},t),a^{\dagger}_{s'\lambda'}(\mathbf{r}',t')\Big] &=& 0,\nonumber\\
\Big[a_{s\lambda}(\mathbf{r},t),a_{s'\lambda'}(\mathbf{r}',t')\Big] &=& 0.
\end{eqnarray}
These expressions hold for all $\mathbf{r}$, $\mathbf{r}'$, $s$, $s'$, $\lambda$, $\lambda'$, $t$ and $t'$.

\subsection{Blips in the momentum representation}
\label{Sec:3Dmomentum}

\subsubsection{Photon states}

An alternative but equivalent representation of these local excitations can be found by expressing the blip operators in their Fourier representations as was done for the one-dimensional blips in Section \ref{Sec:photonHilbertspace}. By applying a Fourier transform to the blip operators $a_{s\lambda}(\mathbf{r},t)$ we obtain the operators $\widetilde{a}_{s\lambda}(\mathbf{k},t)$.  These are defined as
\begin{equation}
\label{3DFourier1}
\widetilde{a}_{s\lambda}(\mathbf{k},t) = \int_{\mathbb{R}^3}\frac{\text{d}^3\mathbf{r}}{(2\pi)^{3/2}}\;e^{-is\mathbf{k}\cdot\mathbf{r}}\,a_{s\lambda}(\mathbf{r},t).
\end{equation}
We assume in the following, and will see later on, that the $\widetilde{a}_{s\lambda}(\mathbf{k},t)$ operators represent the annihilation operators of monochromatic photons.  The inverse transformation is given by
\begin{equation}
\label{3DFourier2}
a_{s\lambda}(\mathbf{r},t) = \int_{\mathbb{R}^3}\frac{\text{d}^3\mathbf{k}}{(2\pi)^{3/2}}\;e^{is\mathbf{k}\cdot\mathbf{r}}\,\widetilde{a}_{s\lambda}(\mathbf{k},t).
\end{equation}
As in Section \ref{Sec:photonHilbertspace}, the parameter $s$ is included in the exponent of the Fourier transformation as a convenient convention so that by reversing the sign of $s$ we simultaneously invert the orientation of the wave vector.  Using Eq.~(\ref{3DFourier1}) we can see that the photon vacuum state coincides with the blip vacuum state. 

\subsubsection{Photon commutation relations}

Since there is a strict invertible relationship between the blip and photon operators given by Eqs.~(\ref{3DFourier1}) and (\ref{3DFourier2}), we can determine the commutation relations for the latter by direct calculation.  Using the blip commutation relations given in Eq.~(\ref{3Dcommutator6}), one finds that  
\begin{equation}
\label{3Dcommutator7}
\Big[\widetilde{a}_{s\lambda}(\mathbf{k},t),\widetilde{a}^\dagger_{s'\lambda'}(\mathbf{k}',t)\Big] = \delta^3(\mathbf{k}-\mathbf{k}')\,\delta_{s,s'}\,\delta_{\lambda,\lambda'}.
\end{equation}
One may also show that the photon creation and annihilation operators commute with all other creation and annihilation operators respectively:
\begin{equation}
\label{3Dcommutator8}
\Big[\widetilde{a}_{s\lambda}(\mathbf{k},t),\widetilde{a}_{s'\lambda'}(\mathbf{k}',t')\Big] = \Big[\widetilde{a}^\dagger_{s\lambda}(\mathbf{k},t),\widetilde{a}^\dagger_{s'\lambda'}(\mathbf{k}',t')\Big] =0
\end{equation}
for all $\mathbf{k}$, $\mathbf{k}'$, $s$, $s'$, $\lambda$, $\lambda'$, $t$ and $t'$.

Using Eq.~(\ref{3DFourier2}), we can now solve Eq.~(\ref{waveequation2}) which determines the time-dependence of the photon operators.  By substituting Eq.~(\ref{3DFourier2}) into Eq.~(\ref{waveequation2}) and solving the resulting equation one finds that
\begin{equation}
\label{3DFourier3}
a_{s\lambda}(\mathbf{r},t) = \int_{\mathbb{R}^3}\frac{\text{d}^3\mathbf{k}}{(2\pi)^{3/2}}\;e^{is\mathbf{k}\cdot\mathbf{r}-i\omega t}\,\widetilde{a}_{s\lambda}(\mathbf{k},0).
\end{equation} 
In the above $\omega$ is a real number satisfying the relation $\omega^2 = |\mathbf{k}|^2c^2$; however, $\omega$ must take both positive and negative values in order to ensure that the transformation $s \mapsto -s$ always reverses the dynamics of the blip. In Chapter $\ref{Chapter:1D}$, when the parameter $s$ is defined as the direction of propagation, there were only two ways in which the frequency $\omega$ could be continuously related to the wave vector $k$.  These were $\omega = \pm kc$.  In three dimensions, however, where single-photon wave packets can propagate in all sorts of directions, there are many different possible choices of how to relate $\omega$ to $\mathbf{k}$ and $s$, and different choices will be more suitable than others depending on the wave packet we are describing.

\subsection{A complete basis of blip states}

\label{Sec:blipstoplanes}

\subsubsection{The relation between plane waves and blips}

In this section we shall look at the relationship between the blip operators $a_{s\lambda}(\mathbf{r},t)$ introduced in Section \ref{Sec:3Dlocalblips} and the plane wave operators $a_{\mathbf{s}\lambda}(q_1,t)$ introduced in Section \ref{Sec:planewaves}.  Both excitations are characterised by an identical number of parameters which suggests that a relationship exists.  First, let us consider the following wave packet:
\begin{equation}
\label{planewavepacket1}
\ket{1_{s\lambda}(q_1,t)} = \int_{-\infty}^{\infty}\frac{\text{d}k_1}{\sqrt{2\pi}}\;e^{isk_1q_1}\,\widetilde{a}^\dagger_{s\lambda}(\mathbf{k},t)\ket{0}.
\end{equation} 
In the above we shall choose the coordinate $q_1$ such that its axis lies either parallel or anti-parallel to the vector $\mathbf{k}$ in such a way that  $\mathbf{k}\cdot \mathbf{r} = k_1q_1$ where $k_1$ takes all real values.  By applying the above definitions and substituting into Eq.~(\ref{planewavepacket1}) the time-dependence of the photon operator given in Eq.~(\ref{3DFourier3}), the state given in Eq.~(\ref{planewavepacket1}) simplifies to
\begin{equation}
\label{planewavepacket2}
\ket{1_{s\lambda}(q_1,t)} = \int_{-\infty}^{\infty}\frac{\text{d}k_1}{\sqrt{2\pi}}\; e^{isk_1(q_1-sct)}\,\widetilde{a}^\dagger_{s\lambda}(\mathbf{k},0)\ket{0}.
\end{equation}
In the above expression we have made the particular choice $\omega = k_1c$ which satisfies $\omega^2 = |\mathbf{k}|^2c^2$; although, as mention earlier, other choices are possible.  This wave packet contains only a single blip in a superposition over the region spanned by the coordinates $q_2$ and $q_3$ and has no dependence on either coordinate.  This region has the structure of a plane whose normal is parallel to the $q_1$ axis.  We can think of this wave packet, therefore, as a plane wave.  By noting the exponent in Eq.~(\ref{planewavepacket2}), one can see that the plane wave excitation above propagates along the $q_1$ axis in a direction determined by the parameter $s$.  This is the exact same dynamics that were displayed by the plane wave solutions in Section (\ref{Sec:planewaves}).  Moreover, in fact, when $\mathbf{s} = s\,\widehat{\mathbf{q}}_1$ where $\widehat{\mathbf{q}}_1$ is a unit vector oriented in the direction of increasing $q_1$, the operator $a_{s\lambda}(q_1,t)$ associated with the state in Eq.~(\ref{planewavepacket2}) satisfies Eq.~(\ref{3Deom2}).  Here $\widehat{\mathbf{q}}_1$ lies on the unit half-sphere.

\subsubsection{Plane wave commutation relations}

We have determined that we can construct plane wave operators $a_{s\lambda}(q_1,t)$ with the correct dynamics by superposing monochromatic photons over a single axis.  With the help of the photon commutation relations in Eq.~(\ref{3Dcommutator7}), we can also determine a set of commutation relations for the plane wave excitations constructed in Eq.~(\ref{planewavepacket1}).  To determine this commutation relation we first look at the inner product between the two plane wave excitations defined at equal times.  Using Eq.~(\ref{planewavepacket2}) we find that the equal-time inner product between two plane wave packets with possibly different normals is given by the expression
\begin{eqnarray}
\label{planewavepacket3}
\braket{1_{s\lambda}(q_1,t)|1_{s'\lambda'}(q_1',t)} &=& \frac{1}{2\pi}\int_{-\infty}^{\infty}\text{d}k_1\int_{-\infty}^{\infty}\text{d}k'_1\;e^{-i(sk_1q_1-s'k'_1q'_1)}\nonumber\\
&&\times\braket{0|\Big[\widetilde{a}_{s\lambda}(\mathbf{k},t), \widetilde{a}^\dagger_{s'\lambda'}(\mathbf{k}',t)\Big]|0}.
\end{eqnarray}
To continue we note that the photon commutator (\ref{3Dcommutator7}), which appears in Eq.~(\ref{planewavepacket3}), can be written in a slightly different form.  As the vector $\mathbf{k}$ satisfies $\mathbf{k} = k_1\widehat{\mathbf{q}}_1$, the photon commutator can be expressed as
\begin{eqnarray}
\label{alternatecommutator1}
\Big[\widetilde{a}_{s\lambda}(\mathbf{k},t),\widetilde{a}^\dagger_{s'\lambda'}(\mathbf{k}',t)\Big] &=& \delta^2(\widehat{\mathbf{q}}_1-\widehat{\mathbf{q}}'_1)\,\delta(k_1-k'_1)\,\delta_{s,s'}\,\delta_{\lambda,\lambda'}\nonumber\\
&=&	\delta^2(\mathbf{s}-\mathbf{s}')\,\delta(k_1-k'_1)\,\delta_{\lambda,\lambda'}.
\end{eqnarray}
In this final line we have used the definition $\mathbf{s} = s\hat{\mathbf{q}}_1$ from the previous section.  Here $\mathbf{s}$ is the direction of propagation.  As $\hat{\mathbf{q}}_1$ lies on the unit half-sphere, $\mathbf{s}$ lies on the unit sphere.  By substituting Eq.~(\ref{alternatecommutator1}) into Eq.~(\ref{planewavepacket3}) one finds that 
\begin{eqnarray}
\label{alternatecommutator2}
\Big[a_{s\lambda}(q_1,t), a^\dagger_{s'\lambda'}(q_1',t)\Big] &=& \int_{-\infty}^{\infty}\frac{\text{d}k_1}{2\pi}\;e^{-isk_1(q_1-q_1')}\delta^2(\mathbf{s}-\mathbf{s}')\,\delta_{\lambda,\lambda'}\nonumber\\
&=& \delta_s(q_1-q_1')\,\delta^2(\mathbf{s}-\mathbf{s}')\,\delta_{\lambda,\lambda'}. 
\end{eqnarray} 
The commutator coincides exactly with that for the plane wave quanta in Section {\ref{Sec:planewaves}}.  Finally, let us add for completeness that planes wave annihilation and creation operators, commute amongst themselves:
\begin{equation}
\label{alternativecommutator3}
\Big[a^{\dagger}_{s\lambda}(q_1,t), a^{\dagger}_{s'\lambda'}(q_1',t')\Big] = \Big[a_{s\lambda}(q_1,t), a_{s'\lambda'}(q_1',t')\Big] = 0
\end{equation}
for all $q_1$, $q_1'$, $s$, $s'$, $\lambda$, $\lambda'$, $t$ and $t'$ in complete analogy to Eq.~(\ref{blipcommutator1}).

\subsubsection{Blips as a complete basis}

The localised blips defined in Section \ref{Sec:3Dlocalblips} can be used to construct single-photon plane wave states that have a well-defined direction of propagation in addition to a polarisation about and position along the photon's trajectory.  These states were constructed in Eq.~\ref{planewavepacket1}.  What is more, the plane wave operators constructed in Section \ref{Sec:blipstoplanes} are characterised by an identical set of parameters, an identical set of commutation relations, and governed by the same equation of motion as the plane wave operators defined in Section \ref{Sec:planewaves}.  Both excitations also share a vacuum state.  As the commutation relations define the Hilbert space of our system, we find that the Hilbert spaces of the two systems are identical, and governed by the same dynamical Hamiltonian.  The main result of this section, then, is that the system of localised blips defined in Section \ref{Sec:blipsin3D}, or alternatively the system of monochromatic photons defined in Section \ref{Sec:3Dmomentum}, provide a complete description of the free quantised EM field in three dimensions.    

Unlike the plane wave energy quanta, blips are localised to a point.  We are now in a position to quantise the free EM field in the position representation.  This will be the topic of Section \ref{Sec:3Dcomplete} of this Chapter.  In the standard theory of the quantised EM field, photon states could be characterised at most by a position and a polarisation.  In this section we have introduced an additional parameter $s = \pm 1$, thus doubling the usual Hilbert space.  In one-dimension, this additional parameter could be interpreted as a direction of propagation.  In three dimensions, however, this is not necessarily the case.  Here $s$ characterises states that are the time-reversal of each other.  As discussed in Section \ref{Sec:3Dmomentum}, there are many different ways to introduce the parameter $s$; for instance, it can be interpreted as the direction of propagation of a plane wave.  What this parameter implies, however, is that for every particle in the Hilbert space, the time-reversal of that particle is also a member of the Hilbert space.  Moreover, these excitations are orthogonal to each other. 

Finally, we would like to mention that it is perhaps not entirely surprising that the plane wave excitations form a complete basis of states.  In Chapter \ref{Chapter:1D}, the one-dimensional blip states were localised to a point along the $x$ axis only.  These excitations are essentially also plane wave solutions as they are totally delocalised in the $y$-$z$ plane.  We can see therefore that the results of this section are consistent with the results of Chapter \ref{Chapter:1D}, the difference being that in one dimension we only consider propagation along the $x$ axis; in three dimensions we consider propagation in all directions.

\subsection{Vector excitations}

\label{Sec:vectorexcitations}
\subsubsection{Vector operators}

\label{Sec:vectorblips}

Blips form a complete basis of Hilbert space and are perfectly localised in space.  We need not go any further; we have what we are looking for, and can construct all possible states from these blip excitations.  For practical reasons, however, in this section we shall introduce one more type of excitation: one that is vector valued.

\begin{figure}
\centering
\includegraphics[width = 0.4\textwidth]{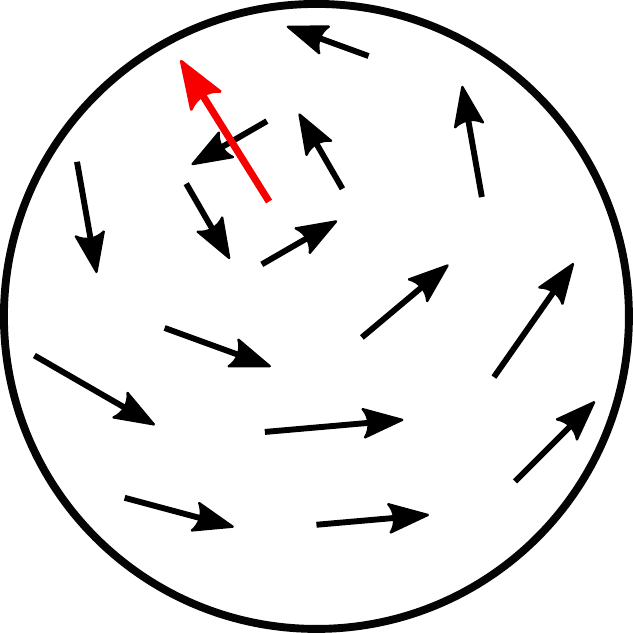}
\caption[The ``Hairy ball" theorem]{For any continuous vector field tangent to the surface of a sphere there must be at least one point where the vector field vanishes.  This is known as the ``Hairy-ball" theorem.  The red arrow in the figure indicates the point at which the vector field disappears. As the EM field is oriented tangentially to its direction of propagation, it is impossible to construct continuous and spherically symmetric solutions to Maxwell's equations.}
\label{Fig:Hairyball}
\end{figure}

The original reason for constructing blips was to describe the quantised solutions of Maxwell's equations.  The solutions of Maxwell's equations are vector fields; the states generated by blips, on the other hand, are not.  In general the orientation of the field will have an effect on the type of wave packets that can be constructed and their resulting dynamics, particularly if the orientation of the fields is not position-independent.  For example, according to the so-called ``Hairy Ball Theorem," perfectly isotropic electromagnetic spherical waves do not exist (see Fig.~\ref{Fig:Hairyball}).  On the other hand, blip states localised at a position $\mathbf{r}$ can be superposed into a sphere without difficulty.  For this reason, it becomes difficult to relate the vector-valued field observables to the scalar-valued blip operators that we have considered so far.

In general, we may wish to consider wave packets that have different structures and propagate in position-dependent directions. To accommodate these possibilities, and more closely relate the field observables and particle excitations, in this section we shall use the photon operators defined in Section \ref{Sec:3Dmomentum} to construct vector-valued blip operators.  This particular choice of notation results in simpler expressions for the field observables later on in this chapter.  The purpose of this section is to introduce tools that will help us when applying this formalism to real applications in future.

Let us introduce a new vector-valued annihilation operator $\boldsymbol{a}_{s}(\mathbf{r},t)$.  In terms of the photon operators $\widetilde{a}_{s\lambda}(\mathbf{k},t)$, the vector operator $\boldsymbol{a}_{s}(\mathbf{r},t)$ is given by the formula
\begin{equation}
\label{vectorblip1}
\boldsymbol{a}_{s}(\mathbf{r},t) = \sum_{\lambda = {\mathsf{H},\mathsf{V}}}\int_{\mathbb{R}^3}\frac{\text{d}^3\mathbf{k}}{(2\pi)^{3/2}}\;e^{i(s\mathbf{k}\cdot\mathbf{r}-\omega t)}\,\widetilde{a}_{s\lambda}(\mathbf{k})\,\boldsymbol{e}_{s\lambda}(\mathbf{k}).
\end{equation}
Here the vectors $\boldsymbol{e}_{s\mathsf{H}}(\mathbf{k})$ and $\boldsymbol{e}_{s\mathsf{V}}(\mathbf{k})$ are two real polarisation vectors orthogonal both to each other and the vector $\mathbf{k}$ for any $\mathbf{k}$, $s$ and $\lambda$.  These operators are closely related to the electric field observable; for instance, one may show that $\boldsymbol{a}_{s}(\mathbf{k},t)$ is, like the electric field, divergence-less: $\mathbf{\nabla}\cdot\boldsymbol{a}_{s}(\mathbf{k},t) = 0$.  To specify the polarisation vectors more completely we demand that the unit vectors $(\widehat{\mathbf{q}}_1,\boldsymbol{e}_{s\mathsf{H}}(\mathbf{k}), \boldsymbol{e}_{s\mathsf{V}}(\mathbf{k}))$ form a right-handed coordinate system.  The horizontally and vertically polarised components of the vector annihilation operator are given by the localised blip operators $a_{s\mathsf{H}}(\mathbf{r},t)$ and $a_{s\mathsf{V}}(\mathbf{r},t)$ respectively.

\subsubsection{A fundamental equation of motion}

\label{Sec:vectoreom}

From Section \ref{Sec:blipsin3D}, we know that the blip operators $a_{s\lambda}(\mathbf{r},t)$ satisfy the second order wave equation for waves propagating at the speed of light (see Eq.~(\ref{waveequation2})).  This equation is unchanged by the transformation $s \mapsto -s$, and therefore does not distinguish a difference in the dynamics of light characterised by excitations with different values of $s$.  As the $\boldsymbol{a}_{s}(\mathbf{r},t)$ operators are closely related to the electric field observable, we can introduce another vector-valued operator $\boldsymbol{b}_{s}(\mathbf{r},t)$ that is closely related to the magnetic field observable.  We define the $\boldsymbol{b}_{s}(\mathbf{r},t)$ operators to be those satisfying the following equation of motion:
\begin{equation}
\label{vectoreom1}
\mathbf{\nabla}\times \boldsymbol{a}_{s}(\mathbf{r},t) = -\frac{s}{c}\frac{\partial}{\partial t}\,\boldsymbol{b}_s(\mathbf{r},t).
\end{equation} 
Unlike the wave equation, this equation contains only first-order derivatives with respect to time and the role of the parameter $s$ can be seen.  The relationship between $\boldsymbol{a}_{s}(\mathbf{r},t)$ and $\boldsymbol{b}_{s}(\mathbf{r},t)$ given in Eq.~(\ref{vectoreom1}) can also be represented by means of a position-independent superoperator $\chi$ where $\boldsymbol{b}_{s}(\mathbf{r},t) = -i\chi [\boldsymbol{a}_{s}](\mathbf{r},t)$.  The operator $\chi$ acts on the vector operators in a distributional sense just like the regularisation operator in Eq.~(\ref{Rsuperposition1}).  The operator $\chi$, however, mixes together the different components of the vector operators.  In Ref.~\cite{BB6}, $\chi$ is referred to as the helicity operator.  By calculating the divergence of both sides of Eq.~(\ref{vectoreom1}) one can show that, like the $\boldsymbol{a}_{s}(\mathbf{r},t)$ operators, the $\boldsymbol{b}_{s}(\mathbf{r},t)$ operators are divergence-less too.

Although different from the $\boldsymbol{a}_{s}(\mathbf{r},t)$ operators we should bear in mind that the $\boldsymbol{b}_{s}(\mathbf{r},t)$ operators are not independent of the $\boldsymbol{a}_{s}(\mathbf{r},t)$ operators.  Both of these operators can be represented in terms of the $a_{s\lambda}(\mathbf{r},t)$ blip operators.  Consequently, any state containing $\boldsymbol{a}_{s}(\mathbf{r},t)$ energy quanta also contains $\boldsymbol{b}_{s}(\mathbf{r},t)$ energy quanta.  Any light pulse containing an electric component also contains a magnetic component.  In particular, whereas the horizontally and vertically polarised components of $\boldsymbol{a}_{s}(\mathbf{r},t)$ are superpositions of horizontally and vertically polarised blips respectively, the horizontally and vertically polarised components of the $\boldsymbol{b}_{s}(\mathbf{r},t)$ operators are superpositions of vertically and horizontally polarised blips respectively.  In addition, by calculating the curl of Eq.~(\ref{vectoreom1}) and using Eq.~(\ref{waveequation2}) one can show that $\boldsymbol{a}_{s}(\mathbf{r},t)$ and $\boldsymbol{b}_{s}(\mathbf{r},t)$ satisfy the equation
\begin{equation}
\label{vectoreom2}
\mathbf{\nabla}\times \boldsymbol{b}_s(\mathbf{r},t) = \frac{s}{c} \frac{\partial}{\partial t}\,\boldsymbol{a}_{s}(\mathbf{r},t).
\end{equation}
Using the $\chi$ notation, the above equation may be written $\boldsymbol{a}_{s}(\mathbf{r},t) = i\chi[\boldsymbol{b}_s](\mathbf{r},t)$.  Hence together Eqs.~(\ref{vectoreom1}) and (\ref{vectoreom2}) form a closed set of first-order equations for the $\boldsymbol{a}_{s}(\mathbf{r},t)$ and $\boldsymbol{b}_{s}(\mathbf{r},t)$ operators and $\chi^2 = -1$.  We shall use these equations in Section \ref{Sec:dynHamiltonian3} to construct a dynamical Hamiltonian in the position representation.

For a given single-blip state $\ket{\psi(t)}$ we may define the following vector wave functions:
\begin{eqnarray}
\boldsymbol{\psi}^{E}_{s}(\mathbf{r},t) &=& \braket{0|\boldsymbol{a}_{s}(\mathbf{r},0)|\psi(t)}\nonumber\\
\boldsymbol{\psi}^{B}_{s}(\mathbf{r},t) &=& \braket{0|\boldsymbol{b}_{s}(\mathbf{r},0)|\psi(t)}.
\end{eqnarray}
We may refer to $\boldsymbol{\psi}^{E}_s(\mathbf{r},t)$ and $\boldsymbol{\psi}^{B}_s(\mathbf{r},t)$ as the electric and magnetic wave functions respectively.  Both are divergence-less and represent the probability distribution of a particular electric and magnetic field configuration.

\subsubsection{Are blips local?}

The blip creation operators $a^\dagger_{s\lambda}(\mathbf{r},t)$ defined at different positions are orthogonal to one another and therefore define perfectly localised particles.  One the other hand, no two components $a_{si}(\mathbf{r},t)$ or $b_{si}(\mathbf{r},t)$ of the $\boldsymbol{a}_{s}(\mathbf{r},t)$ and $\boldsymbol{b}_{s}(\mathbf{r},t)$ operators (where $i = 1,2,3$ refers to a component in a constant Cartesian basis) possess a bosonic commutation relation.  This result can be shown using Eqs.~(\ref{vectorblip1}) and \ref{vectoreom1}.  It seems, therefore, that the components of the vector operators cannot be localised in space.  Does this mean that photons cannot be localised and that we are unable to give an exact description of photons in the position representation?  The answer is emphatically ``no."  

The components of the $\boldsymbol{a}_{s}(\mathbf{r},t)$ and $\boldsymbol{b}_{s}(\mathbf{r},t)$ operators do not have a local bosonic commutation relation precisely because the operators possess only transverse polarisations.  It is sometimes thought that this implies that the photon cannot be localised.  In this quantisation, however, the principle excitations of the electromagnetic field in three-dimensions are the blip operators $a_{s\lambda}(\mathbf{r},t)$ which do have a local commutation relation as all momenta are taken into account for each polarisation (see Eq.~(\ref{3Dcommutator6})).  This is reminiscent of the position operator introduced by Hawton \cite{Haw10} which takes into consideration longitudinally as well as transversely polarised field components.  

By representing any quantum state in a basis of blip states, we can define its position-dependent wave function.  The vector excitations $\boldsymbol{a}_{s}(\mathbf{r},t)$ and $\boldsymbol{b}_{s}(\mathbf{r},t)$ are non-local, but the position of these non-local excitations is well-defined.  The non-local commutation relations for the $\boldsymbol{a}_{s}(\mathbf{r},t)$ and $\boldsymbol{b}_{s}(\mathbf{r},t)$ operators therefore do not represent a quantum problem founded in an apparent inability to localise the photon; it is a classical problem brought about by the fact that electromagnetic waves do not form themselves into point-like objects. In this sense, the energy quanta of the field behave very much like the classical solutions of Maxwell's equations.  Shortly we shall see that the EM field observables are closely related to the vector excitations.

\section{The EM field observables}

\label{Sec:3Dcomplete}

Hitherto we have constructed a complete Hilbert space for the carrier particles of the quantised EM field in position space.  These are the localised blips, which are characterised by a position $\mathbf{r}$, a time $t$, a polarisation $\lambda$ and an additional discrete parameter $s$.  Furthermore, we have shown that these blip excitations can be used to construct vector-valued operators that, like the electric and magnetic field observables, are divergence-less and obey the wave equation for vector fields.  In this next section we shall complete the development of the three-dimensional theory of the free EM field by constructing the electromagnetic field observables that act on this Hilbert space and a dynamical Hamiltonian that determines the evolution of the blips.

\subsection{Observables in the position representation}

\label{Sec:completefields}

\subsubsection{Field observables in the position representation}

Consistent with the field observables given in Eq.~(\ref{3Dfieldobservables1}) and the vector operator defined in Eq.~(\ref{vectorblip1}), we assume that the complex electric and magnetic field observables are a linear superposition of the $\boldsymbol{a}_{s}(\mathbf{r},t)$ and $\boldsymbol{b}_{s}(\mathbf{r},t)$ operators respectively:
\begin{eqnarray}
\label{3Dfieldobservables2}
\mathbf{E}(\mathbf{r},t) &=& \sum_{s = \pm 1} \, c\,\mathcal{R}\left[\boldsymbol{a}_s\right](\mathbf{r},t)\nonumber\\
\mathbf{B}(\mathbf{r},t) &=& \sum_{s = \pm 1} \, s\, \mathcal{R}\left[\boldsymbol{b}_{s}\right](\mathbf{r},t).
\end{eqnarray} 
Here as in previous definitions of the field observables, the blip operators are related to the field observables by a position-independent regularisation operator $\mathcal{R}$.  Using the equations of motion (\ref{vectoreom1}) and (\ref{vectoreom2}) for the vector operators, it can be shown that the field observables defined in Eq.~(\ref{3Dfieldobservables2}) satisfy all four of Maxwell's equations (\ref{fMaxwell1})-(\ref{fMaxwell4}).

\subsubsection{Field observables in the momentum representation}

The field observables in the momentum representation are given by the expressions
\begin{eqnarray}
\label{3Dfieldobservables3}
\hspace*{-0.6cm}
\mathbf{E}(\mathbf{r},t) &=& \sum_{s = \pm 1}\sum_{\lambda = {\mathsf{H},\mathsf{V}}}\int_{\mathbb{R}^3}\frac{\text{d}^3\mathbf{k}}{(2\pi)^{3/2}}\; c\,\Omega(\mathbf{k})e^{i(s\mathbf{k}\cdot\mathbf{r}+\varphi(k))}\,\widetilde{a}_{s\lambda}(\mathbf{k},t)\,\boldsymbol{e}_{s\lambda}(\mathbf{k},t)\nonumber\\
\hspace*{-0.6cm}
\mathbf{B}(\mathbf{r},t) &=& \sum_{s = \pm 1}\sum_{\lambda = {\mathsf{H},\mathsf{V}}}\int_{\mathbb{R}^3}\frac{\text{d}^3\mathbf{k}}{(2\pi)^{3/2}}\; s\,\frac{\Omega(\mathbf{k})}{\omega}e^{i(s\mathbf{k}\cdot\mathbf{r}+\varphi(k))}\,\widetilde{a}_{s\lambda}(\mathbf{k},t)\,\left(\mathbf{k}\times\boldsymbol{e}_{s\lambda}(\mathbf{k},t)\right).
\end{eqnarray} 
This is in agreement with Eqs.~(\ref{vectorblip1}) and (\ref{vectoreom1}) and the expressions for the field observables in Eq.~(\ref{3Dfieldobservables2}).

\subsubsection{The regularisation operator}

Unlike the regularisation operator defined in Section \ref{Sec:3Dposfields}, in Eq.~(\ref{3Dfieldobservables2}) the regularisation operator is not vector valued.  Instead, it acts on the vector blip operators to produce the also vector-valued electric and magnetic field observables.  The regularisation operator can be understood again in a distributional sense such that
\begin{eqnarray}
\label{Regularisationofvectors}
\mathcal{R}\left[\boldsymbol{a}_s\right](\mathbf{r},t) = \int_{\mathbb{R}^3}\text{d}^3\mathbf{r}'\;\mathcal{R}(\mathbf{r}-\mathbf{r}')\,\boldsymbol{a}_{s}(\mathbf{r}',t).
\end{eqnarray}
Due to the isotropy of free space, the regularisation operator $\mathcal{R}(\mathbf{r}-\mathbf{r}')$ is independent of $\mathbf{s}$ and $\lambda$, and symmetric in its two arguments: $\mathcal{R}(\mathbf{r}-\mathbf{r}') = \mathcal{R}(\mathbf{r}'-\mathbf{r})$.

\subsection{The energy observable}

\label{Sec:3Denergy}

\subsubsection{Energy in the position representation}

Now that we have expressions for the electric and magnetic field observables, by substituting these expressions into the classical energy (\ref{classicalenergy4}) we can determine an expression for the energy observable of the system.  After carrying out this substitution one finds that
\begin{eqnarray}
\label{energyobservable3D1}
H_{\text{energy}}(t) &=& \sum_{s = \pm 1}\int_{\mathbb{R}^3}\text{d}^3\mathbf{r}\;\frac{\varepsilon_0c^2}{8}\nonumber\\
&& \hspace{-1cm} \times\left\{\|\mathcal{R}[\boldsymbol{a}_{s}](\mathbf{r},t) + \text{H.c.}\|^2  + \|\mathcal{R}[\boldsymbol{b}_{s}](\mathbf{r},t) + \text{H.c.}\|^2\right\}.
\end{eqnarray}

Due to the complicated commutation relations between the vector blip operators, this operator is not in a form that is of much practical use to us. Fortunately we may simplify this expression further.  We should recall that the horizontally and vertically polarised components of the $\boldsymbol{a}_s(\mathbf{r},t)$ operators are given by the horizontally and vertically polarised blip operators respectively. Consequently, in Eq.~(\ref{energyobservable3D1}), the product $\boldsymbol{a}_{s}(\mathbf{r},t)\cdot\boldsymbol{a}_{s'}(\mathbf{r},t)$ is given by the expression $\sum_{\lambda = \mathsf{H},\mathsf{V}} a_{s\lambda}(\mathbf{r},t)a_{s'\lambda}(\mathbf{r},t)$ where $a_{s\lambda}(\mathbf{r},t)$ are the usual local blip operators. Moreover, using the property $\chi^2 = -1$ it may be shown that the total energy of the magnetic field is equal to the total energy of the electric field. Putting both of these observations together one finds that the energy observable can be written in the simpler form
\begin{equation}
\label{energyobservable3D2}
H_{\text{energy}}(t) = \sum_{s = \pm 1}\sum_{\lambda = {\mathsf{H},\mathsf{V}}}\frac{\varepsilon_0 c^2}{4}\int_{\mathbb{R}^3}\text{d}^3\mathbf{r}\;\left\{\mathcal{R}\left[a_{s\lambda}\right](\mathbf{r},t) + \text{H.c.}\right\}^2.
\end{equation}
This expression is entirely analogous to the energy observable derived for the one dimensional fields in Eq.~(\ref{Heng1}).  Like the expression for the energy observable in Eq.~(\ref{Heng1}), the above energy observable is always positive, which is ensured by the square in the integrand.

\subsubsection{The regularisation operator revisited}

As in one dimension, the regularisation operator plays an important role in specifying the energy of a blip excitation.  In order to determine the exact form of the energy observable, we must now determine what effect the regularisation operator has on a blip, or in other words, what is the exact expression for the distribution $\mathcal{R}(\mathbf{r}-\mathbf{r}')$.  In Chapter \ref{Chapter:1D}, the regularisation operator was determined by ensuring that the field observables transformed correctly under Lorentz boosts.  We proceed analogously in this section and first determine the effect of the regularisation operator on the photon operator $\widetilde{a}_{s\lambda}(\mathbf{k},t)$.  The distribution $\mathcal{R}(\mathbf{r}-\mathbf{r}')$ is then determined by Fourier transforming any multiplicative factor gained in the regularisation of the photon operators.

Consider again a pure Lorentz boost in a given direction denoted by the Greek letter $\Lambda$.  This transformation shifts all momenta $\mathbf{k}$ to the new momenta $\mathbf{p}$. As before, the corresponding transformation on the states of the system is implemented by the unitary transformation $U(\Lambda)$ which ensures that all transition amplitudes are invariant under the change of reference frame.  Using the Lorentz invariant measure $\text{d}^3\mathbf{k}/|\mathbf{k}|$ and the photon commutation relation given in Eq.~(\ref{3Dcommutator7}), one can show that the inner product between two normalised photon wave packets is both Lorentz- and form-invariant only when
\begin{equation}
U(\Lambda)\,\widetilde{a}_{s\lambda}(\mathbf{k},0)\,U^\dagger(\Lambda) = \sqrt{\frac{|\mathbf{p}|}{|\mathbf{k}|}}\,\widetilde{a}_{s\lambda}(\mathbf{p})
\end{equation}
up to a unitary rotation of the polarisation states.  I shall ignore this rotation in the above transformation as the anti-symmetric field tensor transforms correctly to accommodate this rotation and we are only interested in the overall normalisation of the field.  See Section 5.9 of Ref.~\cite{Wei} for more details.  Using this result one can again show that the fields transform correctly only when the regularised photon operator is given by 
\begin{equation}
\label{3Domega}
\mathcal{R}[\widetilde{a}_{s\lambda}](\mathbf{k},t) = \Omega_0\,\sqrt{|\mathbf{k}|}\,\widetilde{a}_{s\lambda}(\mathbf{k},t)
\end{equation}
for some real and constant numerical factor $\Omega_0$. See again Appendix.~\ref{App:Omegaproof} for more details. 

As we shall see in Section~\ref{Sec:dynHamiltonian3}, when restricted to positive-frequency excitations the energy observable and dynamical Hamiltonian coincide when $\Omega_0$ is given by
\begin{equation}
\Omega_0 = \sqrt{\frac{2\hbar}{\varepsilon_0 c}}.
\end{equation}
For this choice of $\Omega_0$ the energy expectation value of a single-photon state in the $(\mathbf{k}, s,\lambda)$ mode is $\hbar |\mathbf{k}| c$.  Moreover, since the regularised blip operator is just the Fourier transform of the regularised photon operator we can determine that 
\begin{equation}
\label{3Dregularisation1}
\mathcal{R}(\mathbf{r}-\mathbf{r}') = \int_{\mathbb{R}^3}\text{d}^3\mathbf{k}\;(2\pi)^{-3}\,e^{i\mathbf{k}\cdot(\mathbf{r}-\mathbf{r}')}\,\sqrt{\frac{2\hbar|\mathbf{k}|}{\varepsilon_0 c}} = -\frac{3}{8\pi^{3/2}}\sqrt{\frac{\hbar}{c\varepsilon_0}}\frac{1}{|\mathbf{r}-\mathbf{
r}'|^{7/2}}.
\end{equation}
The final equality sign has been determined, for example, in Refs.~\cite{Amr, MW} after an exponential regularisation has been used.  The result above again shows that in three dimensions the field observables are non-locally related to the blip operators.  Given Eq.~(\ref{3Dregularisation1}), we have completely determined the field observables and the energy observable in the position representation.

\subsection{The dynamical Hamiltonian}

\label{Sec:dynHamiltonian3}

\subsubsection{An equation of motion for blips}

The two equations of motion (\ref{vectoreom1}) and (\ref{vectoreom2}) describe completely the dynamics of the blip operators in three-dimensions.  As these equations are first-order in time they provide a suitable starting point for our determining a dynamical Hamiltonian that satisfies Heisenberg's equation.  The commutation relations for the vector operators $\boldsymbol{a}_{s}(\mathbf{r},t)$ and $\boldsymbol{b}_{s}(\mathbf{r},t)$ are not simple and it is more appropriate therefore for us to express the dynamical Hamiltonian in terms of the locally bosonic blip operators $a_{s\lambda}(\mathbf{r},t)$.  Hence, to continue, we must find a way to express Eq.~(\ref{vectoreom1}) in terms of the local blip operators.  

In Eqs.~(\ref{vectoreom1}) and (\ref{vectoreom2}) there are two different types of operator: there are the $\boldsymbol{a}_{s}(\mathbf{r},t)$ operators, and there are the $\boldsymbol{b}_{s}(\mathbf{r},t)$ operators.  By comparing Eqs.~(\ref{vectorblip1}) and (\ref{3DFourier2}) one can see that in the $(q_1, q_2, q_3)$ basis the $\boldsymbol{a}_{s}(\mathbf{r},t)$ have components\\ $(0, a_{s\mathsf{H}}(\mathbf{r},t), a_{s\mathsf{V}}(\mathbf{r},t))$.  Hence, it is straightforward to represent the $\boldsymbol{a}_{s}(\mathbf{r},t)$  operators in terms of blip operators.  The $\boldsymbol{b}_{s}(\mathbf{r},t)$ operators on the other hand are defined entirely by the equation Eq.~(\ref{vectoreom1}), or, equivalently, related to the $\boldsymbol{a}_{s}(\mathbf{r},t)$ operators by a helicity operator $\chi$. This relation is not necessarily local as it depends upon the relationship between the wave vector $\mathbf{k}$ and the frequency $\omega$ for which there are many different choices.  I shall proceed as in Section \ref{Sec:blipstoplanes} where the direction of propagation is indicated by the vector $\mathbf{s} = s \widehat{\mathbf{q}}_1$.  As the electric and magnetic fields are therefore related by a rotation about the $q_1$ axis, in the $(q_1, q_2, q_3)$ basis the components of the  $\boldsymbol{a}_s(\mathbf{r},t)$ and $\boldsymbol{b}_{s}(\mathbf{r},t)$ operators are related by a rotation.  In particular, using the right hand rule and the expressions for the field observables given in Eq.~(\ref{3Dfieldobservables2}), the components of the $\boldsymbol{b}_s(\mathbf{r},t)$ operators are given by $(0, -a_{s\mathsf{V}}(\mathbf{r},t), a_{s\mathsf{H}}(\mathbf{r},t))$.

By substituting the above expressions for the vector operators back into Eq.~(\ref{vectoreom1}) one finds that
\begin{equation}
\label{vectoreom3}
\mathbf{\nabla} \times 
\begin{pmatrix}
0 \\ a_{s\mathsf{H}}(\mathbf{r},t) \\ a_{s\mathsf{V}}(\mathbf{r},t)
\end{pmatrix}
 = -\frac{s}{c}\frac{\partial}{\partial t}
\begin{pmatrix}
0 \\ - a_{s\mathsf{V}}(\mathbf{r},t) \\ a_{s\mathsf{H}}(\mathbf{r},t).
\end{pmatrix}
\end{equation}
This equation can be solved component-wise by remembering that $(\widehat{\mathbf{q}}_1, \boldsymbol{e}_\mathsf{H}, \boldsymbol{e}_\mathsf{V})$ forms a right-handed coordinate system and that the blip operators depend on the $q_1$ coordinate only.  The above equation (\ref{vectoreom3}) then reduces to the simpler equation
\begin{equation}
\label{vectoreom4}
\frac{\partial}{\partial t}a_{s\lambda}(\mathbf{r},t) = -sc\frac{\partial}{\partial q_1}a_{s\lambda}(\mathbf{r},t).
\end{equation}
This equation is completely analogous to the equation of motion (\ref{blipmotion1}) for light propagating in one dimension.  In Eq.~(\ref{blipmotion1}), the space derivative is with respect to the $x$ coordinate axis.  Here the space derivative is with respect to the $q_1$ axis lying either parallel or antiparallel to the direction of propagation.  This equation is general when the direction of propagation is given by $\mathbf{s} = s \widehat{\mathbf{q}}_1$.

\subsubsection{The dynamical Hamiltonian}

By using the equation of motion (\ref{vectoreom4}) for localised blip excitations we can now construct a dynamical Hamiltonian for light in three dimensions.  We proceed as before by assuming that the dynamical Hamiltonian is an exchange Hamiltonian for blips.  This operator takes the form
\begin{eqnarray}
\label{Hdyn3D1}
H_{\text{dyn}}(t) &=& i \sum_{s = \pm 1}\sum_{\lambda=\mathsf{H},\mathsf{V}}\int_{\mathbb{R}^3}\text{d}^3\mathbf{r}'\int_{\mathbb{R}^3}\text{d}^3\mathbf{r}''\; \hbar s c\,f_{s\lambda}(\mathbf{r}',\mathbf{r}'')\nonumber\\
&& \times a^\dagger_{s\lambda}(\mathbf{r}',t)a_{s\lambda}(\mathbf{r}'',t).
\end{eqnarray}
In the above $f_{s\lambda}(\mathbf{r}',\mathbf{r}'')$ is complex.  To determine $f_{s\lambda}(\mathbf{r}',\mathbf{r}'')$ we use Eq.~(\ref{Hdyn3D1}) to calculate the time evolution of a blip operator and compare the result with the blip equation of motion: in this case Eq.~(\ref{vectoreom4}).  Using Heisenberg's equation and the blip commutation relations given in Eq.~(\ref{3Dcommutator6}) we find that
\begin{equation}
\frac{\text{d}}{\text{d}t}a_{s\lambda}(\mathbf{r},t) = sc\int_{\mathbb{R}^3}\text{d}^3\mathbf{r}''\;f_{s\lambda}(\mathbf{r},\mathbf{r}'')\,a_{s\lambda}(\mathbf{r}'',t).
\end{equation}
Hence we may verify that the dynamical Hamiltonian is given by
\begin{eqnarray}
\label{Hdyn3D2}
H_{\text{dyn}}(t) &=& -i\sum_{s= \pm 1}\sum_{\lambda = {\mathsf{H},\mathsf{V}}}\int_{\mathbb{R}^3}\text{d}^3\mathbf{r}'\int_{\mathbb{R}^3}\text{d}^3\mathbf{r}''\;\hbar sc\,\delta'(\mathbf{r}'-\mathbf{r}'')\nonumber\\
&& \times a^\dagger_{s\lambda}(\mathbf{r}',t)a_{s\lambda}(\mathbf{r}'',t).
\end{eqnarray}
Here $\delta'(\mathbf{r}'-\mathbf{r}'')$ denotes the derivative of a delta function with respect to $q_1$.  By recalling that $\mathbf{k}\cdot \mathbf{r} = k_1 q_1$, Eq.~(\ref{Hdyn3D2}) can be written in the alternative form
\begin{eqnarray}
\label{Hdyn3D3}
H_{\text{dyn}}(t) &=& \sum_{s=\pm 1} \sum_{\lambda = {\sf H}, {\sf V}} \int_{\mathbb{R}^3}\text{d}^3\mathbf{r}' \int_{\mathbb{R}^3}\text{d}^3\mathbf{r}''  \int_{\mathbb{R}^3}\text{d}^3\mathbf{k} \; \frac{\hbar k_1 c}{(2\pi)^3} \nonumber \\
&& \times {\rm e}^{is\mathbf{k}\cdot(\mathbf{r}'-\mathbf{r}'')} \, a^\dagger_{s\lambda}(\mathbf{r}',t) \,a_{s\lambda}(\mathbf{r}'',t).
\end{eqnarray}
As one can see above, the dynamical Hamiltonian in three dimensions also has positive and negative eigenvalues.  This differentiates it from the dynamical Hamiltonian in standard quantisations that has only positive eigenvalues.  

In Eq.~\ref{Hdyn3D3}, the frequency of a photon with wave vector $s\mathbf{k}$ is given by $k_1$ where $k_1$ is the $q_1$ component of $\mathbf{k}$.  This is a slightly odd choice of frequency because, as mentioned in Section \ref{Sec:blipstoplanes}, the unit vector $\widehat{\mathbf{q}}_1$ lies on the half-sphere, and there are many such half-spheres to choose from.  This choice of dynamical Hamiltonian is therefore most appropriate when there is a clear overall direction of propagation, as there is, for instance, for plane waves.  One may instead wish to redefine $\widehat{\mathbf{q}}_1$ such that it covers the entire unit sphere, which is much more general.  Consider the alternative description where $\widehat{\mathbf{q}}_1$ covers the unit sphere, the wave vector is given by $\mathbf{k} = |k|\widehat{\mathbf{q}}_1$, and the direction of propagation is given by $\mathbf{s} = s\widehat{\mathbf{q}}_1$.  In such circumstances, while Eq.~(\ref{Hdyn3D2}) is still satisfied, the frequency of a monochromatic photon is given by $s|k|c$.  From this alternative form of the dynamical Hamiltonian we can see that the negative frequency contribution to the dynamical Hamiltonian arises due to the introduction of the new parameter $s$.  Furthermore, when restricted to positive frequencies only, the energy observable (\ref{energyobservable3D2}) and dynamical Hamiltonian (\ref{Hdyn3D3}) coincide.  One can also check that the energy observable commutes with the dynamical Hamiltonian and is therefore conserved.

\section{Discussion}

In Chapter \ref{Chapter:1D}, the basic excitations of the free EM field were a set of bosonic particles localised to a position along the $x$ axis that propagate either to the left or the right at the speed of light.  In this chapter we have generalised the results of Chapter \ref{Chapter:1D} in order to determine a complete set of basis states for the free EM field in three dimensions that are both localised and propagate at the speed of light.  After an initial attempt to quantise the field in terms of localised excitations that have a well-defined direction of propagation, which we found to be an over-complete description, we introduced a complete set of blip annihilation operators $a_{s\lambda}(\mathbf{r},t)$.  The corresponding blip states are locally orthogonal to one another and, in addition to polarisation, are characterised by a discrete parameter $s = \pm1$ which doubles the usual Hilbert space. 

Unlike in one dimension, where the localised blips have a clear direction of propagation, in three dimensions a localised blip state contains contributions from photons propagating along all axes through the point of localisation with a forwards or backwards motion determined by the choice of $s$.  This feature is reminiscent of the Huygens-Fresnel principle in which each point on a wave front acts as a source of spherical waves.  This does mean, however, that highly-localised wave packets will spread out, but they do so only in accordance with Maxwell's equations. 

In Section \ref{Sec:completefields} we derived expressions for the electric and magnetic field observables in terms of the blip operators by means of a position-independent regularisation operator $\mathcal{R}$ and a set of non-local and vector-valued operators $\boldsymbol{a}_{s}(\mathbf{r},t)$ and $\boldsymbol{b}_{s}(\mathbf{r},t)$.  By introducing vector operators we are more easily able to relate the EM fields to the blips or carriers of the field.  These vector operators are similar to those defined in Ref.~\cite{BB6}, but differ in their dimension, as we make a clear distinction between the fields and the carriers.   As in Chapter \ref{Chapter:1D}, the regularisation operator allows us to introduce particle excitations without sacrificing the Lorentz covariance of the field observables.  The expression given in Eq.~(\ref{3Dregularisation1}) shows that the relationship between the field observables and blip operators is non-local with infinite tails.  This was also the case for the one-dimensional fields in Chapter \ref{Chapter:1D}.  Hence, a single blip generates a non-local field that can be felt across all of space. 

By expressing the blip operators in terms of monochromatic photon operators $\widetilde{a}_{s\lambda}(\mathbf{k},t)$, we have also shown that our approach is consistent with the standard theory up to the addition of negative-frequency states.  The Hilbert space in three dimensions, as in one dimension, is only doubled, and this doubling takes into account the negative-frequency states that are usually overlooked.  As states of the quantised EM field now evolve with negative as well as positive frequencies, the dynamical Hamiltonian and the energy observable are no longer equal, and the two operators only coincide when restricted to the subset of positive-frequency states.  

The energy observable (\ref{energyobservable3D2}) is strictly positive due to the square in the integrand.  This observable also contains pure annihilation and pure creation components that do not occur in the standard energy observable.  As in one dimension, these extra terms determine the interference that occurs between the more traditional positive-frequency excitations and the new negative-frequency excitations.  The addition of these terms also means that single-photon states are not eigenstates of the energy observable.  Single-photon states are, however, eigenstates of the dynamical Hamiltonian (\ref{Hdyn3D3}).  This is so because the dynamical Hamiltonian is a recreation operator that annihilates exactly one excitation and replaces it at a new location.  Unlike the energy observable, the dynamical Hamiltonian has both positive and negative eigenvalues.  One can see that by changing the value of $s$, the dynamical Hamiltonian changes sign and reverses the apparent flow of time.

\part{Applications}

\chapter{A Fermi problem for light}

\label{Chapter:Fermi}

Fermi's two-atom problem \cite{Fer} provides an interesting model for studying causality in quantum electrodynamics.  The problem is complicated, however, by the precise manner in which the atoms and fields are coupled together. In this chapter we study a simpler but equivalent problem that does not use atoms, involving a beam-splitter and a pair of photon detectors, in order to determine whether the field propagates causally or not. In Section \ref{Sec:alternative} we describe this experiment in more detail and study the propagation of light through the new system.  In Section \ref{Sec:Detection} we construct a pair of detection operators representing each detector in the experiment, and show that a causality violation takes place for light described by the standard theory.  In Section \ref{Sec:resolvingcausality} we show that there is no causality violation for light when we introduce negative-frequency photons. We conclude this chapter with a discussion in Section \ref{Sec:Fermidiscussion}. 

\section{Introduction: Fermi's original two-atom problem}

In empty space, light propagates at a constant and finite speed.  The speed of light is the same for all time-like observers, irrespective of their motion relative to one another or to that of the light source. In order to ensure the validity of relativity theory, it is crucial that the speed of light cannot be exceeded by any signal capable of relaying causal information, or, in other words, any signal capable of relaying to the receiver some knowledge of the actions of the transmitter.  Naturally, this speed limit introduces a lower bound on the time it takes for particles to propagate from one place to another, and it is important that this lower bound be apparent in any quantum theory.  

\begin{figure}[t]
\begin{center}
\includegraphics[width = 0.7 \textwidth]{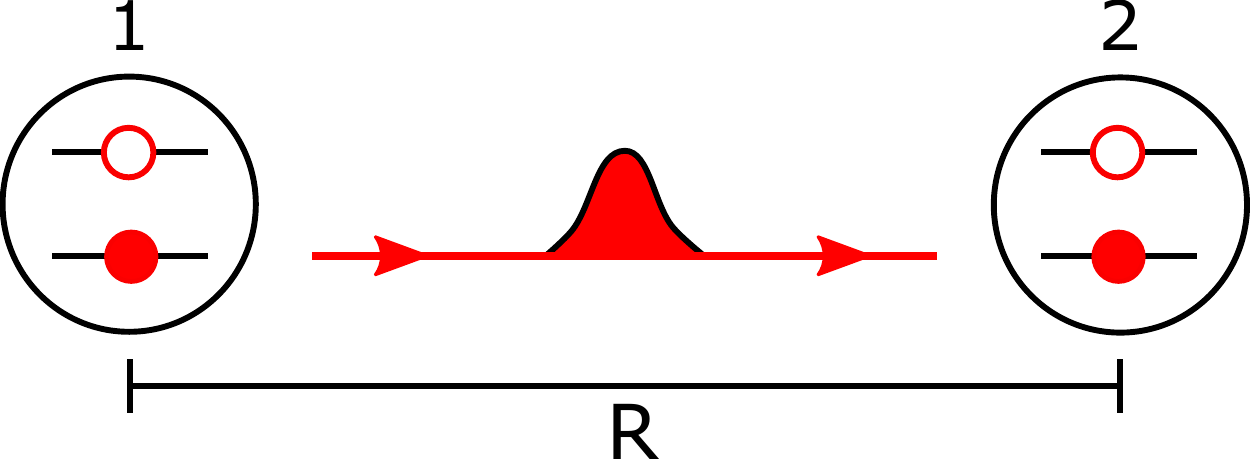}
\end{center}
\caption[Fermi's two-atom problem]{A illustration of the expected behaviour in Fermi's two-atom problem.  An atom initially in its excited state drops to its ground state generating a single photon from the vacuum.  The photon propagates to a distant, unexcited atom causing it to transition into its excited state.  The second atom is expected to be found in an excited state only after a time $R/c$ has passed.} \label{diagram2}
\label{Fig:fermi1}
\end{figure}

In 1932, Fermi published a paper on the quantisation of the EM field and its interactions with matter \cite{Fer}.  Here he considered the following problem. Consider two identical atoms separated by some large distance $R$.  Suppose also that one of these atoms is in its first excited state whilst the other atom and the pervading EM field are in their ground states.  After a negligible time the excited atom will decay emitting a single photon that then propagates towards the second atom.  Being in its ground state, the second atom will absorb the incoming photon and ``jump" to its excited state. How long should one have to wait after the photon is first emitted until it is absorbed again by the second atom?  In an intuitive picture of this interaction, the energy transferred from the first atom is carried by a  ``flying" photon, as illustrated in Fig.~\ref{Fig:fermi1}, that propagates from the first atom to the second.  Taking causality into consideration, the second atom should not become excited until a time $R/c$ has passed.

In Ref.~\cite{Fer}, Fermi used a perturbative approach to determine the probability of finding the second atom in an excited state, and found that this probability remains strictly zero for all times less than $R/c$.  This is the desired and expected result.  In Fermi's calculations, however, an approximation was used that extended the frequency of photon states into the negative frequency domain.  Shirokov \cite{Shi} and others \cite{Pow, Rub, Val, Cr} have since pointed out that without making this assumption the solution to Fermi's problem results in a small but non-zero probability for the second atom to be found in an excited state before the expected time, thereby violating Einstein causality. 

In the years since, many attempts have been made to reconcile the predictions of special relativity and quantum optical models for the experimental scenario illustrated in Fig.~\ref{diagram2}, \cite{BenB1, BY, Mil1, Mil2, Sab1, Sab2, Tjo, For1, Pli, Kau}. In some of these cases, however, a causal result is only found due to various approximations and modifications of the standard theory \cite{BenB1, Pli}, or by looking at the problem from a different view point, for example, Refs. \cite{BY, Tjo} look at the problem from the view point of algebraic QFT rather than of photon wave packets.  Other authors have found that there is no strict Einstein causality \cite{Heg2, Bis, Bor, Heg4, Pow, Rub, Val, Cr}, and that there is always a non-zero probability of immediately exciting the second atom.  In these cases, however, there is no superluminal signalling as this probability does not depend on the presence of the first atom, but is instead attributed to vacuum correlations between the two atoms \cite{Bis, Bor, Pow, Val}.  Hence, there is no causal relationship between measurements at the two atoms.  Such a result, however, is closely related to the particular choice of the coupling between the EM field and the two atoms, or upon the final states of the field and source atom \cite{Bis, Pow}.  For instance, In Ref.~\cite{Bis}, causality is preserved only when one refrains from applying the rotating wave approximation, the validity of which is also investigated in \cite{Pli, Stk}.  Unfortunately, due to the expected non-causal effects in the original two-atom Fermi problem being very weak, it is very difficult to verify their presence experimentally \cite{Set}.

By taking a wider view of the Fermi problem and not focussing solely on the atom-field interaction, Hegerfeldt demonstrated that the violation of strict causality is a result of the fact that the system Hamiltonian is bounded below \cite{Heg2, Heg4}.  This is a direct result of Hegerfeldt's theorem \cite{Heg8}, which states that, under a projective measurement, a single-photon wave packet initially localised to a finite region will spread out immediately.  The work in Ref.~\cite{Heg4} has been criticised in relation to the validity of the measuring process and the fact that possible loopholes must be treated with adequate care \cite{BY, Sab1, Tjo}.  For instance, in Ref.~\cite{Sab1}, the probability of exciting the second atom is not represented as a projective measurement and Hegerfeldt's conclusions are therefore avoided.  Nevertheless, in a position wave function approach to quantum optics, Hegerfeldt's theorem appears to demonstrate that photons can propagate at superluminal speeds.       

In summary, although there appears to be no superluminal signalling between the two atoms in Fig.~{\ref{Fig:fermi1}}, it is unclear whether there is any evidence of a causality violation or not.  The results on this topic vary and are intimately related to the particular way in which the atom and field are coupled together; however, the existence of superluminal correlations would seem to suggest that photon wave packets may be able to propagate at speeds exceeding the speed of light.  The work of Hegerfeldt suggests that the presence of atoms is not crucial to the calculations that predict non-causal behaviour. In this chapter, therefore, we analyse a simpler but analogous problem to Fermi's two-atom problem involving a beam-splitter and two spatially localised detectors.  We find in this chapter that single-photon wave packets described in the standard quantum theory do not propagate as would be expected for a localised light pulse. When expressed as a superposition of blip operators, however, localised single-photon wave packets propagate causally.

\section{An alternative experiment}

\label{Sec:alternative}

The alternative experimental setup that we shall consider in this chapter is illustrated in Fig.~\ref{Fig:fermi2}.  In this section we shall construct an analogy between Fermi's original two-atom system and our experiment in order to determine how a causality violation may be verified.  We shall also describe the propagation of light through the beam-splitter.  Initially we shall only consider positive-frequency photon states, as are present in standard quantisations, but later in this chapter we shall consider the propagation of both the positive-frequency and negative-frequency photons described in Chapter \ref{Chapter:1D}.

\subsection{Our experiment}   

\begin{figure}[h]
	\begin{center}
		\includegraphics[width= 0.85 \textwidth]{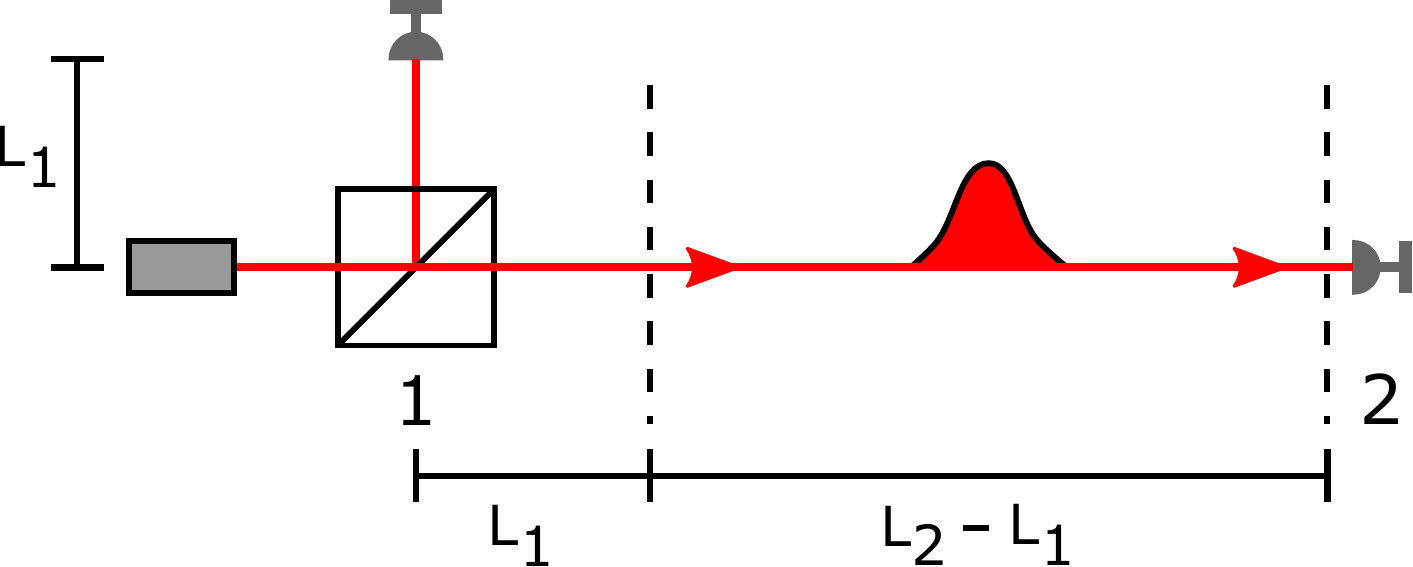}
	\end{center}		
	\caption[The experimental setup for a test of causality]{An interferometer consisting of a beam-splitter and two photon detectors.  A short pulse of light approaches the beam-splitter from the left and is then diverted towards the two detectors. The distance from the beam-splitter to Detector 1 is $L_1$ and the distance from the beam-splitter to Detector 2 is $L_2 \gg L_1$.  In relation to Fig.~\ref{Fig:fermi1}, region 1 corresponds to the first atom whereas region 2 corresponds to the second atom.}
	\label{Fig:fermi2}
\end{figure}

The interferometer in Fig.~\ref{Fig:fermi2} consists of a source of spatially-localised ultra-broadband light pulses \cite{Nasr, Ok, Jav, Tak} aimed at a beam-splitter, and a pair of photon detectors positioned along the output trajectories of the beam-splitter.  We denote the detector at the end of the vertical arm Detector 1 and the detector at the end of the horizontal arm Detector 2.  We assume that the separation between the beam-splitter and Detector 2 is much greater than that between the beam-splitter and Detector 1.

In the original formulation of Fermi's two-atom problem, the photon leaves the first atom at the time the system is turned on and the atom begins to decay.  The photon then arrives at the second atom the moment we first register a non-zero probability of finding the second atom in an excited state.  The difference between these times is the photon travel time from which a speed is calculated.  In the interferometer system illustrated in Fig.~\ref{Fig:fermi2}, all photons will initially be emitted by a source to the left of the beam-splitter.  Upon reaching the beam-splitter, some photons will be directed towards Detector 1 and the rest towards Detector 2.  Due to the presence of the beam-splitter, the positions of photons propagating along the vertical and horizontal arms of the interferometer are highly correlated, and a signal at Detector 1 will notify us of the position of the remaining photons in the horizontal arm to within a width of the wave packet. 

After passing through the beam-splitter, the two output wave packets have the same wave function as the initial wave packet entering the beam-splitter.  If Detector 1 has a particular probability of clicking at a time $t_1$, then, assuming that all wave packets propagate at the speed of light, the probability for a subsequent click at Detector 2 at a time $t_2$ must be directly proportional to the initial probability for Detector 1 to click after taking into account a short time delay caused by the difference in the distances from the beam-splitter to the detectors.  More specifically, when the photon wave packets propagate at the speed of light, the two probabilities must be directly proportional when 
\begin{equation}
t_2 = t_1 +(L_2 - L_1)/c
\end{equation}
Furthermore, there must be a zero probability for Detector 2 to click before a sufficient time has elapsed after Detector 1 has clicked.  

This type of causality, in which correlations cannot be generated at speeds exceeding the speed of light, is known as Einstein Causality.  From the point of view of localised photon wave packets, if a wave packet propagates at the speed of light, correlations cannot be generated at speeds exceeding the speed of light. This experiment specifically tests Einstein causality, hence a verification of causality is possible by ensuring that the probabilities for Detectors 1 and 2 to click obey the relationship described above.  In this chapter we carry out these calculations for photons in both the standard and blip descriptions of light.  

The model we propose here offers a simpler theoretical model for studying possible causality violations in a way that can be realised in a laboratory.  In this experiment there is a clear signature for the beginning and the end of the photon's journey in the form of the detector clicks.  This is well defined in the theory and in a possible experiment.  Furthermore, the nature of the coupling between photons and atoms does not need to be discussed or justified.

\subsection{Light in the interferometer}

\label{Sec:Interferometer}

We next describe the propagation of a short photon wave packet through the beam-splitter illustrated in Fig.~\ref{Fig:fermi2} by introducing a unitary scattering operator $S(t)$.  We use the standard description of photon wave packets in this section, but a generalisation to our new quantisation is easily made later on.

\subsubsection{One-dimensional systems}

In the experimental setup illustrated in Fig.~\ref{Fig:fermi2}, we consider light propagating along two perpendicular arms of an interferometer.  In each arm of the interferometer, light propagates either towards or away from the beam-splitter independently of what is going on in the other; if we ignore the effect of the beam-splitter that is.  In reality, the photon wave packets travelling along each arm would be localised in the plane orthogonal to their direction of propagation, but as there is no interference between light in the two arms of the interferometer except at the beam-splitter, it is sufficient for our purposes to treat light in the interferometer as two distinct one-dimensional systems: one system describing light propagating along the vertical arm, and the other describing light propagating along the horizontal arm.  States in different arms occupy distinct regions of the total Hilbert space.  We shall proceed next to construct wave packets of light in one dimension in preparation for our discussion on the beam-splitter in Section \ref{Sec:beamsplitter}.

\subsubsection{Single-photon wave packets}

In the quantised theory of the free EM field, the stationary solutions of Maxwell's equations are associated with a set of bosonic particles which form the basic energy quanta of the system.  In Section \ref{Sec:Quantisation1}, these energy quanta are associated with the running wave solutions of Maxwell's equations in one dimension.  This approach is particularly well suited for modelling light in the experimental setup considered in Fig.~\ref{Fig:fermi2} which is concerned with light propagating in one direction only.  In this chapter we shall consider again the set of creation and annihilation operators associated with light propagating in one dimension with a single frequency and a given polarisation, $a^\dagger_{k\lambda}$ and $a_{k\lambda}$.  The magnitude and sign of $k$ denote the frequency divided by the speed of light and the direction of propagation respectively.

As in Section \ref{Sec:wavepackets}, a single-photon wave packet is a linear superposition of the single-photon excitations generated by applying the $a^\dagger_{k\lambda}$ creation operators to the vacuum state.  Hence,
\begin{equation}
\label{wavepacket5}
\ket{\psi} = \sum_{\lambda = {\mathsf{H},\mathsf{V}}}\int_{-\infty}^{\infty}\text{d}k\;\psi_{k\lambda}\, a^\dagger_{k\lambda}\ket{0}
\end{equation}
with complex coefficients $\psi_{k\lambda}$.  This state is normalised when the following condition holds:
\begin{equation}
\label{normalise1}
\sum_{\lambda = {\mathsf{H},\mathsf{V}}}\int_{-\infty}^{\infty}\text{d}k\;|\psi_{k\lambda}|^2 = 1.
\end{equation}

\subsubsection{Many-photon wave packets}

The single-photon creation operator that generates the above wave packet from the vacuum is given by $a^\dagger$ where 
\begin{equation}
	\label{wavepacket6}
a = \sum_{\lambda = {\mathsf{H},\mathsf{V}}}\int_{-\infty}^{\infty}\text{d}k\;\psi^*_{k\lambda}\,a_{k\lambda}
\end{equation} 
Using Eq.~(\ref{wavepacket6}) and the commutation relation (\ref{standardcommutator1}) one can check that $a$ satisfies the single-photon commutation relation $\big[ a, a^\dagger\big] = 1$ when the state is properly normalised according to Eq.~(\ref{normalise1}).  To construct a many-photon state we proceed as in Section \ref{Sec:wavepackets}.  In this chapter we consider wave packets for any general set of coefficients $c_n$.  We can then afterwards examine particular choices of these coefficients.

\subsection{The dynamics of photon wave packets} 
\label{SecFD}

The dynamics of a state in the Hilbert space can be determined by proceeding as in Section \ref{Sec:shortQM} and solving the Schr\"odinger or Heisenberg equation.  In standard descriptions of the quantised EM field, the Hamiltonian is given by the energy observable (\ref{energyobs1}).  This is in contrast to the quantisations developed in Chapters \ref{Chapter:1D} and \ref{Chapter:3D} where the dynamical Hamiltonian determines the time-evolution of the system.  For simplicity, we work in the Heisenberg picture that was described in Section \ref{Sec:Dynamics1}.

By Solving Heisenberg's equation with the Hamiltonian given in Eq.~(\ref{energyobs1}), the single-photon annihilation operator for a time-dependent state is given by
\begin{eqnarray}
	\label{wavepacket3}
	a(t) = U(t,0) \, a \, U^\dagger(t,0) &=& \sum_{\lambda = {\sf H}, {\sf V}} \int_{-\infty}^{\infty}\text{d}k\, \psi^*_{k\lambda} \, {\rm e}^{{\rm i} c|k|t} \, a_{k\lambda}.
\end{eqnarray}
Here $U(t,0) = \exp (-{\rm i}H t/\hbar)$ is the usual free-space time-evolution operator of the quantised EM field.  

Suppose that the EM field has initially been prepared in the many-particle state $\ket{\psi}$, which we introduced in Eq.~(\ref{manyphoton}) for the single-photon operators defined in Eq.~(\ref{wavepacket3}).  The state at a later time $t$ is found in an equivalent superposition of number states constructed in an analogous way using the time-evolved annihilation operator $a(t)$ defined in Eq.~(\ref{wavepacket3}).  To show this we substitute the identity $\text{id} = U^\dagger(t,0)U(t,0)$ between each time-independent creation operator in the many-particle state and use the relation $U(t,0)\ket{0} = \ket{0}$.  This leads to the expression 
\begin{eqnarray}
	\label{extra}
	U(t,0) \, |\psi \rangle &=& \sum_{n=0}^\infty {c_n \over n!} \left( U(t,0) \, a^\dagger \, U^\dagger(t,0) \right)^n |0 \rangle \nonumber \\
	&=& \sum_{n=0}^\infty {c_n \over n!} \left(a^\dagger(t) \right)^n |0 \rangle.
\end{eqnarray}

\subsection{The beam-splitter}

\label{Sec:beamsplitter}

\subsubsection{The Hilbert space of the beam-splitter system} 

The purpose of the beam splitter in Fig.~\ref{Fig:fermi2} is to split any incoming wave packets into two parts that will propagate along the vertical and horizontal arms of the interferometer respectively.  To distinguish the photons in the vertical arm from those in the horizontal arm, we denote the annihilation operators of the latter by $b_{k\lambda}$ and the former by $c_{k\lambda}$.  The total single-photon Hilbert space is then given by the direct sum of $\mathcal{H}^b$ and $\mathcal{H}^c$, $\mathcal{H}^b \bigoplus \mathcal{H}^c$, which are the single-photon Hilbert spaces for photons in the horizontal and vertical arms respectively.  Since the single-photon states $|1^{(b)}_{k\lambda}\rangle = b^\dagger_{k\lambda}|0\rangle$ and $|1^{(c)}_{k\lambda}\rangle = c^\dagger_{k\lambda}|0\rangle$ evolve independently, they must be orthogonal to one another.  This implies that
\begin{eqnarray}
\label{Prod1}
\langle 1^{i}_{k\lambda}|1^{i}_{k'\lambda'} \rangle = \big[ b_{k\lambda}, b^\dagger_{k'\lambda'} \big] =  \big[c_{k\lambda}, c^\dagger_{k'\lambda'} \big] = \delta_{\lambda,\lambda'} \, \delta(k-k')
\end{eqnarray} 
where $i = b,c$.  All other commutators vanish. Thus the $b_{k\lambda}$ and $c_{k\lambda}$ operators obey the usual bosonic commutation relations.  

Next, in analogy to Eq.~(\ref{wavepacket6}), we construct annihilation operators $b$ and $c$ for normalised wave packets travelling along each arm of the interferometer, 
\begin{eqnarray}
	\label{wavepacket4}
	b &=& \sum_{\lambda = {\sf H}, {\sf V}} \int_{-\infty}^{\infty}\text{d}k \, \psi^{(b)*}_{k\lambda} \, b_{k\lambda} \, , \nonumber \\
	c &=& \sum_{\lambda = {\sf H}, {\sf V}} \int_{-\infty}^{\infty}\text{d}k \, \psi^{(c)*}_{k\lambda} \, c_{k\lambda} \, .
\end{eqnarray} 
The time evolution of the corresponding single-photon operators can be determined by solving Heisenberg's equation (\ref{Heisenberg's equation}). The  Hamiltonian of the total system is given by the sum of two harmonic oscillator Hamiltonians of the type given in Eq.~(\ref{energyobs1}): one in terms of the $b_{k\lambda}$ operators, and the other in terms of the $c_{k\lambda}$ operators.

\subsubsection{Scattering operators} 

As an initial state passes through the beam-splitter, a superposition of $b$ and $c$ states will be produced.  The effect of this splitting on an initial state $|\psi_{\text{in}} (0) \rangle$ can be described with the help of a scattering operator $S(t,0)$.  After passing through the beam-splitter, the state of the outgoing light at a time $t$ is given by
\begin{eqnarray}
	\label{scatter1}
	\ket{\psi_{\text{out}}(t)} &=& S(t,0) \, \ket{\psi_{\text{in}} (0)}. 
\end{eqnarray} 
As the time-evolution of a state before and after passing through the beam splitter is the same, it does not matter when the wave packet is split into two parts. Without any restrictions, we can therefore assume that 
\begin{eqnarray}
	\label{scatter2}
	S(t,0) &=& S U(t,0) \, ,
\end{eqnarray} 
where $S$ satisfying $SS^\dagger = S^\dagger S = 1$ is a unitary scattering operator.  In the case of a 50:50 beam-splitter, by denoting the annihilation operators for a photon that enters the beam-splitter along the $x$ axis $a_{k\lambda}$, $S$ changes the single-photon excitations such that \cite{Lim,Kok1}
\begin{eqnarray}
	\label{scatter3}
	S a_{k\lambda} S^\dagger &=& {1 \over \sqrt{2}} \left( b_{k \lambda} + {\rm i} c_{k\lambda} \right) \, .
\end{eqnarray} 
Taking this into account we find that the annihilation operator $a$ with the complex coefficients $\psi^*_{k\lambda}$ introduced in Eq.~(\ref{wavepacket3}) transforms such that
\begin{eqnarray}
	\label{scatter4}
	S a(t) S^\dagger &=& {1 \over \sqrt{2}} \sum_{\lambda = {\sf H}, {\sf V}} \int_{-\infty}^{\infty}\textrm{d}k \, \psi^*_{k \lambda} \, {\rm e}^{{\rm i} c|k|t} \left( b_{k \lambda} + {\rm i} c_{k\lambda} \right) \nonumber \\
	&=& {1 \over \sqrt{2}} \left( b(t) + {\rm i} c(t) \right).
\end{eqnarray}

Suppose that the light source in Fig.~\ref{Fig:fermi2} prepares the light in a superposition state such that $|\psi_{\text{in}} (0) \rangle = |\psi \rangle$.  We shall assume that the second input port remains empty.  Using Eq.~(\ref{scatter4}), the light exiting the beam-splitter at a time $t$ will be given by
\begin{eqnarray}
	\label{scatter5}
	\ket{\psi_{\text{out}}(t)} &=& \sum_{n=0}^\infty {c_n \over n!} \left( {1 \over \sqrt{2}} \left( b^\dagger(t) + {\rm i} c^\dagger(t) \right) \right)^n \ket{0}. 
\end{eqnarray}
The right hand side of Eq.~(\ref{scatter5}) is determined using the unitarity of $S$ and substituting $S^\dagger\,S$ between each input creation operator.

\section{Photon detection in the interferometer}

\label{Sec:Detection}

In this section we construct two detection operators that represent the measurement of a photon by Detectors 1 and 2 in Fig.~\ref{Fig:fermi2}.  Interpreting the photon wave function as the probability amplitude for detection, Detectors 1 and 2 are represented by a pair of projection operators. Later in this section we shall calculate and compare the probabilities for Detector 1 to click at a time $t_1$, and for Detector 2 to click at a later time $t_2$ conditional to the earlier click at Detector 1.  We then show that a causality violation takes place in the standard theory of the quantised EM field.  For simplicity, and since it does not affect our conclusions, from now on we shall only consider light with a single fixed polarisation $\lambda$.  This could be achieved by placing a polarising filter in front of the detectors or behind the source.

\subsection{Detection operators}

\label{Sec:detectionoperators}

When light enters a detection device, it will be absorbed by the device and cause a click at a given time $t$ with a particular probability $P_\text{click}(t)$.  The act of detecting a photon is represented by a detection operator acting on the Hilbert space.  In this chapter we shall assume that the detection operator represents an ideal measurement on the field and that we therefore do not need to pay any attention to the internal mechanism of the detecting device.  Furthermore, we shall assume that our photon detector is a perfect detector, although the experiment would work equally well with a less efficient detector.  This implies that a light pulse entering the detector will be detected with absolute certainty, and that the detector will never produce a false reading.  When a detector makes a measurement on a field, it does so with respect to a particular measurable characteristic of the field.  In this chapter, the characteristic of interest is the position of a photon.  The detector is designed to respond to any photon that comes within the bounds of the detecting equipment and not otherwise.

As localised particles are represented by the locally bosonic position eigenstates of a system, the detector responds to the position eigenstates of the field that lie within the detecting region. In spite of this, the detection operator does not usually coincide with the position observable in that region.  Rather the detection operator projects onto the eigenstates of the position observable with a particular probability.  This probability represents the belief that the projection onto a position eigenstate indicates a photon at that position \cite{Hel}.  Since we are assuming that our detectors are perfect detectors, this belief is an absolute certainty.  We therefore associate a unit probability with each projector.  The resulting mathematical structure of our detection operator is a projection valued measure associated with every possible region of space.  A projection valued measure is suitable for a detection operator because, for any excited state, the probability of a detection is always positive and there is a unit probability of finding a photon somewhere in space.

In the following the probability for a detector to click $P_{\text{click}}(t)$ at a time $t$ is determined by calculating the expectation value of the projection operator $P_{\text{detect}}(t)$ also at a time $t$ with respect to the state of the light incident on the detector.  As the detection operator is positive, any chance to detect a photon always results in a non-zero expectation value.  Using the idempotent property of the projection operator, $P^2 = P$, the expectation value of the detection operator with respect to a state $\ket{\psi}$ is equal to the square magnitude of the projection operator acting on that state.  Hence, the probability of the detector clicking at a time $t$ is given by
\begin{eqnarray} 
\label{clickprob1}
	P_{\rm click} (t) &=& \left\| P_{\rm detect}(t) \, \ket{\psi(0)}  \right\|^2 = \left\| P_{\rm detect} \, \ket{\psi(-t)}  \right\|^2 
\end{eqnarray}
where the projection operator $P_{\text{detect}}$ is defined at $t=0$.  In the remainder of this section we identify appropriate detection operators $P^{(1)}_{\rm detect}$ and $P^{(2)}_{\rm detect}$ for Detectors 1 and 2 in the setup shown in Fig.~\ref{Fig:fermi2} and calculate their respective detection rates.

\subsection{Detection operators in the interferometer}

\label{Sec:detectionoperators2}

\subsubsection{Localised single-photon wave packets}

The first step in constructing the appropriate detection operators is to identify a set of position eigenstates for light in the two arms of the interferometer onto which states will be projected.  As discussed in the introduction of this thesis, the task of constructing localised photon states in quantum theory has been beset with difficulties.  Indeed, Part \ref{Part:theory} of this thesis is dedicated to constructing a local theory of the free quantised EM field.  In this section,  we shall construct localised states in the standard Hilbert space (Section \ref{Sec:Quantisation1}) in the manner of Mandel \cite{Man1, MW} and Cook \cite{Ck1, Ck2}.  There are several reasons for considering these states as the localised photon states.  First of all, these states are locally orthogonal to one another, which is consistent with our definition of localised states in Chapters \ref{Chapter:1D} and \ref{Chapter:3D}. Furthermore, their inner product has the correct dimensions to be interpreted as a probability density, and, in one dimension, the position-dependent number operator represents exactly the photon number density operator \cite{Man1}.  

According to this definition, we define a pair of localised single-excitation states $\ket{1_b(x)}$ and $\ket{1_c(y)}$ in the following way:
\begin{eqnarray} 
\label{standardlocalstates1}
|1_b (x) \rangle &=& {1 \over \sqrt{2 \pi}} \int_{-\infty}^\infty \textrm{d}k \, {\rm e}^{ikx} \, b^\dagger_{k \lambda} \, \ket{0}\, , \nonumber \\ 
|1_c (y) \rangle &=& {1 \over \sqrt{2 \pi}} \int_{-\infty}^\infty \textrm{d}k \, {\rm e}^{iky} \, c^\dagger_{k \lambda} \, \ket{0}\,
\end{eqnarray}
Here $\ket{0}$ is the vacuum state for the total combined system $\ket{0} = \ket{0_b}\bigotimes\ket{0_c}$.  These states relate to the non-local single-photon states $b^\dagger_{k \lambda} \ket{0}$ and $c^\dagger_{k \lambda} \ket{0}$ via a Fourier transform and describe localised single-photon wave packets at positions $x$ and $y$ along the horizontal and vertical axes respectively.  There are no other field excitations anywhere else in the system.

\subsubsection{Number-resolving detectors}

For simplicity, in the setup shown in Fig.~\ref{Fig:fermi2}
we shall assume that our detectors are number-resolving detectors and designed such that they only react to single-excitation states, i.e.~states which can be obtained by applying a local photon creation operator, Eq.~(\ref{standardlocalstates1}), to the vacuum state.  This experiment could also be done with non-number resolving detectors, but since contributions from different number states are purely additive and positive, any violation of causality that occurs for the number-resolving detectors would also be present for more general single-photon detection operators. 

We shall define a pair of detectors that each have a finite length $D$ and respond only to photons within this range. The detection operators project onto the single-photon subspace of photons spanned by the states (\ref{standardlocalstates1}) that lie within the region of the detectors.  For convenience we also suppose that the beam-splitter is positioned at the origin of both the $x$ and $y$ coordinate axes.  Consequently, Detector 1 extends from $L_1$ to $L_1 + D$ and Detector 2 extends from $L_2$ to $L_2 + D$.  The detection operators may therefore be written as
\begin{eqnarray} 
\label{detectionoperators1}
P^{(1)}_{\rm detect} &=& \int_{L_1}^{L_1 + D}\textrm{d}y \, |1_c(y) \rangle \langle 1_c (y)| \, , \nonumber \\
P^{(2)}_{\rm detect} &=& \int_{L_2}^{L_2 + D}\textrm{d}x \, |1_b(x) \rangle \langle 1_b (x)|  \, .
\end{eqnarray}
Using the definition of the localised excitations in Eq.~(\ref{standardlocalstates1}), one may check that  $(P^{(i)}_{\rm detect})^2 = P^{(i)}_{\rm detect}$ for both $i = 1,2$.

\subsubsection{Signalling between detectors}

Using the commutation relations in Eq.~(\ref{Prod1}) one may verify that any operator that acts only on the $b$-photons in the horizontal arm will always commute with any operator that acts only on the $c$-photons in the vertical arm.  It is important that observables on the different arms of the interferometer commute with each other as this implies that the probability of measuring a particular outcome of an experiment on the $b$-photons is statistically independent of any operation carried out on the $c$-photons and vice-versa.  In other words, the frequency of a particular result occurring when the $b(c)$-photons are measured does not depend on any experiment performed on the $c(b)$-photons.  This includes the measurements performed by Detectors $1$ and $2$.

The statistical independence of measurements on the two arms of the interferometer does not, however, prohibit correlations existing between the $b$- and $c$-photons.  In fact, this experiment relies on the strong correlations between photons in each arm as these correlations provide a relation between the expected times of arrival of a photon in each of the detectors.  When one detector clicks, if the light pulse is very short, a click at the second detectors cannot be far behind.  The fact that the position observables in the two arms of the interferometer commute is important because, although there are correlations between the $b$- and $c$-photons, it means that a detection in one detector cannot either hurry along or hold off a detection in the other: the probability of detection in each will be unchanged.  Therefore, by making a measurement at Detector $1$ we set for ourselves an expected time of arrival for a photon to arrive at Detector $2$ (assuming propagation at $c$), but do not change the independent trajectory of light towards that detector.  This ensures that we are calculating the speed of a light pulse when unobserved.

\subsection{Probability of a detection}

\label{Sec:detectioncalculations}

\subsubsection{Detector 1}

We now consider the probability for Detector 1 to detect exactly one photon at some time $t_1$ and the probability for Detector 2 to detect exactly one photon at a later time $t_2$ conditional on the earlier click at Detector 1 at time $t_1$.  In the following, the probability for Detector 1 to click will be denoted $P^{(1)}_{\text{click}}(t_1)$.  The probability for Detector 2 to click will be denoted $P^{(2)}_{\text{click}}(t_2)$.

As the scattered light leaves the beam-splitter, the quantum state for the component of light seen by Detector 1 at a time $t_1$ is given by the projection of the detection operator $P^{(1)}_\text{detect}$ onto the output state $\ket{\psi_{\text{out}}(-t_1)}$ defined in Eq.~(\ref{scatter5}).  The probability for Detector 1 to click is then, according to Eq.~(\ref{clickprob1}), given by the square magnitude of the resulting state.  Using the expression for the output state $\ket{\psi_{\text{out}}(-t_1)}$ in Eq.~(\ref{scatter5}) and the detection operator for Detector 1 in Eq.~(\ref{detectionoperators1}), one finds that the probability for Detector 1 to click at a time $t_1$ is given by
\begin{equation} 
\label{detect1}
P^{(1)}_{\rm click} (t_1) = {1 \over 2} \sum_{n=0}^\infty {|c_{n+1}|^2 \over 2^n n!} \, \int_{L_1}^{L_1 + D} \textrm{d}y \, \| \braket{1_c(y)| c^\dagger(-t_1) |0_c} \|^2.
\end{equation}
When the coefficients $c_n$ are chosen for a coherent state (see Eq.~(\ref{coherent1})), one may show that $\sum_{n=0} |c_{n+1}|^2/2^n n! = \alpha^2\, \text{exp}[-|\alpha|^2/2]$.

\subsubsection{Detector 2}

The state seen by Detector 1 was given by $\ket{\psi_{\rm out} (-t_1)}$ specified in Eq.~(\ref{scatter5}).  On the other hand, once a click has been registered by Detector 1, the state seen by Detector 2 is given by
\begin{eqnarray}
\label{output1}
\ket{\psi_{\rm out} (-t_2)} &=&U(-t_2,-t_1) \, {P^{(1)}_{\rm detect} \, \ket{\psi_{\rm out}(-t_1)} \over \| P^{(1)}_{\rm detect} \, \ket{\psi_{\rm out}(-t_1)} \|^{1/2}}.  
\end{eqnarray}
This state is normalised.  To determine the probability for Detector 2 to click, one proceeds analogously to the previous calculation by applying the detection operator $P^{(2)}_{\text{detect}}$ to the output state $\ket{\psi_{\text{out}}(-t_2)}$.  The squared magnitude of the resulting state gives the probability of detection.  Using Eqs.~(\ref{output1}) and (\ref{scatter5}), and both detection operators in Eq.~(\ref{detectionoperators1}) one finds that
\begin{equation}
\label{detect2}
P^{(2)}_{\rm click} (t_2) = \frac{|c_2|^2}{4} \left( \sum_{n=0}^\infty {|c_{n+1}|^2 \over 2^n n!} \right)^{-1} \, \int_{L_2}^{L_2 + D} \textrm{d}x \, \| \braket{1_b(x)| b^\dagger(-t_2) |0_b} \|^2.
\end{equation}
This probability depends solely on the two-photon contribution of the initial output state in Eq.~(\ref{scatter5}).  This is because both detectors will only register a photon if there is exactly one photon in each arm of the interferometer.  It is therefore only the two-photon part of the output state that contributes to Eq.~(\ref{detect2}).

\subsubsection{Transition amplitudes}

By expressing the single-photon creation operators and the localised photon states in their Fourier representations, we may show that the transition amplitudes within the modulus signs in Eqs.~(\ref{detect1}) and (\ref{detect2}) are equivalent to the integral expressions
\begin{eqnarray} 
	\label{transition1}
	\braket{1_b(x)| b^\dagger(-t_2) |0_b} &=& {1 \over \sqrt{2 \pi}} \int_{- \infty}^\infty {\rm d} k \, {\rm e}^{-i(kx-c|k|t_2)} \, \psi_{k \lambda} \, , \nonumber \\
	\label{transition2}
	\braket{1_c(y)| c^\dagger(-t_1) |0_c} &=& {1 \over \sqrt{2 \pi}} \int_{- \infty}^\infty {\rm d} k \, {\rm e}^{-i(ky-c|k|t_1)} \, \psi_{k \lambda}  
\end{eqnarray}
for the particular $\lambda$ that we have chosen for this output state.  If the initially prepared incoming wave packet contains photons with positive $k$ only, the above transition amplitudes depend on $y - ct_1$ and $x - c t_2$ respectively.  This is as one would expect for light moving towards the two detectors.

\subsection{Causality violations in the standard theory}

\label{Sec:Causality}

We now study our expressions for the detection probabilities (\ref{detect1}) and (\ref{detect2}) and show that a causality violation takes place when using the standard description of single-photon wave packets.

\subsubsection{Expected results}

The probability for Detector 2 to click at a time $t_2$ conditional on a previous click by Detector 1 at a time $t_1$ was given in Eq.~(\ref{detect2}).  In order to determine whether causality is satisfied, let us look more closely at the first transition amplitude in Eq.~(\ref{transition1}) that appears in Eq.~(\ref{detect2}).  By decomposing the transition amplitude into its Fourier components, and making the substitutions $x=y+L_2-L_1$ and $t_2 = t_1 + (L_2-L_1)/c$, we find that
\begin{equation} 
\label{transition3}
\braket{1_b(x)| b^\dagger(-t_2) |0_b}  = {1 \over \sqrt{2 \pi}} \int_{- \infty}^\infty {\rm d} k \, {\rm e}^{-i(ky-c|k|t_1)}\,e^{-i(k-|k|)(L_2-L_1)} \, \psi_{k \lambda}
\end{equation}
By comparing this result with the transition amplitude in Eq.~(\ref{transition2}) we can see that the two expressions are almost identical.  The only difference here is that the amplitude in Eq.~(\ref{transition3}) contains an additional exponential term.  

Recall here that causality applies when the probabilities $P^{(1)}_{\text{click}}(t_1)$ and $P^{(2)}_{\text{click}}(t_1 + (L_2-L_1)/c)$ are proportional to each other.  This occurs only when the second line of Eq.~(\ref{transition2}) and Eq.~(\ref{transition3}) are also proportional to each other.  As the Fourier transform of any absolutely integrable function is unique \cite{How}, we can conclude that the two transition amplitudes in Eq.~(\ref{transition2}) are proportional, and in fact equal, only when the second exponential in Eq.~(\ref{transition3}) vanishes altogether. This occurs on only two occasions: when $L_1 = L_2$, or when $k = |k|$.  Next we shall examine these conditions in more detail in order to determine the characteristics of the system or state under which causality is preserved.

\subsubsection{An inconsistency with Einstein's relativity}

Let us first consider the condition that $L_2 = L_1$.  It was asserted at the beginning of the chapter that the horizontal arm of the interferometer was much longer than the vertical arm.  This implies that $L_2 \gg L_1$ and therefore that the detectors do not click at the same time. Hence, we can immediately disregard this condition and assume that $L_1 \neq L_2$.  The second condition states that $k = |k|$ and, therefore, that $k$ takes positive values only.  In the standard theory of the quantised EM field, the sign of $k$ indicates the direction of propagation.  In this experiment, light propagates in a single direction away from the beam-splitter and towards the detectors.  The above condition, then, is only the statement that the state under consideration ought to contain contributions only from photons propagating towards rather than away from the detectors.  On first inspection this condition seems to be one that we can accept perfectly well.  We should not, however, be too hasty with our conclusions.  Let us have a closer look at Eq.~(\ref{wavepacket5}) which is the general expression for a single-photon wave packet.  

The photon wave packet propagates causally only when we neglect the negative $k$ terms.  These wave packets are defined in Eq.~(\ref{wavepacket5}) by changing the limits of the integral from $(-\infty, \infty)$ to $[0, \infty)$ or by replacing the momentum-dependent wave function $\psi_{k\lambda}$ with $\theta(k)\,\psi_{k\lambda}$ where $\theta(k)$ is the Heaviside step-function.  Using the definition of perfectly localised states given in Eq.~(\ref{standardlocalstates1}), the position-dependent wave function of a state is given by the Fourier transform of the momentum-dependent wave function.  The Fourier transform of the wave function $\theta(k)\,\psi_{k\lambda}$ is non-zero at all positions for all allowed choices of $\psi_{k\lambda}$.  Consequently, a single-photon wave packet that propagates in a single direction cannot be localised completely to a finite region, but must be spread out across all available space.

At the beginning of this chapter we asserted that the photons entering the interferometer were initially localised somewhere to the left of the beam-splitter.  This is a condition that could be imposed by introducing a shutter mechanism in front of the beam-splitter.  Since this assumption implies that the photon wave packet contains contributions from all wave numbers, we must also disregard the second causality condition.  Thus, we find that, in the standard description of the quantised EM field, localised wave packets do not propagate at the speed of light, and correlations do form at superluminal speeds.  In the language of single-photon wave functions, this implies that photon wave packets may propagate at speeds exceeding the speed of light, and consequently that either causality is violated or there does not exist a wave function that represents the position of a single photon propagating through space.  In the following section we shall examine whether this dilemma may be resolved by describing the light propagating along each arm of the interferometer as a system of blips.

\section{Resolving causality issues with blips}

\label{Sec:resolvingcausality}

By carrying out similar calculations as in the previous section, in this section we demonstrate that localised photon wave packets propagate causally when described as a system of blips.  We also discuss causality of the field observables.  

\subsection{An argument for negative-frequency photons}

In the previous section we have shown that, in the standard theory of the quantised EM field, single-photon wave packets cannot simultaneously be localised and propagate at the speed of light.  This problem arises because causality applies only when the frequency of all photons involved is equal to their wave number multiplied by the speed of light.  Since localised wave packets necessarily contain photons with both positive and negative wave numbers, and frequency is strictly positive, this can never be the case.  In Chapter \ref{Chapter:1D} of this thesis we introduced blip excitations in one dimension.  We discovered, however, that blip excitations contain contributions from photons that evolve with negative as well as positive frequencies.  This suggests that, if the quantised field is described in terms of blip excitations, causality may be preserved for local excitations. 

Hence, in this section, we shall consider again the interferometer experiment described in the introduction of this chapter.  This time, however, we shall model the EM field propagating along the arms of the interferometer as two distinct systems of one-dimensional blips.  To continue we shall first describe the scattering of blip excitations by the beam-splitter and construct detection operators appropriate for detecting blips at the desired locations.  We shall then repeat the calculations carried out in Section \ref{Sec:detectioncalculations} in order to determine whether the blip wave packets propagate causally or not.

\subsection{Scattering operators for blips}

In this experiment we shall, as before, assume that the photon is localised somewhere to the left of the interferometer.  The initial state of the system is given by
\begin{equation}
\ket{\psi_{\text{in}}(0)} = \int_{-\infty}^{\infty}\text{d}x\;\psi_{s\lambda}(x)\,a^\dagger_{s\lambda}(x,0)\ket{0}
\end{equation}
Since the wave packet approaches the beam-splitter from the left we shall assume that $\psi_{s\lambda}(x) = 0$ when $s=-1$, or, in other words, the initial wave packet only contains blips that propagate to the right.  We shall also assume that the wave function $\psi_{s\lambda}(x)$ is non-zero only in a small region to the left of the beam-splitter and for a single polarisation $\lambda$.

As in Eq.~(\ref{scatter1}), the output state of the beam-splitter is determined by acting upon the input state with a unitary scattering operator $S$ that splits the incoming wave packet into two components.  The annihilation operators for blips propagating along the horizontal arm are denoted $b_{s\lambda}(y)$, whereas those propagating along the vertical arm are denoted $c_{s\lambda}(x)$.  As before, these excitations are distinct, and all commutators between $b$ and $c$ creation and annihilation operators are zero.  The scattering operator is modelled completely analogously to Eq.~(\ref{scatter3}) which was defined for monochromatic photon states.  This transformation leaves both the wave number and polarisation unchanged, but we shall now also assume that the parameter $s$ is unchanged by the scattering transformation.  The two output wave packets therefore propagate away from the beam-splitter. The output state is then given by Eq.~(\ref{scatter5}).  In this case, however, the time-dependent excitations are given by
\begin{eqnarray}
\label{scatteredblips1}
b(t) &=& \int_{-\infty}^{\infty}\text{d}x\;\psi_{s\lambda}^*(x)\,b_{s\lambda}(x,t)\nonumber\\
c(t) &=& \int_{-\infty}^{\infty}\text{d}y\;\psi_{s\lambda}^*(y)\,c_{s\lambda}(y,t)
\end{eqnarray}
respectively.  The $b_{s\lambda}(x,t)$ and $c_{s\lambda}(x,t)$ operators independently satisfy the blip commutation relations given in Eq.~(\ref{blipcommutator2}).

\subsection{Detecting blips in the interferometer}

\subsubsection{Detection operators}

We consider now a pair of number-resolving detection operators that respond to single-blip states located within the detecting regions $L_1$ to $L_1 + D$ up the vertical arm and $L_2$ to $L_2 + D$ along the horizontal arm.  Here we shall drop the $s$ and $\lambda$ subscripts because we assume that all the light reaching the detector propagates away from the beam-splitter and is polarised in a single direction.  The assumption that light reaching the detectors propagates in only one direction is based upon the fact that the photon source is to the left of the beam-splitter.  As the blip excitations are locally orthogonal to one another, the appropriate detection operators are given just as in Eq.~(\ref{detectionoperators1}).  Here, however, the single-excitation states $\ket{1_b(x)}$ and $\ket{1_c(y)}$ are generated by acting upon the vacuum state with the blip operators $b_{s\lambda}^\dagger(x,0)$ and $c_{s\lambda}^\dagger(y,0)$ for the values $s$ and $\lambda$ selected by the wave function $\psi_{s\lambda}(x)$.

\subsubsection{Detection probabilities}

We are now in a position to calculate the probabilities for Detectors 1 and 2 to click at the times $t_1$ and $t_2$ respectively.  As before, the probability for Detector 1 to click is evaluated by first evolving the beam-splitter output state backwards in time by a time $t_1$, and then taking the expectation value of the new projection operator $P^{(1)}_{\text{detect}}$.  The resulting probability $P^{(1)}_{\text{click}}$ is given by Eq.~(\ref{detect1}).  In the new expression, however, the local states and operators within the transition amplitude are the local blip states and operators, and not the local states defined in Eq.~(\ref{standardlocalstates1}).  Using the time-dependent inner product given in Eq.~(\ref{blipproduct1}) and the annihilation operators given in Eq.~(\ref{scatteredblips1}), the probability of finding a blip at Detector 1 at a time $t_1$ is now given by
\begin{equation}
\label{detect3}
P^{(1)}_{\rm click} (t_1) = {1 \over 2} \sum_{n=0}^\infty {|c_{n+1}|^2 \over 2^n n!} \, \int_{L_1}^{L_1 + D} \textrm{d}y \, \| \psi(y - ct_1) \|^2.
\end{equation}
We can now see, by looking at the square of the wave function, that the probability of detecting a blip is equal to the probability of finding a blip in a region of length $D$ shifted along from Detector 1 by a length $ct_1$.  Thus, there is only a non-zero probability to detect a blip at Detector 1 if the wave packet is initially a distance of $ct_1$ away from the detector, as we would expect.

Next, let us calculate the probability of finding a second blip at Detector 2 at a time $t_2$.  After a detection has been made at Detector 1, the remaining output state is again given by Eq.~(\ref{output1}).  After evolving the state backwards to a time $-t_2$, we then calculate the expectation value of the second detection operator $P^{(2)}_{\text{detect}}$.  The conditional probability for Detector 2 to click at a time $t_2$ is given by
\begin{equation}
\label{detect4}
P^{(2)}_{\rm click} (t_2) = \frac{|c_2|^2}{4} \left( \sum_{n=0}^\infty {|c_{n+1}|^2 \over 2^n n!} \right)^{-1} \, \int_{L_2}^{L_2 + D} \textrm{d}x \, \| \psi(x-ct_2) \|^2.
\end{equation}
This expression can be written in an alternative form by making the substitution $y = x - L_2 + L_1$, which is always true due to our choice of detector placement.  We shall also examine the particular case where $t_2 = t_1 + (L_2-L_1)/c$.  We choose this time because it is the time we would expect a light signal to take to travel from a distance $L_1$ to a distance $L_2$.  The resulting expression for $P^{(2)}_{\text{click}}(t_2)$ is given by
\begin{equation}
\label{detect5}
P^{(2)}_{\rm click} (t_2) = \frac{|c_2|^2}{4} \left( \sum_{n=0}^\infty {|c_{n+1}|^2 \over 2^n n!} \right)^{-1} \, \int_{L_1}^{L_1 + D} \textrm{d}x \, \| \psi(y-ct_1) \|^2.
\end{equation}

One can see by cross-examining Eqs.~(\ref{detect3}) and (\ref{detect5}), that the probabilities of finding a blip in Detector 1 at a time $t_1$ and a blip in Detector 2 at a time $t_2 = t_1 + (L_2-L_1)/c$ are directly proportional to each other.  In both cases the overlap between the wave function and the detector regions are identical, as we would expect after taking causality into consideration.  Thus, when the quantised EM field in one dimension is described using blips, the position-dependent wave functions all propagate at the speed of light.

\subsection{Causality of field observables}

In Section \ref{Sec:Causality} we demonstrated that, when described using the standard theory (see Section \ref{Sec:Quantisation1}), single-photon wave functions cannot be simultaneously localised to a finite region of space and propagate at the speed of light.  This is a slightly puzzling and apparently contradictory result considering that the electric and magnetic field observables are solutions of the wave equation for waves propagating at $c$ (Eq.~(\ref{waveequation1})).  In fact, some authors claim that, in the context of algebraic quantum field theory, the fact that the field observables obey this wave equation is enough to demonstrate that no causality issues can occur \cite{BY, Tjo}.  The problem of infinite propagation speeds, however, is not directly related to the propagation of the electric and magnetic field observables.

In the Heisenberg picture, the electric and magnetic field observables in one dimension are given, as in Eqs.~(\ref{Efield2}) and (\ref{Bfield2}), by the expressions
\begin{eqnarray}
\label{EB}
\mathbf{E}(x) &=& c \int_{-\infty}^{\infty}\text{d}k \, \sqrt{\frac{\hbar |k|}{4\pi A c\varepsilon_0}}\, e^{ikx-i|k| ct} \, \left[a_{k\mathsf{H}} \, \widehat{\boldsymbol{y}} + a_{k\mathsf{V}} \, \widehat{\boldsymbol{z}} \right] + \text{H.c.} \, , \nonumber\\
\mathbf{B}(x) &=& \int_{-\infty}^{\infty}\text{d}k \, \sqrt{\frac{\hbar |k|}{4\pi A c \varepsilon_0}} \, e^{ikx-i|k| ct}\, \text{sgn}(k) \left[a_{k\mathsf{H}} \, \widehat{\boldsymbol{z}} - a_{k\mathsf{V}} \, \widehat{\boldsymbol{y}} \right] + \text{H.c.}
\end{eqnarray}
Let us now calculate the expectation value of these field observables with respect to a horizontally polarised coherent state containing on average $|\alpha|^2$ photons defined by the wave function $\psi_{k\mathsf{H}}$.
Taking into account Eqs.~(\ref{manyphoton}) and (\ref{coherent1}), one can show that
\begin{eqnarray}
	\label{EB3}
	\Braket{ \mathbf{E}(x,t) } &=& c \alpha \int_{-\infty}^{\infty}\textrm{d}k\, \sqrt{\frac{\hbar |k|}{4\pi A c \varepsilon_0}}\, {\rm e}^{ikx-i|k| ct} \, \psi^*_{k\mathsf{H}}\, \widehat{\boldsymbol{y}} + \text{c.c.} \, , \nonumber\\
	\Braket{ \mathbf{B}(x,t)} &=&  \alpha  \int_{-\infty}^{\infty}\textrm{d}k\, \sqrt{\frac{\hbar |k|}{4\pi A c \varepsilon_0}} \, {\rm e}^{ikx-i|k| ct} \, \text{sgn}(k) \,\psi^*_{k\mathsf{H}}\,\widehat{\boldsymbol{z}} + \text{c.c.}
\end{eqnarray}
Restricting the wave function to a single polarisation does not affect the generality of this result.  

In these expressions, as before, positive and negative $k$ correspond to left- and right-travelling waves.  In these expressions one can see that at a time $t=0$ the expectation values of the field observables can have any distribution in space we like, even localised ones.  This is the case because there are no restrictions on $\psi_{k\mathsf{H}}$ for any $k$.  What is less obvious is that these expectation values can have any distribution in space at any time and propagate at the speed of light even if $k$ is restricted to only positive or negative values.  Unlike the expectation values of the projection operators, when restricted to only positive values of $k$ we find that the additional complex conjugate term completes the missing frequency terms to create localised wave packets with a definite direction of propagation that do not disperse as time evolves.  Hence, the expectation values of the electric and magnetic fields are complete solutions of Maxwell's equations that propagate both to the left and the right at a speed $c$.

Although the electric and magnetic field expectation values propagate causally for positive frequency states, the expectation values of the detection operators do not because they measure only the complex part of the field.  They do, however, propagate causally for states containing both positive- and negative-frequency photons.  To see that this is so consider the position operator acting on the single-blip Hilbert space below:
\begin{equation}
\label{blipposition1}	
X = \sum_{s = \pm 1}\sum_{\lambda = \mathsf{H},\mathsf{V}}\int_{-\infty}^{\infty}\text{d}x\;x\,\ket{1_{s\lambda}(x)}\bra{1_{s\lambda}(x)}\, . 
\end{equation}
The blip states are the eigenvalues of this observable.  After evolving this operator in time with respect to the dynamical Hamiltonian (\ref{Hdyn2}), the time-dependent position operator is given by 
\begin{equation}
\label{blipposition2}
X(t) = \sum_{s = \pm 1}\sum_{\lambda = {H,V}}\int_{\infty}^{\infty}\text{d}x\;(x+sct)\,\ket{1_{s\lambda}(x)}\bra{1_{s\lambda}(x)}.
\end{equation}
The eigenvalues of the position operator are the same at all times.  As time advances they are displaced by a distance $sct$ as would be expected for a localised photon.

If the position states in Eq.~(\ref{blipposition1}) are replaced by the localised states given in Eq.~(\ref{standardlocalstates1}) and the system evolved forward in time using the one-dimensional generalisation of Eq.~(\ref{energyobs1}), the time-dependent position operator is given by
\begin{equation}
X(t) = \sum_{\lambda = {H,V}}\int_{-\infty}^{\infty}\text{d}k_1\int_{-\infty}^{\infty}\text{d}k_2\;g(k_1,k_2;t)\ket{1_\lambda(k_1)}\bra{1_{\lambda}(k_2)}.
\end{equation}
where
\begin{equation}
g(k_1,k_2;t) = \int_{-\infty}^{\infty}\text{d}x\;x\,e^{i(k_1-k_2)x+ i(|k_1|-|k_2|)ct}.
\end{equation}
The localised states $\ket{1_\lambda(x)}$ are not the eigenstates of the above operator for any non-zero time.

\section{Discussion}

\label{Sec:Fermidiscussion}

When two identical atoms, one in an excited state and one in its ground state, are coupled to the electromagnetic field, the excited atom will emit a photon that is absorbed at a later time by the second atom.  It is often found that there is a non-zero probability for the second atom to become excited immediately after the coupling is turned on, but as this probability is independent of the first atom, it is not usually interpreted as evidence of superluminal propagation from the first atom to the second.  This result, however, does not depend on the use of atoms.  For this reason, in this chapter we have considered an alternative causality experiment that uses a beam-splitter and a pair of number-resolving photon detectors positioned along the output ports of the beam-splitter as illustrated in Fig.~\ref{Fig:fermi2}.

The results of this chapter show that in the standard theory of the quantised EM field a state initially localised to the left of the beam-splitter has a non-zero probability of causing a click at either of the detectors at a later time.  In fact, Detector 2 may register a click long before a photon is received by Detector 1.  In this experiment a click at either detector must correspond to a measurement of the state initially localised to the left of the beam-splitter.  In the standard quantisation of the free EM field, therefore, localised photon states propagate at speeds exceeding the speed of light.  This is in agreement with the results of Hegerfeldt \cite{Heg8} and Malament \cite{Mal}.  As the two detection operators defined in Eq.~(\ref{detectionoperators1}) commute with one another it is impossible for an observer at Detector 1 to send a message to an observer at Detector 2 by taking measurements with his detector.  It is then possible to say that there is no causality violation because this experiment does not examine the propagation of causal information.  Nevertheless, the results in this chapter clearly demonstrate that the localised states defined in Eq.~(\ref{standardlocalstates1}) disperse filling all space immediately.  This result cannot be accepted if we are to construct a local theory of photons. 

We have also shown in this chapter that when the field is modelled in terms of blips, localised photon wave packets propagate at the speed of light.  In particular, the probability of detecting a photon at Detector 2 conditioned on an earlier detection made by Detector 1 is directly proportional to the initial probability of Detector 1 clicking once an appropriate delay has been taken into consideration.  In Section \ref{Sec:Causality} we found that causality only applies when the photon wave number (multiplied by $c$) and frequency are equal. The problem of infinite propagation speeds can therefore be overcome by removing the lower bound on the Hamiltonian, as was done in Chapters \ref{Chapter:1D} and \ref{Chapter:3D} of this thesis, but also in Refs.~\cite{Haw11,Jake2, Jake1, Al1, Al2, Al3, For2, Haw8, CP}.  In spite of this, theories that are restricted to positive field Hamiltonians are by far the most prevalent in standard quantum physics textbooks \cite{Hei,Coh, Wei, Lou, RB2}. In Ref.~\cite{Bis}, the superluminal contribution to the excitation probability of the second atom was attributed, not to a ``flying" photon, but to the vacuum fluctuations of the EM field.  Our results are in partial agreement with this conclusion, as we have also shown that such correlations have nothing to do with the use of atoms but with the current description of the EM field.  Here, however, we do not see this as a reason to excuse these correlations, but as a justification that an alternative description of the field is needed.   

This chapter has provided an important verification that the blip theory offers a more intuitive means of describing photon wave packets than the standard theory in realistic experimental scenarios.  In particular the results of this chapter confirm that negative-frequency states are a necessary part of the construction of photon wave packets.  The existence of negative-frequency modes is fundamental to a variety of amplification effects including Penrose superradiance \cite{Pen}, which has recently been measured in a non-linear optical system \cite{Brai}, and the production of Hawking radiation by a black hole \cite{Hawk} or Unruh radiation in an accelerating reference frame \cite{Unruh}.  In these instances, however, amplification occurs due to a conversion from modes with a negative norm to those with a positive norm.  The positive-frequency and negative-frequency parts of the blip's Fourier spectrum discussed in this chapter both have a positive norm and so do not relate to these effects.  Modes with a negative norm do still exist in our new theory in addition to the newly introduced negative-frequency photon states.  In order to study effects such as the Unruh effect we would need to study whether negative-norm states can be converted to positive-norm states in this new theory.  

The results of this chapter provide a good indication that the blip theory will prove a useful tool for studying quantum effects in the position representation.  In this next chapter we shall investigate the behaviour of blips in a one-dimensional cavity and a three-dimensional cavity.  By studying light inside a cavity in the position representation we hope to gain a new perspective on the origin of the Casimir effect.

\chapter{The Casimir effect}

\label{Chapter:Casimir}

In this chapter we provide an alternative derivation of the Casimir force between two parallel, perfectly reflecting mirrors in both one and three dimensions using the local theory of photons constructed in Chapters \ref{Chapter:1D} and \ref{Chapter:3D}. We recover the usual textbook expressions for the Casimir force as an exact result without the need for regularisation procedures. In Section \ref{Sec:Lightincav} we investigate the propagation of blips near a perfect cavity in one dimension and construct a corresponding set of field observables.  In Section \ref{Sec:1Dforce} we show that the zero-point energy of the field inside the cavity differs from its usual expression by a convergent term that depends on the width of the cavity.  This additional term leads to an attractive force between the mirrors.  In Section \ref{Sec:3Dforce} we repeat these calculations for the case of a perfectly reflecting cavity in three dimensions.  We similarly find that there is an attractive force between the two plates in agreement with standard results.  We conclude this chapter with a discussion in Section \ref{Sec:Casimirdiscussion}.

\section{Introduction: The Casimir effect}

The Casimir effect refers to the attractive force between two parallel and highly reflecting mirrors in a vacuum \cite{Cas}.  Since its initial discovery the Casimir effect has received a lot of attention in the
literature \cite{Lif, Rey, Em, Gol, Dal, RB3}, and, despite not having a counterpart in classical
electrodynamics, recent experiments confirm its predictions \cite{Lam, Moh,Ed, Aks, Bre, Dec}. Nevertheless, there is still some controversy surrounding the origin of this effect \cite{Ki, Buh1, Shah}.  For example, the standard derivation of the Casimir effect requires regularisation procedures to be used before identifying a finite contribution to the zero-point energy of the EM field that depends on the mirror separation $D$ \cite{Mil3}. 

Moreover, the standard derivation simply assumes that the plates restrict the quantised EM field inside the cavity to standing waves with a discrete set of resonant frequencies. In these standing wave mode models one cannot take into account from which direction light enters an optical cavity and therefore cannot reproduce the typical behaviour of Fabry-Perot cavities which have maximum transmission rates for resonant light \cite{Bnt4}. Discrete mode models also imply that no light is permitted inside a cavity with mirror separations well below typical optical wavelengths, which would appear to contradict recent experiments with nano-cavities \cite{Bau}.  Special care must be taken in these circumstances, however, as the modes excited inside the cavity are often described as nano-gap plasmonic modes, which lie on the surface of the cavity walls. The plasmon wavelengths are always smaller than the free-space optical wavelengths allowing an enhancement of light inside very small cavities.  To fully capture all aspects of optical cavities, an alternative approach to the Casimir effect is needed.

In Chapters \ref{Chapter:1D} and \ref{Chapter:3D} of this thesis we introduced a quantisation of the one- and three-dimensional EM field in the position representation.  A description such as this lends itself naturally to the modelling of locally interacting quantum systems, like a cavity, and other quantum optics experiments \cite{Jav, Alv}.  In the blip descriptions, the EM field is viewed as a collection of localised energy quanta described by the local operators $a_{s\lambda}(x,t)$ and $a_{s\lambda}(\mathbf{r},t)$.  These operators avoid the previous localisability issues as the associated energy quanta are completely localised in space (see Eqs.~(\ref{blipcommutator2}) and (\ref{3Dcommutator6})).  Moreover, as was shown in Chapter \ref{Chapter:Fermi} for the one-dimensional excitations, because these excitations evolve according to a dynamical Hamiltonian that has both positive and negative eigenvalues, these states propagate at the speed of light without dispersion.  

The purpose of this chapter is to use the local theories developed in Chapters \ref{Chapter:1D} and \ref{Chapter:3D} to model the behaviour of the quantised EM field between two parallel mirrors and re-derive the Casimir effect in both one and three dimensions. Instead of imposing boundary conditions by demanding that the electric field amplitudes vanish at the mirror surfaces and restricting the EM field inside the cavity to standing waves, in this chapter we realise boundary conditions in a dynamical fashion by looking at how blips propagate near a cavity.  The results of this chapter will provide a more intuitive description of the Casimir effect and will support the validity of the local field quantisation approach to light-matter systems.  As the calculations involved in the derivation of the Casimir effect in one and three dimensions are very similar, the most part of this chapter will be dedicated to the one-dimensional effect.  The derivation of the three-dimensional effect will be shorter as many of the relevant discussions carry over.

 \section{Light inside an optical cavity}
 
In this section we describe the propagation of blips both inside and outside a one-dimensional cavity using the mirror image method of classical optics.  By applying the regularisation operator of Chapter \ref{Chapter:1D} to a blip near the cavity and its mirror images, we determine an expression for the field observables near the cavity.
 
 \label{Sec:Lightincav}

\subsection{The propagation of blips near a one-dimensional mirror}

\label{Sec:blipsnearmirror1D}

\begin{figure}[h]
\centering
\includegraphics[width = 0.6\textwidth]{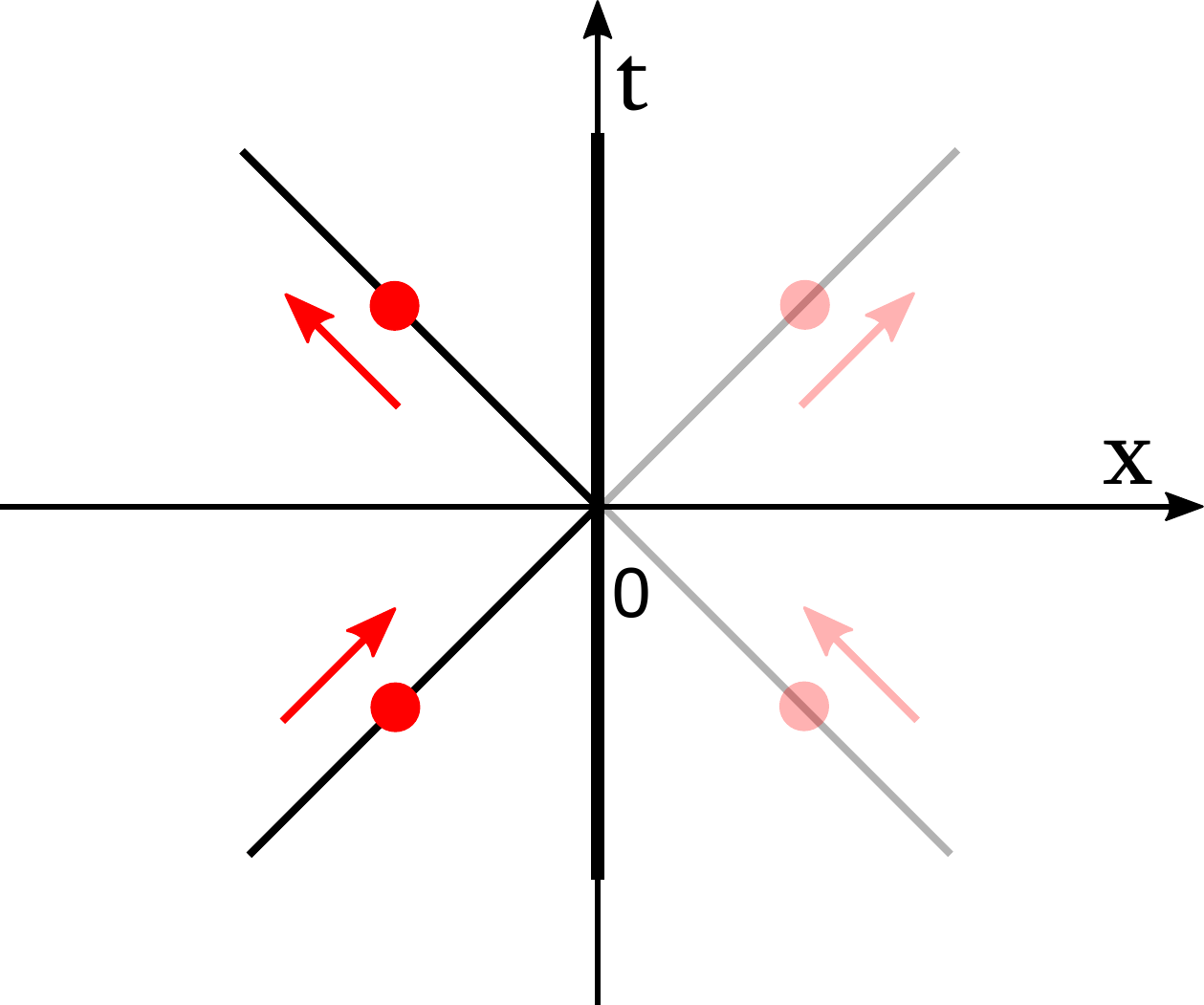}
\caption[The world-line of a blip near a one-dimensional mirror]{The figure depicts the world-line of a single blip interacting with a perfect mirror placed at the origin.  The blip, initially approaching from the left, propagates freely until it reaches the mirror at which point the direction of propagation is reversed.  While the incoming blip is characterised by $s = 1$, the outgoing blip is characterised by $s = -1$.  A blip initially approaching from the right would behave analogously, the reflection implementing the transition from an $s = -1$ blip to an $s = 1$ blip.  The translucent trajectory on the right-hand side of the mirror shows the mirror image of the reflected blip.}
\label{Fig:mirror1}
\end{figure}

In free space, blips propagate along the boundaries of the light-cones as illustrated in Fig.~\ref{Fig:Lightcone1}.  In such a system the blips propagate away from the source at the speed of light and never change their direction.  Let us now consider a single mirror placed at some point along the $x$ axis.  For convenience we shall choose this point to be the origin $x=0$.  We shall name the region to the left of the mirror Region I, and the region to the right of the mirror Region II.  In this chapter we shall treat all mirrors as classical objects that interact locally with light.  The mirrors we are describing are perfect mirrors that reflect all incident light.  As blips are the local energy quanta of the EM field, it is the blips that interact locally with the mirror.  Therefore, when away from the mirror the blips evolve as if in a vacuum; when in contact with the mirror, the blips are reflected by the mirror and their direction of propagation is reversed ($s \mapsto -s$).  The trajectory of a blip initially approaching the mirror from the left along with its mirror image is illustrated in Fig.~\ref{Fig:mirror1}.

Since the mirror is perfect and light therefore cannot pass through it, a blip that is initially localised to the left of the mirror will remain to the left of the mirror for all time.  The reflection of light from the left-hand side of the mirror can therefore be modelled entirely with the use of blips localised in Region I.  Since the light changes direction when it interacts with the mirror, blips characterised by $s = 1$ and $s = -1$ are required to give a full account of this interaction.  In particular, the reflection process would be described by an interaction Hamiltonian of the kind studied in Ref.~\cite{Jake1} that would transform a blip characterised by a direction $s$ into another characterised by $-s$.  In this chapter, however, for simplicity we shall describe the trajectory of the blip using the mirror image method of classical electrodynamics.  

Looking more closely at Fig.~\ref{Fig:mirror1}, one can see that the combination of the blip being reflected from the left and its mirror image are equivalent to the superposition of two blips propagating in free space, that is, as if the mirror were not there.  The two trajectories cross at the origin.  Whereas the blip propagating towards the mirror is defined by the operator $a_{s\lambda}(x,t)$, its mirror image is defined by the operator $a_{s\lambda}(-x,t)$.  Denoting the operators for blips to the left and the right of the mirror $a^\text{I}_{s\lambda}(x,t)$ and $a^\text{II}_{s\lambda}(x,t)$ respectively, it is therefore possible to express these operators conveniently as
\begin{eqnarray}
\label{blipmirror1}
a^\text{I}_{s\lambda}(x,t) &=& \mathcal{X}^\text{I}\left(a_{s\lambda}(x,t) + a_{s\lambda}(-x,t)\right)\nonumber\\
a^\text{II}_{s\lambda}(x,t) &=& \mathcal{X}^\text{II}\left(a_{s\lambda}(x,t) + a_{s\lambda}(-x,t)\right).
\end{eqnarray}
In the above the $a_{s\lambda}(x,t)$ operators are the free-space blip operators.  Expressing the $a^\text{I}_{s\lambda}(x,t)$ and $a^\text{II}_{s\lambda}(x,t)$ operators as a superposition of the free-space blip operators is particularly useful as we already know the commutation relations for the free operators.  In Eq.~(\ref{blipmirror1}), the operator $\mathcal{X}^\text{I}$ tells us to neglect all blips located to the right of the mirror for all times.  This effectively removes the translucent trajectory in Fig.~\ref{Fig:mirror1} leaving only the physical trajectory.  The operator $\mathcal{X}^\text{II}$ tells us to neglect all blips located to the left of the mirror for the same reason.

Note finally that the mirror image method is not needed to describe the propagation of light near a double-sided mirror in the blip formalism \cite{Jake1}.  We use the mirror image method only for convenience.  However, whereas one would usually need both parameters $s =\pm 1$ to describe the behaviour of light at either side of the mirror, using the mirror image method, the entire world-line of the blip near the mirror is described using only a single value of $s$.  Therefore, from now only we need only make use of one value of $s$.  Which value doesn't matter so we shall just leave this parameter as $s$.

\subsection{The propagation of blips near a one-dimensional cavity}

\label{Sec:blipsincav1}

To construct a cavity system we place two double-sided mirrors at points along the $x$ axis a width $D$ apart.  For symmetry purposes we assume that the leftmost mirror is placed at $x = -D/2$ whilst the rightmost is placed at $x = D/2$.  The two mirrors partition the real line into three regions.  The region to the left of the cavity spanning all $x <-D/2$ shall be denoted Region I, the region between the mirrors spanning $-D/2<x<D/2$ shall be denoted Region II, and the region to the right of the cavity spanning all $x>D/2$ shall be denoted Region III.  As there are now two mirrors, there will be many more mirror images to take into consideration in order to describe blips both inside and outside the cavity.  We now look at how to construct the appropriate blip operators in terms of the free-space blip operators just like we did in Eq.~(\ref{blipmirror1})

\subsubsection{Blips outside the cavity}

Consider a single blip propagating towards the cavity from the left.  The blip will propagate freely until it comes into contact with the leftmost mirror.  As these mirrors are perfect there is no chance of a blip penetrating the mirror.  Consequently all blips in Region I are unaffected by the mirror forming the right-hand wall of the cavity.  Therefore, the blips to the left of the cavity propagate as if there were only one mirror, just as illustrated in Fig.~\ref{Fig:mirror1}.  Blips in region III behave analogously.  We can therefore express the blips in Regions I and III in a manner very similar to Eq.~(\ref{blipmirror1}).  Denoting blips in Regions I and III $a^{\text{I}}_{s\lambda}(x,t)$ and $a^\text{III}_{s\lambda}(x,t)$ respectively, and taking into account the new position of the mirrors we find that 
\begin{eqnarray}
\label{blipcavity1}
a^{\text{I}}_{s\lambda}(x,t) &=& \mathcal{X}^\text{I}\left(a_{s\lambda}(x,t) + a_{s\lambda}(-D-x,t)\right)\nonumber\\
a^{\text{III}}_{s\lambda}(x,t) &=& \mathcal{X}^\text{III}\left(a_{s\lambda}(x,t) + a_{s\lambda}(D-x,t)\right).
\end{eqnarray} 
As before the operators $\mathcal{X}^\text{I}$ and $\mathcal{X}^{\text{III}}$ remove contributions that are outside their designated regions.

\subsubsection{Blips inside the cavity}

\begin{figure}[h]
\centering
\includegraphics[width = 0.6\textwidth]{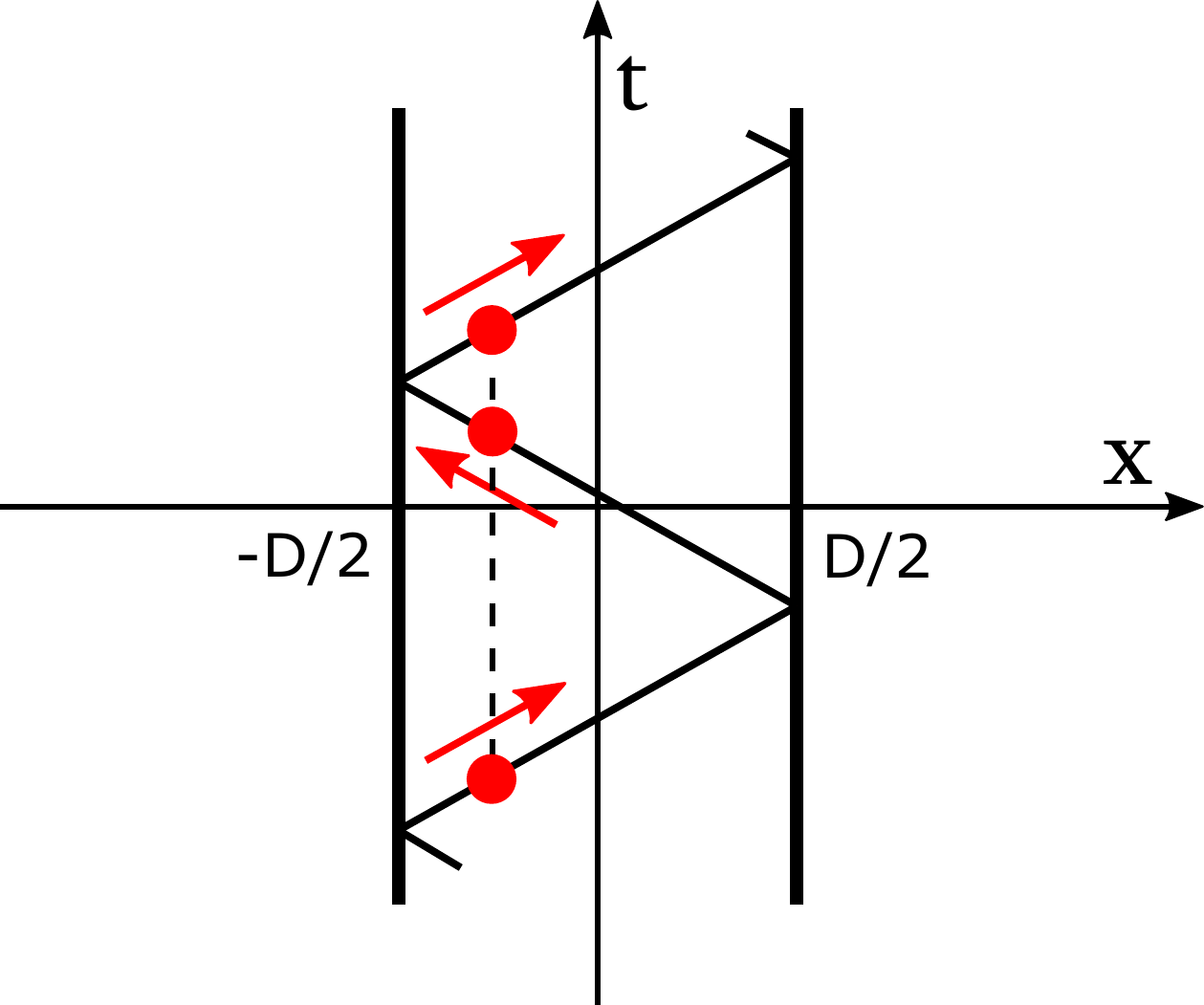}
\caption[The world-line of a blip inside a one-dimensional cavity]{The figure shows the world-line of a single blip placed inside a cavity.  A blip initially propagating to the left will meet the right-hand mirror and be reflected to the left.  At a later time the same blip will meet the left-hand mirror and be reflected again causing it to propagate to the right.  This process continues indefinitely.  Light initially inside the cavity remains inside for all time, and light initially outside the cavity remains outside for all time.}
\label{Fig:mirror2}
\end{figure}

Inside the cavity a single blip will bounce backwards and forwards between the two mirrors for all time.  This motion is illustrated in Fig.~\ref{Fig:mirror2}.  This is true for any size cavity as blips are infinitely well localised and therefore guaranteed to fit inside the cavity.  The trajectory of blips inside the cavity differs therefore from that of those on the outside.  As a blip inside the cavity is reflected an infinite number of times there is an infinite number of mirror images that must be taken into consideration.  These can be imagined by extending the world-lines of the blips in Fig.~\ref{Fig:mirror2} through the cavity walls.  Taking these reflections into consideration we find that the blip operator for a single blip placed inside the cavity is given by the expression
\begin{eqnarray}
\label{blipcavity2}
a^\text{II}_{s\lambda}(x,t) = \mathcal{X}^{\text{II}}\sum_{n = -\infty}^{\infty}\left(a_{s\lambda}(x + 2nD, t) + a_{s\lambda}((2n-1)D-x,t)\right).
\end{eqnarray} 
The operation $\mathcal{X}^{\text{II}}$ disregards all blips located outside Region II for all times.  The infinite sum is representative of the infinite number of reflections made by a blip inside the cavity.

\subsection{The propagation of the EM field near a cavity}

\label{Sec:fieldincav1}

The Casimir effect arises due to changes in the zero-point energy of the EM field as the cavity width is varied.  The energy of the system is determined by the fields, so in this section we shall investigate how the field observables are altered by the presence of a cavity.

In Chapters \ref{Chapter:1D} and \ref{Chapter:3D}, the bosonic blip operators are related to the field observables by means of a regularisation operator $\mathcal{R}$.  In both cases the regularisation operator is highly non-local (see Eqs.~(\ref{explicitR}) and (\ref{3Dregularisation1})).  The field measured at a single point therefore depends on the distribution of blips across all of space.  We have seen, however, that when a mirror is introduced to the system all blips are reflected, which prevents them from reaching the other side.  A mirror, therefore, will effectively shield an observer on one side of the mirror from all blips present on the other.  When such an observer measures the field observables on their side of the mirror, they ought not to be aware of any blips on the other side.  For this reason, all field observables measured on one side of a mirror cannot depend on blips on the other.  This is one consideration that must be taken into account.  We must also remember that the cavity walls are not dull but perfectly reflecting.  Therefore it is not only the blips on the correct side of the mirror that contribute to the field observables, but all of their mirror images too.

\begin{figure}[h]
\centering
\includegraphics[width=\textwidth]{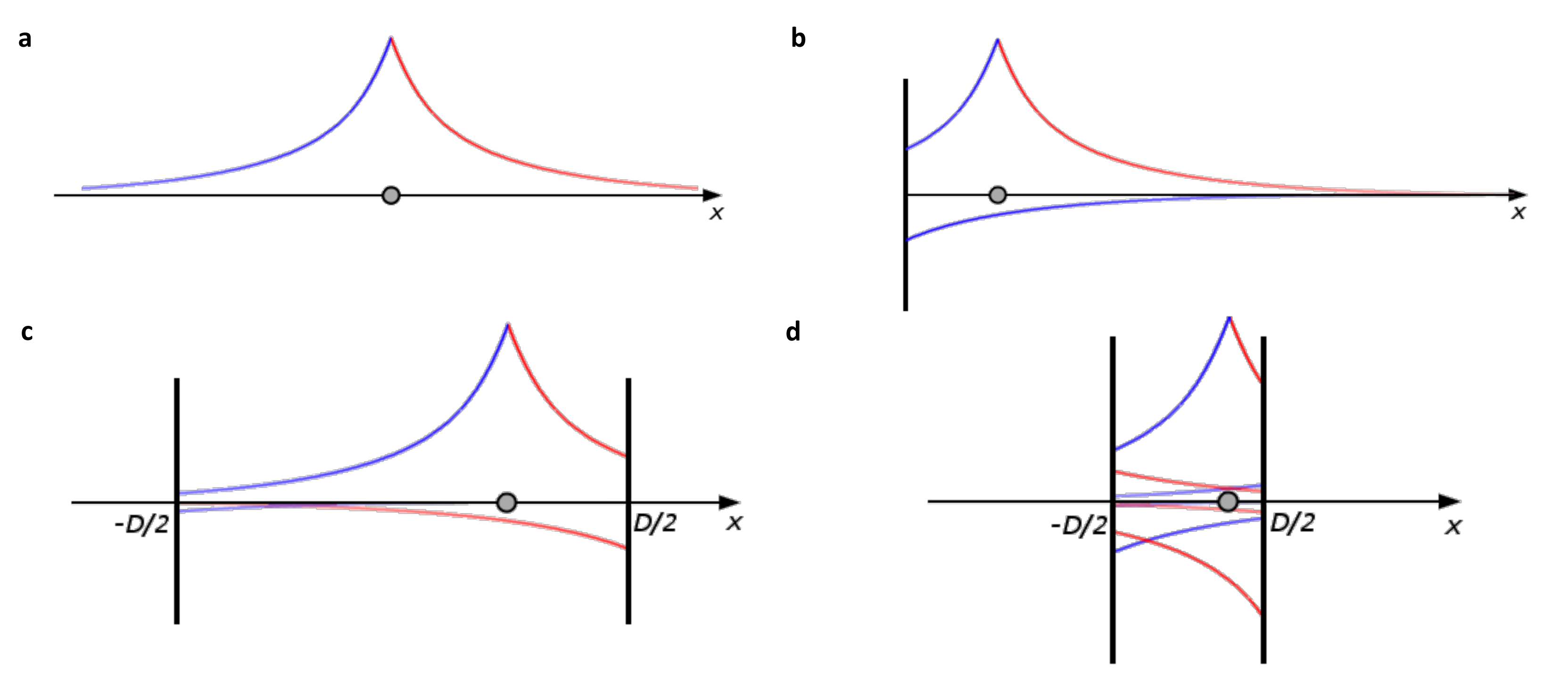}  
\caption[The reflection of field amplitudes]{\cite{CP} {\bf a.} The diagram shows the amplitude of the electric field observable generated by a single blip shown by the grey dot.  As the regularisation operator ${\cal R}$ is non-local, local blip excitations contribute to local electric and magnetic field expectation values everywhere along the $x$ axis. {\bf b.} Since a blip on one side of a highly reflecting mirror cannot contribute to the field expectation value on the other side, its field contribution must be reflected back on itself. This effect alters the electric and magnetic field observables in the presence of a mirror. {\bf c.} In the presence of two highly-reflecting mirrors, blips outside the cavity cannot contribute to field expectation values on the inside. Moreover, the field contributions of blips on the inside need to be reflected as in the case of one mirror. Now, however, the field contributions must be folded infinitely many times. {\bf d.} By comparing two cavities of different sizes we see that the behaviour of the fields inside the cavity is dependent on the cavity width.} 
\label{Fig:foldedfields}
\end{figure}

As the field observables defined in either Region I, II or III can only depend on blips defined in that same region, and must also depend on all of their mirror images, the field operators defined in either one of these regions is simply the regularisation of the photon operators $a^{\text{I}, \text{II}, \text{III}}_{s\lambda}(x,t)$ that we defined earlier.  Conveniently, as the regularisation operators in both one and three dimensions are symmetric, the regularisation of the mirror images is equal to the mirror images of the regularisation.  Hence, the field observables near a cavity are determined  by taking the mirror images of the free field observables.  Unlike the blip operators that only become aware of the mirror when the two come into contact, as the field observables are non-local, the fields at all locations are changed quite significantly by the presence of the two mirrors.  In particular, by taking into account the mirror images of the fields both inside and outside the cavity, the non-local field observables are folded over onto themselves many times forming a concertina shape.  This is illustrated for the field observables both inside and outside the cavity in Fig.~\ref{Fig:foldedfields}.  In the remainder of the section we shall determine explicit expressions for the field observables near a one-dimensional cavity. 

\subsubsection{Outside the cavity}

In the previous section we failed to mention that every time light is reflected by a perfect mirror, the field observables gain a unitary phase factor.  In this section we choose a mirror that applies a factor of $-1$ and $+1$ to the electric and magnetic fields respectively.  As we calculate the mirror images of the fields we must remember to include these additional phase factors.  Taking into account all mirror images, the field observables in Region I are given by
\begin{equation}
\label{FieldsI}
\mathbf{O}^{\text{I}}_{s}(x,t) = \mathcal{X}^{\text{I}} \left(\mathbf{O}_{s}(x, t) \mp \mathbf{O}_{s}(-D -x, t) \right).
\end{equation}
Here $\mathbf{O}_s = \mathbf{E}_s, \mathbf{B}_s$ refers to the free field observables defined in Eqs.~(\ref{fieldobservables1}) and (\ref{fieldobservables3}), and the $\mp$ sign applies to the electric and magnetic fields respectively.  

In Eq.~(\ref{FieldsI}), the operation $\mathcal{X}^\text{I}$ restricts the Hilbert space to blips characterised by $x$ in Region I only.  This ensures that blips outside of Region I cannot contribute in any way to the fields in Region I as both are separated by a perfect mirror.  After substituting into Eq.~(\ref{FieldsI}) the expressions for the free field observables defined in Eq.~(\ref{fieldobservables1}) and using the regularisation operator given in Eq.~(\ref{explicitR}), one finds that 
\begin{eqnarray}
\label{EfieldI}
\mathbf{E}^{\text{I}}_{s}(x,t) &=& -\int^{-D/2}_{-\infty} {\rm d}x'  \left({\hbar c \over 16\pi\varepsilon_0 A} \right)^{1/2}  \left[\left|x-x' \right|^{-3/2} - \left| x+x'+D \right|^{-3/2}\right]\nonumber\\
&&  \hspace*{2.2cm} \times \left[ a_{s\mathsf{H}}(x',t)\,\widehat{\boldsymbol{y}} + a_{s\mathsf{V}}(x',t)\,\widehat{\boldsymbol{z}} \right] + \text{H.c.}
\end{eqnarray}
and
\begin{eqnarray}
\label{BfieldI}
\mathbf{B}^{\text{I}}_{s}(x,t) &=& -\int^{-D/2}_{-\infty} {\rm d}x'\;\frac{s}{c}  \left({\hbar c \over 16\pi\varepsilon_0 A} \right)^{1/2} \left[\left|x-x' \right|^{-3/2} + \left| x+x'+D \right|^{-3/2}\right]\nonumber\\
&& \hspace*{2.2cm} \times  \left[ a_{s\mathsf{H}}(x',t)\,\widehat{\boldsymbol
z} - a_{s\textsf{V}}(x',t)\,\widehat{\boldsymbol{y}} \right] + \text{H.c.}
\end{eqnarray}
respectively.  

The field observables to the right of the cavity are similarly defined by taking into account the mirror images of the field about the mirror on the right.  The field observables in Region III are given by
\begin{equation}
\label{FieldsIII}
\mathbf{O}^{\text{III}}_{s}(x,t) = {\cal X}^{\text{III}} \left(\mathbf{O}^{\text{free}}_{s}(x, t) \mp \mathbf{O}^{\text{free}}_{s}(D -x, t) \right).
\end{equation}
Here the operation $\mathcal{X}^{\text{III}}$ restricts the Hilbert space to blip excitations that are to the right of the mirror.  By again expanding out the free field observables one finds that 
\begin{eqnarray}
\label{EfieldIII}
\mathbf{E}^{\text{III}}_{s}(x,t) &=& -\int^{\infty}_{D/2} {\rm d}x'  \left({\hbar c \over 16\pi\varepsilon_0 A} \right)^{1/2} \left[\left|x-x' \right|^{-3/2} - \left| x+x'-D \right|^{-3/2}\right]\nonumber\\
&& \hspace*{1.8cm} \times  \left[ a_{s\mathsf{H}}(x',t)\,\widehat{\boldsymbol{y}} + a_{s\mathsf{V}}(x',t)\,\widehat{\boldsymbol{z}} \right] + \text{H.c.}
\end{eqnarray}
and
\begin{eqnarray}
\label{BfieldIII}
\mathbf{B}^{\text{III}}_{s}(x,t) &=& -\int^{\infty}_{D/2} {\rm d}x'\;\frac{s}{c}  \left({\hbar c \over 16\pi\varepsilon_0 A} \right)^{1/2} \left[\left|x-x' \right|^{-3/2} + \left| x+x'-D \right|^{-3/2}\right]\nonumber\\
&& \hspace*{1.8cm} \times  \left[ a_{s\mathsf{H}}(x',t)\,\widehat{\boldsymbol{z}} - a_{s\mathsf{V}}(x',t)\,\widehat{\boldsymbol{y}} \right] + \text{H.c.}
\end{eqnarray}
respectively.

\subsubsection{Inside the cavity}

Inside the cavity we must take into consideration an infinite number of mirror images that each contribute to a section of the zig-zagging trajectory illustrated in Fig.~\ref{Fig:mirror2}.  Analogous to Eq.~(\ref{blipcavity2}), the expressions for the total field observables inside the cavity are given by  
\begin{equation}
	\label{FieldsII}
\mathbf{O}^{\text{II}}_{s}(x,t) = \mathcal{X}^{\text{II}}\sum_{n=-\infty}^{\infty}\left(\mathbf{O}_{s}(x+2nD,t) \mp\mathbf{O}_{s}((2n-1)D-x,t)\right)
\end{equation}
Here $\mathcal{X}^\text{II}$ reduces the Hilbert space to the Hilbert space of blips located within Region $\text{II}$.  By substituting in the expressions for the free field observables, we arrive at the following expressions for the electric and magnetic field observables: 
\begin{eqnarray} 
\label{EfieldII}
\mathbf{E}_s^{\text{II}}(x,t) &=& -\sum_{n=-\infty}^\infty \int^{D/2}_{-D/2} {\rm d}x'  \left({\hbar c \over 16\pi\varepsilon_0 A} \right)^{1/2} \notag \\
&& \hspace*{-1.3cm} \left[ \left|x-x'+2nD \right|^{-3/2} - \left| x+x' +(2n-1)D \right|^{-3/2} \right] \nonumber \\
&& \hspace*{-1.3cm} \times \left[ a_{s\mathsf{H}}(x',t)\,\widehat{\boldsymbol{y}} + a_{s\mathsf{V}}(x',t)\,\hat{\boldsymbol{z}} \right] + \text{H.c.}
\end{eqnarray}
and
\begin{eqnarray}
\label{BfieldII}
\mathbf{B}_s^{\text{II}}(x,t) &=&- \sum_{n=-\infty}^\infty \int^{D/2}_{-D/2} {\rm d}x'  \, {s \over c} \left({\hbar c \over 16\pi\varepsilon_0 A} \right)^{1/2} \notag \\ 
&& \hspace*{-1.3cm} \left[ \left| x-x'+2nD \right|^{-3/2} + \left| x + x' +(2n-1)D \right|^{-3/2} \right] \notag \\
&& \hspace*{-1.3cm} \times \left[ a_{s\mathsf{H}}(x',t)\,\widehat{\boldsymbol{z}} - a_{s\mathsf{V}}(x',t)\,\widehat{\boldsymbol{y}} \right] + \text{H.c.} 
\end{eqnarray} 

We now have a complete set of operators for the electric and magnetic field observables inside and outside the cavity.

\section{The Casimir Effect in one dimension}

In this section we use the field observables derived in the previous section to calculate the zero-point energy of the field near the cavity.  We find that the zero-point energy of the field inside the cavity is given by the usual free-space expression plus a $D$-dependent term caused by the interference of the field's infinite tails.  This leads to the usual expression for the Casimir force in one dimension.

\label{Sec:1Dforce}

\subsection{The vacuum energy of the cavity}

Now that we have determined the field observables in the presence of a cavity we can calculate the total energy of the EM field inside and outside the cavity.  The purpose of this section is to determine whether the zero-point energy of the field changes as the cavity width $D$ varies and, if so, determine an expression for the accompanying Casimir force.  As the field observables now contain one or more mirror images, the expression for the energy observable will become very complicated.  Fortunately many of these expressions disappear when we take the vacuum expectation value.  Therefore in this section we shall calculate the zero-point energy of the fields directly.   The first step is to calculate the zero-point energy of the EM field both inside and outside the cavity.

\subsubsection{Outside the cavity}

The electric and magnetic fields to the left of the cavity are given by the observables $\mathbf{E}_s^{\text{I}}(x,t)$ and $\mathbf{B}_s^\text{I}(x,t)$ expressed in Eqs.~(\ref{EfieldI}) and (\ref{BfieldI}).  We only use the field observables defined for one value of $s$ in this expression, as these field observables describe the entire world-line of the blips.  The energy observable in Region I is obtained by substituting the above field observables into the classical expression for the field energy given in Eq.~(\ref{classicalenergy4}).  As the fields defined in Region I only contain contributions from blips also in Region I, this energy observable is indifferent to blips in Regions II and III as we planned. 

After taking the vacuum expectation value of the resulting energy observable and using the blip commutation relation in Eq.~(\ref{blipcommutator2}), one finds that the zero-point energy of the field to the left of the mirror is given by
\begin{eqnarray}
\label{VEVI1}
H^\text{I}_{\text{ZPE}} &=&\frac{\hbar c}{8\pi}\int_{-\infty}^{-D/2}\text{d}x\int_{-\infty}^{-D/2}\text{d}x'\; \left[|(x-x')(x-x')|^{-3/2}\right. \nonumber\\
&&\left.  + |(x+x'+D)(x+x'+D)|^{-3/2}\right]\nonumber\\
&=& \frac{\hbar c}{8\pi}\int_{-\infty}^{\infty}\text{d}x\int_{-\infty}^{-D/2}\text{d}x'\; |(x-x')(x-x')|^{-3/2}.
\end{eqnarray}
The final equality occurs after making the substitution $x' \mapsto -x' - D$ in the second line.  To simplify this expression further we make use of the following identity
\begin{eqnarray}
\label{Casimiridentity1}
\int_{-\infty}^{\infty}\text{d}x \, |(x'-x)(x-x'')|^{-3/2} &=& 4\int_{-\infty}^{\infty}\text{d}k\;|k|\,e^{ik(x'-x'')}\nonumber\\
&=&  - {8 \over |x'-x''|^2}.
\end{eqnarray}
The final equality holds for $x' \neq x''$ only.  Using the identity above we can show that Eq.~(\ref{VEVI1}) is equal to
\begin{equation}
\label{VEVI2}
H^{\text{I}}_{\text{ZPE}} = -\frac{\hbar c }{\pi}\int_{-\infty}^{-D/2}\text{d}x_1\;\left. \frac{1}{|x_1-x|^2}\right|_{x = x_1}.
\end{equation}
The above expression is properly defined as the Fourier transform given in the first line of Eq.~(\ref{Casimiridentity1}). This term is divergent and the factor under the integral sign is independent of $D$.  This term corresponds to the infinite zero-point energy of the field in Region I.

The total vacuum energy of the fields to the right of the cavity is found in the same way by calculating the expectation value of the energy observable defined for the $\mathbf{E}^{\text{III}}(x,t)$ and $\mathbf{B}^{\text{III}}(x,t)$ field observables given by Eqs.~(\ref{EfieldIII}) and (\ref{BfieldIII}).  By an almost identical set of calculations one finds that
\begin{equation}
\label{VEVIII1}
H^{\text{III}}_{\text{ZPE}} = -\frac{\hbar c }{\pi}\int_{D/2}^{\infty}\text{d}x_1\;\left. \frac{1}{|x_1-x|^2}\right|_{x = x_1}.
\end{equation}
The zero-point energy density to the right of the mirror is exactly the same as the zero-point energy density to the left of the mirror.  This would be expected due to the symmetry of the system.

\subsubsection{The vacuum energy inside the cavity}

Inside the cavity the process is just the same.  After calculating the energy observable for the $\mathbf{E}_s^\text{II}(x,t)$ and $\mathbf{B}_s^\text{II}(x,t)$ fields defined in Eqs.~(\ref{EfieldII}) and (\ref{BfieldII}), one finds that the total zero-point energy between the cavity walls is given by
\begin{eqnarray}
\label{VEVII1}
H_{\text{ZPE}}^{\text{II}} &=& \frac{\hbar c}{8 \pi} \sum_{n,m = -\infty}^{\infty} \int_{-D/2}^{D/2}\text{d}x \int_{-D/2}^{D/2}\text{d}x' \notag \\
&& \hspace*{-0.8cm} \times\left[  \left|(x+ x' + (2n-1)D)(x+ x' + (2m-1)D) \right|^{-3/2} \right. \nonumber \\
&& \hspace*{-0.8cm} \left. + \left|(x- x' +2nD )(x- x' +2mD) \right|^{-3/2} \right] \, .
\end{eqnarray}
This term can be simplified to a single line by a series of substitutions.  There are several steps to this substitution, so for clarity in this section we set the full calculation apart in Appendix \ref{App:ProofofVEVII}.  After performing the mentioned substitutions we find that  
\begin{eqnarray}
\label{VEVII2}
H_{\text{ZPE}}^{\text{II}} &=& \frac{\hbar c}{8 \pi} \sum_{n,m = -\infty}^{\infty} \int_{-D}^{D}\text{d}x \int_{-D/2}^{D/2}\text{d}x' \, \left|(x- x' + 2nD)(x- x' + 2mD) \right|^{-3/2} \nonumber \\
&=& \frac{\hbar c}{8 \pi} \sum_{n,m = -\infty}^{\infty} \int_{-D+2nD}^{D+2nD}\text{d}x \int_{-D/2}^{D/2}\text{d}x' \, \left| ( x- x' ) (x- x' + 2(m-n)D) \right|^{-3/2} \nonumber \\
&=& \frac{\hbar c}{8 \pi} \sum_{m = -\infty}^{\infty} \int_{-\infty}^\infty \text{d}x \int_{-D/2}^{D/2}\text{d}x' \, \left|(x- x')(x- x' + 2mD) \right|^{-3/2} \, .
\end{eqnarray} 
The second line in Eq.~(\ref{VEVII2}) follows from the first by making the substitution $x \mapsto x +2nD$.  To make the transition from the second to the third line we should first look at the expression $m-n$.  As both $n$ and $m$ run between $-\infty$ and $\infty$, for each fixed value of $n$, $m-n$ also takes all values between $-\infty$ and $\infty$.  Therefore, it does no harm to simply replace $m-n$ with $m$.  At this point, the sum over all $n$ has the sole effect of stretching the limits of the $x$ integral, which now covers the entire real line.  We are now in a position to apply the identity given in Eq.~(\ref{Casimiridentity1}) to Eq.~(\ref{VEVII2}).  We then find that
\begin{eqnarray}
\label{VEVII3}
H_{\text{ZPE}}^{\text{III}} &=& -\frac{\hbar c}{4\pi D}\sum_{m=-\infty}^{\infty}\frac{1}{m^2}\, .
\end{eqnarray}
We shall examine this expression further in the next section.

\subsection{Divergent contributions to the total zero-point energy}

The expression in Eq.~(\ref{VEVII3}) is the total zero-point energy of the quantised EM field inside the cavity.  As might be expected, this expression is divergent.  The divergent term, however is easily identified as the term for which $m = 0$.  We can separate this divergent contribution from the remaining terms which allows us to express the total zero-point energy in the form
\begin{equation}
\label{VEVII4}
H_{\text{ZPE}}^{\text{II}} = \left.-\frac{\hbar c}{4\pi Dm^2}\right|_{m=0} - \frac{\hbar c}{2\pi D}\sum_{m = 1}^\infty\frac{1}{m^2}.
\end{equation}
The second term in this expression is convergent and will be examined shortly.  For now we shall investigate more closely the divergent contribution to the total zero-point energy inside the cavity.  Let us look again at the final line of Eq.~(\ref{VEVII2}).  If we compare the $m=0$ contribution to this expression  with the final line of Eq.~(\ref{VEVI1}) we can see that the two are identical apart from the limits of the integration over $x$.  The divergent zero-point energy density inside the cavity is therefore just the usual zero-point energy of the free fields in the region between $x = -D/2$ and $x=D/2$.  We can therefore rewrite Eq.~(\ref{VEVII4}) in the form
\begin{equation}
\label{VEVII5}
H_{\text{ZPE}}^{\text{II}} = H_{\text{ZPE}}^{\text{II}(\text{free})} - \frac{\hbar c}{2\pi D}\sum_{m = 1}^\infty\frac{1}{m^2}.
\end{equation}

A rough physical argument can be given as to why the $m=0$ contribution corresponds to the zero-point energy of the free fields if we look again at Eq.~(\ref{VEVII2}).  The $m$ that appears in Eq.~(\ref{VEVII4}) and the final line of Eq.~(\ref{VEVII2}) corresponds to the difference between $n$ and $m$ in the first line of Eq.~(\ref{VEVII2}).  As $n$ and $m$ here refer to the components of the fields that have been reflected $2n$ and $2m$ times respectively, the $m=0$ contribution to the zero-point energy of the field given in Eq.~(\ref{VEVII4}) is the contribution to the total vacuum energy caused by the interference between components of the fields that have been reflected an equal number of times.  In part (d) of Fig.~\ref{Fig:foldedfields}  one can see that the effect of the mirrors is to fold the non-local field amplitudes back in on themselves when they meet the cavity walls.  If this diagram were unfolded we would recover part (a) of the same figure.  The interference, then, between components of the field amplitudes that have been folded an equal number of times corresponds exactly to the interference of the free fields with themselves.  The difference between the former and the latter is only that the first set of fields have been scrunched up into a cavity.

The $m \neq 0$ contributions to Eq.~(\ref{VEVII5}), however, correspond to interference between field amplitudes when one of the fields has been folded at least twice more than the other; this is only possible inside a cavity.  Moreover, the greater the number of reflections that separate two contributions to the EM field observables, the larger $m$ and the smaller their contribution to the total zero-point energy.    Next we examine more closely the remaining correction to the free zero-point energy.

\subsection{The Casimir force}

Using our expressions for the zero-point energy of the EM field inside and outside the cavity we can now determine whether a Casimir force exists.  To do so we calculate how the total zero-point energy of the system changes as the width of the cavity is varied.  From Eqs.~(\ref{VEVI2}) and (\ref{VEVIII1}) we know that the vacuum energy density in Regions I and III, which is given by the usual expression for the free field zero-point energy density, is independent of the cavity width and therefore does not contribute to an overall Casimir force.  In Region II the zero-point energy density is given by the usual cavity-independent expression with a $D$-dependent correction. In the position representation, as can be seen in Eq.~(\ref{VEVII5}), it is easy to separate the $D$-independent free field contribution from the $D$-dependent correction.  Hence, the total Casimir force on the mirrors is given by the formula
\begin{equation}
\label{Casimirforce1}
F_{\text{Casimir}} = -\frac{\partial E_{\text{Casimir}}}{\partial D}
\end{equation}
where $E_\text{Casimir}$ is the correction to the zero-point energy inside the cavity.  By substituting in our value for the Casimir energy $E_{\text{Casimir}}$ in Eq.~(\ref{VEVII5}), we find that
\begin{equation}
\label{Casimirforce2}
F_{\text{Casimir}} = -\frac{\hbar c}{2\pi D^2}\sum_{m=1}^{\infty}\frac{1}{m^2}.
\end{equation}
This force is $D$-dependent and finite.  The sum in Eq.~(\ref{Casimirforce2}) is the convergent Basel sum equal to $\pi^2/6$.  Hence we find that 
\begin{equation}
\label{Casimirforce3}
F_{\text{Casimir}} = -\frac{\hbar c\pi}{12 D^2}.
\end{equation}
This is exactly the predicted Casimir force for photons with two possible polarisations.  This force is attractive and varies with the second negative power of $D$ \cite{Far}.

\section{The Casimir effect in three dimensions}

\label{Sec:3Dforce}

In this section we investigate the propagation of three-dimensional blips near two infinite, flat and parallel mirrors. In a similar manner to Section \ref{Sec:Lightincav}, we determine the appropriate expressions for the field observables near the cavity, and then use these expressions to calculate the zero-point energy of the system.  The geometry of the cavity in this system is very similar to the one-dimensional cavity described in Section \ref{Sec:blipsincav1}.  For this reason many of the methods and results used in Sections \ref{Sec:Lightincav} and \ref{Sec:1Dforce} will be used again in this section.  We find that the zero-point energy of the field inside the cavity is given by its free-space value with the addition of a $D$-dependent correction.  This result leads to the usual expression for the Casimir force in three dimensions.

\subsection{The propagation of the EM field near a cavity}

\label{Sec:fieldnearcav2}

In three dimensions, the basic energy quanta of the EM field are the localised $a_{s\lambda}(\mathbf{r},t)$ excitations.  These excitations are completely localised and propagate without obstruction until they come into contact with the mirrors.  At this point the blips will be reflected changing their direction of propagation.  To define this motion we shall again use the method of mirror images.  This also means that only one value of $s$ is required to describe the entire trajectory of the blips near a cavity.  

The field observables, on the other hand, are not localised, but depend upon the configuration of blips across all space.  In this section we must therefore make sure that blips hidden at one side of a mirror cannot influence the measurement of the fields at the other side.  Moreover, the mirror images of the blips in one particular region must also contribute to the field observables in that region.  As in Section \ref{Sec:fieldincav1}, the resulting fields can be determined by superposing the contributions from all the mirror images of the field observables in each particular region.  This will be our starting point in this section.

\subsubsection{Outside the cavity}

Consider two infinite, flat and perfectly reflecting mirrors positioned parallel to each other a distance $D$ apart.  We shall work in an orthogonal Cartesian system of coordinates such that the two mirrors lie in the $y$-$z$ plane.  For symmetry purposes we shall assume that one mirror is positioned at $x= -D/2$ and the other at $x = D/2$.  It is convenient again to refer to the regions to the left of the cavity, inside the cavity, and to the right of the cavity as Regions I, II, and III respectively.  As the mirrors are perfectly reflecting, no blips can penetrate the mirrors.  Consequently, blips initially found in Regions I, II, or III will remain in those regions for all time.  As before we must remember that upon reflection the electric field observables gain a phase factor $-1$ whereas the magnetic fields gain a phase factor $+1$.

In Region I the fields have only a single mirror image reflected about the mirror placed at $x = -D/2$.  Analogous to the field observables given in Eq.~(\ref{FieldsI}) one finds that the field observables in Region I are given by
\begin{equation}
\label{3DFieldsI}
\mathbf{O}^{\text{I}}_{s}(x,y,z,t) = {\cal X}^{\text{I}} \left(\mathbf{O}_{s}(x,y,z, t) \mp \mathbf{O}_{s}(-D -x,y,z, t) \right).
\end{equation}
In the above $\mathbf{O}_s = \mathbf{E}_s, \mathbf{B}_s$ refer to the free field observables defined in Eq.~(\ref{3Dfieldobservables2}) and the operator $\mathcal{X}^\text{I}$ restricts the Hilbert space to blips defined in Region I only.  Here also the negative sign applies to electric fields and the positive sign to magnetic fields.  After substituting in the expressions for the free field observables (\ref{3Dfieldobservables2}) and the explicit form of the regularisation operator in Eq.~(\ref{3Dregularisation1}) one finds that 
\begin{eqnarray}
\label{3DEfieldI}
\mathbf{E}^\text{I}_s(\mathbf{r},t) &=& - \int_{-\infty}^{-D/2}\text{d}x\int_{-\infty}^{\infty}\text{d}y\int_{-\infty}^{\infty}\text{d}z\; \left(\frac{9\hbar c}{(4\pi)^3\varepsilon_0}\right)^{1/2}\nonumber\\
&&\left[|\mathbf{r}-\mathbf{r}'|^{-7/2} - |\mathbf{r}-\mathbf{r}'_{-1}|^{-7/2}\right]\, \boldsymbol{a}_s(\mathbf{r}') + \text{H.c.} 
\end{eqnarray}
and
\begin{eqnarray}
\label{3DBfieldI}
\mathbf{B}^\text{I}_s(\mathbf{r},t) &=& - \int_{-\infty}^{-D/2}\text{d}x\int_{-\infty}^{\infty}\text{d}y\int_{-\infty}^{\infty}\text{d}z\; \left(\frac{9\hbar c}{(4\pi)^3\varepsilon_0}\right)^{1/2}\nonumber\\
&& \left[|\mathbf{r}-\mathbf{r}'|^{-7/2} + |\mathbf{r}-\mathbf{r}'_{-1}|^{-7/2}\right]\, \boldsymbol{b}_s(\mathbf{r}') + \text{H.c.}
\end{eqnarray}
In the above $\mathbf{r}' = x' \,\widehat{\boldsymbol{x}} + y'\, \widehat{\boldsymbol{y}} + z'\, \widehat{\boldsymbol{z}}$ is the position vector of a point to the left of the cavity.  The vector $\mathbf{r}'_{-1} =  (-x'-D)\, \widehat{\boldsymbol{x}} + y'\, \widehat{\boldsymbol{y}} + z'\, \widehat{\boldsymbol{z}}$ is the position vector of the mirror image of $\mathbf{r}'$ in the left-hand mirror.

In Region III the fields are similarly defined once we have taken into account the new position of the mirror.  In this case one finds that
\begin{equation}
\label{3DFieldsIII}
\mathbf{O}^{\text{III}}_{s}(x,y,z,t) = \mathcal{X}^{\text{III}} \left(\mathbf{O}_{s\lambda}(x,y,z, t) \mp \mathbf{O}_{s}(D -x,y,z, t) \right).
\end{equation}
Here also the negative sign applies to the electric field and the positive sign to the magnetic field, and the operator $\mathcal{X}^\text{III}$ restricts the Hilbert space to blips defined in Region III.  After substituting the free field observables into the above expression one finds that
\begin{eqnarray}
	\label{3DEfieldIII}
	\mathbf{E}^\text{III}_s(\mathbf{r},t) &=& - \int_{D/2}^{\infty}\text{d}x\int_{-\infty}^{\infty}\text{d}y\int_{-\infty}^{\infty}\text{d}z\; \left(\frac{9\hbar c}{(4\pi)^3\varepsilon_0}\right)^{1/2}\nonumber\\
	&&\left[|\mathbf{r}-\mathbf{r}'|^{-7/2} - |\mathbf{r}-\mathbf{r}'_{1}|^{-7/2}\right]\, \boldsymbol{a}_s(\mathbf{r}') + \text{H.c.} 
\end{eqnarray}
and
\begin{eqnarray}
	\label{3DBfieldIII}
	\mathbf{B}^\text{III}_s(\mathbf{r},t) &=& - \int_{D/2}^{\infty}\text{d}x\int_{-\infty}^{\infty}\text{d}y\int_{-\infty}^{\infty}\text{d}z\; \left(\frac{9\hbar c}{(4\pi)^3\varepsilon_0}\right)^{1/2}\nonumber\\
	&& \left[|\mathbf{r}-\mathbf{r}'|^{-7/2} + |\mathbf{r}-\mathbf{r}'_{1}|^{-7/2}\right]\, \boldsymbol{b}_s(\mathbf{r}') + \text{H.c.}
\end{eqnarray}
In the above $\mathbf{r}'$ is defined as before and $\mathbf{r}'_{1} =  (-x'+D)\, \widehat{\boldsymbol{x}} + y'\, \widehat{\boldsymbol{y}} + z'\, \widehat{\boldsymbol{z}}$ is now the position vector of the mirror image of $\mathbf{r}'$ in the right-hand mirror.

\subsubsection{Inside the cavity}

Finally, in region II the field observables are given by the expression
\begin{equation}
\label{3DFieldsII}
\mathbf{O}^\text{II}_{s}(x,t) = \mathcal{X}^\text{II}\sum_{n=-\infty}^{\infty}\left(\mathbf{O}_{s}(x+2nD,y,z,t)\mp\mathbf{O}_{s}((2n-1)D-x,y,z,t)\right).
\end{equation}
The operator $\mathcal{X}^\text{III}$ restricts the Hilbert space to blips defined in Region III only. The field observables inside the cavity contain infinitely many image terms due to the infinitely many mirror images each blip has.  In this instance, after substituting in the free field observables one finds that 
\begin{eqnarray}
	\label{3DEfieldII}
	\mathbf{E}^\text{II}_s(\mathbf{r},t) &=& - \sum_{n=-\infty}^{\infty} \int_{-D/2}^{D/2}\text{d}x\int_{-\infty}^{\infty}\text{d}y\int_{-\infty}^{\infty}\text{d}z\; \left(\frac{9\hbar c}{(4\pi)^3\varepsilon_0}\right)^{1/2}\nonumber\\
	&&\left[|\mathbf{r}-\mathbf{r}'_{2n}|^{-7/2} - |\mathbf{r}-\mathbf{r}'_{2n-1}|^{-7/2}\right]\, \boldsymbol{a}_s(\mathbf{r}') + \text{H.c.} 
\end{eqnarray}
and
\begin{eqnarray}
	\label{3DBfieldII}
	\mathbf{B}^\text{II}_s(\mathbf{r},t) &=& - \sum_{n=-\infty}^{\infty} \int_{-D/2}^{D/2}\text{d}x\int_{-\infty}^{\infty}\text{d}y\int_{-\infty}^{\infty}\text{d}z\; \left(\frac{9\hbar c}{(4\pi)^3\varepsilon_0}\right)^{1/2}\nonumber\\
	&& \left[|\mathbf{r}-\mathbf{r}'_{2n}|^{-7/2} + |\mathbf{r}-\mathbf{r}'_{2n-1}|^{-7/2}\right]\, \boldsymbol{b}_s(\mathbf{r}') + \text{H.c.}
\end{eqnarray}
In the above $\mathbf{r}'$ is defined as before.  The vector $\mathbf{r}'_{2n} =  (x'-2nD)\, \widehat{\boldsymbol{x}} + y'\, \widehat{\boldsymbol{y}} + z'\, \widehat{\boldsymbol{z}}$ is the position vector of the mirror image of a blip that has been reflected $2|n|$ times.  Positive $n$ applies when the first reflection is off the right-hand mirror, and negative $n$ when the first reflection is off the left-hand mirror.  The vector $\mathbf{r}'_{2n-1} =  (-x'-(2n-1)D)\, \widehat{\boldsymbol{x}} + y'\, \widehat{\boldsymbol{y}} + z'\, \widehat{\boldsymbol{z}}$ is the position vector of the mirror image of a blip that has been reflected an odd number of times.  In this case, $n>0$ applies to blips that have been reflected $2n-1$ times, initially off the left-hand mirror.  Terms for which $n \leq 0$ applies to blips that have been reflected $-(2n-1)$ times, the first reflection being off the right-hand mirror.

\subsection{The vacuum energy of the cavity}

We are now in a position to calculate the vacuum energy both inside and outside the cavity.  As before, the energy observable in each region is determined by substituting the field observables into the classical expression of energy for that region.  We shall also calculate the vacuum expectation value of the energy observable directly without writing out the full operator itself. 

\subsubsection{Outside the cavity}

After substituting the field observables in Region I (Eqs.~(\ref{3DEfieldI}) and (\ref{3DBfieldI})) into the expression for the total field energy to the left of the cavity and taking the vacuum expectation value, one finds that
\begin{eqnarray}
\label{3DVEVI1}
H_{\text{ZPE}}^{\text{I}} &=& \int_{-\infty}^{-D/2}\text{d}x\int_{-\infty}^{-D/2}\text{d}x'\int_{-\infty}^{\infty}\text{d}y'\int \text{d}y\int_{-\infty}^{\infty}\text{d}z'\int\text{d}z\;\left(\frac{9\hbar c }{2(4\pi)^3}\right)\nonumber\\
&&\times\left[(|\mathbf{r}- \mathbf{r}'||\mathbf{r}-\mathbf{r}'|)^{-7/2} +  (|\mathbf{r}- \mathbf{r}'_{-1}||\mathbf{r}-\mathbf{r}'_{-1}|)^{-7/2}\right]\nonumber\\
&=& \int_{-\infty}^{-D/2}\text{d}x\int\text{d}y\int\text{dz}\int_{\mathbb{R}^3}\text{d}^3\mathbf{r}'\;\left(\frac{9\hbar c}{2(4\pi)^3}\right)\,(|\mathbf{r}- \mathbf{r}'||\mathbf{r}-\mathbf{r}'|)^{-7/2} .
\end{eqnarray} 
In Eq.~(\ref{3DVEVI1}), the final equality holds by making the substitution $x' \mapsto -x' - D$.  In order to perform the integral over $\mathbf{r}'$ we make use of another identity.  In particular we use the identity
\begin{eqnarray}
\label{3Didentity}
\int_{\mathbb{R}^3}\text{d}^3\mathbf{r}\;(|\mathbf{r}-\mathbf{r}'||\mathbf{r}-\mathbf{r}''|)^{-7/2} &=& \int_{\mathbb{R}^3}\text{d}^3\mathbf{k}\;\frac{16|\mathbf{k}|}{9}\,e^{i\mathbf{k}\cdot(\mathbf{r}'-\mathbf{r}'')}\nonumber\\
&=& -\frac{128\pi}{9|\mathbf{r}'-\mathbf{r}''|^4}.
\end{eqnarray}
The final equality holds when $\mathbf{r}' \neq \mathbf{r}''$.  By using Eq.~(\ref{3Didentity}) we can see that the zero-point energy to the left of the cavity is given by
\begin{equation}
\label{3DVEVI2}
H_{\text{ZPE}}^{\text{I}} = -\int_{-\infty}^{-D/2}\text{d}x\;\left.\frac{8\pi A\hbar c}{(2\pi)^3}\frac{1}{|\mathbf{r}-\mathbf{r}'|^4}\right|_{\mathbf{r} = \mathbf{r}'}.
\end{equation}
The area $A$ is the area inhabited by the field in the $y$-$z$ plane.  This term is divergent, as can be seen by substituting $\mathbf{r}' = \mathbf{r}''$ into the first line of Eq.~(\ref{3Didentity}), and the term under the integral sign is independent of $D$.  This term corresponds to the infinite zero-point energy of the field to the left of the cavity and is completely analogous to the one-dimensional result found in Eq.~(\ref{VEVI2}).

In Region III the calculation follows almost identically.  In this region we find the similar result
\begin{equation}
\label{3DVEVIII1}
H_{\text{ZPE}}^{\text{I}} = -\int_{D/2}^{\infty}\text{d}x\;\left.\frac{8\pi A\hbar c}{(2\pi)^3}\frac{1}{|\mathbf{r}-\mathbf{r}'|^4}\right|_{\mathbf{r} = \mathbf{r}'}.
\end{equation}
As would be expected from symmetry arguments, the zero-point energy density of the field to the left and the right of the cavity is the same.

\subsubsection{Inside the cavity}

Inside the cavity we calculate the zero-point energy in the same way.  After substituting the field observables given in Eqs.~(\ref{3DEfieldII}) and (\ref{3DBfieldII}) into the classical expression for the energy and calculating the vacuum expectation value, we find that 
\begin{eqnarray}
\label{3DVEVII1}
H_{\text{ZPE}}^{\text{II}} &=& \sum_{n,m = -\infty}^{\infty} \int_{-D/2}^{D/2}\text{d}x \int_{-D/2}^{D/2}\text{d}x'\int_{-\infty}^{\infty}\text{d}y'\int_{-\infty}^{\infty}\text{d}z\; \left(\frac{9A\hbar c }{2(4\pi)^3}\right) \nonumber \\
&& \hspace*{-0.8cm} \times\left[(|\mathbf{r}-\mathbf{r}'_{2n}||\mathbf{r}-\mathbf{r}'_{2m}|)^{-7/2} + (|\mathbf{r}-\mathbf{r}'_{2n-1}||\mathbf{r}-\mathbf{r}'_{2m-1}|)^{-7/2}\right].
\end{eqnarray} 
In the above, $A$ has the same interpretation as before.  The result above in Eq.~(\ref{3DVEVII1}) is analogous to the expression for the zero-point energy inside a one-dimensional cavity given in Eq.~(\ref{VEVII1}).  The two terms are alike due to the similar geometries of both cavities.  To simplify Eq.~(\ref{3DVEVII1}) we can therefore proceed exactly as before by making the substitutions described in Appendix \ref{App:ProofofVEVII} and the paragraph beneath Eq.~(\ref{VEVII2}).  The resulting expression for the zero-point energy inside the three-dimensional cavity is then given by
\begin{eqnarray}
\label{3DVEVII2}
H_{\text{ZPE}}^{\text{II}} &=& \sum_{m = -\infty}^{\infty} \int_{-D/2}^{D/2}\text{d}x \int_{\mathbb{R}^3}\text{d}^3\mathbf{r}'\;\left(\frac{9\hbar c A}{2(4\pi)^3}\right) \nonumber \\
&& \left[(|\mathbf{r}-\mathbf{r}'||\mathbf{r}-\mathbf{r}'_{2m}|)^{-7/2}\right].
\end{eqnarray}
This expression is analogous to the final line of Eq.~(\ref{VEVII2}).  To simplify this expression further we again use the identity given in Eq.~(\ref{3Didentity}).  As $\mathbf{r}'$ and $\mathbf{r}'_{2m}$ differ only by a distance $2mD$ along the $x$ axis, we find that 
\begin{equation}
\label{3DVEVII3}
H_{\text{ZPE}}^{\text{II}} = -\sum_{m = -\infty}^{\infty} \int_{-D/2}^{D/2}\text{d}x\;\frac{\hbar c A}{\pi^2}\frac{1}{|2mD|^4}.
\end{equation}

\subsection{The Casimir force}

The zero-point energy derived in Eq.~(\ref{3DVEVII3}) contains the addition of two terms: one divergent term and one convergent term.  The divergent term can be immediately identified as the one for which $m = 0$; let us look at this term more closely.  Compare the final line of Eq.~(\ref{3DVEVI1}) and the $m=0$ contribution to Eq.~(\ref{3DVEVII2}).  By noting that $\mathbf{r}'_{2m} = \mathbf{r}'$ when $m = 0$, we can see that the energy density in both terms are equal.  The $m = 0$ contribution to Eq.~(\ref{3DVEVII3}) is just the divergent zero-point energy of the free fields.  We may therefore rewrite Eq.~(\ref{3DVEVII3}) in the form
\begin{equation}
\label{3DVEVII4}
H_{\text{ZPE}}^{\text{II}} = H_{\text{ZPE}}^{\text{II(free)}} - \frac{\hbar c A}{8\pi^2 D^3}\sum_{m = 1}^{\infty}\frac{1}{m^4}.
\end{equation}
This expression is equal to the sum of the usual $D$-independent free field contribution to the zero-point energy and an additional $D$-dependent correction.  As before, the negative correction to the zero-point energy is caused by the interference between two components of the electric and magnetic fields where one component has been reflected at least twice more than the other.  

To determine the total Casimir force on the mirrors we use the formula given in Eq.~(\ref{Casimirforce1}).  As the only $D$-dependent contribution to the total zero-point energy density is the correction term to the zero-point energy inside the cavity, the Casimir energy in this calculation is given by the second term on the right-hand side of Eq.~\ref{3DVEVII4}.  This generates a force acting from the inside of the cavity.  Using this expression we find that the Casimir force per unit area is given by
\begin{equation}
\label{3DCasimir1}
F_{\text{Casimir}} = -\frac{1}{A}\frac{\partial E_{\text{Casimir}}}{\partial D} = -\frac{3\hbar c}{8\pi^2 D^4}\sum_{m=1}^{\infty}\frac{1}{m^4}.
\end{equation}
The infinite sum is well known and corresponds to the Riemann zeta function of argument $4$.  This sum is given by $\pi^4/90$.  Hence, we find that the Casimir force per unit area between two infinite parallel mirrors is given by 
\begin{equation}
F_{\text{Casimir}} = - \frac{\hbar c \pi^2}{240 D^4}. 
\end{equation}
This value is identical to the usual result found by other methods for light with two distinct polarisations \cite{Buh1, Bor1, Bor2}.  The resulting expression describes an attractive force that pulls the two mirrors together and increases with the negative fourth power of the mirror separation.

\section{Discussion}

\label{Sec:Casimirdiscussion}

The Casimir effect is usually calculated by examining changes to the standing modes formed inside a perfect cavity.  From this point of view only those modes with a wavelength less than twice the width of the cavity are permitted inside the cavity.  Moreover, additional regularisation procedures must be applied in order to wrestle the expression for the zero-point energy of the field into a form where corrections caused by the cavity can be examined.  In this chapter we have shown that the usual well-known Casimir forces in one- and three-dimensional cavities are an exact result that can be calculated by considering the propagation of localised energy quanta both inside and outside the cavities.

In both one and three dimensions the primary excitations of the EM field, blips, are entirely localised.  The blips fit perfectly inside cavities of any width and move around as if in a vacuum until coming into direct contact with the mirrors.  From this point of view, light is treated as a collection of small particles that interact with the cavity like classical point-like particles.  As blips are constructed by superposing monochromatic photons of all frequencies, there are no restrictions on the frequency of light that can be injected into the cavity, even if the cavity is exceptionally narrow. By viewing the monochromatic photons as only a tool for constructing localised wave packets we can see that more appropriate boundary conditions can be applied just by specifying the trajectory of a blip.

Although the blip excitations are localised, the electric and magnetic field observables have a non-local dependence on the configuration of blips, which is implemented by the regularisation operator $\mathcal{R}$.  As a result, the electric and magnetic fields are altered by the presence of the cavity at positions far removed from the mirrors themselves.  In particular, as the two mirrors are highly reflecting, a single blip inside the cavity contributes to the electric and magnetic fields many times due to the many mirror images that that blip has.  These changes were implemented by taking the mirror images of the field observables both inside and outside the cavity (see Sections \ref{Sec:fieldincav1} and \ref{Sec:fieldnearcav2}).  In this description, the Casimir effect arises due to interference between the reflected components of the highly non-local field observables that must be folded back into the cavity many times over.

This result also helps us to understand certain characteristics of the resulting force.  For example, looking again at Fig.~\ref{Fig:foldedfields}, we can see that, since the fields decrease the further one moves from the unregularised blip, contributions that have been folded the most contribute the least to changes of the field observables.  This can be seen in Eqs.~(\ref{Casimirforce2}) and (\ref{3DCasimir1}) where the greater $m$ becomes the smaller the contribution to the overall force.  Moreover, by comparing parts (c) and (d) of Fig.~\ref{Fig:foldedfields}, one can see that the closer together the mirrors become, the more the reflected components contribute to the field inside the cavity.  Hence, as the mirrors are moved closer the force between the mirrors increases.  

The methods used in this chapter provide a particularly intuitive way of deriving the Casimir effect.  By piecing together the classical world-line of localised particles inside a cavity we can calculate the zero-point energy of the field in a way that easily distinguishes any divergent contributions without the need for regularisation procedures or approximations \cite{Buh1}, and gives a more natural understanding of the physics at play.  This approach is therefore very well suited to investigations into Casimir forces in cavities with different geometries, lossy or partially transparent mirrors, and even inhomogeneous materials or curved space-times.

\chapter{Discussion}

\label{Chap:conclusions}

In this thesis we refer to single-photon wave packets as the set of excitations associated with the normalised annihilation operators $a$ satisfying the commutation relation $[a,a^\dagger]=1$.  It is usual to construct these wave packets by superposing linear combinations of the orthogonal and non-normalisable excitations associated with the monochromatic plane wave solutions of Maxwell's equations.  These excitations are non-localised and are characterised by a single wave vector $\mathbf{k}$ and polarisation $\lambda = \mathsf{H}, \mathsf{V}$, each carrying an energy $\hbar|\mathbf{k}|c$.  In this thesis we have taken an alternative approach and constructed single-photon wave packets from a set non-normalisable bosonic particles that have a well-defined position in space-time, so-called blips.  In the past it has been thought that localised excitations of the quantised EM field do not exist \cite{Jor2}; however, here we overcome existing no-go theorems by distinguishing between localised particles and the field observables themselves.

In the classical theory of the free EM field in one dimension, localised wave packets propagate in both directions at the speed of light without dispersion.  By considering a set of highly localised single-photon wave packets in one dimension we have pointed out that an additional discrete parameter $s = \pm1$ is needed in order to describe the unitary evolution of short light pulses that behave in an analogous way.  States characterised by different values of $s$ must be orthogonal to one another in order to distinguish between the different dynamics of left- and right-propagating photons.   As we demonstrated in Section \ref{Sec:Causality}, in standard quantisations of the EM field that neglect this additional parameter, single-photon wave packets do not propagate causally but spread out filling all of space immediately.  When our quantisation scheme is used, however, we find that single-photon wave packets propagate at the speed of light.  In three-dimensions, despite a much larger Hilbert space, light is also only characterised by an additional parameter $s =\pm 1$.  As demonstrated in Chapter \ref{Chapter:3D}, in three dimensions the localised solutions of Maxwell's equations do not propagate in straight lines.  The role of the parameter $s$ in this case is to specify the overall motion of the system, for example, the direction of a plane wave.

Following closely the quantisation scheme of Bennett \textit{et al.}~\cite{RB2}, we were able to derive a set of field observables in terms of the orthogonal blip operators.  In order to ensure that the field observables possessed the correct dimensions and transformation properties, the electric and magnetic field observables were related to the blip operators by means of a non-local regularisation operator $\mathcal{R}$.  In this thesis, in contrast to many other works, we differentiate between the basic excitations of the EM field, which are completely localised, and the corresponding field observables, which are spread out across all of space (see Figs.~\ref{Fig:regularisation}.b and \ref{Fig:foldedfields}.a).  In this sense our views are more in line with those of Fleming \cite{Fle3} who states that locally orthogonal particle states also have some fundamental meaning.  By distinguishing between the particles and fields we were able to demonstrate how the Casimir effect can be shown to arise entirely due to the non-local characteristics of the field observables.  More specifically, although the localised blip excitations are unaffected by the mirror when the two are not in contact, the non-local fields are necessarily reflected many times (see Fig.~\ref{Fig:foldedfields}.d) causing a large amount of interference inside the cavity. The effect of this interference on the zero-point energy of the field varies with the width of the cavity causing an attractive force between the mirrors.  

In both one and three dimensions our quantisation differs from standard approaches by introducing negative-frequency states.  These states are often disregarded as they do not usually possess a positive norm \cite{Wald}; however, in this thesis we have found that negative-frequency states are necessary for the construction of wave packets corresponding to all possible solutions of Maxwell's equations.  In quantum mechanics, the positive energy observable and the dynamical Hamiltonian would usually be equal to each other.  When both positive and negative frequencies are taken into account, however, the dynamical Hamiltonian cannot be bounded below. 

To avoid this problem we no longer insist that the positive energy observable and the generator of time translations be the same.  In this way we avoid the consequences of Hegerfeldt's theorem \cite{Heg8} without the unnerving issue of having an energy observable without a lower bound.  Unlike the energy observable, the dynamical Hamiltonian takes the form of an exchange operator.  The role of this operator is to continually recreate localised blips at new positions at a rate equal to the energy expectation value of that state.  Without an energy observable to guide us, the exact form of the dynamical Hamiltonian is determined from a constraint on the dynamics of the system.  In free space, light is restricted to the boundaries of the light-cones as illustrated in Fig.~\ref{Fig:Lightcone1}, but similar constraints can be found for any other trajectory. 

The quantisation schemes presented in Chapters \ref{Chapter:1D} and \ref{Chapter:3D} not only provide solutions to existing problems; for example, in Chapter \ref{Chapter:Fermi} we found a possible resolution to the Fermi problem and in Ref.~\cite{Jake1} we constructed locally-acting mirror Hamiltonians, but can also be used, as we have seen in Chapter \ref{Chapter:Casimir}, to predict experimentally verifiable results.  In spite of these successes, there are still important aspects of this quantisation that we do not yet fully understand and that need to be examined further.  For example, in classical treatments, the momentum of a light pulse is given by its energy multiplied by a unit vector oriented in the direction of propagation.  Although the energy of a blip is undetermined we appear to know something about its direction, and therefore its momentum, whilst simultaneously having complete knowledge of its position.  In our case, when the position of a blip is known, there is complete uncertainty in the wave vector $\mathbf{k}$ and therefore complete uncertainty in the blip's dynamics.  Nevertheless, we're not yet sure if this can be related to the uncertainty of a measurable quantity and, if so, what that quantity would be.  

There is also the related question of, if our  description of the quantised EM field is approached from the point of view of canonical quantisation, what would the canonical variables be and how would their commutation relation be altered.  This question has been asked in Refs.~\cite{Haw6, Haw8, Bab} that look at a canonical quantisation approach to real fields containing negative-frequency modes.  In standard QED the canonical momentum is given by the electric field observable which commutes non-trivially with the vector field $\mathbf{A}(\mathbf{r},t)$ and therefore also $\mathbf{B}(\mathbf{r},t)$.  In our scheme the electric and magnetic fields are always related by a rotation and therefore commute.  This is an important result that would need to be investigated and lends itself to be measured via uncertainties.

Other questions which could be addressed experimentally are, for instance, can the negative-frequency photons be measured in a laboratory?  As both positive- and negative-frequency waves are required to localise a wave packet to any length it is practically impossible to separate the two, and so a clever experiment must be devised. The authors of Ref.~\cite{Fac} claim to have already detected negative-frequency photons, but the real relation between what we both call negative-frequency photons would need to be examined more carefully before we can draw final conclusions.  One possibility for detecting negative-frequency photons could be to construct a transformation between the positive-frequency and negative-frequency modes, and then probe this transformation by means of an interference experiment. 

In standard quantum electrodynamics a conversion from the positive-frequency to negative-frequency modes is imposed by a non-unitary Bogoliubov transformation between the annihilation and creation operators of the original field operators.  This results in an amplification effect such as the Unruh effect for uniformly accelerating observers.  In our case, however, it is possible that we would require an anti-unitary transformation on the Hilbert space (similar to time reversal) that converts between the positive- and negative-frequency but positive-norm photon states.  Materials with negative or temporally oscillating refractive indices have been studied, for example, by Pendry \cite{Pend} and others \cite{Smi, Vezz, Har}, which reverse the phase evolution of states, essentially inverting their frequencies. These experiments create the possibility for realising the required transformation in a laboratory.  

In this thesis we have also supposed that interactions take place by means of a local exchange of blips.  As it is the field observables that couple to charged matter and currents (see Eqs.~(\ref{Maxwell1}) and (\ref{Maxwell4})), many other authors hold the view that interactions take place by means of an exchange of field excitations.  An interesting extension to this work would therefore be to couple blips to matter.  This would not only increase the spectrum of problems to which this theory could be applied, but also open an investigation into the validity of the local blip approach to interactions by studying known effects.  It would also open the path to modelling experimental setups and photonic devices which have not yet been accessible \cite{Bau, Kyo}.  This is a major motivation for the work presented here.   

Although there is still much to be understood and investigated, the work in this thesis is beginning to open up new possibilities for studying complex or non-intuitive systems.  For example, this new formalism would also be very well suited to studying gravitational systems in which the metric or curvature of space-time varies from place to place.  The Unruh effect is one effect that is usually discussed in terms of frequency modes, which makes it difficult to visualise the underlying physical processes.  A blip description of this effect is very likely to produce a more lucid explanation \cite{Haw7}.  Casimir cosmology is also an interesting field in gravitational physics \cite{Leo1, Leo2}.  Regularisation procedures, however, become increasingly difficult to manage in curved space-times.  As our approach has no need of regularisation procedures, this theory may offer a satisfying resolution to this outstanding problem.

Our local theory of light also offers solutions to other long-standing problems.  For instance, the problem of time in quantum theory refers to the impossibility of defining a time operator \cite{Bauer, Del, VH}.  The Pauli objection prohibits such an operator on the basis that energy is bounded from below \cite{Tol, Gro, Mac, Alt1} (and Refs.~therein).  As discussed in Ref.~\cite{Alt1}, this theory evades this objection by defining a distinct dynamical Hamiltonian, which is unbounded below, leading to the possibility of introducing time observables or a quantum clock.  This is perhaps not so surprising as our Dirac-like equations of motion for the blips, Eqs.~(\ref{blipmotion1}) and (\ref{vectoreom4}), closely resemble the Wheeler-DeWitt equation of quantum gravity \cite{Page} which places a constraint on the dynamics of a system.  

In spite of remaining open questions, the results of this thesis and the way in which our local theory appears to be so beautifully clearing the fog around many other problems encourages me to believe that solutions will be found.  By looking at electromagnetic phenomena from the point of view of monochromatic photons we only see half the picture; by introducing local photons we might in the future not only be able to gain unique perspectives on a wide range of existing problems, but create all the necessary tools for studying new ones.

\singlespacing

\doublespacing

\appendix

\chapter{Derivation of $\Omega(\mathbf{k})$}
\label{App:Omegaproof}

Consider the pure boost that takes us from an initial reference frame $\mathcal{O}$ into a new reference frame $\mathcal{O}'$ moving in a straight line at a speed $v$ relative to the original frame.  This transformation is imposed by the unitary operator $U(\Lambda)$ where $\Lambda$ deonotes the transformation.  If we define a Cartesian coordinate system such that the reference frame $\mathcal{O}'$ is moving in the positive $x$ direction with respect to the frame $\mathcal{O}$, the electric and magnetic fields will undergo a transformation that is given by, for example, Eq.~(12.108) of Ref.~\cite{Grif}.  Under this transformation the orientation of the field vectors transforms in a way that ensures the new field observables are divergence-less.  One can check this using the field transformations in Ref.~\cite{Grif}, the usual Lorentz transformations for the space and time coordinates, and Maxwell's equations (\ref{fMaxwell1})-(\ref{fMaxwell4}).  The transformation also preserves helicity. A detailed description of the transformation can be found in Section 5.9 of Ref.~\cite{Wei}.  We shall ignore any transformation of the polarisation states here as it is not relevant to determining the overall normalisation of the field.

Under the Lorentz boost an initial wave vector $\mathbf{k}$ will be transformed into the new wave vector $\mathbf{p}$.  As we cannot boost into a frame in which light travels in the opposite direction to its original trajectory, the parameter $s$ can always be chosen such that it remains constant under a boost.  When we boost from the $\mathcal{O}$ frame to the $\mathcal{O}'$ frame the fields also undergo a Doppler-shift by a factor of $|\mathbf{k}|/|\mathbf{p}|$.  By transforming the field observables given in Eq.~(\ref{3Dfieldobservables3}) under the unitary operation $U(\Lambda)$, one finds that   
\begin{eqnarray}
\mathbf{E}'(\mathbf{r}') &=&\int_{\mathbb{R}^3}\frac{\text{d}^3\mathbf{k}}{(2\pi)^{3/2}}\;c\,\Omega(\mathbf{k})\,e^{is\mathbf{k}\cdot\mathbf{r}'} U(\Lambda) \widetilde{a}_{s\lambda}(\mathbf{k}) U^\dagger(\Lambda)\,\boldsymbol{e}_{s\lambda}(\mathbf{k})\nonumber\\
&=&
\int_{\mathbb{R}^3}\frac{\text{d}^3\mathbf{k}}{(2\pi)^{3/2}}\;c\,\Omega(\mathbf{k})\,e^{is\mathbf{k}\cdot\mathbf{r}'} \sqrt{\frac{|\mathbf{p}|}{|\mathbf{k}|}} \,\widetilde{a}_{s\lambda}(\mathbf{p})\,\boldsymbol{e}'_{s\lambda}(\mathbf{p})\nonumber\\
&=&
\int_{\mathbb{R}^3}\frac{\text{d}^3\mathbf{p}}{(2\pi)^{3/2}}\;\frac{|\mathbf{k}|}{|\mathbf{p}|}\,c\,\Omega(\mathbf{k})\,e^{is\mathbf{p}\cdot\mathbf{r}} \sqrt{\frac{|\mathbf{p}|}{|\mathbf{k}|}}\, \widetilde{a}_{s\lambda}(\mathbf{p})\,\boldsymbol{e}'_{s\lambda}(\mathbf{p})\nonumber\\ 
&=& \int_{\mathbb{R}^3}\frac{\text{d}^3\mathbf{p}}{(2\pi)^{3/2}}\;\frac{|\mathbf{k}|}{|\mathbf{p}|}\,c\,\Omega(\mathbf{p})\,e^{is\mathbf{p}\cdot\mathbf{r}}\, \widetilde{a}_{s\lambda}(\mathbf{p})\,\boldsymbol{e}'_{s\lambda}(\mathbf{p}). 
\end{eqnarray} 
In the above we made use of the relation $\mathbf{k}\cdot\mathbf{r}' = \mathbf{p}\cdot \mathbf{r}$.  The additional factor of $|\mathbf{k}|/|\mathbf{p}|$ in the last line is the blue-shift factor.  By equating the last two lines of the above calculation one finds that 
\begin{equation}
\frac{\Omega(\mathbf{k})}{\Omega(\mathbf{p})} = \sqrt{\frac{|\mathbf{k}|}{|\mathbf{p}|}}
\end{equation} 
which is in agreement with Eq.~(\ref{3Domega}).  In one dimension the argument follows analogously which leads us to Eq.~(\ref{Omega1}).

\chapter{Divergence of the field expectation values}
\label{Rorthogonaltopsi}

Consider again the regularised wave function given in Eq.~(\ref{regularisedwavefunction1}).  In the following we shall define a set of Cartesian coordinates $x_i$ for $i\in\{1,2,3\}$.  Furthermore, the derivative $\partial_i$ shall be used as shorthand for the partial derivative with respect to the coordinate $x_i$.  The divergence of the regularised wave function is then given by
\begin{eqnarray}
\label{wavefunctiondivergence1}
\mathbf{\nabla}\cdot \int_{\mathbb{R}^3}\text{d}^3\mathbf{r}'\;\boldsymbol{\mathcal{R}}_{\mathbf{s}\lambda}(\mathbf{r}-\mathbf{r}')\,\psi_{\mathbf{s}\lambda}(\mathbf{r}')
&=&\sum_{i=1}^3\int_{\mathbb{R}^3}\text{d}^3\mathbf{r}'\;\partial_i\,\mathcal{R}^i_{\mathbf{s}\lambda}(\mathbf{r}-\mathbf{r}')\,\psi_{\mathbf{s}\lambda}(\mathbf{r}')\nonumber\\
&=& - \sum_{i=1}^3\int_{\mathbb{R}^3}\text{d}^3\mathbf{r}'\;\left[\partial'_i\, \mathcal{R}^i(\mathbf{r}-\mathbf{r}')\right]\psi_{\mathbf{s}\lambda}(\mathbf{r}')\nonumber\\
&=&
\sum_{i=1}^3\int_{\mathbb{R}^3}\text{d}^3\mathbf{r}'\;\mathcal{R}^i_{\mathbf{s}\lambda}(\mathbf{r}-\mathbf{r}')\,\partial'_i\,\psi_{\mathbf{s}\lambda}(\mathbf{r}')\nonumber\\
&=& \int_{\mathbb{R}^3}\text{d}^3\mathbf{r}'\;\boldsymbol{\mathcal{R}}_{\mathbf{s}\lambda}(\mathbf{r}-\mathbf{r}')\cdot \mathbf{\nabla}'\psi_{\mathbf{s}\lambda}(\mathbf{r}').
\end{eqnarray}
In the above, the third line follows from the second by integration by parts.  We have assumed in this step that our wave function vanishes at large distances.  One finds, therefore, that the regularised wave function is divergence-less only when the gradient of the wave function is orthogonal to $\boldsymbol{\mathcal{R}}_{\mathbf{s}\lambda}$.  A similar calculation will show that the divergence of the vector field
\begin{equation}
\boldsymbol{\mathcal{R}}_{\mathbf{s}\lambda}[\psi_{\mathbf{s}\lambda}](\mathbf{r}) \times \mathbf{s}
\end{equation}
vanishes only when the gradient of the wave function is orthogonal to $\boldsymbol{\mathcal{R}}_{\mathbf{s}\lambda} \times \mathbf{s}$.  Using the expressions for the field expectation values in Eq.~(\ref{fieldexpectationvalues1}), one can see that Gauss's laws for electric and magnetic fields are satisfied when the above conditions holds.  This proves our statement in Section \ref{Sec:Physicalstates}.

\chapter{Derivation of Eq.~(\ref{VEVII2})}

\label{App:ProofofVEVII}

This appendix derives Eq.~(\ref{VEVII2}) from Eq.~(\ref{VEVII1}) by making a series of substitutions.  Taking as our starting point the second line of Eq.~(\ref{VEVII1}), we first make the substitution $x_2 \mapsto -x_2$.  The resulting expression is given by
\begin{eqnarray}
\label{CasApp1}
H_{\text{ZPE}}^{\text{II}} &=& \frac{\hbar c}{4 \pi} \sum_{n,m = -\infty}^{\infty} \int_{-D/2}^{D/2}\text{d}x_1 \int_{-D/2}^{D/2}\text{d}x_2\; \nonumber\\
&& \hspace*{-1.3cm}\left|(x_1 - x_2 + (2n-1)D)(x_1 - x_2 + (2m-1)D) \right|^{-3/2}. 
\end{eqnarray}

In this next step we shall first separate the $x_1$ integral into two halves.  One of these integrals shall run between $-D/2$ and $0$, covering the negative values of $x_1$, and the other shall cover the remaining positive values between $0$ and $D/2$.  The resulting integral is expressed
\begin{eqnarray}
\label{CasApp2}
H_{\text{ZPE}}^{\text{II}} &=& \frac{\hbar c}{4 \pi} \sum_{n,m = -\infty}^{\infty} \left[\int_{-D/2}^{0}\text{d}x_1 + \int_{0}^{D/2}\text{d}x_1\right] \int_{-D/2}^{D/2}\text{d}x_2\; \nonumber\\
&& \hspace*{-1.3cm}\left|(x_1 - x_2 + (2n-1)D)(x_1 - x_2 + (2m-1)D) \right|^{-3/2}. 
\end{eqnarray}
We shall next make a separate substitution for each of these two integrals.  For the integral over negative values of $x_1$ we shall make the substitution $x_1 \mapsto x_1 - D$.  For the integral over the positive values of $x_1$ we shall make the substitution $x_1 \mapsto x_1 + D$.  The first integral then takes the form
\begin{eqnarray}
\label{CasApp3}
H_{\text{ZPE}}^{\text{II}} &=& \frac{\hbar c}{4 \pi} \sum_{n,m = -\infty}^{\infty} \int_{D/2}^{D}\text{d}x_1 \int_{-D/2}^{D/2}\text{d}x_2\; \nonumber\\
&& \hspace*{-1.3cm}\left|(x_1 - x_2 + (2n-2)D)(x_1 - x_2 + (2m-2)D) \right|^{-3/2}. 
\end{eqnarray}
The second integral becomes
\begin{eqnarray}
\label{CasApp4}
H_{\text{ZPE}}^{\text{II}} &=& \frac{\hbar c}{4 \pi} \sum_{n,m = -\infty}^{\infty} \int_{-D}^{-D/2}\text{d}x_1 \int_{-D/2}^{D/2}\text{d}x_2\; \nonumber\\
&& \hspace*{-1.3cm}\left|(x_1 - x_2 + 2nD)(x_1 - x_2 + 2mD) \right|^{-3/2}. 
\end{eqnarray}

As one final step, we make the substitution $n \mapsto n+1$ and $m \mapsto m+1$ in Eq.~(\ref{CasApp3}).  This substitution does not change the limits of either sum.  The integrands in both Eqs.~(\ref{CasApp3}) and (\ref{CasApp4}) are then equal both to each other and the integrand in the third line of Eq.~(\ref{VEVII1}).  Putting both lines of Eq.~(\ref{VEVII1}) together we have three integrals with the same integrand: the third line of Eq.~(\ref{VEVII1}), Eq.~(\ref{CasApp3}) and Eq.~(\ref{CasApp4}).  In each of these three integrals the limits of $x_2$ are identical so can be factorised out of this sum.  The integrals over $x_1$ contribute to different sections of the real line.  Eq.~(\ref{CasApp4}) integrates over the region between $-D$ and $-D/2$, the third line of Eq.~(\ref{VEVII1}) over the region between $-D/2$ and $D/2$, and Eq.~(\ref{CasApp3}) over the region $D/2$ to $D$.  Altogether the integral over $x_1$ covers the region $[-D,D]$.  Eq.~(\ref{VEVII1}) is therefore equal to Eq.~(\ref{VEVII2}).
\end{document}